\documentclass[11pt,urlcolor=blue, linkcolor=blue]{article} 
\usepackage{cite}
\usepackage{tablefootnote}
\usepackage{amsmath, amsthm, amssymb,slashed}
\usepackage{ifpdf}
\ifpdf
  \usepackage[pdftex]{graphicx}
  \usepackage{epstopdf}
\else
  \usepackage[dvips]{graphicx}
\fi
\parskip=\baselineskip

\usepackage[usenames, dvipsnames]{color}

\allowdisplaybreaks[1]

\numberwithin{equation}{section}

\usepackage{booktabs}

\theoremstyle{definition}

\newcommand{\cblue}[1]{\textcolor{black}{#1}}
\newcommand{\cred}[1]{\textcolor{black}{#1}}

\definecolor{mygray}{gray}{0.6}

\usepackage{upgreek}
\usepackage{bbm}

\newenvironment{myfont}[2][]{\csname#2\endcsname[#1]}{}

\newcommand{\nn}{\nonumber}

\usepackage{slashed}
\usepackage[makeroom]{cancel}
\usepackage[normalem]{ulem}
\usepackage{soul}
\newcommand{\stkout}[1]{\ifmmode\text{\sout{\ensuremath{#1}}}\else\sout{#1}\fi}

\usepackage{sseq}
\usepackage[all,cmtip]{xy}
\usepackage{tikz-cd}
\usepackage{tikz}
\usetikzlibrary{matrix}

\newcommand{\bea}{\begin{eqnarray}}
\newcommand{\eea}{\end{eqnarray}}
\def\be{\begin{equation}}
\def\ee{\end{equation}}

\newcommand{\ii}{\hspace{1pt}\mathrm{i}\hspace{1pt}}

\def\RP{{\mathbb{RP}}}
\def\CP{{\mathbb{CP}}}

\usepackage{color}
\usepackage[colorlinks,citecolor=blue]{hyperref}
\definecolor{red}{rgb}{1,0,0}
\definecolor{blue}{rgb}{0,0,1}
\definecolor{dblue}{rgb}{0,0,0.4}
\definecolor{green}{rgb}{0,1,0}
\definecolor{black}{rgb}{0,0,0}
\definecolor{white}{rgb}{1,1,1}

\definecolor{brn}{rgb}{.8,.4,.0}
\definecolor{redo}{rgb}{1,.5,.0}
\definecolor{ddgrn}{rgb}{0,0.4,0}
\definecolor{dgrn}{rgb}{0,0.55,0}
\definecolor{dbl}{rgb}{0,0,0.5}

\usepackage[bbgreekl]{mathbbol}
\usepackage{amscd}

\newcommand{\Z}{\mathbb{Z}}
\newcommand{\C}{\mathbb{C}}
\newcommand{\R}{\mathbb{R}}

\newcommand{\dd}{\hspace{1pt}\mathrm{d}}
\newcommand{\<}{\langle} 
\renewcommand{\>}{\rangle} 

\newcommand{\Ref}[1]{Ref.~\cite{#1}}
\newcommand{\Eq}[1]{(\ref{#1})} 
\newcommand{\eq}[1]{(\ref{#1})} 
\newcommand{\eqn}[1]{eqn.~(\ref{#1})}

\newcommand{\Tr}{{\rm Tr}}

\newcommand{\prt}{\partial}

\newcommand{\etc}{{\it etc~}}

\newcommand{\bpm}{\begin{pmatrix}}
\newcommand{\epm}{\end{pmatrix}}
\newcommand{\bmm}{\begin{matrix}}
\newcommand{\emm}{\end{matrix}}

\newcommand{\cB}{ {\cal B} }

\newcommand{\cH}{ {\cal H} }

\newcommand{\cP}{ {\cal P} }

\newcommand{\cT}{ {\cal T} }

\def\Z{{\mathbb{Z}}}

\def\R{{\mathbb{R}}}
\def\C{{\mathbb{C}}}

\def\Tr{{\mathrm{Tr}}}

\def \To{\longrightarrow}
\def \Hom{\operatorname{Hom}}
\def \Tor{\operatorname{Tor}}

\def \H{\operatorname{H}}

\def \F{\mathbb{F}}

\def \Z{\mathbb{Z}}
\def \A{\mathcal{A}}
\def \RP{\mathbb{RP}}
\def \CP{\mathbb{CP}}

\def\Ext{\operatorname{Ext}}
\newcommand{\RZ}{{\mathbb{R}/\mathbb{Z}}}

\newcommand\hcup[1]{\underset{{\scriptscriptstyle #1}}{\cup}}

\newcommand {\emptycomment}[1]{}

\def\TP{\mathrm{TP}}
\def\Sq{\mathrm{Sq}}
\def\B{\mathrm{B}}
\newcommand{\tO}{{\text O}}

\newcommand{\SO}{{\rm SO}}
\newcommand{\Spin}{{\rm Spin}}
\newcommand{\U}{{\rm U}}
\newcommand{\PU}{{\rm PU}}
\newcommand{\SU}{{\rm SU}}
\newcommand{\PSU}{{\rm PSU}}
\newcommand{\Pin}{{\rm Pin}}

\newcommand{\W}{{\rm W}}
\newcommand{\rN}{{\rm N}}

\newtheorem{theorem}{Theorem}%[section]
\newtheorem{lemma}[theorem]{Lemma}

\newtheorem{definition}[theorem]{Definition}
\newtheorem{example}[theorem]{Example}

\usetikzlibrary{decorations.markings}
\usetikzlibrary{positioning}
\usetikzlibrary{shadings}

\usepackage{datetime}

\newcommand{\symfootnote}[1]{%
\let\oldthefootnote=\thefootnote%
\stepcounter{mpfootnote}%
\addtocounter{footnote}{-1}%
\renewcommand{\thefootnote}{\fnsymbol{mpfootnote}}%
\footnote{#1}%
\let\thefootnote=\oldthefootnote%
}

\tikzset{
particle/.style={thick,draw=blue, postaction={decorate},
    decoration={markings,mark=at position .5 with {\arrow[blue]{triangle 45}}}},
gluon/.style={decorate, draw=black,
    decoration={coil,aspect=0.3,segment length=3pt,amplitude=3pt}},
photon/.style={thick, decorate, draw=black,
decoration={coil,aspect=0}}
 }

\usepackage{enumitem} 
\usepackage{moreenum}

\newcommand{\middlewave}[1]{\raisebox{0.5em}{\uwave{\hspace{#1}}}}  

\def\GL{{\mathrm{GL}}}
\def\SL{{\mathrm{SL}}}
\newcommand{\Sec}[1]{Sec.~\ref{#1}} 
\def\bZ{{\mathbf{Z}}}
\newcommand{\onlinecite}[1]{Ref.~\cite{#1}}

\DeclareRobustCommand\sWan{\includegraphics[height=4.85ex]{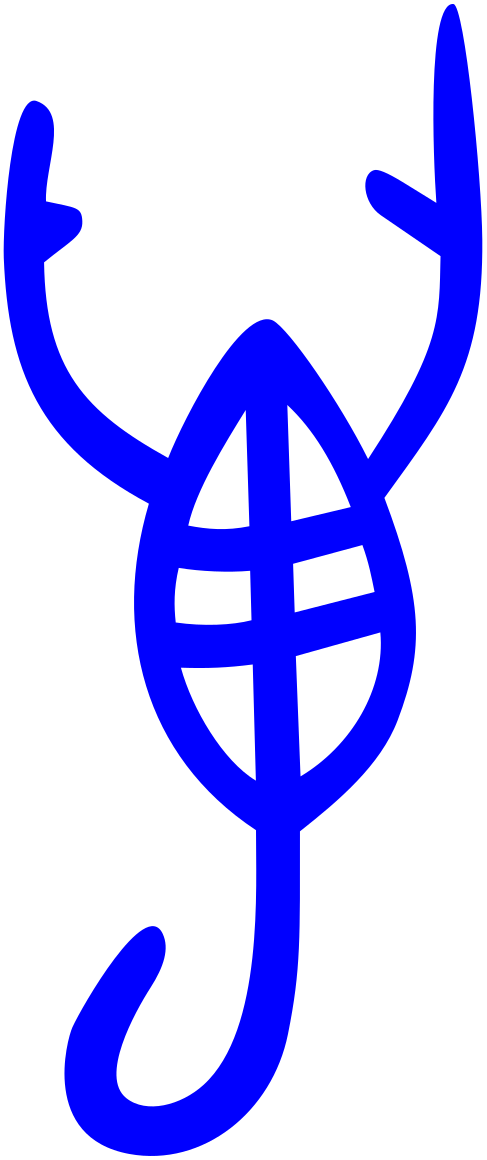}}
\DeclareRobustCommand\sWang%{{\includegraphics[height=4.25ex]{1280px-Wang}}}
{\cblue{\includegraphics[height=4.55ex]{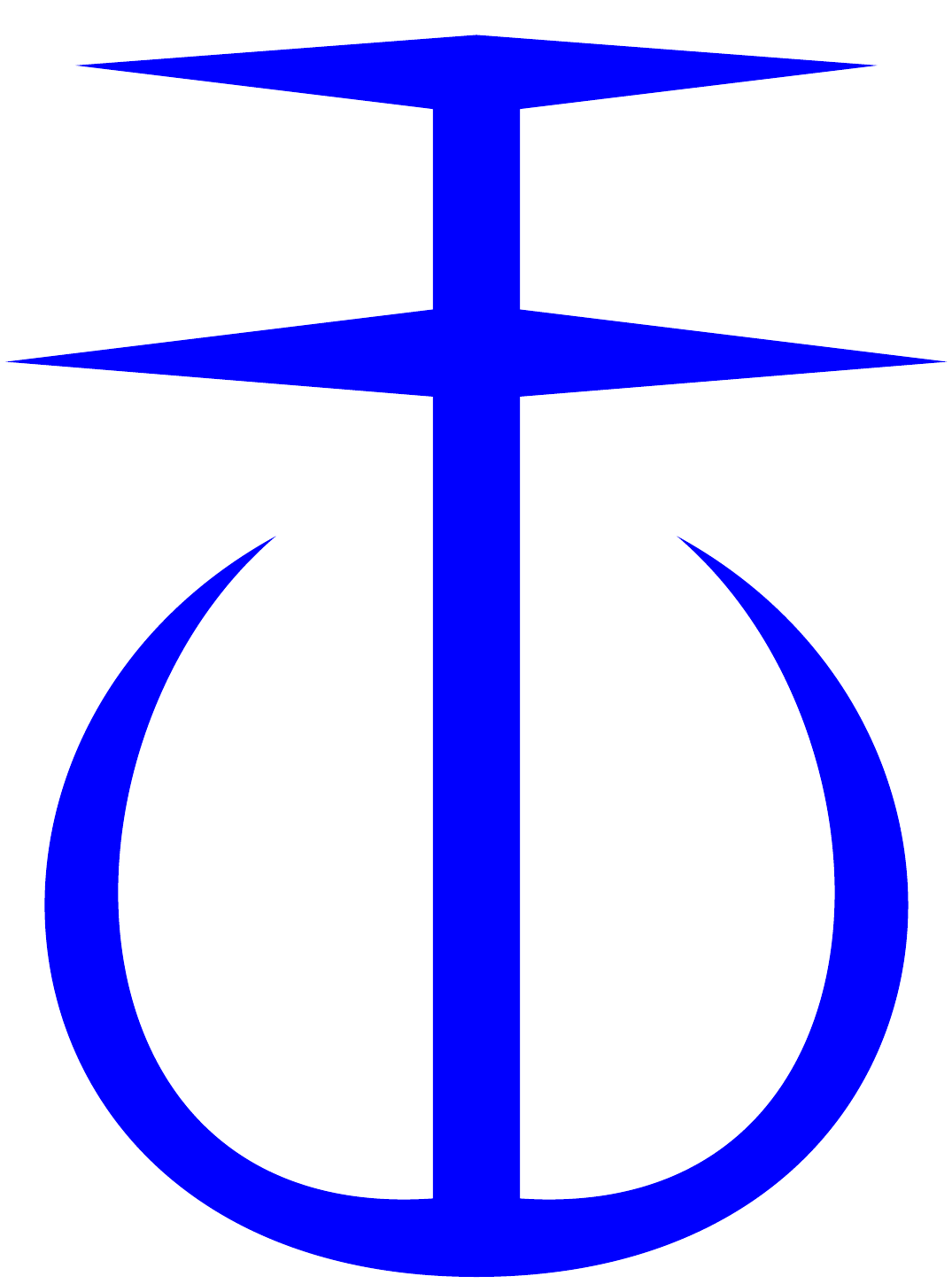}}}

\newcommand{\Wanfootnote}[1]{%
\let\oldthefootnote=\thefootnote%
\stepcounter{mpfootnote}%
\addtocounter{footnote}{-1}%
\renewcommand{\thefootnote}{\sWan}
\footnote{#1}%
\let\thefootnote=\oldthefootnote%
}

\newcommand{\Wangfootnote}[1]{%
\let\oldthefootnote=\thefootnote%
\stepcounter{mpfootnote}%
\addtocounter{footnote}{-1}%
\renewcommand{\thefootnote}{\sWang}
\footnote{#1}%
\let\thefootnote=\oldthefootnote%
}

\begin{document}
\begin{titlepage}
\begin{flushright}
%cond-mat/yymm.nnnn
\end{flushright}
\vskip .25in
\begin{center}

{\bf
%\LARGE{ 
%Yang-Mills 
%Non-Abelian Gauge
%Theories},  %\\[2.75mm]
%\LARGE{ 
%Sigma Models}, \\[2.75mm]
 \LARGE{Higher Anomalies, Higher Symmetries, 
 and\\[3.75mm]  
 Cobordisms I:}\\[4.75mm]
 \large{Classification of Higher-Symmetry-Protected Topological States and Their}\\[4.75mm]
\large{Boundary Fermionic/Bosonic Anomalies} %\\[4.75mm]
  \large{via a Generalized %higher-
  Cobordism Theory}\\[5.75mm]
}

\vskip0.5cm 
\Large{Zheyan Wan$^{1,2}$ {\Wanfootnote{e-mail: {\tt wanzy@mail.ustc.edu.cn} (Corresponding Author)}}
%\symfootnote{e-mail: {\tt wanzy@mail.ustc.edu.cn}}
and Juven Wang$^{3,4}${\Wangfootnote{e-mail: {\tt juven@ias.edu} (Corresponding Author)
}}
%\symfootnote{e-mail: {\tt juven@ias.edu}}
\\[3.75mm] 
} 
\vskip.5cm
{\small{\textit{$^1${Yau Mathematical Sciences Center, Tsinghua University, Beijing 100084, China}\\}}
}
 \vskip.2cm
 {\small{\textit{$^2${School of Mathematical Sciences, USTC, Hefei 230026, China}}\\}}
 \vskip.2cm
 {\small{\textit{$^3$School of Natural Sciences, Institute for Advanced Study, Einstein Drive, Princeton, NJ 08540, USA}\\}}
 \vskip.2cm
 {\small{\textit{$^4${Center of Mathematical Sciences and Applications, Harvard University,  Cambridge, MA 02138, USA} \\}}
}
% \vskip.2cm
%{\small{\textit{$^4${Department of Physics, Harvard University,  Cambridge, MA 02138, USA} \\}}
%}
 %\vskip.2cm
% {\small{\textit{$^4$Department of Physics, Massachusetts Institute of Technology, Cambridge, MA 02139, USA \\}}
%}

\end{center}

\vskip1.5cm
\baselineskip 16pt
\begin{abstract}

%[Draft for authors ONLY: NOT for distributions]\\

By developing a generalized cobordism theory, 
we explore the higher global symmetries
and higher anomalies of quantum field theories  
and interacting fermionic/bosonic systems in condensed matter.
Our essential math %mathematical 
input is a generalization of
Thom-Madsen-Tillmann spectra,  Adams spectral sequence, and Freed-Hopkins's theorem, to incorporate 
higher-groups and higher classifying spaces.
%the classifying space of higher-groups. 
%
We provide many examples of bordism groups with a generic $H$-structure manifold with a higher-group $\mathbb{G}$, and their bordism invariants
 --- e.g. perturbative anomalies of chiral fermions [originated from Adler-Bell-Jackiw] or bosons with U(1) symmetry in any even spacetime dimensions;
 non-perturbative global anomalies such as Witten anomaly and the new SU(2) anomaly in 4d and 5d.
Suitable $H$ such as SO/Spin/O/Pin$^\pm$ enables the study 
of quantum vacua of general bosonic or fermionic systems 
with time-reversal or reflection symmetry on (un)orientable spacetime.
Higher 't Hooft anomalies of $d$d live on the boundary of $(d+1)$d higher-Symmetry-Protected Topological states (SPTs) or 
symmetric invertible topological orders (i.e., invertible topological quantum field theories at low energy);
thus our cobordism theory also classifies and characterizes higher-SPTs.
Examples of higher-SPT's anomalous boundary theories include strongly coupled non-Abelian Yang-Mills (YM) gauge theories and sigma models,
complementary to physics obtained in [arXiv:1810.00844, 1812.11955, 1812.11968, 1904.00994].\\

\noindent
This article is a companion with further detailed calculations supporting other shorter articles.

\end{abstract}

\date{December, 2018}
\end{titlepage}

  \pagenumbering{arabic}
    \setcounter{page}{2}

\tableofcontents   

%\newpage

\section{Introduction and Summary}

\label{sec:intro}

\subsection{Preliminaries}

\label{sec:intro-pre}

Thom, as the pioneer of bordism theory, studied the criteria when the disjoint union of two closed $n$-manifolds is the boundary of a compact $(n+1)$-manifold  \cite{thom1954quelques}. Thom found that this relation is an equivalence relation on the set of closed $n$-manifolds. Moreover, the disjoint union operation defines an abelian group structure on the set of equivalence classes. This group is called the unoriented bordism group, it is denoted by $\Omega_n^{\tO}$. Furthermore, Thom found that the Cartesian product defines a graded ring structure on $\Omega_*^{\tO}:=\bigoplus_{n\ge0}\Omega_n^{\tO}$, which is called the unoriented bordism ring.
Thom also found that the bordism invariants of $\Omega_n^{\tO}$ are the Stiefel-Whitney numbers. Namely, 
 two manifolds are unorientedly bordant if and only if they have identical sets of Stiefel-Whitney characteristic numbers. This yields many interesting consequences. For example, 
the real projective space $\RP^2$ is not a boundary while $\RP^3$ is; also 
the complex projective space  $\CP^2$ and $\RP^2\times\RP^2$ are unorientedly bordant.

Many generalizations are made to bordism theories so far.
\cred{For example, we can consider manifolds which are equipped with an $H$-structure,
 we follow the definition of $H$-structure given in \cite{freed2013bordism}.}
Our work is inspired by a cobordism theory from the Madsen-Tillmann spectrum \cite{MadsenTillmann4} and from Freed-Hopkins \cite{Freed2016}.
Freed and Hopkins propose a cobordism theory \cite{Freed2016} to classify the
Symmetry-Protected Topological states (SPTs) \cite{1106.4772} in condensed matter physics \cite{Wen1610.03911} with ordinary internal global symmetries of group $G$
and their classifying space $\B G$. Examples of SPTs include the famous topological insulators and topological superconductors \cite{2010RMP_HasanKane, 2011_RMP_Qi_Zhang}.

The major motivation of our work is to generalize the calculations and the cobordism theory of Freed-Hopkins \cite{Freed2016} ---
such that, instead of the ordinary group $G$ or ordinary classifying space $\B G$, we consider a generalized cobordism theory studying manifolds (i.e., spacetime manifolds) 
endorsed with $H$ structure, with additional higher group $\mathbb{G}$ (i.e., generalized as principal-$\mathbb{G}$ bundles) 
and \emph{higher classifying spaces} $\B \mathbb{G}$. 
We consider this particular generalized  cobordism theory in order to study, characterize and classify: 
\begin{enumerate}
\item
 Generalized higher global symmetries of $\mathbb{G}$ \cite{Gaiotto2014kfa1412.5148} in physics \footnote{Generalized higher global symmetries may or may not be 
higher-differential form global symmetries. For example, 
there exist certain fermionic SPTs
whose higher global symmetries whose
charged objects are not in terms of higher-differential forms, see \Ref{1812.11959} and References therein.} and their higher classifying spaces $\B\mathbb{G}$. 
\item Higher-Symmetry-Protected Topological states (higher-SPTs), which are nontrivial quantum vacua protected by  higher global symmetries of $\mathbb{G}$. 
Higher SPTs are characterized by (co)bordism invariants obtained from (co)bordism groups of higher classifying spaces $\B\mathbb{G}$.
For example, in $(d+1)$-dimensional spacetime, denoted $(d+1)$d, 
consider the quantum vacua of internal global symmetry $\mathbb{G}$ on a $(d+1)$d spacetime manifold with $H$-structure,  
we will propose a bordism group $\Omega_{(d+1)}^H(\B\mathbb{G})$ and a related cobordism group $\TP_{(d+1)}(H\times\mathbb{G})$
to classify higher-SPTs in $(d+1)$d. See the earlier pioneer work on higher-SPTs in \cite{Kapustin2013uxa1309.4721, Thorngren2015gtw1511.02929}.
\item Higher quantum anomalies, e.g. higher 't Hooft anomalies: The ordinary 't Hooft anomalies \cite{tHooft:1980xss} of global symmetry $G$ is the anomaly for
QFT of the ordinary global symmetry  $G$.
In comparison, given the internal higher global symmetry $\mathbb{G}$ and the $d$d spacetime manifold with $H$-structure,
we can ask what are the possible higher quantum anomalies in the $d$d physical theories? 
The associated higher anomalies, given by the data $\mathbb{G}$ and $H$, in the $d$d spacetime,
via a generalization of the anomaly-inflow picture \cite{Callan:1984sa},
turns out to relate to the anomalies of the $d$d boundary theory (called the boundary anomalies) of $(d+1)$d higher-SPTs (given by the same data $\mathbb{G}$ and $H$).
So the characterization and classification of $(d+1)$d higher-SPTs in the previous remark turns out to help on the characterization and classification of 
$d$d higher 't Hooft anomalies.

Modern examples of higher 't Hooft anomalies are found 
in quantum field theories (QFTs) including Yang-Mills gauge theories \cite{PhysRev.96.191-YM} and sigma models.
The first example of higher 't Hooft anomalies is discovered by a remarkable work \Ref{Gaiotto2017TTT1703.00501} 
for a pure 4d SU(N) Yang-MiIls gauge theory of even integer N 
with a second Chern-class topological term (called the $\theta$-term or $\theta \Tr[F \wedge F]$-term in particle physics.)
Further new higher 't Hooft anomalies are found in \cite{2018arXiv180609592C, Wan:2018zql,Wan:2018djl,Wan:2019oyr}.
\end{enumerate}
In summary, as we have said, we aim to study higher global symmetries, characterize and classify higher-SPTs and higher quantum anomalies.
\begin{itemize}
\item By \emph{characterization}, we mean that given certain physics phenomena or theories (here, higher-SPTs and higher quantum anomalies),
we like to write down their mathematical invariants (here, we mean the bordism invariant) to
fully describe or capture their essences/properties. Hopefully, we can further compute their physical observables.
\item By \emph{classification}, we mean that given the spacetime dimensions (here $d+1$d for higher-SPTs or $d$d for higher quantum anomalies),
their $H$-structure and the internal higher global symmetry $\mathbb{G}$,
we aim to know how \emph{many classes} (a number to count them) there are?
Also, we aim to determine the mathematical structures of classes (i.e. here group structure as for  (co)bordism groups:
would the classes be a finite group $\Z_n$ or an infinite group $\Z$ or their mixing, etc.).
\end{itemize}

Another purpose of this article is a companion article with further detailed mathematical calculations in order to support other shorter articles \cite{Wan:2018zql,Wan:2018djl,Wan:2019oyr}.

In this Introduction, we will provide some basic physics preliminaries in \Sec{sec:intro-phy}
and {mathematical preliminaries} in \Sec{sec:intro-math}.
Since the concepts of higher symmetries and higher anomalies are crucial, we will also clarify what precisely we mean by
higher symmetries/anomalies in condensed matter, in QFTs and in mathematics, in \Sec{sec:intro-anom}.

After some additional introduction to mathematical background in \Sec{sec:Background},
we will provide explicit interpretations of familiar examples (to QFT-ist and physicists)  of
\emph{perturbative} anomalies in \Sec{sec:p-anom-ex-U1}:
\begin{enumerate} [label=\textcolor{blue}{(\arabic*)}:, ref={(\arabic*)}]
%$\bullet$
\item
Perturbative fermionic anomalies from chiral fermions with U(1) symmetry, originated from Adler-Bell-Jackiw (ABJ) anomalies \cite{Adler1969gkABJ,Bell1969tsABJ}. \\
%$\bullet$
\item Perturbative bosonic anomalies from bosonic systems with U(1) symmetry.
\end{enumerate}
We will also provide more exotic \emph{non-perturbative global} anomalies in \Sec{sec:np-anom-ex-SU2}:
\begin{enumerate} [label=\textcolor{blue}{(\arabic*)}:, ref={(\arabic*)}]
\setcounter{enumi}{2}
%$\bullet$
\item The SU(2) anomaly of Witten \cite{Witten:1982fp}. 
%$\bullet$
\item A new SU(2) anomaly \cite{Wang:2018qoyWWW},
\end{enumerate}
matching the physics results of $d$d anomalies  to mathematical cobordism group calculations in $(d+1)$d.

%\ccred
{We briefly comment the difference between a previous cobordism theory \cite{Freed2016, 2017arXiv171111587GPW} and this work:
In all Adams charts of the computation in \cite{Freed2016, 2017arXiv171111587GPW}, there are no nonzero differentials, 
while in this paper we encounter nonzero differentials $d_n$ due to the $(p,p^n)$-Bockstein homomorphisms in the computation involving $\B ^2\Z_{p^n}$ and $\B \Z_{p^n}$.}

\subsection{Physics preliminaries}
\label{sec:intro-phy}

Freed-Hopkins's work \cite{Freed2016} is motivated by the development of cobordism theory classification \cite{Kapustin2014tfa1403.1467,Kapustin1406.7329} of so-called the 
Symmetry Protected Topological (SPT) state in condensed matter physics \cite{Wen1610.03911}.
In a very short summary, 
Freed-Hopkins's work \cite{Freed2016} applies the theory of
Thom-Madsen-Tillmann spectra \cite{thom1954quelques,MadsenTillmann4},
to prove a theorem  relating the 
``Topological Phases'' (which later will be abbreviated as TP) or certain 
deformation classes of reflection positive invertible $n$-dimensional extended
topological field theories (iTQFT) with
symmetry group (or in short, symmetric iTQFT),
to Madsen-Tillmann spectrum \cite{MadsenTillmann4} of the symmetry group.

%\cred
{
Here an $n$-dimensional extended
topological field theory is a symmetric monoidal functor $F$ from the $(\infty,n)$-category of extended cobordisms 
$\text{Bord}_n(H_n)$ to a symmetric monoidal $(\infty,n)$-category $\mathcal{C}$
where $\text{Bord}_n(H_n)$ is defined as follows (all manifolds are equipped with $H$-structures, see definition \ref{H-structure}):
\begin{itemize}
\item
objects are 0-manifolds;
\item
1-morphisms are 1-cobordisms between objects;
\item
2-morphisms are 2-cobordisms between 1-morphisms;
\item
\dots
\item 
$n$-morphisms are $n$-cobordisms between $(n-1)$-morphisms;
\item
$(n+1)$-morphisms are diffeomorphisms between $n$-morphisms;
\item
$(n+2)$-morphisms are smooth homotopies between $(n+1)$-morphisms;
\item
\dots
\end{itemize}
An $n$-dimensional extended
topological field theory is called invertible if $F$ factor through the Picard groupoid $\mathcal{C}^{\times}$.
By a theorem of Galatius-Madsen-Tillmann-Weiss \cite{MadsenTillmann4}, the classifying space
of $\text{Bord}_n(H_n)$ is exactly the 0-th space of the Madsen-Tillmann spectrum $\Sigma^nMTH_n$.
}

In this work, we will consider the generalization of \cite{Freed2016} 
to include higher symmetries \cite{Gaiotto2014kfa1412.5148}, for example, including both 0-form symmetry of group $G_{(0)}$ and 1-form symmetry of group $G_{(1)}$,
or in certain cases, as higher symmetry group of higher $n$-group.\footnote{
For the physics application of our result,  please see \cite{Wan:2018zql,Wan:2018djl,Wan:2019oyr}.
Some of these 4d non-Abelian $\SU(N)$ Yang-Mills\cite{PhysRev.96.191-YM}-like gauge theories can be obtained from 
gauging the time-reversal symmetric $\SU(N)$-SPT generalization of topological insulator/superconductor (TI/SC) \cite{2017arXiv171111587GPW}.
We can understand their anomalies of 0-form symmetry of group $G_{(0)}$ and 1-form symmetry of group $G_{(1)}$,
as the obstruction to regularize the global symmetries locally in its own dimensions (4d for YM theory).
Instead, in order  to regularize the global symmetries locally and onsite,
the 4d gauge theories need to be placed on the boundary of 5d higher SPTs. 
The 5d higher SPTs corresponds to the nontrivial generators of %$\Omega^{d=5}(\B G_{(0)} \times \B ^2G_{(1)})$ 
cobordism groups of higher {classifying} spaces.
We write $G_{(0)}$ or $G_{a}$ to indicate some 0-form symmetry probed by 1-form $a$ field. 
We write $G_{(1)}$ or $G_{b}$ to indicate some 1-form symmetry probed by 2-form $b$ field. 
}
Other physics motivations to study higher group can be found in 
\cite{Cordova:2018cvg2group, Benini:2018reh, Delcamp2018wlb1802.10104, 2018arXiv180809394Z} and references therein.

We generalize the work of Freed-Hopkins \cite{Freed2016}: there is a 1:1 correspondence
\bea \label{eq:propose-1-thm}
\left\{\begin{array}{ccc}\text{deformation classes of reflection positive}\\\text{invertible }n\text{-dimensional extended topological}\\\text{field theories with a symmetry group }H_n\times\mathbb{G}\end{array}\right\}\cong[MT(H\times\mathbb{G}),\Sigma^{n+1}I\Z]_{\text{tors}}.
\eea
where $H$ is the space time symmetry, $\mathbb{G}$ is the internal symmetry which is possibly a higher group,
$MT(H\times\mathbb{G})$ is the Madsen-Tillmann spectrum \cite{MadsenTillmann4} of the group $H\times\mathbb{G}$,
$\Sigma$ is the suspension, $I\Z$ is the Anderson dual spectrum, and ${\text{tors}}$ means the torsion part.

Since there is an exact sequence
\bea
0\to\Ext^1(\pi_n\mathcal{B},\Z)\to[\mathcal{B},\Sigma^{n+1}I\Z]\to\Hom(\pi_{n+1}\mathcal{B},\Z)\to0
\eea
for any spectrum $\mathcal{B}$, especially for $MT(H\times\mathbb{G})$.
The torsion part $[MT(H\times\mathbb{G}),\Sigma^{n+1}I\Z]_{\text{tors}}$ is $\Ext^1((\pi_nMT(H\times\mathbb{G}))_{\text{tors}},\Z)=\Hom((\pi_nMT(H\times\mathbb{G}))_{\text{tors}},\U(1))$.

By the generalized Pontryagin-Thom isomorphism \eqref{ponthom}, $\pi_nMT(H\times\mathbb{G})=\Omega_n^{H\times\mathbb{G}}=\Omega_n^H(\B\mathbb{G})$ which is the bordism group defined in definition \ref{bordism-def}.

Namely, we can classify the deformation classes of symmetric iTQFTs and also symmetric invertible topological orders (iTOs), via
the particular group
\bea\label{eq:TP}
\TP_n(H\times\mathbb{G})\equiv[MT(H\times\mathbb{G}),\Sigma^{n+1}I\Z].
\eea
 Here TP means the abbreviation of ``Topological Phases''
classifying the above symmetric iTQFT,
the torsion part of $\TP_n(H\times\mathbb{G})$ and $\Omega_n^H(\B\mathbb{G})$ are the same.

In this work, we compute the (co)bordism groups $\Omega_d^H(\B\mathbb{G})$ ($\TP_d(H\times\mathbb{G})$) for $H=\tO/\SO/\Spin/\Pin^{\pm}$ and several $\mathbb{G}$, we also consider $\Omega_d^{\mathbb{G}}$ where $\B\mathbb{G}$ is the total space of the nontrivial fibration with base space $\B\tO$ and fiber $\B^2\Z_2$ in section \ref{(BG_a,B^2G_b):(BO,B^2Z_2)}.

%\ccred
{
If there is a nontrivial group action between $H$ and $\mathbb{G}$ (let us denote the action as the semi-direct product $\ltimes$),
or if there is a shared common normal subgroup $N_{\text{shared}}$ or sub-higher-group between $H$ and $\mathbb{G}$,
then we can generalize our above proposal \eqn{eq:propose-1-thm} and \eqn{eq:TP} to
\bea
\left\{\begin{array}{ccc}\text{deformation classes of reflection positive}\\\text{invertible }n\text{-dimensional extended topological}\\\text{field theories  with a symmetry group }
(\frac{H_n \ltimes\mathbb{G}}{N_{\text{shared}}})
\end{array}\right\}\cong[MT(\frac{H \ltimes\mathbb{G}}{N_{\text{shared}}}),\Sigma^{n+1}I\Z]_{\text{tors}}.
\eea
and
\bea\label{eq:TP-2}
\TP_n(\frac{H \ltimes\mathbb{G}}{N_{\text{shared}}})\equiv[MT(\frac{H \ltimes\mathbb{G}}{N_{\text{shared}}}),\Sigma^{n+1}I\Z].
\eea
}

For readers who wishes to explore other physics stories and some introduction materials, we suggest to look at the introduction of \cite{2017arXiv171111587GPW}
and other shorter articles \cite{Wan:2018zql,Wan:2018djl,Wan:2019oyr, Tanizaki:2017qhf1710.08923, Wen2018zux1812.02517}. 
In particular,  \Ref{Wen2018zux1812.02517} provides a condensed matter interpretation of higher symmetries.
We also encourage readers to read the Section I to III of  \cite{Wan:2018zql}. 

We will explore the generic manifold with $H$-structure, including the orientable $H=\SO$, $\Spin$, etc., 
or unorientable $H=\tO$, $\Pin^{\pm}$. 
In physics, the quantum system that can be put on an unorientable $H$ manifold implies that 
there is a time-reversal symmetry or a reflection symmetry (commonly termed the parity symmetry in an odd dimensional space). 
Physicists can find the introduction materials on the reflection symmetry and unorientable manifolds in \Ref{Witten2016cio1605.02391, Barkeshli2016mew1612.07792}.

For readers who wishes to explore other mathematical introductory materials, we suggest to look at the \cite{Freed2016, 2018arXiv180107530B} and 
Appendices of \cite{2017arXiv171111587GPW}.

Readers may be also interested in other recent work along the cobordism theory applications to physics 
\cite{Kapustin2017jrc1701.08264} \cite{2018PTEP1801.05416} \cite{Yonekura2018arXiv180310796Y} \cite{Montero2018arXiv180800009G}.

\subsection{Mathematical preliminaries}

\label{sec:intro-math}

\cred{In this subsection, we review the basics of bordism theory and possible generalizations.}

\begin{definition}\label{H-structure}
If $H$ is a group with a group homomorphism $\rho:H\to\tO$, $V$ is a vector bundle over $M$ with a metric, then an $H$-structure on $V$ is a principal
$H$-bundle $P$ over $M$, together with an isomorphism of bundles $P\times_H\tO\xrightarrow{\sim}\cB_{\tO}(V)$ where $P\times_H\tO$ is the quotient $(P\times\tO)/H$ where $H$ acts freely on right of $P\times\tO$ by 
$$(p,g)\cdot h=(p\cdot h,\rho(h)^{-1}g),\;\;\;p\in P,\;\;\;g\in\tO,\;\;\;h\in H$$
and $\cB_{\tO}(V)$ is the orthonormal frame bundle of $V$.
\end{definition}

In particular, if $V=TM$, then an $H$-structure on $TM$ is also called a tangential $H$-structure (or an $H$-structure) on $M$.
Here we assume the $H$-structures are defined on the tangent bundles instead of normal bundles.

Below we consider manifolds with a metric.

Any manifold admits an $\tO$-structure,
a manifold $M$ admits an $\SO$-structure if and only if $w_1(TM)=0$, a manifold admits a Spin structure if and only if $w_1(TM)=w_2(TM)=0$,
a manifold admits a Pin$^+$ structure if and only if $w_2(TM)=0$, a manifold admits a Pin$^-$ structure if and only if $w_2(TM)+w_1(TM)^2=0$. \cred{Here $w_i(TM)$ is the $i$-th Stiefel-Whitney class of the tangent bundle of $M$.}

\cred{ Moreover, we can consider manifolds equipped with a map to a fixed topological space $X$,
 we are interested in the case when $X$ is an Eilenberg-MacLane space since 
 $$[M,K(G,n)]=\H^n(M,G)$$
 where the left hand side is the group of homotopy classes of maps from $M$ to $K(G,n)$, the right hand side is the $n$-th cohomology group of $M$ with coefficients in $G$.}
 
\begin{definition}\label{bordism-def}
\cred{Let $H$ be a group, $X$ be a fixed topological space, we can define an abelian group}
\bea
&&\Omega^{H}_{n}(X):=\{(M,f)\vert M\text{ is a closed }\notag\\
&&n\text{-manifold with $H$-structure, }f:M\to X\text{ is a map}\}/\text{bordism}.
\eea
where bordism is an equivalence relation, namely, $(M,f)$ and $(M',f')$ are bordant if there exists a compact $n+1$-manifold $N$ with $H$-structure and a map $h:N\to X$ such that the boundary of $N$ is the disjoint union of $M$ and $M'$,
the $H$-structures on $M$ and $M'$ are induced from the $H$-structure on $N$ and $h|_M=f$, $h|_{M'}=f'$.
\end{definition}

In particular, when $X=\B^2\Z_n$, $f:M\to \B^2\Z_n$ is a cohomology class in $\H^2(M,\Z_n)$.
When $X=\B G$, with $G$ is a Lie group or a finite group (viewed as a Lie group with discrete topology), then $f:M\to \B G$ is a principal $G$-bundle over $M$.

To explain our notation, here $\B G$ is a classifying space of $G$, and $\B^2\Z_n$ is a higher classifying space (Eilenberg-MacLane space $K(\Z_n,2)$) of $\Z_n$. 

In the particular case that $H=\tO$ and $X$ is a point, this definition \ref{bordism-def} coincides with Thom's original definition.

In this article, we study the cases in which $H=\tO/\SO/\Spin/\Pin^{\pm}$, and $X$ is a higher classifying space, or more complicated cases.

\cred{
If $\Omega_n^H(X)=G_1\times G_2\times\cdots\times G_r$ where $G_i$ are cyclic groups, then 
group homomorphisms $\phi_i:\Omega_n^H(X)\to G_i$ form a complete set of \emph{bordism invariants} if 
$\phi=(\phi_1,\phi_2,\dots,\phi_r):\Omega_n^H(X)\to G_1\times G_2\times\cdots\times G_r$ is a group isomorphism.
}

\cred{
Elements of $\Omega_n^H(X)$ are \emph{manifold generators} if their images in $G_1\times G_2\times\cdots\times G_r$ under $\phi$ generate $G_1\times G_2\times\cdots\times G_r$.
}

%\cred{
%Convention:
%For a top degree cohomology class 
% we often suppress explicit integration over the manifold (i.e. pairing with the fundamental class $[M]$).
% If $M$ is orientable, then $[M]$ has coefficients in $\Z$.
% If $M$ is non-orientable, then $[M]$ has coefficients in $\Z_2$.
%}

We first introduce several concepts which are important for bordism theory:

Thom space: Let $V\to Y$ be a real vector bundle, and fix a Euclidean metric. The Thom space $\text{Thom}(Y;V)$ is the quotient $D(V)/S(V)$ where $D(V)$ is the unit disk bundle and $S(V)$ is the unit sphere bundle. Thom spaces satisfy
\bea\label{thomsum}
\text{Thom}(X\times Y;V\times W)&=&\text{Thom}(X;V)\wedge \text{Thom}(Y;W),\notag\\
\text{Thom}(X,V\oplus \underline{\R^n})&=&\Sigma^n\text{Thom}(X;V),\notag\\
\text{Thom}(X,\underline{\R^n})&=&\Sigma^nX_+
\eea
where $V\to X$ and $W\to Y$ are real vector bundles, $\underline{\R^n}$ is the trivial real vector bundle of dimension $n$, $\Sigma$ is the suspension, $X_+$ is the disjoint union of $X$ and a point.
 
We follow the definition of Thom spectrum and Madsen-Tillmann spectrum given in \cite{freed2013bordism}.
 
Thom spectrum \cite{thom1954quelques}: 
 $MH$ is the Thom spectrum of the group $H$, 
 \cred{it is the spectrification (see \ref{sec:spectra}) of the prespectrum whose $n$-th space is}
 %its 0-th space is the colimit of $\Omega^nMH(n)$, where 
 $MH(n)=\text{Thom}(\B H(n);V_n)$, and $V_n$ is the induced vector bundle (of dimension $n$) by the map $\B H(n)\to \B \tO(n)$.
 
In other words, 
$MH=\text{Thom}(\B H;V)$, where $V$ is the induced virtual bundle (of dimension $0$) by the map $\B H\to \B \tO$.

Madsen-Tillmann spectrum \cite{MadsenTillmann4}: 
$MTH$ is the Madsen-Tillmann spectrum of the group $H$, it is the colimit of $\Sigma^nMTH(n)$, where 
$MTH(n)=\text{Thom}(\B H(n);-V_n)$, and $V_n$ is the induced vector bundle (of dimension $n$) by the map $\B H(n)\to \B \tO(n)$.
\cred{
The %0-th space of the 
virtual Thom spectrum
$MTH(n)$ is the spectrification (see \ref{sec:spectra}) of the prespectrum whose $(n+q)$-th space is 
$\text{Thom}(\B H(n,n+q),Q_q)$
}
%colimit of $\Omega^{n+q}\text{Thom}(\B H(n,n+q),Q_q)$
 where  $\B H(n,n+q)$ is the pullback 
\bea
\xymatrix{
\B H(n,n+q)\ar@{-->}[r]\ar@{-->}[d]&\B H(n)\ar[d]\\
Gr_n(\R^{n+q})\ar[r]&\B\tO(n)
}
\eea
and there is a direct sum $\underline{\R^{n+q}}=V_n\oplus Q_q$ of vector bundles over $Gr_n(\R^{n+q})$ and, by pullback, over $\B H(n,n+q)$ where $\underline{\R^{n+q}}$ is the trivial real vector bundle of dimension $n+q$.

In other words, 
$MTH=\text{Thom}(\B H;-V)$, where $V$ is the induced virtual bundle (of dimension $0$) by the map $\B H\to \B \tO$.

Here $\Omega$ is the loop space, $\Sigma$ is the suspension.

Note: ``$T$'' in $MTH$ denotes that the $H$-structures are on tangent bundles instead of normal bundles.

(Co)bordism theory is a generalized (co)homology theory which is represented by a spectrum by the Brown representability theorem.

In fact, it is represented by Thom spectrum due to the Pontryagin-Thom isomorphism:

\bea
\pi_n(MTH)&=&\Omega_n^H\text{   the cobordism group of }n\text{-manifolds with tangential }H\text{-structure},\notag\\
\pi_n(MH)&=&\Omega_n^{\nu H}\text{   the cobordism group of }n\text{-manifolds with normal }H\text{-structure}
\eea

In the case when tangential $H$-structure is the same as normal $H'$-structure, the relevant Thom spectra are weakly equivalent. In particular,
$MT\tO\simeq M\tO$, $MT\SO\simeq M\SO$, $MT\Spin\simeq M\Spin$, $MT\Pin^+\simeq M\Pin^-$, $MT\Pin^-\simeq M\Pin^+$.

$\Pin^{\pm}$ cobordism groups are not rings, though they are modules over the $\Spin$ cobordism ring.

By the generalized Pontryagin-Thom construction, for $X$ a topological space, then the group of $H$-bordism classes of $H$-manifolds in $X$ is isomorphic to the generalized homology of $X$ with coefficients in $MTH$:
\bea\label{ponthom}
\Omega_d^H(X)=\pi_d(MTH\wedge X_+)=MTH_d(X)
\eea
where 
$\pi_d(MTH\wedge X_+)$ is the $d$-th stable homotopy group of the spectrum $MTH\wedge X_+$. 
The $d$-th stable homotopy group of a spectrum $M$ is
\bea
\pi_d(M)=\text{colim}_{k\to\infty}\pi_{d+k}M_k.
\eea

So the computation of the bordism group $\Omega_d^H(X)$ is the same as the computation of the stable homotopy group of the spectrum $MTH\wedge X_+$ which can be computed by Adams spectral sequence method.

Next, we introduce the
Thom isomorphism \cite{thom1954quelques}:
Let $p:E\to B$ be a real vector bundle of rank $n$. Then there is an isomorphism, called Thom isomorphism 
\bea\label{thomiso}
\Phi:\H^k(B,\Z_2)\to\tilde{\H}^{k+n}(T(E),\Z_2)
\eea
where $\tilde{\H}$ is the reduced cohomology, $T(E)=\text{Thom}(E;B)$ is the Thom space and
\bea
\Phi(b)=p^*(b)\cup U
\eea
where $U$ is the Thom class.
We can define the $i$-th Stiefel-Whitney class of the vector bundle $p:E\to B$ by
\bea\label{thom}
w_i(p)=\Phi^{-1}(\Sq^iU)
\eea
where $\Sq$ is the Steenrod square.

%\subsection{A Preliminary Introduction}

%\subsection{Bordism theory}

\subsection{Basics of Higher Symmetries and Higher Anomalies of Quantum Field Theory for Physicists and Mathematicians}
\label{sec:intro-anom}

In order to obtain a complete classification of 't Hooft anomalies of quantum field theories (QFTs), we aim to first identify the relevant (if not all of) global symmetry $G$ 
(here we will abuse the notation to have $G$ including the higher symmetry $\mathbb{G}$)
of QFTs.
Then we couple the QFTs to classical background-symmetric gauge field of $G$. Then we try to detect the possible obstructions of such
coupling \cite{tHooft:1980xss}. Such obstructions, known as the obstruction of gauging the global symmetry, are termed 
`` 't Hooft anomalies'' in QFT.
In the literature, when people refer to ``anomalies,'' however, they can means several related but different issues. 
To fix our terminology, we refer ``anomalies'' to be one of the followings:
\begin{enumerate}
\item Classical global symmetry is violated at the quantum theory, such that the 
classical global symmetry fails to survive as a quantum global symmetry, e.g. the original Adler-Bell-Jackiw (ABJ) anomaly \cite{Adler1969gkABJ,Bell1969tsABJ}.

\item Quantum global symmetry is well-defined and preserved (for the Hamiltonian or path integral Lagrangian formulation of quantum theory). 
Namely, global symmetry is sensible, not only at a classical theory (if there is any classical description), but also for a quantum theory.
However, there is an obstruction to gauge the global symmetry.
Specifically, we can detect a certain obstruction to even \emph{weakly gauge}
the symmetry or couple the symmetry to a \emph{non-dynamical background probed gauge field}.
(We may abbreviate this background field as ``bgd.field.'')
%\footnote{
%We will refer this %kind of field simply 
%as a background 
%(non-dynamical gauge) 
%field, abbreviated as ``bgd.field.''}
This is known as ``'t Hooft anomaly,'' or sometimes
regarded as a ``weakly gauged anomaly'' in condensed matter.
Namely, the partition function $\bZ$ does not sum over background gauge connections, but only fix a background gauge connection
and only depend on the background gauge connection as a classical field (as a classical coupling constant).
Say if the background gauge connection is $A$, the partition function is $\bZ[A]$ depending on $A$.
Normally, the $\bZ[A]$ on a closed manifold in its own dimension is an invertible topological QFT (iTQFT), such that
$\bZ[A]= \exp(\ii \theta(A))$ is a complex phase (thus physically meaningfully \emph{invertible}) while its absolute value $|\bZ[A]|=1$ for any choice of background $A$. 
%\cred{ [do not sum over gauge connection: fix a gauge connection]}

\item Quantum global symmetry is well-defined and preserved (for the Hamiltonian or path integral Lagrangian formulation of quantum theory). 
However, once we promote the global symmetry to a gauge symmetry of the dynamical gauge theory,
then the gauge theory becomes inconsistent. Some people call this as 
a ``dynamical gauge anomaly'' which makes a quantum theory inconsistent.
Namely, the partition function $\bZ$ after summing over dynamical gauge connections becomes inconsistent or ill-defined.
%\cred{[sum over gauge connection]. Related to 2 by dynamical gauging.}

\end{enumerate}

From now on,  when we simply refer to ``anomalies,'' we mean mostly ``'t Hooft anomalies,'' 
which still have several intertwined interpretations:\\

\noindent
Interpretation (1): In condensed matter physics, ``'t Hooft anomalies''  are known as the obstruction to \emph{lattice-regularize} the global symmetry's quantum operator in a strictly 
local manner. By claiming local on a lattice or on a simplicial complex, 
we mean:\\ 
$\bullet$ on-site (e.g. on 0-simplex) for which 0-form symmetry operator acts on.  \\
$\bullet$ on-link (e.g. on 1-simplex) for which 1-form symmetry operator acts on.  \\
$\bullet$ on-plaquette (e.g. on 2-simplex) for which 2-form symmetry operator acts on.  \\
$\dots$\\
$\bullet$ on $n$-simplex for which $n$-form symmetry operator acts on.  \\
This obstruction is due to the symmetry-twists (See [\onlinecite{Wen2013oza1303.1803, 1405.7689, Wang2017locWWW1705.06728}] for QFT-oriented discussion and references therein).
This obstruction can be detected at high energy lattice scale (known as the ultraviolet [UV] in QFT).
This ``non-onsite symmetry'' viewpoint is generically applicable to both, \emph{perturbative} anomalies, and  \emph{non-perturbative global} anomalies:\\
$\bullet$ \emph{Perturbative} anomalies --- Characterized and captured by perturbative Feynman diagram calculations. Classified by an infinite integer $\Z$ class, 
known as the free (sub)group.\\ 
$\bullet$ \emph{Non-perturbative or global} anomalies --- 
Examples of global anomalies include the old and the new SU(2) anomalies \cite{Witten:1982fp, Wang:2018qoyWWW} 
(here we mean their 't Hooft anomaly analogs if we view the SU(2) gauge field as a non-dynamical classical background field)
and the global gravitational anomalies \cite{Witten1985xe}.
These are classified by a finite group $\Z_n$ class, known as the torsion (sub)group.

%\newpage

These anomalies are sensitive to the underlying UV-completion 
not only of fermionic systems, but also of bosonic systems \cite{WangSantosWen1403.5256, 1405.7689, Kapustin1404.3230, JWangthesis}.
We term the anomalies of QFT whose UV-completion requires only the bosonic degrees of freedom as bosonic anomalies \cite{WangSantosWen1403.5256}.
While we term those must require fermionic degrees of freedom as fermionic anomalies.\\

\noindent
Interpretation (2): In QFTs, the obstruction is on the impossibility of adding any counter term 
in its own dimension ($d$-d) in order to absorb a one-higher-dimensional counter term (e.g. $(d+1)$d topological term) due to background $G$-field \cite{Kapustin:2014gua}.
This is named the ``anomaly-inflow \cite{Callan:1984sa}.''
The  $(d+1)$d topological term is known as the $(d+1)$d SPTs in condensed matter physics \cite{1106.4772, Senthil1405.4015}.\\

\noindent
Interpretation (3): In math, the $d$d anomalies can be systematically captured by $(d+1)$d topological invariants \cite{Witten:1982fp} known as 
bordism invariants \cite{DaiFreed1994kq9405012,Kapustin2014tfa1403.1467,Kapustin1406.7329,Freed2016}.\\

\noindent
$\bullet$ Bosonic anomalies or bosonic SPTs are normally characterized by topological terms detected via manifolds with $H=\SO$ (orientable) or $\tO$ (unorientable) structures.\\
$\bullet$ Fermionic anomalies or fermionic SPTs 
are normally characterized by topological terms detected via manifolds with $H=\Spin$ (orientable) or $\Pin^{\pm}$ (unorientable) structures.

Below we summarize the higher symmetry $\mathbb{G}$ systematically introduced in  \cite{Gaiotto2014kfa1412.5148}.

\noindent
($i$) Higher symmetries and higher anomalies:
The ordinary 0-form global symmetry has a \emph{charged} object of 0d measured by the \emph{charge} operator of $(d-1)$d.
The generalized $q$-form global symmetry is introduced by \Ref{Gaiotto2014kfa1412.5148}.  
A \emph{charged} object of $q$d is measured by the \emph{charge} operator of $(d-q-1)$d (i.e. codimension-$(q+1)$). This concept turns out to be powerful to
detect new anomalies, e.g. the pure SU($\rN$)-YM at $\theta=\pi$ 
has a mixed anomaly between 0-form time-reversal symmetry $\Z_2^T$
and 1-form center symmetry $\Z_{\rN,[1]}$ at an even integer $\rN$, firstly discovered in a remarkable work [\onlinecite{Gaiotto2017TTT1703.00501}].

\noindent
($ii$) Relate (higher)-SPTs to (higher)-topological invariants: 
In the condensed matter literature, based on the earlier discussion on the \emph{symmetry twist},
it has been recognized that the classical background-field partition function under the {symmetry twist}, called $\bZ_{{\text{sym.twist}}}$
in $(d+1)$d
can be regarded as the partition function of  $(d+1)$d SPTs $\bZ_{{\text{SPTs}}}$.
 These descriptions are applicable to both low-energy infrared (IR) field theory, but also to the UV-regulated SPTs on a lattice,
see [\onlinecite{Wen2013oza1303.1803, 1405.7689, Kapustin2014tfa1403.1467}]
and References therein. Schematically, we follow the framework of \cite{1405.7689}, 
\begin{multline}
\label{eq:Zpart}
\bZ^{\text{$(d+1)$d}}_{{\text{sym.twist}}}=\bZ^{\text{$(d+1)$d}}_{{\text{SPTs}}}=\bZ^{\text{$(d+1)$d}}_{{\text{topo.inv}}}=\bZ^{\text{$(d+1)$d}}_{{\text{Cobordism.inv}}}  
\longleftrightarrow \text{$d$d-(higher) 't Hooft anomaly}.\;\;\;\;\;
\end{multline}
 In general,
the partition function $\bZ_{{\text{sym.twist}}}=\bZ_{{\text{SPTs}}}[A_1,B_2, w_i, \dots]$
is a functional containing background gauge fields of 1-form $A_1$, 2-form $B_2$ or higher forms;
and can contain characteristic classes \cite{milnor1974characteristic} such as the $i$-th Stiefel-Whitney class ($w_i$)
and other geometric probes such as gravitational background fields, e.g. a gravitational Chern-Simons 3-form CS$_3(\Gamma)$ involving the Levi-Civita connection or
the spin connection $\Gamma$. 
For our convention, we use the capital letters ($A,B,...$) to denote \emph{non-dynamical background gauge} fields (which, however, later they may or may not be dynamically gauged),  while 
the little letters ($a,b,...$) to denote \emph{dynamical gauge} fields.\\
More generally, \\
$\bullet$ For the ordinary 0-form symmetry, we may couple the charged 0d point operator to 1-form background gauge field (so the symmetry-twist occurs
in the Poincar\'e dual codimension-1 sub-spacetime [$d$d] of SPTs).\\
$\bullet$ For the 1-form symmetry, we may couple the charged 1d line operator to 2-form background gauge field (so the symmetry-twist occurs in the Poincar\'e dual codimension-2 sub-spacetime [$(d-1)$d] of SPTs).\\
$\bullet$ For the $q$-form symmetry, we may couple the \emph{charged} $q$d extended operator to $(q+1)$-form background gauge field. 
The  \emph{charged} $q$d extended operator can be measured by another \emph{charge} operator of codimension-$(q+1)$ [i.e. $(d-q)$d].\\
In summary, for the $q$-dimensional symmetry, we use the following terminology: 
\begin{enumerate} [label=\textcolor{blue}{($\diamondsuit$ \arabic*)}:, ref={($\diamondsuit$ \arabic*)}]
%$\bullet$
\item  \label{1:charged}
\emph{Charged} object: The \emph{charged} $q$-dimensional extended operator as the $q$-dimensional-symmetry generator which is being measured
by a symmetry generator.
\item  \label{2:charge}
\emph{Charge} operator: The corresponding  \emph{charge} operator of codimension-$(q+1)$ [i.e. $(d-q)$-dimension] which measures the  $q$-dimensional-symmetry
charged object.
\end{enumerate} 
So the symmetry-twist can be interpreted as the occurrence of the codimension-$(q+1)$ \emph{charge} operator.
In other words, the symmetry-twist happens at a Poincar\'e dual codimension-$(q+1)$ sub-spacetime [$(d-q)$d] of SPTs. 
We shall view the measurement of a \emph{charged} $q$d extended object, happening at any $q$-dimensional intersection 
between the $(q+1)$d form background gauge field
and the codimension-$(q+1)$ symmetry-twist or \emph{charge} operator of this SPT vacua.

By {higher-SPTs}, we mean SPTs protected by higher symmetries (for generic $q$, especially for any SPTs with at least a symmetry of $q>0$). 
So our principle above is applicable to higher-SPTs \cite{Thorngren2015gtw1511.02929, Delcamp2018wlb1802.10104}.
In the following of this article, thanks to \eq{eq:Zpart}, we can interchange the usages and interpretations of
``higher SPTs $\bZ_{{\text{SPTs}}}$,'' ``higher topological terms due to symmetry-twist $\bZ^{\text{$(d+1)$d}}_{{\text{sym.twist}}}$,'' ``higher topological invariants $\bZ^{\text{$(d+1)$d}}_{{\text{topo.inv}}}$'' or ``bordism invariants 
$\bZ^{\text{$(d+1)$d}}_{{\text{Cobordism.inv}}} $'' in $(d+1)$d.
They are all physically equivalent, and can uniquely determine a $d$d higher anomaly: if we study the anomaly of any
boundary theory of the $(d+1)$d higher SPTs living on a manifold with $d$d boundary.
Thus, we regard all of them as physically tightly-related given by \eq{eq:Zpart}.
By turning on the classical background probed field (denoted as ``{bgd.field}'' in \eq{eq:QFT-SPT}) coupled to $d$d QFT, under the symmetry transformation (i.e. symmetry twist),
its partition function $ \bZ^{\text{$d$d}}_{{\text{QFT}}}$
can be \emph{shifted} 
\bea \label{eq:QFT-SPT}
\left.
 \bZ^{\text{$d$d}}_{{\text{QFT}}}   \right|_{\text{bgd.field}=0} 
 \longrightarrow%\to  
 \left. \bZ^{\text{$d$d}}_{{\text{QFT}}}
   \right|_{\text{bgd.field}\neq 0} \cdot
   \bZ^{\text{$(d+1)$d}}_{{\text{SPTs}}}(\text{bgd.field}) , 
\eea
to detect the underlying {$(d+1)$d} topological terms/counter term/SPTs, namely the $(d+1)$d partition function  $\bZ^{\text{$(d+1)$d}}_{{\text{SPTs}}}$.
To check whether the underlying {$(d+1)$d} SPTs really specifies a true $d$d 't Hooft anomaly unremovable from $d$d \emph{counter term},
it means that $\bZ^{\text{$(d+1)$d}}_{{\text{SPTs}}}(\text{bgd.field})$ cannot be absorbed by a lower-dimensional SPTs 
$\bZ^{\text{$d$d}}_{{\text{SPTs}}}(\text{bgd.field})$, namely 
\bea 
 \left. \bZ^{\text{$d$d}}_{{\text{QFT}}}
   \right|_{\text{bgd.field}} \cdot
   \bZ^{\text{$(d+1)$d}}_{{\text{SPTs}}}(\text{bgd.field})
\neq % \centernot\longrightarrow
  \left. \bZ^{\text{$d$d}}_{{\text{QFT}}}
   \right|_{\text{bgd.field}} \cdot
      \bZ^{\text{$d$d}}_{{\text{SPTs}}}(\text{bgd.field}).
\eea
Readers can find related materials in \cite{Wan:2018zql, Tanizaki:2017qhf1710.08923}.

\subsection{The convention of notations}

\label{sec:plan-convention}

We explain the convention for our notations and terminology below. 
Most of our conventions follow \cite{Freed2016} and \cite{2017arXiv171111587GPW}.

\begin{itemize}

\item {We denote $\tO$ \cred{the stable} orthogonal group,
$\SO$ \cred{the stable} special orthogonal group,
$\Spin$ the \cred{stable} spin group, 
and $\Pin^{\pm}$  the two ways of $\Z_2$ extension (related to the time reversal symmetry) of $\Spin$ group.}
\item $\Z_n$ is the finite cyclic group of order $n$, \cred{$n$ is a positive integer}.
%\item $A \times_{\Z_2} B\equiv (A\times B)/\Z_2$ where both $A$ and $B$ have a central $\Z_2$ subgroup, the quotient is with respect to the diagonal central $\Z_2$ subgroup.
\item
A map between topological spaces is always assumed to be continuous.
\item For a (pointed) topological space $X$, $\Sigma$ denotes a suspension $\Sigma X=S^1\wedge X=(S^1\times X)/(S^1\vee X)$ where $\wedge$ and $\vee$ are smash product and wedge sum (one point union) of pointed topological spaces respectively. 
For a graded algebra $A$, %$\Sigma A$ is obtained from $A$ by shifting its degree by 1. 
\cred{$A=\bigoplus_i A_i$, $\Sigma A$ is the graded algebra defined by $\Sigma A=\bigoplus_i(\Sigma A)_i$ where $(\Sigma A)_i=A_{i-1}$.}
\item For a (pointed) topological space $X$ \cred{with the base point $x_0$}, $\Omega X$ is the loop space of $X$: 
\bea
\Omega X=\{\gamma:I\to X\text{ continuous}|\gamma(0)=\gamma(1)\cred{=x_0}\}.
\eea
%\item
%A spectrum \cred{$M_{\bullet}$} is a collection of (pointed) topological spaces $M_n$ together with structure maps $\Sigma M_n\to M_{n+1}$ such that the adjoints $M_n\to \Omega M_{n+1}$ of the structure maps are homeomorphisms.
\item
\cred{Let $R$ be a ring, $M$ a topological space,
$\H^*(M,R)$ is the cohomology ring of $M$ with coefficients in $R$.}
\item
We will abbreviate the cup product $x\cup y$ by $xy$.

% \item The term ``spin'' can have different meanings in our article, including the $\SU(2)$-spin rotation, or the spin group, or the spin manifold.
%\item  For condensed matter realization, fTI/fTSC stand for fermionic topological insulator/superconductor, bTI/bTSC stand for their bosonic counterparts.
%\item The fSPTs/bSPTs stand for fermionic/bosonic Symmetry Protected Topological states (SPTs) respectively. 
%\item Non-orientable and unorientable manifolds both mean that the manifolds \emph{cannot} be oriented.
%An unoriented manifold means that an orientation has not been chosen (that is, even though it might be possible to orient the manifold, the transition functions in the atlas do not necessarily preserve orientation).
%\item {$(-1)^{F}$ is the generator of $\Z_2^F$ fermionic parity, where $F$ is the fermion number (or $\hat{N}$)}.
%\item Mathematically $\Z_2^F$, $\Z_2$, $\Z/2$, $\Z/2\Z$, $\{\pm 1\}$ all mean the same, the cyclic group of order 2. Notation $\Z_2^F$ is used when we want to emphasize its physical meaning as fermionic parity symmetry. Notation $\{\pm 1\}$ is sometimes used to emphasize that is considered as  a multiplicative group.
%\item $M_d$ (or simply $M$) is a $d$-dimensional (possibly non-orientable) manifold.
\item \cred{If $M_d$ (or simply $M$) is a $d$-dimensional manifold, then $TM_d$ (or simply $TM$) is the tangent bundle over $M_d$ (or $M$).}

\item Rank $r$ real (complex) vector bundle $V$ is a bundle with fibers being real (complex) vector spaces of real (complex) dimension $r$.
\item $w_i(V)$ is the $i$-th Stiefel-Whitney class of a  real vector bundle $V$ (which may be also complex rank $r$ but considered as real rank $2r$). 
%We will use the notation $w_i$ for the $i$-th Stiefel-Whitney class of the virtual bundle (of dimension 0) over $\B \tO$ which is the colimit of $E_n-n$ and $E_n$ is the universal $n$-bundle over $\B \tO(n)$.
%Note that the pullback of the virtual bundle (of dimension 0) over $\B \tO$ along the map $g:M\to \B \tO$ is just $TM-d$ where $M$ is a $d$-dimensional manifold and $TM$ is the tangent bundle of $M$, $g$ is given by the $\tO$-structure on $M$.
%We will use the notation $w_i'$ for the $i$-th Stiefel-Whitney class of the universal principal $\PSU(2)$-bundle over $\B \PSU(2)$.
\item $p_i(V)$ is the $i$-th Pontryagin class of a real vector bundle $V$. 
%We will use the notation $p_i$ for the $i$-th Pontryagin class of the virtual bundle (of dimension 0) over $\B \tO$, and $p_i'$ for the $i$-th Pontryagin class of the universal principal $\PSU(2)$-bundle over $\B \PSU(2)$.
\item $c_i(V)$ is the $i$-th Chern class of a complex vector bundle $V$. 
Pontryagin classes are closely related to Chern classes via complexification:
\bea
p_i(V)=(-1)^ic_{2i}(V\otimes_{\R}\C)
\eea
where $V\otimes_{\R}\C$ is the complexification of the real vector bundle $V$. 
The relation between Pontryagin classes and Stiefel-Whitney classes is 
\bea
p_i(V)=w_{2i}(V)^2\mod2.
\eea
%We will use the notation $c_i$ for the $i$-th Chern class of the universal principal $\PSU(3)$-bundle over $\B \PSU(3)$.
\item
%For a top degree cohomology class with coefficients $\Z_2$ we often suppress explicit integration over the manifold (i.e. pairing with the fundamental class $[M]$ with coefficients $\Z_2$), for example: 
%$w_2(TM)w_3(TM)\equiv\int_Mw_2(TM)w_3(TM)$ where $M$ is a 5-manifold.
\cred{For a top degree cohomology class 
 we often suppress explicit integration over the manifold (i.e. pairing with the fundamental class $[M]$).
 If $M$ is orientable, then $[M]$ has coefficients in $\Z$.
 If $M$ is non-orientable, then $[M]$ has coefficients in $\Z_2$.
}

\item If $x$ is an element of a graded vector space, $|x|$ denotes the degree of $x$.
\item For an odd prime $p$ and a non-negatively and integrally graded vector space $V$ over $\Z_p$, let $V^{\text{even}}$ and $V^{\text{odd}}$ be even and odd graded parts of $V$ . The free algebra $F_{\Z_p}[V]$ generated by the graded vector space $V$ is the tensor product of the polynomial algebra on $V^{\text{even}}$ and the exterior algebra on $V^{\text{odd}}$:
\bea
F_{\Z_p}[V]=\Z_p[V^{\text{even}}]\otimes\Lambda_{\Z_p}(V^{\text{odd}}).
\eea
We sometimes replace the vector space with a set of bases of it.
\item $\A_p$ denotes the mod $p$ Steenrod algebra where $p$ is a prime.
\item $\Sq^n$ is the $n$-th Steenrod square, it is an element of $\A_2$.
%The Steenrod squares were introduced by Steenrod in \cite{Steenrod1947}, they are defined using higher cup product as 
%$\Sq^kx_n=x_n\hcup{n-k}x_n$. Later, Cartan \cite{Cartan1955} defined the Steenrod algebra $\A_p$ to be the algebra of stable cohomology operations for mod 
%$p$ cohomology. In particular, $\A_2$ is generated by the Steenrod squares.
\item $\A_2(1)$ denotes the subalgebra of $\A_2$ generated by $\Sq^1$ and $\Sq^2$.
\item $\beta_{(n,m)}:\H^*(-,\Z_{m})\to\H^{*+1}(-,\Z_{n})$ is the Bockstein homomorphism associated to the extension $\Z_n\stackrel{\cdot m}{\to}\Z_{nm}\to\Z_m$, when $n=m=p$ is a prime, it is an element of $\A_p$. If $p=2$, then $\beta_{(2,2)}=\Sq^1$. 
%\item
%$\beta_n:\H^*(-,\Z_n)\to\H^{*+1}(-,\Z)$ is the Bockstein homomorphism associated to the extension $\Z\stackrel{\cdot n}{\to}\Z\to\Z_n$.
\item $P^n_p:\H^*(-,\Z_p)\to\H^{*+2n(p-1)}(-,\Z_p)$ is the $n$-th Steenrod power, it is an element of $\A_p$ where $p$ is an odd prime. For odd primes $p$, we only consider $p=3$, so we abbreviate $P^n_3$ by $P^n$.

\item $\mathcal{P}_2$ is the Pontryagin square operation $\H^{2i}(M,\Z_{2^k})\to \H^{4i}(M,\Z_{2^{k+1}})$.
% It was introduced by Pontryagin in \cite{Pontryagin1942}. 
Explicitly,
$\mathcal{P}_2$
is given by
\bea
\mathcal{P}_2(x)=x\cup x+x\hcup{1}\delta x\mod2^{k+1}
\eea
and it satisfies
\bea
\mathcal{P}_2(x)=x\cup x\mod2^k.
\eea
Here $\hcup{1}$ is the higher cup product.
%which was studied by Steenrod in \cite{Steenrod1947}.
\item 
Postnikov square
$\mathfrak{P}_3:\H^2(-,\Z_{3^k})\to\H^5(-,\Z_{3^{k+1}})$
is given by 
\bea
\mathfrak{P}_3(u)=\beta_{(3^{k+1},3^k)}(u\cup u)
\eea
where $\beta_{(3^{k+1},3^k)}$ is the Bockstein homomorphism associated to $0\to\Z_{3^{k+1}}\to\Z_{3^{2k+1}}\to\Z_{3^k}\to0$.

\item For a finitely generated abelian group $G$ and a prime $p$, $G_p^{\wedge}=\lim_nG/p^nG$ is the $p$-completion of $G$.
\item \cred{For a topological space $M$, $\pi_d(M)$ is the $d$-th (ordinary) homotopy group of $M$.
%For a spectrum $M_{\bullet}$, $\pi_d(M_{\bullet})$ is the $d$-th stable homotopy group of $M_{\bullet}$: 
%\bea
%\pi_d(M_{\bullet})=\text{colim}_{k\to\infty}\pi_{d+k}M_k.
%\eea
}
%The colimit above can be understood as a limiting group in the sequence $\pi_dM_0\to\pi_{d+1}M_1\to\pi_{d+2}M_2\to\cdots$.
\item For an abelian group $G$, the Eilenberg-MacLane space $K(G,n)$ is a space with homotopy groups satisfying
\bea
\pi_iK(G,n)=\left\{\begin{array}{ll}G,&i=n.\\0,&i\ne n.\end{array}\right.
\eea
%The Eilenberg-MacLane spectrum $HG$ is the spectrum whose $n$-th space is $K(G,n)$.
\item Let $X$, $Y$ be topological spaces, $[X,Y]$ is the set of homotopy classes of maps from $X$ to $Y$.
\item Let $G$ be a group, the classifying space of $G$, $\B G$ is a topological space such that
\bea
[X,\B G]=\{\text{isomorphism classes of principal }G\text{-bundles over }X\}
\eea
for any topological space $X$. In particular, if $G$ is an abelian group, then $\B G$ is a group.

\item There is a vector bundle associated to a principal $G$-bundle $P_G$: $P_G\times_GV=(P_G\times V)/G$ which is the quotient of $P_G\times V$ by the right $G$-action
\bea
(p,v)g=(pg,g^{-1}v)
\eea
where $V$ is the vector space which $G$ acts on.
For characteristic classes of a principal $G$-bundle, we mean the characteristic classes of the associated vector bundle.

\end{itemize}

\subsection{Tables and Summary of Some Co/Bordism Groups}

\cred{Below we use the following notations, all cohomology class are pulled back to the $d$-manifold $M$ along the maps given in the definition of cobordism groups:\\
$\bullet$ $w_i$ is the Stiefel-Whitney class of the tangent bundle of $M$,\\
$\bullet$ $a$ is the generator of $\H^1(\B\Z_2,\Z_2)$,\\
$\bullet$ $a'$ is the generator of $\H^1(\B\Z_3,\Z_3)$, $b'=\beta_{(3,3)}a',$\\
$\bullet$ $x_2$ is the generator of $\H^2(\B^2\Z_2,\Z_2)$, $x_3=\Sq^1x_2$, $x_5=\Sq^2x_3$,\\
$\bullet$ $x_2'$ is the generator of $\H^2(\B^2\Z_3,\Z_3)$, $x_3'=\beta_{(3,3)}x_2'$,\\
$\bullet$ $w_i'=w_i(\PSU(2))\in\H^i(\B\PSU(2),\Z_2)$ is the Stiefel-Whitney class of the principal $\PSU(2)$ bundle,\\
$\bullet$ $c_i=c_i(\PSU(3))\in\H^{2i}(\B\PSU(3),\Z)$ is the Chern class of the principal $\PSU(3)$ bundle,\\
$\bullet$ $z_2=w_2(\PSU(3))\in\H^2(\B\PSU(3),\Z_3)$ is the generalized Stiefel-Whitney class of the principal $\PSU(3)$ bundle, $z_3=\beta_{(3,3)}z_2$.\\
$\bullet$ $\cP_2$ is the Pontryagin square (see \ref{sec:plan-convention}).\\
$\bullet$ $\mathfrak{P}_3$ is the Postnikov square (see \ref{sec:plan-convention}).\\
Conventions:
All product between cohomology classes are cup product, product between a cohomology class $x$ and $\tilde{\eta}$ (or Arf, ABK, etc) means the value of $\tilde{\eta}$ (or Arf, ABK, etc) on the submanifold of $M$ which represents the Poincar\'e dual of $x$.
}

%%%%%%%%%%
%%%%%%%%%%

%%%%%%%%%%
%%%%%%%%%%

%%%%%%%%%%
%%%%%%%%%%
\begin{table}[!h] %[!h] %[tb]
\centering
 \makebox[\textwidth][r]{
 \begin{tabular}{ |c| c | c|  c| c | c| c| c| c| c|}
\hline
$\Omega_d^H(-)$ & $\B ^2\Z_2$ &  $\B^2\Z_3$ & $\B\PSU(2)$  &   $\B\PSU(3)$ & 
\begin{minipage}[c]{.7in} 
$\B\Z_2\times \B^2\Z_2$
\end{minipage}
& 
\begin{minipage}[c]{.7in} 
$\B\Z_3\times \B^2\Z_3$
\end{minipage}
&  
\begin{minipage}[c]{.7in} 
$\B\PSU(2)\times \B^2\Z_2$
\end{minipage}
&  
\begin{minipage}[c]{.7in} 
$\B\PSU(3)\times \B^2\Z_3$
\end{minipage}
\\
\hline
$2$ SO& 
\begin{minipage}[c]{.5in} 
$\Z_2:$\\
$x_2$
\end{minipage}
&
\begin{minipage}[c]{.5in} 
$\Z_3:$\\
$x_2'$
\end{minipage}
&
\begin{minipage}[c]{.5in} 
$\Z_2:$\\
$w_2'$
\end{minipage}
&
\begin{minipage}[c]{.5in} 
$\Z_3:$\\
$z_2$
\end{minipage}
&
\begin{minipage}[c]{.7in} 
$\Z_2:$\\
$x_2$
\end{minipage}
&
\begin{minipage}[c]{.7in} 
$\Z_3:$\\
$x_2'$
\end{minipage}
&
\begin{minipage}[c]{.7in} 
$\Z_2^2:$\\
$w_2',x_2$
\end{minipage}
&
\begin{minipage}[c]{.7in} 
$\Z_3^2:$\\
$x_2',z_2$
\end{minipage}
\\
\hline
$2$ Spin&
\begin{minipage}[c]{.5in} 
$\Z_2^2:$\\
$x_2,\text{Arf}\tablefootnote{\cred{Arf is the Arf invariant of Spin 2-manifolds.}}$
\end{minipage}
&
\begin{minipage}[c]{.5in} 
$\Z_2\times\Z_3:$\\
$\text{Arf},x_2'$
\end{minipage}
&
\begin{minipage}[c]{.5in} 
$\Z_2^2:$\\
$w_2',\text{Arf}$
\end{minipage}
&
\begin{minipage}[c]{.5in} 
$\Z_2\times\Z_3:$\\
$\text{Arf},z_2$
\end{minipage}
&
\begin{minipage}[c]{.7in} 
$\Z_2^3:$\\
$x_2,\text{Arf},$\\
$a{\tilde{\eta}}\tablefootnote{
$\tilde{\eta}$ is the ``mod 2
index'' of the 1d Dirac operator (\#zero eigenvalues mod 2, no
contribution from spectral asymmetry).}$
\end{minipage}
&
\begin{minipage}[c]{.7in} 
$\Z_2\times\Z_3:$\\
$\text{Arf},x_2'$
\end{minipage}
&
\begin{minipage}[c]{.7in} 
$\Z_2^3:$\\
$w_2',x_2,$\\
$\text{Arf}$
\end{minipage}
&
\begin{minipage}[c]{.7in} 
$\Z_2\times\Z_3^2:$\\
$\text{Arf},x_2',$\\
$z_2$
\end{minipage}
\\
\hline
$2$ O&
\begin{minipage}[c]{.5in} 
$\Z_2^2:$\\
$x_2,w_1^2$
\end{minipage}
&
\begin{minipage}[c]{.5in} 
$\Z_2:$\\
$w_1^2$
\end{minipage}
&
\begin{minipage}[c]{.5in} 
$\Z_2^2:$\\
$w_2',w_1^2$
\end{minipage}
&
\begin{minipage}[c]{.5in} 
$\Z_2:$\\
$w_1^2$
\end{minipage}
&
\begin{minipage}[c]{.7in} 
$\Z_2^3:$\\
$a^2,x_2,$\\
$w_1^2$
\end{minipage}
&
\begin{minipage}[c]{.7in} 
$\Z_2:$\\
$w_1^2$
\end{minipage}
&
\begin{minipage}[c]{.7in} 
$\Z_2^3:$\\
$w_2',x_2,$\\
$w_1^2$
\end{minipage}
&
\begin{minipage}[c]{.7in} 
$\Z_2:$\\
$w_1^2$
\end{minipage}
\\
\hline
$2$ Pin$^+$&
\begin{minipage}[c]{.5in} 
$\Z_2^2:$\\
$x_2,{w_1\tilde{\eta}}$
\end{minipage}
&
\begin{minipage}[c]{.5in} 
$\Z_2:$\\
${w_1\tilde{\eta}}$
\end{minipage}
&
\begin{minipage}[c]{.5in} 
$\Z_2^2:$\\
$w_2',{w_1\tilde{\eta}}$
\end{minipage}
&
\begin{minipage}[c]{.5in} 
$\Z_2:$\\
${w_1\tilde{\eta}}$
\end{minipage}
&
\begin{minipage}[c]{.7in} 
$\Z_2^3:$\\
$w_1a=a^2,x_2,$\\
${w_1\tilde{\eta}}$
\end{minipage}
&
\begin{minipage}[c]{.7in} 
$\Z_2:$\\
${w_1\tilde{\eta}}$
\end{minipage}
&
\begin{minipage}[c]{.7in} 
$\Z_2^3:$\\
$w_2',x_2,$\\
${w_1\tilde{\eta}}$
\end{minipage}
&
\begin{minipage}[c]{.7in} 
$\Z_2:$\\
${w_1\tilde{\eta}}$
\end{minipage}
\\
\hline
$2$ Pin$^-$&
\begin{minipage}[c]{.5in} 
$\Z_2\times\Z_8:$\\
$x_2,\text{ABK}\tablefootnote{\cred{ABK is the Arf-Brown-Kervaire invariant of $\Pin^-$ 2-manifolds.}}$
\end{minipage}
&
\begin{minipage}[c]{.5in} 
$\Z_8:$\\
$\text{ABK}$
\end{minipage}
&
\begin{minipage}[c]{.5in} 
$\Z_2\times\Z_8:$\\
$w_2',\text{ABK}$
\end{minipage}
&
\begin{minipage}[c]{.5in} 
$\Z_8:$\\
$\text{ABK}$
\end{minipage}
&
\begin{minipage}[c]{.7in} 
$\Z_2\times\Z_4\times\Z_8:$\\
$x_2,{q(a)}\tablefootnote{\cred{
Any 2-manifold $\Sigma$ always admits a $\Pin^-$ structure. $\Pin^-$ structures are in one-to-one correspondence with quadratic enhancement
\begin{equation}
 {q}:\H^1(\Sigma,\Z_2) \rightarrow \Z_4
\end{equation} 
such that
\begin{equation}
{q}(x+y)-{q}(x)-{q}(y) = 2\int_\Sigma x\cup y\ \mod 4.
\end{equation} 
In particular:
\begin{equation}
{q}(x)=\int_\Sigma x\cup x \mod 2.
\end{equation} 
}
},$\\
$\text{ABK}$
\end{minipage}
&
\begin{minipage}[c]{.7in} 
$\Z_8:$\\
$\text{ABK}$
\end{minipage}
&
\begin{minipage}[c]{.7in} 
$\Z_2^2\times\Z_8:$\\
$w_2',x_2,$\\
$\text{ABK}$
\end{minipage}
&
\begin{minipage}[c]{.7in} 
$\Z_8:$\\
$\text{ABK}$
\end{minipage}
\\
\hline
 \end{tabular}
 }\hspace*{-10mm}
\caption{$2d$ bordism groups.}
 \label{2d bordism groups}
\end{table}

\begin{table}[!h] %[!h] %[tb]
\centering
 \makebox[\textwidth][r]{
 \begin{tabular}{ |c| c | c|  c| c | c| c| c| c| c|}
\hline
$\Omega_d^H(-)$ & $\B^2\Z_2$ &  $\B^2\Z_3$ & $\B\PSU(2)$  &   $\B\PSU(3)$ & 
\begin{minipage}[c]{.7in} 
$\B\Z_2\times \B^2\Z_2$
\end{minipage}
& 
\begin{minipage}[c]{.7in} 
$\B\Z_3\times \B^2\Z_3$
\end{minipage}
&  
\begin{minipage}[c]{.7in} 
$\B\PSU(2)\times \B^2\Z_2$
\end{minipage}
&  
\begin{minipage}[c]{.7in} 
$\B\PSU(3)\times \B^2\Z_3$
\end{minipage}
\\
\hline
$3$ SO& 
\begin{minipage}[c]{.5in} 
$0$
\end{minipage}
&
\begin{minipage}[c]{.5in} 
$0$
\end{minipage}
&
\begin{minipage}[c]{.5in} 
$0$
\end{minipage}
&
\begin{minipage}[c]{.5in} 
$0$
\end{minipage}
&
\begin{minipage}[c]{.7in} 
$\Z_2^2:$\\
$ax_2,a^3$
\end{minipage}
&
\begin{minipage}[c]{.7in} 
$\Z_3^2:$\\
$a'b',a'x_2'$
\end{minipage}
&
\begin{minipage}[c]{.7in} 
$0$
\end{minipage}
&
\begin{minipage}[c]{.7in} 
$0$
\end{minipage}
\\
\hline
$3$ Spin&
\begin{minipage}[c]{.5in} 
$0$
\end{minipage}
&
\begin{minipage}[c]{.5in} 
$0$
\end{minipage}
&
\begin{minipage}[c]{.5in} 
$0$
\end{minipage}
&
\begin{minipage}[c]{.5in} 
$0$
\end{minipage}
&
\begin{minipage}[c]{.78in} 
$\Z_2\times\Z_8:$\\
$ax_2,a\text{ABK}$
\end{minipage}
&
\begin{minipage}[c]{.7in} 
$\Z_3^2:$\\
$a'b',a'x_2'$
\end{minipage}
&
\begin{minipage}[c]{.7in} 
$0$
\end{minipage}
&
\begin{minipage}[c]{.7in} 
$0$
\end{minipage}
\\
\hline
$3$ O&
\begin{minipage}[c]{.5in} 
$\Z_2:$\\
$x_3=w_1x_2$
\end{minipage}
&
\begin{minipage}[c]{.5in} 
$0$
\end{minipage}
&
\begin{minipage}[c]{.5in} 
$\Z_2:$\\
$w_3'=w_1w_2'$
\end{minipage}
&
\begin{minipage}[c]{.5in} 
$0$
\end{minipage}
&
\begin{minipage}[c]{.7in} 
$\Z_2^4:$\\
$x_3=w_1x_2,ax_2,$\\
$aw_1^2,a^3$
\end{minipage}
&
\begin{minipage}[c]{.7in} 
$0$
\end{minipage}
&
\begin{minipage}[c]{.7in} 
$\Z_2^2:$\\
$x_3=w_1x_2,w_3'=w_1w_2'$
\end{minipage}
&
\begin{minipage}[c]{.7in} 
$0$
\end{minipage}
\\
\hline
$3$ Pin$^+$&
\begin{minipage}[c]{.5in} 
$\Z_2^2:$\\
$w_1x_2=x_3,$\\
$w_1\text{Arf}$
\end{minipage}
&
\begin{minipage}[c]{.5in} 
$\Z_2:$\\
$w_1\text{Arf}$
\end{minipage}
&
\begin{minipage}[c]{.5in} 
$\Z_2^2:$\\
$w_1w_2'=w_3',$\\
$w_1\text{Arf}$
\end{minipage}
&
\begin{minipage}[c]{.5in} 
$\Z_2:$\\
$w_1\text{Arf}$
\end{minipage}
&
\begin{minipage}[c]{.7in} 
$\Z_2^5:$\\
$a^3,w_1x_2=x_3,$\\
$ax_2,w_1a{\tilde{\eta}},$\\
$w_1\text{Arf}$
\end{minipage}
&
\begin{minipage}[c]{.7in} 
$\Z_2:$\\
$w_1\text{Arf}$
\end{minipage}
&
\begin{minipage}[c]{.7in} 
$\Z_2^3:$\\
$w_1w_2'=w_3',w_1x_2=x_3,$\\
$w_1\text{Arf}$
\end{minipage}
&
\begin{minipage}[c]{.7in} 
$\Z_2:$\\
$w_1\text{Arf}$
\end{minipage}
\\
\hline
$3$ Pin$^-$&
\begin{minipage}[c]{.5in} 
$\Z_2:$\\
$w_1x_2$\\
$=x_3$
\end{minipage}
&
\begin{minipage}[c]{.5in} 
$0$
\end{minipage}
&
\begin{minipage}[c]{.5in} 
$\Z_2:$\\
$w_1w_2'$\\
$=w_3'$
\end{minipage}
&
\begin{minipage}[c]{.5in} 
$0$
\end{minipage}
&
\begin{minipage}[c]{.7in} 
$\Z_2^4:$\\
$a^3,w_1^2a,$\\
$x_3=$\\
$w_1x_2$,\\ 
$ax_2$
\end{minipage}
&
\begin{minipage}[c]{.7in} 
$0$
\end{minipage}
&
\begin{minipage}[c]{.7in}
$\Z_2^2:$\\
$w_1w_2'$$=w_3',$\\
$w_1x_2=x_3$ 
\end{minipage}
&
\begin{minipage}[c]{.7in} 
$0$
\end{minipage}
\\
\hline
 \end{tabular}
 }\hspace*{-10mm}
\caption{$3d$ bordism groups.}
 \label{3d bordism groups}
\end{table}
%%%%%%%%%%
%%%%%%%%%%

%%%%%%%%%%
%%%%%%%%%%
\begin{table}[!h] %[!h] %[tb]
\centering
 \makebox[\textwidth][r]{
 \begin{tabular}{ |c| c | c|  c| c | c| c| c| c| c|}
\hline
$\Omega_d^H(-)$ & $\B^2\Z_2$ &  $\B^2\Z_3$ & $\B\PSU(2)$  &   $\B\PSU(3)$ & 
\begin{minipage}[c]{.7in} 
$\B\Z_2\times \B^2\Z_2$
\end{minipage}
& 
\begin{minipage}[c]{.7in} 
$\B\Z_3\times \B^2\Z_3$
\end{minipage}
&  
\begin{minipage}[c]{.7in} 
$\B\PSU(2)\times \B^2\Z_2$
\end{minipage}
&  
\begin{minipage}[c]{.7in} 
$\B\PSU(3)\times \B^2\Z_3$
\end{minipage}
\\
\hline
$4$ SO& 
\begin{minipage}[c]{.5in} 
$\Z\times\Z_4$:\\
${\sigma},{\mathcal{P}_2(x_2)}$
\end{minipage}
&
\begin{minipage}[c]{.5in} 
$\Z\times\Z_3$:\\
${\sigma},x_2'^2$
\end{minipage}
&
\begin{minipage}[c]{.5in} 
$\Z^2$:\\
${\sigma},p_1'$
\end{minipage}
&
\begin{minipage}[c]{.5in} 
$\Z^2$:\\
${\sigma},c_2$
\end{minipage}
&
\begin{minipage}[c]{.7in} 
$\Z\times\Z_2\times\Z_4$:\\
${\sigma}$,\\
${ax_3=a^2x_2}$,\\
${\mathcal{P}_2(x_2)}$
\end{minipage}
&
\begin{minipage}[c]{.7in} 
$\Z\times\Z_3^2$:\\
${\sigma}$,\\
${a'x_3'=b'x_2'}$,\\
$x_2'^2$
\end{minipage}
&
\begin{minipage}[c]{.75in} 
$\Z^2\times\Z_2\times\Z_4$:\\
${\sigma},p_1'$,\\
$w_2'x_2,{\mathcal{P}_2(x_2)}$
\end{minipage}
&
\begin{minipage}[c]{.7in} 
$\Z^2\times\Z_3^2$:\\
${\sigma},c_2$,\\
$x_2'^2,x_2'z_2$
\end{minipage}
\\
\hline
$4$ Spin&
\begin{minipage}[c]{.55in} 
$\Z\times\Z_2$:\\
$\frac{\sigma}{16}$,\\
${\frac{\mathcal{P}_2(x_2)}{2}}$
\end{minipage}
&
\begin{minipage}[c]{.5in} 
$\Z\times\Z_3$:\\
$\frac{\sigma}{16},x_2'^2$
\end{minipage}
&
\begin{minipage}[c]{.5in} 
$\Z^2$:\\
$\frac{\sigma}{16},\frac{p_1'}{2}$
\end{minipage}
&
\begin{minipage}[c]{.5in} 
$\Z^2$:\\
$\frac{\sigma}{16},c_2$
\end{minipage}
&
\begin{minipage}[c]{.7in} 
$\Z\times\Z_2^2$:\\
$\frac{\sigma}{16}$,\\
${ax_3=a^2x_2}$,\\
${\frac{\mathcal{P}_2(x_2)}{2}}$
\end{minipage}
&
\begin{minipage}[c]{.7in} 
$\Z\times\Z_3^2$:\\
$\frac{\sigma}{16}$,\\
${a'x_3'=b'x_2'}$,\\
$x_2'^2$
\end{minipage}
&
\begin{minipage}[c]{.7in} 
$\Z^2\times\Z_2^2$:\\
$\frac{\sigma}{16},\frac{p_1'}{2}$,\\
$w_2'x_2$,\\
${\frac{\mathcal{P}_2(x_2)}{2}}$
\end{minipage}
&
\begin{minipage}[c]{.7in} 
$\Z^2\times\Z_3^2$:\\
$\frac{\sigma}{16},c_2$,\\
$x_2'^2,x_2'z_2$
\end{minipage}
\\
\hline
$4$ O&
\begin{minipage}[c]{.5in} 
$\Z_2^4$:\\ 
$x_2^2,w_1^4$,\\
$w_1^2x_2,w_2^2$
\end{minipage}
&
\begin{minipage}[c]{.5in} 
$\Z_2^2$:\\
$w_1^4$,\\
$w_2^2$
\end{minipage}
&
\begin{minipage}[c]{.5in} 
$\Z_2^4$:\\
$w_2'^2,w_1^4$,\\
$w_1^2w_2',w_2^2$
\end{minipage}
&
\begin{minipage}[c]{.65in} 
$\Z_2^3$:\\
$w_1^4,w_2^2$,\\
$c_2$(mod 2)
\end{minipage}
&
\begin{minipage}[c]{.7in} 
$\Z_2^8$:\\
$w_1^4,w_2^2$,\\
$a^4,a^2x_2$,\\
$ax_3,x_2^2$,\\
$w_1^2a^2,w_1^2x_2$
\end{minipage}
&
\begin{minipage}[c]{.7in} 
$\Z_2^2$:\\
$w_1^4,w_2^2$
\end{minipage}
&
\begin{minipage}[c]{.7in} 
$\Z_2^7$:\\
$w_1^4,w_2^2$,\\
$x_2^2,w_2'^2$,\\
$x_2w_1^2,w_2'w_1^2$,\\
$w_2'x_2$
\end{minipage}
&
\begin{minipage}[c]{.7in} 
$\Z_2^3$:\\
$w_1^4,w_2^2$,\\
$c_2$(mod 2)
\end{minipage}
\\
\hline
$4$ Pin$^+$&
\begin{minipage}[c]{.5in} 
$\Z_4\times\Z_{16}$:\\
${q_s(x_2)}\tablefootnote{
$q_s: \H^2(M,\Z_2) \to \Z_4$ is a $\Z_4$ valued quadratic refinement (dependent on
the choice of $\Pin^+$ structure $s\in \Pin^+(M)$) of the intersection form
$$\langle,\rangle: \H^2(M,\Z_2) \times \H^2(M,\Z_2) \to \Z_2$$
i.e. so that $q_s(x+y)-q_s(x)-q_s(y) = 2\langle x,y\rangle  \in \Z_4$ (in particular
$q_s(x) = \langle x,x\rangle \mod 2$)

The space of $\Pin^+$ structures is acted upon freely and transitively
by $\H^1(M,\Z_2)$, and the dependence of $q_s$ on the $\Pin^+$ structure should
satisfy

$$q_{s+h}(x)-q_s(x)=2w_1(TM)hx,\text{ for any }h \in \H^1(M,\Z_2)$$
(note that any two quadratic functions differ by a linear function)

If $w_1(TM)=0$, then $q_s(x)$ is independent on the $\Pin^+$ structure $s\in \Pin^+(M)$,
it reduces to $\mathcal{P}_2(x)$ where $\mathcal{P}_2(x)$ is the Pontryagin square of $x$.
},$\\
$\eta\tablefootnote{Here $\eta$ is the usual Atiyah-Patodi-Singer eta-invariant of the 4d Dirac operator (=``\#zero eigenvalues + spectral asymmetry'').}$
\end{minipage}
&
\begin{minipage}[c]{.5in} 
$\Z_{16}$:\\
$\eta$
\end{minipage}
&
\begin{minipage}[c]{.5in} 
$\Z_4\times\Z_{16}$:\\
${q_s(w_2')}\tablefootnote{
one can also define this $\Z_4$ invariant as
$$(\eta_{\SO(3)} - 3\eta)/4 \in \Z_4 \;\;\;(*)$$
where $\eta \in \Z_{16}$ is the (properly normalized) eta-invariant of the
ordinary Dirac operator, and $\eta_{\SO(3)} \in \Z_{16}$ is the eta
invariant of the twisted Dirac operator acting on the $S \otimes V_3$
where $S$ is the spinor bundle and $V_3$ is the bundle associated to 3-dim
representation of $\SO(3)$. 
Note that $(*)$ is well defined because $\eta_{\SO(3)} = 3\eta  \mod 4$.

{Note that on non-orientable manifold, if $w_2(V_3)=0$, then since $w_1(V_3)=0$, we also have $w_3(V_3)=0$, hence $V_3$ is stably trivial, $\eta_{\SO(3)}=3\eta$.}

Also note that on oriented manifold one can use Atiyah-Patodi-Singer index theorem to show that (here the normalization of eta-invariants is such that $\eta$ is an integer mod 16 on a general non-oriented $4$-manifold)
$$\eta=-\frac{\sigma(M)}{2},$$
$$\eta_{\SO(3)}=-\frac{3\sigma(M)}{2}+4p_1(\SO(3)).$$
So
$$(\eta_{\SO(3)} - 3\eta)/4=p_1(\SO(3))\mod4=\mathcal{P}_2(w_2(\SO(3))).$$
$q_s(w_2(\SO(3)))$ also reduces to $\mathcal{P}_2(w_2(\SO(3)))$ in the oriented case.
},$\\
$\eta$
\end{minipage}
&
\begin{minipage}[c]{.7in} 
$\Z_2\times\Z_{16}$:\\
$c_2${(mod 2)},\\
$\eta$
\end{minipage}
&
\begin{minipage}[c]{.78in} 
$\Z_2^2\times\Z_4\times\Z_8\times\Z_{16}$:\\
$ax_3,w_1ax_2=$\\
${a^2x_2+ax_3}$,\\
${q_s(x_2)},$\\
$w_1a {\text{ABK}}$,\\
$\eta$
\end{minipage}
&
\begin{minipage}[c]{.7in} 
$\Z_{16}$:\\
$\eta$
\end{minipage}
&
\begin{minipage}[c]{.7in} 
$\Z_4^2\times\Z_{16}\times\Z_2$:\\
${q_s(w_2')},$\\
${q_s(x_2)}$,\\
$\eta,w_2'x_2$
\end{minipage}
&
\begin{minipage}[c]{.7in} 
$\Z_2\times\Z_{16}$:\\
$c_2${(mod 2)},\\
$\eta$
\end{minipage}
\\
\hline
$4$ Pin$^-$&
\begin{minipage}[c]{.5in} 
$\Z_2$:\\
$w_1^2x_2$
\end{minipage}
&
\begin{minipage}[c]{.5in} 
$0$
\end{minipage}
&
\begin{minipage}[c]{.5in} 
$\Z_2$:\\
$w_1^2w_2'$
\end{minipage}
&
\begin{minipage}[c]{.65in} 
$\Z_2$:\\
$c_2${(mod $2$)}
\end{minipage}
&
\begin{minipage}[c]{.8in} 
$\Z_2^3$:\\
$w_1^2x_2,ax_3$,\\
$w_1ax_2=$\\
$a^2x_2+ax_3$
\end{minipage}
&
\begin{minipage}[c]{.7in} 
$0$
\end{minipage}
&
\begin{minipage}[c]{.7in} 
$\Z_2^3$:\\
$w_1^2w_2'$,\\
$w_1^2x_2,w_2'x_2$
\end{minipage}
&
\begin{minipage}[c]{.7in} 
$\Z_2$:\\
$c_2${(mod 2)}
\end{minipage}
\\
\hline
 \end{tabular}
 }\hspace*{-15mm}
\caption{$4d$ bordism groups.}
 \label{4d bordism groups}
\end{table}

%%%%%%%%%%
%%%%%%%%%%
\begin{table}[!h] %[!h] %[tb]
\centering
 \makebox[\textwidth][r]{
 \begin{tabular}{ |c| c | c|  c| c | c| c| c| c| c|}
\hline
$\Omega_d^H(-)$ & $\B^2\Z_2$ &  $\B^2\Z_3$ & $\B\PSU(2)$  &   $\B\PSU(3)$ & 
\begin{minipage}[c]{.7in} 
$\B\Z_2\times \B^2\Z_2$
\end{minipage}
& 
\begin{minipage}[c]{.7in} 
$\B\Z_3\times \B^2\Z_3$
\end{minipage}
&  
\begin{minipage}[c]{.7in} 
$\B\PSU(2)\times \B^2\Z_2$
\end{minipage}
&  
\begin{minipage}[c]{.7in} 
$\B\PSU(3)\times \B^2\Z_3$
\end{minipage}
\\
\hline
$5$ SO&
\begin{minipage}[c]{.5in} 
$\Z_2^2$:\\
${x_5}=x_2x_3$,\\
$w_2w_3$
\end{minipage}
&
\begin{minipage}[c]{.5in} 
$\Z_2$:\\
$w_2w_3$
\end{minipage}
&
\begin{minipage}[c]{.5in} 
$\Z_2^2$:\\
$w_2w_3$,\\
$w_2'w_3'$
\end{minipage}
&
\begin{minipage}[c]{.5in} 
$\Z_2$:\\
$w_2w_3$
\end{minipage}
&
\begin{minipage}[c]{.7in} 
$\Z_2^6$:\\
$ax_2^2,a^5$,\\
${x_5}=x_2x_3,a^3x_2$,\\
$w_2w_3,aw_2^2$
\end{minipage}
&
\begin{minipage}[c]{.7in} 
$\Z_2\times\Z_3^2\times\Z_9$:\\
$w_2w_3$,\\
$a'b'x_2'$,\\
$a'x_2'^2$,\\
$\mathfrak{P}_3(b')$
\end{minipage}
&
\begin{minipage}[c]{.7in} 
$\Z_2^4$:\\
$w_2'w_3',{x_5}=x_2x_3$,\\
$w_3'x_2=$\\
$w_2'x_3$,\\
$w_2w_3$
\end{minipage}
&
\begin{minipage}[c]{.7in} 
$\Z_2\times\Z_3$:\\
$w_2w_3$,\\
$z_2x_3'=$\\
$-z_3x_2'$
\end{minipage}
\\
\hline
$5$ Spin&
\begin{minipage}[c]{.5in} 
$0$
\end{minipage}
&
\begin{minipage}[c]{.5in} 
$0$
\end{minipage}
&
\begin{minipage}[c]{.5in} 
$0$
\end{minipage}
&
\begin{minipage}[c]{.5in} 
$0$
\end{minipage}
&
\begin{minipage}[c]{.7in} 
$\Z_2$:\\
$a^3x_2$
\end{minipage}
&
\begin{minipage}[c]{.7in} 
$\Z_3^2\times\Z_9$:\\
$a'b'x_2'$,\\
$a'x_2'^2$,\\
$\mathfrak{P}_3(b')$
\end{minipage}
&
\begin{minipage}[c]{.7in} 
$\Z_2$:\\
$w_3'x_2$\\
$=w_2'x_3$
\end{minipage}
&
\begin{minipage}[c]{.7in} 
$\Z_3$:\\
$z_2x_3'$\\
$=-z_3x_2'$
\end{minipage}
\\
\hline
$5$ O&
\begin{minipage}[c]{.5in} 
$\Z_2^4$:\\ 
$x_2x_3$,\\
$x_5$,\\
$w_1^2x_3,$\\
$w_2w_3$
\end{minipage}
&
\begin{minipage}[c]{.5in} 
$\Z_2$: \\
$w_2w_3$
\end{minipage}
&
\begin{minipage}[c]{.5in} 
$\Z_2^3$: \\
$w_2w_3$,\\
$w_1^2w_3'$,\\
$w_2'w_3'$
\end{minipage}
& 
\begin{minipage}[c]{.5in} 
$\Z_2$:\\ 
$w_2w_3$
\end{minipage}
&
\begin{minipage}[c]{.7in} 
$\Z_2^{12}$: 
$a^5,a^2x_3$,\\
$a^3x_2,a^3w_1^2$,\\
$ax_2^2,aw_1^4$,\\
$ax_2w_1^2,aw_2^2$,\\
$x_2x_3,w_1^2x_3$,\\
$x_5,w_2w_3$
\end{minipage}
&
\begin{minipage}[c]{.7in} 
$\Z_2$:\\ 
$w_2w_3$
\end{minipage}
& 
\begin{minipage}[c]{.7in} 
$\Z_2^{8}$:\\ 
$w_2'w_3',x_2w_3'$,\\
$w_1^2w_3',w_2'x_3$,\\
$x_2x_3,w_1^2x_3$,\\
$x_5,w_2w_3$
\end{minipage}
&
\begin{minipage}[c]{.7in} 
$\Z_2$:\\
$w_2w_3$
\end{minipage}
\\
\hline
$5$ Pin$^+$&
\begin{minipage}[c]{.5in} 
$\Z_2^2$:\\
$x_2x_3$,\\
$w_1^2x_3$\\
${=x_5}$
\end{minipage}
&
\begin{minipage}[c]{.5in} 
$0$
\end{minipage}
&
\begin{minipage}[c]{.5in} 
$\Z_2$:\\
$w_1^2w_3'$\\
${=w_2'w_3'}$
\end{minipage}
&
\begin{minipage}[c]{.5in} 
$0$
\end{minipage}
&
\begin{minipage}[c]{.7in} 
$\Z_2^7$:\\
$w_1^4a$,\\
$a^5=$$w_1^2a^3$,\\
$w_1^2x_3{=x_5}$,\\
$x_2x_3$,\\
$w_1^2ax_2=$\\
$ax_2^2+a^2x_3$,\\
$w_1ax_3=$\\
$a^2x_3$,\\
$a^3x_2$
\end{minipage}
&
\begin{minipage}[c]{.7in} 
$0$
\end{minipage}
&
\begin{minipage}[c]{.73in} 
$\Z_2^5$:\\
$w_1^2w_3'$\\
${=w_2'w_3'}$,\\
$w_1^2x_3=$$x_5$,\\
$x_2x_3,w_3'x_2$,\\
$w_1w_2'x_2=$\\
$w_2'x_3+$$w_3'x_2$
\end{minipage}
&
\begin{minipage}[c]{.7in} 
$0$
\end{minipage}
\\
\hline
$5$ Pin$^-$&
\begin{minipage}[c]{.5in} 
$\Z_2$:\\
$x_2x_3$
\end{minipage}
&
\begin{minipage}[c]{.5in} 
$0$
\end{minipage}
&
\begin{minipage}[c]{.5in} 
$0$
\end{minipage}
&
\begin{minipage}[c]{.5in} 
$0$
\end{minipage}
&
\begin{minipage}[c]{.7in} 
$\Z_2^5$:\\
$w_1^2a^3,x_2x_3$,\\
$w_1^2ax_2$,\\
$w_1ax_3=$\\
$a^2x_3$,\\
$a^3x_2$
\end{minipage}
&
\begin{minipage}[c]{.7in} 
$0$
\end{minipage}
&
\begin{minipage}[c]{.73in} 
$\Z_2^3$:\\
$x_2x_3,w_3'x_2$,\\
$w_1w_2'x_2=$\\
$w_2'x_3+$$w_3'x_2$
\end{minipage}
&
\begin{minipage}[c]{.7in} 
$0$
\end{minipage}
\\
\hline
 \end{tabular}
 }\hspace*{-15mm}
\caption{$5d$ bordism groups.}
 \label{5d bordism groups}
\end{table}

%%%%%%%%%%
%%%%%%%%%%
\begin{table}[!h] %[!h] %[tb]
\centering
 \makebox[\textwidth][r]{
 \begin{tabular}{ |c| c | c|  c| c | c| c| c| c| c|}
\hline
$\TP_d(H\times-)$ & $\B\Z_2$ &  $\B\Z_3$ & $\PSU(2)$  &   $\PSU(3)$ & 
\begin{minipage}[c]{.7in} 
$\Z_2\times \B\Z_2$
\end{minipage}
& 
\begin{minipage}[c]{.7in} 
$\Z_3\times \B\Z_3$
\end{minipage}
&  
\begin{minipage}[c]{.7in} 
$\PSU(2)\times \B\Z_2$
\end{minipage}
&  
\begin{minipage}[c]{.7in} 
$\PSU(3)\times \B\Z_3$
\end{minipage}
\\
\hline
$2$ SO& 
\begin{minipage}[c]{.5in} 
$\Z_2:$\\
$x_2$
\end{minipage}
&
\begin{minipage}[c]{.5in} 
$\Z_3:$\\
$x_2'$
\end{minipage}
&
\begin{minipage}[c]{.5in} 
$\Z_2:$\\
$w_2'$
\end{minipage}
&
\begin{minipage}[c]{.5in} 
$\Z_3:$\\
$z_2$
\end{minipage}
&
\begin{minipage}[c]{.7in} 
$\Z_2:$\\
$x_2$
\end{minipage}
&
\begin{minipage}[c]{.7in} 
$\Z_3:$\\
$x_2'$
\end{minipage}
&
\begin{minipage}[c]{.7in} 
$\Z_2^2:$\\
$w_2'${,}$x_2$
\end{minipage}
&
\begin{minipage}[c]{.7in} 
$\Z_3^2:$\\
$x_2',z_2$
\end{minipage}
\\
\hline
$2$ Spin&
\begin{minipage}[c]{.5in} 
$\Z_2^2:$\\
$x_2,\text{Arf}$
\end{minipage}
&
\begin{minipage}[c]{.5in} 
$\Z_2\times\Z_3:$\\
$\text{Arf},x_2'$
\end{minipage}
&
\begin{minipage}[c]{.5in} 
$\Z_2^2:$\\
$w_2',\text{Arf}$
\end{minipage}
&
\begin{minipage}[c]{.5in} 
$\Z_2\times\Z_3:$\\
$\text{Arf},z_2$
\end{minipage}
&
\begin{minipage}[c]{.7in} 
$\Z_2^3:$\\
$x_2,\text{Arf},$\\
$a{\tilde{\eta}}$
\end{minipage}
&
\begin{minipage}[c]{.7in} 
$\Z_2\times\Z_3:$\\
$\text{Arf},x_2'$
\end{minipage}
&
\begin{minipage}[c]{.7in} 
$\Z_2^3:$\\
$w_2',x_2,$\\
$\text{Arf}$
\end{minipage}
&
\begin{minipage}[c]{.7in} 
$\Z_2\times\Z_3^2:$\\
$\text{Arf},x_2',$\\
$z_2$
\end{minipage}
\\
\hline
$2$ O&
\begin{minipage}[c]{.5in} 
$\Z_2^2:$\\
$x_2,w_1^2$
\end{minipage}
&
\begin{minipage}[c]{.5in} 
$\Z_2:$\\
$w_1^2$
\end{minipage}
&
\begin{minipage}[c]{.5in} 
$\Z_2^2:$\\
$w_2',w_1^2$
\end{minipage}
&
\begin{minipage}[c]{.5in} 
$\Z_2:$\\
$w_1^2$
\end{minipage}
&
\begin{minipage}[c]{.7in} 
$\Z_2^3:$\\
$a^2,x_2,$\\
$w_1^2$
\end{minipage}
&
\begin{minipage}[c]{.7in} 
$\Z_2:$\\
$w_1^2$
\end{minipage}
&
\begin{minipage}[c]{.7in} 
$\Z_2^3:$\\
$w_2',x_2,$\\
$w_1^2$
\end{minipage}
&
\begin{minipage}[c]{.7in} 
$\Z_2:$\\
$w_1^2$
\end{minipage}
\\
\hline
$2$ Pin$^+$&
\begin{minipage}[c]{.5in} 
$\Z_2^2:$\\
$x_2,{w_1\tilde{\eta}}$
\end{minipage}
&
\begin{minipage}[c]{.5in} 
$\Z_2:$\\
${w_1\tilde{\eta}}$
\end{minipage}
&
\begin{minipage}[c]{.5in} 
$\Z_2^2:$\\
$w_2',{w_1\tilde{\eta}}$
\end{minipage}
&
\begin{minipage}[c]{.5in} 
$\Z_2:$\\
${w_1\tilde{\eta}}$
\end{minipage}
&
\begin{minipage}[c]{.7in} 
$\Z_2^3:$\\
$w_1a=a^2,x_2,$\\
${w_1\tilde{\eta}}$
\end{minipage}
&
\begin{minipage}[c]{.7in} 
$\Z_2:$\\
${w_1\tilde{\eta}}$
\end{minipage}
&
\begin{minipage}[c]{.7in} 
$\Z_2^3:$\\
$w_2',x_2,$\\
${w_1\tilde{\eta}}$
\end{minipage}
&
\begin{minipage}[c]{.7in} 
$\Z_2:$\\
${w_1\tilde{\eta}}$
\end{minipage}
\\
\hline
$2$ Pin$^-$&
\begin{minipage}[c]{.5in} 
$\Z_2\times\Z_8:$\\
$x_2,\text{ABK}$
\end{minipage}
&
\begin{minipage}[c]{.5in} 
$\Z_8:$\\
$\text{ABK}$
\end{minipage}
&
\begin{minipage}[c]{.5in} 
$\Z_2\times\Z_8:$\\
$w_2',\text{ABK}$
\end{minipage}
&
\begin{minipage}[c]{.5in} 
$\Z_8:$\\
$\text{ABK}$
\end{minipage}
&
\begin{minipage}[c]{.7in} 
$\Z_2\times\Z_4\times\Z_8:$\\
$x_2,{q(a)},$\\
$\text{ABK}$
\end{minipage}
&
\begin{minipage}[c]{.7in} 
$\Z_8:$\\
$\text{ABK}$
\end{minipage}
&
\begin{minipage}[c]{.7in} 
$\Z_2^2\times\Z_8:$\\
$w_2',x_2,$\\
$\text{ABK}$
\end{minipage}
&
\begin{minipage}[c]{.7in} 
$\Z_8:$\\
$\text{ABK}$
\end{minipage}
\\
\hline
 \end{tabular}
 }\hspace*{-10mm}
\caption{$\TP_2$.}
 \label{TP_2}
\end{table}

\begin{table}[!h] %[!h] %[tb]
\centering
 \makebox[\textwidth][r]{
 \begin{tabular}{ |c| c | c|  c| c | c| c| c| c| c|}
\hline
$\TP_d(H\times-)$ & $\B\Z_2$ &  $\B\Z_3$ & $\PSU(2)$  &   $\PSU(3)$ & 
\begin{minipage}[c]{.7in} 
$\Z_2\times \B\Z_2$
\end{minipage}
& 
\begin{minipage}[c]{.7in} 
$\Z_3\times \B\Z_3$
\end{minipage}
&  
\begin{minipage}[c]{.7in} 
$\PSU(2)\times \B\Z_2$
\end{minipage}
&  
\begin{minipage}[c]{.7in} 
$\PSU(3)\times \B\Z_3$
\end{minipage}
\\
\hline
$3$ SO& 
\begin{minipage}[c]{.5in} 
$\Z:$\\
$\frac{1}{3}\text{CS}_3^{(TM)}$\tablefootnote{$\text{CS}_3(TM)\equiv\text{CS}_3^{(TM)}$ is the Chern-Simons 3-form of the tangent bundle.}
\end{minipage}
&
\begin{minipage}[c]{.5in} 
$\Z:$\\
$\frac{1}{3}\text{CS}_3^{(TM)}$
\end{minipage}
&
\begin{minipage}[c]{.7in} 
$\Z^2:$\\
$\frac{1}{3}\text{CS}_3^{(TM)}$,\\
$\text{CS}_3^{(\SO(3))}$\tablefootnote{$\text{CS}_3(\SO(3))\equiv\text{CS}_3^{(\SO(3))}$ is the Chern-Simons 3-form of the $\SO(3)$ gauge bundle.}
\end{minipage}
&
\begin{minipage}[c]{.7in} 
$\Z^2:$
$\frac{1}{3}\text{CS}_3^{(TM)}$,\\
$\text{CS}_3^{(\PSU(3))}$\tablefootnote{$\text{CS}_3(\PSU(3))\equiv\text{CS}_3^{(\PSU(3))}$ is the Chern-Simons 3-form of the $\PSU(3)$ gauge bundle.}
\end{minipage}
&
\begin{minipage}[c]{.7in} 
$\Z\times\Z_2^2:$\\
$\frac{1}{3}\text{CS}_3^{(TM)}$,\\
$ax_2,a^3$
\end{minipage}
&
\begin{minipage}[c]{.7in} 
$\Z\times\Z_3^2:$\\
$\frac{1}{3}\text{CS}_3^{(TM)}$,\\
$a'b',a'x_2'$
\end{minipage}
&
\begin{minipage}[c]{.7in} 
$\Z^2:$\\
$\frac{1}{3}\text{CS}_3^{(TM)}$,\\
$\text{CS}_3^{(\SO(3))}$
\end{minipage}
&
\begin{minipage}[c]{.7in} 
$\Z^2:$\\
$\frac{1}{3}\text{CS}_3^{(TM)}$,\\
$\text{CS}_3^{(\PSU(3))}$
\end{minipage}
\\
\hline
$3$ Spin&
\begin{minipage}[c]{.55in} 
$\Z:$\\
$\frac{1}{48}\text{CS}_3^{(TM)}$
\end{minipage}
&
\begin{minipage}[c]{.55in} 
$\Z:$\\
$\frac{1}{48}\text{CS}_3^{(TM)}$
\end{minipage}
&
\begin{minipage}[c]{.6in} 
$\Z^2:$\\
$\frac{1}{48}\text{CS}_3^{(TM)}$,\\
$\frac{1}{2}\text{CS}_3^{(\SO(3))}$
\end{minipage}
&
\begin{minipage}[c]{.6in} 
$\Z^2:$\\
$\frac{1}{48}\text{CS}_3^{(TM)}$,\\
$\text{CS}_3^{(\PSU(3))}$
\end{minipage}
&
\begin{minipage}[c]{.7in} 
$\Z\times\Z_2\times\Z_8:$\\
$\frac{1}{48}\text{CS}_3^{(TM)}$,\\
$ax_2,a\text{ABK}$
\end{minipage}
&
\begin{minipage}[c]{.7in} 
$\Z\times\Z_3^2:$\\
$\frac{1}{48}\text{CS}_3^{(TM)}$,\\
$a'b',a'x_2'$
\end{minipage}
&
\begin{minipage}[c]{.7in} 
$\Z^2:$\\
$\frac{1}{48}\text{CS}_3^{(TM)}$,\\
$\frac{1}{2}\text{CS}_3^{(\SO(3))}$
\end{minipage}
&
\begin{minipage}[c]{.7in} 
$\Z^2:$\\
$\frac{1}{48}\text{CS}_3^{(TM)}$,\\
$\text{CS}_3^{(\PSU(3))}$
\end{minipage}
\\
\hline
$3$ O&
\begin{minipage}[c]{.65in} 
$\Z_2:$\\
$x_3=$\\
${w_1 x_2}$  
\end{minipage}
&
\begin{minipage}[c]{.5in} 
$0$
\end{minipage}
&
\begin{minipage}[c]{.66in} 
$\Z_2:$\\
$w_3'=$\\
${w_1 w_2'}$ 
\end{minipage}
&
\begin{minipage}[c]{.5in} 
$0$
\end{minipage}
&
\begin{minipage}[c]{.7in} 
$\Z_2^4:$\\
$x_3=w_1x_2,ax_2,$\\
$aw_1^2,a^3$
\end{minipage}
&
\begin{minipage}[c]{.7in} 
$0$
\end{minipage}
&
\begin{minipage}[c]{.7in} 
$\Z_2^2:$\\
$x_3=w_1x_2,w_3'=w_1w_2'$
\end{minipage}
&
\begin{minipage}[c]{.7in} 
$0$
\end{minipage}
\\
\hline
$3$ Pin$^+$&
\begin{minipage}[c]{.5in} 
$\Z_2^2:$\\
$w_1x_2=x_3,$\\
$w_1\text{Arf}$
\end{minipage}
&
\begin{minipage}[c]{.5in} 
$\Z_2:$\\
$w_1\text{Arf}$
\end{minipage}
&
\begin{minipage}[c]{.5in} 
$\Z_2^2:$\\
$w_1w_2'=w_3',$\\
$w_1\text{Arf}$
\end{minipage}
&
\begin{minipage}[c]{.5in} 
$\Z_2:$\\
$w_1\text{Arf}$
\end{minipage}
&
\begin{minipage}[c]{.7in} 
$\Z_2^5:$\\
$a^3,w_1x_2=x_3,$\\
$ax_2,w_1a{\tilde{\eta}},$\\
$w_1\text{Arf}$
\end{minipage}
&
\begin{minipage}[c]{.7in} 
$\Z_2:$\\
$w_1\text{Arf}$
\end{minipage}
&
\begin{minipage}[c]{.7in} 
$\Z_2^3:$\\
$w_1w_2'=w_3',w_1x_2=x_3,$\\
$w_1\text{Arf}$
\end{minipage}
&
\begin{minipage}[c]{.7in} 
$\Z_2:$\\
$w_1\text{Arf}$
\end{minipage}
\\
\hline
$3$ Pin$^-$&
\begin{minipage}[c]{.5in} 
$\Z_2:$\\
$w_1x_2=x_3$
\end{minipage}
&
\begin{minipage}[c]{.5in} 
$0$
\end{minipage}
&
\begin{minipage}[c]{.5in} 
$\Z_2:$\\
$w_1w_2'=w_3'$
\end{minipage}
&
\begin{minipage}[c]{.5in} 
$0$
\end{minipage}
&
\begin{minipage}[c]{.7in} 
$\Z_2^4:$\\
$a^3,w_1^2a,$\\
$x_3=$\\
$w_1x_2$,\\
$ax_2$
\end{minipage}
&
\begin{minipage}[c]{.7in} 
$0$
\end{minipage}
&
\begin{minipage}[c]{.7in}
$\Z_2^2:$\\
$w_1w_2'=$\\
$w_3',$\\
$w_1x_2=$\\
$x_3$ 
\end{minipage}
&
\begin{minipage}[c]{.7in} 
$0$
\end{minipage}
\\
\hline
 \end{tabular}
 }\hspace*{-26mm}
\caption{$\TP_3$. 
%\cred{In which case, do we gain a 3d topological term
%$w_1 x_2+ w_1^3$?
%$a x_2+ a^3$?
%$w_1 w_2+ w_1^3$?
%$a w_2+ a^3$?
%Descend from 5d?
%}
}
 \label{TP_3}
\end{table}
%%%%%%%%%%
%%%%%%%%%%

%%%%%%%%%%
%%%%%%%%%%
\begin{table}[!h] %[!h] %[tb]
\centering
 \makebox[\textwidth][r]{
 \begin{tabular}{ |c| c | c|  c| c | c| c| c| c| c|}
\hline
$\TP_d(H\times-)$ & $\B\Z_2$ &  $\B\Z_3$ & $\PSU(2)$  &   $\PSU(3)$ & 
\begin{minipage}[c]{.7in} 
$\Z_2\times \B\Z_2$
\end{minipage}
& 
\begin{minipage}[c]{.7in} 
$\Z_3\times \B\Z_3$
\end{minipage}
&  
\begin{minipage}[c]{.7in} 
$\PSU(2)\times \B\Z_2$
\end{minipage}
&  
\begin{minipage}[c]{.7in} 
$\PSU(3)\times \B\Z_3$
\end{minipage}
\\
\hline
$4$ SO& 
\begin{minipage}[c]{.5in} 
$\Z_4$:\\
${\mathcal{P}_2(x_2)}$
\end{minipage}
&
\begin{minipage}[c]{.5in} 
$\Z_3$:\\
$x_2'^2$
\end{minipage}
&
\begin{minipage}[c]{.5in} 
$0$
\end{minipage}
&
\begin{minipage}[c]{.5in} 
$0$
\end{minipage}
&
\begin{minipage}[c]{.7in} 
$\Z_2\times\Z_4$:\\
$ax_3=$\\
$a^2x_2$,\\
${\mathcal{P}_2(x_2)}$
\end{minipage}
&
\begin{minipage}[c]{.7in} 
$\Z_3^2$:\\
$a'x_3'=$\\
$b'x_2'$,\\
$x_2'^2$
\end{minipage}
&
\begin{minipage}[c]{.75in} 
$\Z_2\times\Z_4$:\\
$w_2'x_2,{\mathcal{P}_2(x_2)}$
\end{minipage}
&
\begin{minipage}[c]{.7in} 
$\Z_3^2$:\\
$x_2'^2,x_2'z_2$
\end{minipage}
\\
\hline
$4$ Spin&
\begin{minipage}[c]{.55in} 
$\Z_2$:\\
${\frac{\mathcal{P}_2(x_2)}{2}}$
\end{minipage}
&
\begin{minipage}[c]{.5in} 
$\Z_3$:\\
$x_2'^2$
\end{minipage}
&
\begin{minipage}[c]{.5in} 
$0$
\end{minipage}
&
\begin{minipage}[c]{.5in} 
$0$
\end{minipage}
&
\begin{minipage}[c]{.7in} 
$\Z_2^2$:\\
$ax_3=$\\
$a^2x_2$,\\
${\frac{\mathcal{P}_2(x_2)}{2}}$
\end{minipage}
&
\begin{minipage}[c]{.7in} 
$\Z_3^2$:\\
$a'x_3'=$\\
$b'x_2'$,\\
$x_2'^2$
\end{minipage}
&
\begin{minipage}[c]{.7in} 
$\Z_2^2$:\\
$w_2'x_2$,\\
${\frac{\mathcal{P}_2(x_2)}{2}}$
\end{minipage}
&
\begin{minipage}[c]{.7in} 
$\Z_3^2$:\\
$x_2'^2,x_2'z_2$
\end{minipage}
\\
\hline
$4$ O&
\begin{minipage}[c]{.5in} 
$\Z_2^4$:\\ 
$x_2^2,w_1^4$,\\
$w_1^2x_2,w_2^2$
\end{minipage}
&
\begin{minipage}[c]{.5in} 
$\Z_2^2$:\\
$w_1^4$,\\
$w_2^2$
\end{minipage}
&
\begin{minipage}[c]{.5in} 
$\Z_2^4$:\\
$w_2'^2,w_1^4$,\\
$w_1^2w_2',w_2^2$
\end{minipage}
&
\begin{minipage}[c]{.6in} 
$\Z_2^3$:\\
$w_1^4,w_2^2$,\\
$c_2{(\text{mod }2)}$
\end{minipage}
&
\begin{minipage}[c]{.7in} 
$\Z_2^8$:\\
$w_1^4,w_2^2$,\\
$a^4,a^2x_2$,\\
$ax_3,x_2^2$,\\
$w_1^2a^2,w_1^2x_2$
\end{minipage}
&
\begin{minipage}[c]{.7in} 
$\Z_2^2$:\\
$w_1^4,w_2^2$
\end{minipage}
&
\begin{minipage}[c]{.7in} 
$\Z_2^7$:\\
$w_1^4,w_2^2$,\\
$x_2^2,w_2'^2$,\\
$x_2w_1^2,w_2'w_1^2$,\\
$w_2'x_2$
\end{minipage}
&
\begin{minipage}[c]{.7in} 
$\Z_2^3$:\\
$w_1^4,w_2^2$,\\
$c_2{(\text{mod }2)}$
\end{minipage}
\\
\hline
$4$ Pin$^+$&
\begin{minipage}[c]{.5in} 
$\Z_4\times\Z_{16}$:\\
${q_s(x_2)},$\\
$\eta$
\end{minipage}
&
\begin{minipage}[c]{.5in} 
$\Z_{16}$:\\
$\eta$
\end{minipage}
&
\begin{minipage}[c]{.5in} 
$\Z_4\times\Z_{16}$:\\
${q_s(w_2')},$\\
$\eta$
\end{minipage}
&
\begin{minipage}[c]{.62in} 
$\Z_2\times\Z_{16}$:\\
$c_2{(\text{mod }2)},$\\
$\eta$
\end{minipage}
&
\begin{minipage}[c]{.7in} 
$\Z_2^2\times\Z_4\times\Z_8\times\Z_{16}$:\\
$ax_3,$\\
$w_1ax_2=$\\
$a^2x_2+ax_3$,\\
${q_s(x_2)},$\\
$w_1a {\text{ABK}}$,\\
$\eta$
\end{minipage}
&
\begin{minipage}[c]{.7in} 
$\Z_{16}$:\\
$\eta$
\end{minipage}
&
\begin{minipage}[c]{.7in} 
$\Z_4^2\times\Z_{16}\times\Z_2$:\\
${q_s(w_2')},$\\
${q_s(x_2)}$,\\
$\eta,w_2'x_2$
\end{minipage}
&
\begin{minipage}[c]{.7in} 
$\Z_2\times\Z_{16}$:\\
$c_2{(\text{mod }2)},$\\
$\eta$
\end{minipage}
\\
\hline
$4$ Pin$^-$&
\begin{minipage}[c]{.5in} 
$\Z_2$:\\
$w_1^2x_2$
\end{minipage}
&
\begin{minipage}[c]{.5in} 
$0$
\end{minipage}
&
\begin{minipage}[c]{.5in} 
$\Z_2$:\\
$w_1^2w_2'$
\end{minipage}
&
\begin{minipage}[c]{.6in} 
$\Z_2$:\\
$c_2{(\text{mod }2)}$
\end{minipage}
&
\begin{minipage}[c]{.7in} 
$\Z_2^3$:\\
$w_1^2x_2,ax_3$,\\
$w_1ax_2=$\\
$a^2x_2+ax_3$
\end{minipage}
&
\begin{minipage}[c]{.7in} 
$0$
\end{minipage}
&
\begin{minipage}[c]{.7in} 
$\Z_2^3$:\\
$w_1^2w_2'$,\\
$w_1^2x_2,w_2'x_2$
\end{minipage}
&
\begin{minipage}[c]{.7in} 
$\Z_2$:\\
$c_2{(\text{mod }2)}$
\end{minipage}
\\
\hline
 \end{tabular}
 }\hspace*{-15mm}
\caption{$\TP_4$.}
 \label{TP_4}
\end{table}

%%%%%%%%%%
%%%%%%%%%%
\begin{table}[!h] %[!h] %[tb]
\centering
 \makebox[\textwidth][r]{
 \begin{tabular}{ |c| c | c|  c| c | c| c| c| c| c|}
\hline
${\TP}_d(H\times-)$ & $\B\Z_2$ &  $\B\Z_3$ & $\PSU(2)$  &   $\PSU(3)$ & 
\begin{minipage}[c]{.7in} 
$\Z_2\times \B\Z_2$
\end{minipage}
& 
\begin{minipage}[c]{.7in} 
$\Z_3\times \B\Z_3$
\end{minipage}
&  
\begin{minipage}[c]{.7in} 
$\PSU(2)\times \B\Z_2$
\end{minipage}
&  
\begin{minipage}[c]{.7in} 
$\PSU(3)\times \B\Z_3$
\end{minipage}
\\
\hline
$5$ \SO&
\begin{minipage}[c]{.5in} 
$\Z_2^2$:\\
${x_5}=x_2x_3$,\\
$w_2w_3$
\end{minipage}
&
\begin{minipage}[c]{.5in} 
$\Z_2$:\\
$w_2w_3$
\end{minipage}
&
\begin{minipage}[c]{.5in} 
$\Z_2^2$:\\
$w_2w_3$,\\
$w_2'w_3'$
\end{minipage}
&
\begin{minipage}[c]{.7in} 
$\Z\times\Z_2$:\\
$\text{CS}_5^{(\PSU(3))}\tablefootnote{$\text{CS}_5(\PSU(3))\equiv \text{CS}_5^{(\PSU(3))}$ is the Chern-Simons 5-form of the $\PSU(3)$ gauge bundle.},$\\
$w_2w_3$
\end{minipage}
&
\begin{minipage}[c]{.7in} 
$\Z_2^6$:\\
$ax_2^2,a^5$,\\
${x_5}=x_2x_3,a^3x_2$,\\
$w_2w_3,aw_2^2$
\end{minipage}
&
\begin{minipage}[c]{.7in} 
$\Z_2\times\Z_3^2\times\Z_9$:\\
$w_2w_3$,\\
$a'b'x_2'$,\\
$a'x_2'^2$,\\
$\mathfrak{P}_3(b')$
\end{minipage}
&
\begin{minipage}[c]{.7in} 
$\Z_2^4$:\\
$w_2'w_3',{x_5}=x_2x_3$,\\
$w_3'x_2=$\\
$w_2'x_3$,\\
$w_2w_3$
\end{minipage}
&
\begin{minipage}[c]{.7in} 
$\Z\times\Z_2\times\Z_3$:\\
$\text{CS}_5^{(\PSU(3)})$,\\
$w_2w_3$,\\
$z_2x_3'=$\\
$-z_3x_2'$
\end{minipage}
\\
\hline
$5$ Spin&
\begin{minipage}[c]{.5in} 
$0$
\end{minipage}
&
\begin{minipage}[c]{.5in} 
$0$
\end{minipage}
&
\begin{minipage}[c]{.5in} 
$0$
\end{minipage}
&
\begin{minipage}[c]{.5in} 
$\Z$:\\
$\frac{1}{2}\text{CS}_{5}^{(\PSU(3))}$
\end{minipage}
&
\begin{minipage}[c]{.7in} 
$\Z_2$:\\
$a^3x_2$
\end{minipage}
&
\begin{minipage}[c]{.7in} 
$\Z_3^2\times\Z_9$:\\
$a'b'x_2'$,\\
$a'x_2'^2$,\\
$\mathfrak{P}_3(b')$
\end{minipage}
&
\begin{minipage}[c]{.7in} 
$\Z_2$:\\
$w_3'x_2=$\\
$w_2'x_3$
\end{minipage}
&
\begin{minipage}[c]{.7in} 
$\Z\times\Z_3$:\\
$\frac{1}{2}\text{CS}_{5}^{(\PSU(3))}$,\\
$z_2x_3'=$\\
$-z_3x_2'$
\end{minipage}
\\
\hline
$5$ O&
\begin{minipage}[c]{.85in} 
$\Z_2^4$:\\ 
$x_2x_3$,\\
$x_5=$\\
$(w_2+w_1^2)x_3$\\
$=$\\
$(w_3+w_1^3)x_2$,\\
$w_1^2x_3=$\\
$w_1^3x_2,$\\
$w_2w_3$
\end{minipage}
&
\begin{minipage}[c]{.5in} 
$\Z_2$: \\
$w_2w_3$
\end{minipage}
&
\begin{minipage}[c]{.5in} 
$\Z_2^3$: \\
$w_2w_3$,\\
$w_1^2w_3'=$\\
\cblue{$w_1^3w_2',$}\\
$w_2'w_3'$.\\
\end{minipage}
& 
\begin{minipage}[c]{.5in} 
$\Z_2$:\\ 
$w_2w_3$
\end{minipage}
&
\begin{minipage}[c]{.7in} 
$\Z_2^{12}$: 
$a^5,a^2x_3$,\\
$a^3x_2,a^3w_1^2$,\\
$ax_2^2,aw_1^4$,\\
$ax_2w_1^2,aw_2^2$,\\
$x_2x_3,w_1^2x_3$,\\
$x_5,w_2w_3$
\end{minipage}
&
\begin{minipage}[c]{.7in} 
$\Z_2$:\\ 
$w_2w_3$
\end{minipage}
& 
\begin{minipage}[c]{.7in} 
$\Z_2^{8}$:\\ 
$w_2'w_3',x_2w_3'$,\\
$w_1^2w_3',w_2'x_3$,\\
$x_2x_3,w_1^2x_3$,\\
$x_5,w_2w_3$
\end{minipage}
&
\begin{minipage}[c]{.7in} 
$\Z_2$:\\
$w_2w_3$
\end{minipage}
\\
\hline
$5$ Pin$^+$&
\begin{minipage}[c]{.55in} 
$\Z_2^2$:\\
$x_2x_3$,\\
$x_5 =$\\
$w_1^2x_3=$\\
${w_1^3x_2}$
\end{minipage}
&
\begin{minipage}[c]{.5in} 
$0$
\end{minipage}
&
\begin{minipage}[c]{.5in} 
$\Z_2$:\\
$w_1^2w_3'=$\\
${w_2'w_3'}$
\end{minipage}
&
\begin{minipage}[c]{.5in} 
$0$
\end{minipage}
&
\begin{minipage}[c]{.7in} 
$\Z_2^7$:\\
$w_1^4a$,\\
${a^5=w_1^2a^3}$,\\
$w_1^2x_3{=x_5}$,\\
$x_2x_3$,\\
$w_1^2ax_2=$\\
$ax_2^2+a^2x_3$,\\
$w_1ax_3=$\\
$a^2x_3$,\\
$a^3x_2$
\end{minipage}
&
\begin{minipage}[c]{.7in} 
$0$
\end{minipage}
&
\begin{minipage}[c]{.7in} 
$\Z_2^5$:\\
$w_1^2w_3'=$\\
$w_2'w_3'$,\\
$w_1^2x_3=$\\
$x_5$,\\
$x_2x_3,w_3'x_2$,\\
$w_1w_2'x_2=$\\
$w_2'x_3+$$w_3'x_2$
\end{minipage}
&
\begin{minipage}[c]{.7in} 
$0$
\end{minipage}
\\
\hline
$5$ Pin$^-$&
\begin{minipage}[c]{.5in} 
$\Z_2$:\\
$x_2x_3$
\end{minipage}
&
\begin{minipage}[c]{.5in} 
$0$
\end{minipage}
&
\begin{minipage}[c]{.5in} 
$0$
\end{minipage}
&
\begin{minipage}[c]{.5in} 
$0$
\end{minipage}
&
\begin{minipage}[c]{.7in} 
$\Z_2^5$:\\
$w_1^2a^3,x_2x_3$,\\
$w_1^2ax_2$,\\
$w_1ax_3=$\\
$a^2x_3$,\\
$a^3x_2$
\end{minipage}
&
\begin{minipage}[c]{.7in} 
$0$
\end{minipage}
&
\begin{minipage}[c]{.7in} 
$\Z_2^3$:\\
$x_2x_3,w_3'x_2$,\\
$w_1w_2'x_2=$\\
$w_2'x_3+$$w_3'x_2$
\end{minipage}
&
\begin{minipage}[c]{.7in} 
$0$
\end{minipage}
\\
\hline
 \end{tabular}
 }\hspace*{-28mm}
\caption{$\TP_5$.
} \label{TP_5}
\end{table}

In Section \ref{(BG_a,B^2G_b):(BO,B^2Z_2)}, we compute the topological terms (involving the cohomology classes of $\B ^2\Z_2$) of $\Omega_5^{\mathbb{G}}$ where $\mathbb{G}$ is a 2-group with $G_a=\tO$, $G_b=\Z_2$  
We find that the term $x_2w_3$ (or $x_3w_2$) survives only for $\beta=0,w_1^3$ (the Postnikov class $\beta\in\H^3(\B \tO,\Z_2)=\Z_2^3$ which is generated by $w_1^3,w_1w_2,w_3$). This term also appears in eq. 2.57 of \cite{2018arXiv180609592C}.

\clearpage
%\section{Difference between a previous cobordism theory and this work}

%\appendix

%\begin{center}
%{\bf\LARGE{Appendix}}
%\end{center}

\newpage
\section{Background information}

\label{sec:Background}

For more information, see \cite{hatcher, Wen2014zga1410.8477, wen2017exactly, 2018arXiv180809394Z}.

\subsection{Cohomology theory}
\subsubsection{Cup product}

Let $X$ be a topological space, an $n$-simplex of $X$ is a map $\sigma:\Delta^n\to X$ where 
\bea
\Delta^n=\{(t_0,t_1,\dots,t_n)\in\R^{n+1}|t_0+t_1+\cdots+t_n=1,t_i\ge0\},
\eea
it is denoted by $[v_0,\dots,v_n]$ where $v_i$ are vertices of $\Delta^n$.

$n$-simplexes of $X$ generates an abelian group $C_n(X)$, the elements of $C_n(X)$ are called $n$-chains.
$\Delta^{n-1}$ embeds in $\Delta^n$ in the canonical way, define $\partial:C_n(X)\to C_{n-1}(X)$ by 
\bea
\partial(\sigma)=\sum_{i=0}^n(-1)^i\sigma|_{[v_0,\dots,\hat{v}_i,\dots,v_n]}.
\eea
It is easy to verify that $\partial^2=0$, so $(C_{\bullet}(X),\partial)$ is a chain complex. 

Let $G$ be an abelian group, let $C^n(X,G):=\Hom(C_n(X),G)$, the elements of $C^n(X,G)$ are called $n$-cochains with coefficients $G$. Define $\delta:C^n(X,G)\to C^{n+1}(X,G)$ by 
$\delta(\alpha)(\sigma)=\alpha(\partial(\sigma))$, then $\delta^2=0$, so $(C^{\bullet}(X,G),\delta)$ is a cochain complex.

$\H^n(X,G)$ is defined to be $\frac{\text{Ker}\delta:C^n(X,G)\to C^{n+1}(X,G)}{\text{Im}\delta:C^{n-1}(X,G)\to C^n(X,G)}$.
It is an abelian group, called the $n$-th cohomology group of $X$ with coefficients $G$, the elements of the abelian group $Z^n(X,G):=\text{Ker}\delta:C^n(X,G)\to C^{n+1}(X,G)$ are called $n$-cocycles, the elements of $B^n(X,G):=\text{Im}\delta:C^{n-1}(X,G)\to C^n(X,G)$ are called $n$-coboundaries.

By abusing the notation, we also use $[v_0,\dots,v_n]$ to denote an $n$-chain.

If $G$ is additionally a ring $R$, then we can define a cup product such that $\H^*(X,R)$ is a graded ring.
First we define the cup product of two cochains:
\bea
C^n(X,R)\times C^m(X,R)&\to&C^{n+m}(X,R)\notag\\
(\alpha,\beta)&\mapsto&\alpha\cup\beta
\eea
\bea
\alpha\cup\beta([v_0,\dots,v_{n+m}]):=\alpha([v_0,\dots,v_n])\cdot\beta([v_n,\dots,v_{n+m}])
\eea
where $\cdot$ is the multiplication in $R$.

The cup product satisfies 
\begin{align}
\label{cupprop}
 \delta(\alpha \cup \beta) &= 
(\delta \alpha) \cup \beta 
+ (-1)^n \alpha \cup (\delta \beta) 
\end{align}
$\alpha \cup \beta$ is a cocycle if both $\alpha$ and $\beta$ are cocycles.
If both $\alpha$ and $\beta$ are  cocycles, then $\alpha \cup \beta$ is a coboundary if
one of $\alpha$ and $\beta$ is a coboundary.  So the cup product is also an
operation on cohomology groups $\hcup{} : \H^n(X,R)\times \H^m(X,R) \to
\H^{n+m}(X,R)$.  The cup product of two cocycles satisfies
\begin{align}
 \alpha \cup \beta &= (-1)^{nm} \beta \cup \alpha + \text{coboundary}
\end{align}

For the convenience of defining higher cup product, we use the notation $i\to j$ for the consecutive sequence from $i$ to $j$
\begin{align}
i\to j\equiv i,i+1,\cdots,j-1,j. 
\end{align}
We also denote an $n$-chain by $(0\to n)$. We use $\<\alpha,\sigma\>$ to denote the value of $\alpha(\sigma)$ for $n$-cochain $\alpha$ and $n$-chain $\sigma$.

Let $f_m$ be an $m$-cochain, $h_n$ be an $n$-cochain, we define higher cup product $f_m \hcup{k}
h_n$ which yields an $(m+n-k)$-cochain:
\begin{align}
&\ \ \ \
 \<f_m \hcup{k} h_n, (0,1,\cdots,m+n-k)\> 
\nonumber\\
&
 = \hskip -1em \sum_{0\leq i_0<\cdots< i_k \leq n+m-k} \hskip -2em  (-1)^p
\<f_m,(0 \to i_0, i_1\to i_2, \cdots)\>\times
\<h_n,(i_0\to i_1, i_2\to i_3, \cdots)\>,
\end{align} 
and $f_m \hcup{k} h_n =0$ for  $k>m \text{ or } n$ or $k<0$.
Here
$i\to j$ is the sequence $i,i+1,\cdots,j-1,j$, and
$p$ is the number of transpositions (it is not unique but its parity is unique) in the decomposition of the permutation to bring the sequence
\begin{align}
 0 \to i_0, i_1\to i_2, \cdots; i_0+1\to i_1-1, i_2+1\to i_3-1,\cdots
\end{align}
to the sequence
\begin{align}
 0 \to m+n-k.
\end{align}
For example
\begin{align}
&
 \<f_m \hcup1 h_n, (0,1,\cdots,m+n-1)\> 
 = \sum_{i=0}^{m-1} (-1)^{(m-i)(n+1)}\times
\nonumber\\
&
\<f_m,(0 \to i, i+n\to m+n-1)\>
\<h_n,(i\to i+n)\>.
\end{align} 
We can see that $\hcup0 =\cup$.  
Unlike cup product at $k=0$, the higher cup product of two
cocycles may not be a cocycle.

Steenrod studied the higher cup product of cochains and found a formula \cite[Theorem 5.1]{Steenrod1947}:
\bea \label{eq:Steenrod's}
\delta(u\hcup{i}v)=(-1)^{p+q-i}u\hcup{i-1}v+(-1)^{pq+p+q}v\hcup{i-1}u+\delta u\hcup{i}v+(-1)^pu\hcup{i}\delta v
\eea
where $u$ is a $p$-cochain, $v$ is a $q$-cochain.

Also Steenrod defined Steenrod square using higher cup product:
\begin{align}
 \Sq^{n-k}(z_n) \equiv z_n\hcup{k} z_n
\end{align}

\subsubsection{Universal coefficient theorem and K\"unneth formula}
If $X$ is a topological space,
$R$ is a principal ideal domain ($\Z$ or a field), $G$ is an $R$-module, then the homology version of universal coefficient theorem is 
\bea
\H_n(X,G)=\H_n(X,R)\otimes_RG\oplus\Tor_1^R(\H_{n-1}(X,R),G).
\eea

The cohomology version of universal coefficient theorem is 
\bea\label{uctcohomology1}
\H^n(X,G)=\Hom_R(\H_n(X,R),G)\oplus\Ext_R^1(\H_{n-1}(X,R),G).
\eea

We will abbreviate $\Tor_1^{\Z}$ by $\Tor$, $\Ext_{\Z}^1$ by $\Ext$.

If $X$ and $X'$ are topological spaces,
$R$ is a principle ideal domain and $G,G'$ are $R$-modules such that
$\Tor_1^R(G,G')=0$.  
We also require either\\
(1) $\H_n(X;\Z)$ and  $\H_n(X';\Z)$ are finitely generated, or\\
(2) $G'$ and $H_n(X';\Z)$ are
finitely generated.\\

The homology version of K\"unneth formula is 

\begin{align}
\label{kunnhomology}
&\ \ \ \ \H_d(X\times X',G\otimes_R G')
\nonumber\\
&\simeq \Big[\oplus_{k=0}^d \H_k(X,G)\otimes_R \H_{d-k}(X',G')\Big]\oplus
\Big[\oplus_{k=0}^{d-1}
\Tor_1^R(\H^k(X,G), \H_{d-k-1}(X',G'))\Big]  .
\end{align}
The cohomology version of K\"unneth formula is 
\begin{align}
\label{kunncohomology}
&\ \ \ \ \H^d(X\times X',G\otimes_R G')
\nonumber\\
&\simeq \Big[\oplus_{k=0}^d \H^k(X,G)\otimes_R \H^{d-k}(X',G')\Big]\oplus
\Big[\oplus_{k=0}^{d+1}
\Tor_1^R(\H^k(X,G), \H^{d-k+1}(X',G'))\Big]  .
\end{align}

 Note that $\Z$ and $\R$ are principal ideal
domains, while $\RZ$ is not.  Also, $\R$ and $\RZ$ are not finitely
generate $R$-modules if $R=\Z$.

Special cases: 1. $R=G'=\Z$.

In this case, the
condition $\Tor_1^R(G,G')=\Tor_1^{\Z}(G, \Z)=0$ is always
satisfied. $G$ can be $\RZ$, $\Z$, $\Z_n$ \etc. So we have
\begin{align}
\label{kunnZ}
&\ \ \ \ \H^d(X\times X',G)
\nonumber\\
&\simeq \Big[\oplus_{k=0}^d \H^k(X,G)\otimes_{\Z} \H^{d-k}(X';\Z)\Big]\oplus
\Big[\oplus_{k=0}^{d+1}
\Tor(\H^k(X,G), \H^{d-k+1}(X';\Z))\Big]  .
\end{align}

Take $X$ to be the space of one point in \eqref{kunnZ},
and use
\begin{align}
\H^{n}(X,G))=
\begin{cases}
G, & \text{ if } n=0,\\
0, & \text{ if } n>0,
\end{cases}
\end{align}
to reduce \eqref{kunnZ} to
\begin{align}
\label{uctcohomology2}
 \H^d(X,G)
&\simeq \H^d(X;\Z) \otimes_{\Z} G 
\oplus
\Tor(\H^{d+1}(X;\Z),G)  .
\end{align}
where $X'$ is renamed as $X$.  This is also called the universal coefficient
theorem which can be used to calculate $\H^*(X,G)$ from $\H^*(X;\Z)$ and the
module $G$.   
Here $\Tor=\Tor_1^{\Z}$.

Homology version of \eqref{uctcohomology2} is just the universal coefficient theorem for homology with $R=\Z$.

2. $R=G=G'=\F$ is a field, $\Tor_1^R(G,G')=0$.

\bea
\H^*(X\times X',\F)=\H^*(X,\F)\otimes\H^*(X',\F),
\eea
This is called the K\"unneth formula.

There is also a relative version of K\"unneth formula \cite[Theorem 3.18]{hatcher}: 
\bea\label{kunneth}
\tilde{\H}^*(X\wedge X',\F)=\tilde{\H}^*(X,\F)\otimes\tilde{\H}^*(X',\F).
\eea
Here $X\wedge X'$ is the smash product, $\tilde{\H}$ is the reduced cohomology.

\subsection{Spectra}\label{sec:spectra}

\begin{definition}
$\bullet$
A prespectrum $T_{\bullet}$ is a sequence $\{T_q\}_{q\in\Z^{\ge0}}$ of pointed spaces and maps $s_q:\Sigma T_q\to T_{q+1}$.

$\bullet$
An $\Omega$-prespectrum is a prespectrum $T_{\bullet}$ such that the adjoints $t_q: T_q\to\Omega T_{q+1}$ of the structure maps are weak homotopy equivalences.

$\bullet$
A spectrum is a prespectrum $T_{\bullet}$ such that the adjoints $t_q: T_q\to\Omega T_{q+1}$ of the structure maps are homeomorphisms.

\end{definition}

\begin{example}
$\bullet$ Let $X$ be a pointed space, $T_q=\Sigma^qX$ for $q\ge0$, then $T_{\bullet}$ is a prespectrum.

$\bullet$ $T_q=S^q$, $T_{\bullet}$ is a prespectrum.

$\bullet$ Let $G$ be an abelian group, $T_q=K(G,q)$ the Eilenberg-MacLane space, $T_{\bullet}$ is an $\Omega$-prespectrum.

\end{example}

Spectrification: Let $T_{\bullet}$ be a prespectrum, define
$(LT)_q$ to be the colimit of 
$$T_q\xrightarrow{t_q}\Omega T_{q+1}\xrightarrow{\Omega t_{q+1}}\Omega^2T_{q+2}.$$
Namely,
$$(LT)_q=\text{colim}_{l\to\infty}\Omega^lT_{q+l},$$
then $(LT)_{\bullet}$ is a spectrum.

\begin{example}

$\bullet$ $T_q=S^q$, $(LT)_{\bullet}$ is a spectrum $\mathbb{S}$.

$\bullet$ Let $G$ be an abelian group, $T_q=K(G,q)$ the Eilenberg-MacLane space, $(LT)_{\bullet}$ is a spectrum $HG$ (the Eilenberg-MacLane spectrum).

\end{example}

Stable homotopy groups of spectra:
Let $M_{\bullet}$ be a spectrum, define
$\pi_dM_{\bullet}$ to be the colimit of 
$$\pi_{d+n}M_n\xrightarrow{\pi_{d+n}t_n}\pi_{d+n}\Omega M_{n+1}\xrightarrow{\text{adjunction}}\pi_{d+n+1}M_{n+1}.$$
Namely,
$$\pi_dM_{\bullet}=\text{colim}_{n\to\infty}\pi_{d+n}M_n.$$

Maps between spectra: If $M_{\bullet},N_{\bullet}$ are two spectra, then for any integer $k$, the abelian group of homotopy classes of maps from $M_{\bullet}$ to $N_{\bullet}$ of degree $-k$: $[M_{\bullet},N_{\bullet}]_{-k}$ is defined as follows: a map in 
$[M_{\bullet},N_{\bullet}]_{-k}$ is a sequence of maps $M_n\to N_{n+k}$ such that the following diagram commutes
\bea
\xymatrix{
\Sigma M_n\ar[r]\ar[d]&\Sigma N_{n+k}\ar[d]\\
M_{n+1}\ar[r] &N_{n+k+1}
}
\eea
where the columns are the structure maps of the spectra $M_{\bullet}$ and $N_{\bullet}$.
If in addition the spectrum $N_{\bullet}$ is a ring spectrum, then the abelian groups 
$[M_{\bullet},N_{\bullet}]_{-k}$ form a graded ring $[M_{\bullet},N_{\bullet}]_{-*}$.

\begin{example}
$\pi_dM_{\bullet}=[\mathbb{S},M_{\bullet}]_d$.
\end{example}

Cohomology rings of spectra:
\begin{definition}
A ring spectrum is a spectrum $E$ along with a unit map $\eta:\mathbb{S}\to E$ and a multiplication map $\mu:E\wedge E\to E$.
\end{definition}
\begin{example}
Let $R$ be a ring, then the Eilenberg-MacLane spectrum $HR$ is a ring spectrum.
\end{example}
The cohomology ring of a spectrum $M_{\bullet}$ with coefficients in $R$ is defined to be $[M_{\bullet},HR]_{-*}$.

\subsection{Spectral sequences}

In this paper, we use three kinds of spectral sequence: Adams spectral sequence, Atiyah-Hirzebruch spectral sequence, and Serre spectral sequence.

\subsubsection{Adams spectral sequence}\label{sec:Adams}

The Adams spectral sequence is a spectral sequence introduced by Adams in \cite{Adams1958}, it is of the form
\bea
E_2^{s,t}=\Ext_{\A_p}^{s,t}(\H^*(Y,\Z_p),\Z_p)\Rightarrow\pi_{t-s}(Y)_p^{\wedge}
\eea
\cred{where $Y$ is any spectrum.}
We consider $Y=MTH\wedge X_+$ and focus on $p=2$ {and $p=3$}.

We introduce the notions used in Adams spectral sequence:

$p$-completion: For any finitely generated abelian group $G$, $G_p^{\wedge}=\lim_nG/p^nG$ is the $p$-completion of $G$. If $G$ is finite, then $G_p^{\wedge}$ is the Sylow $p$-subgroup of $G$. If $G=\Z$, $G_p^{\wedge}$ is the ring of $p$-adic integers.

Steenrod algebra:
The mod $p$ Steenrod algebra is $\A_p:=[H\Z_p,H\Z_p]_{-*}$ where $H\Z_p$ is the mod $p$ Eilenberg-MacLane spectrum.
\cred{For any spectrum $Y$, the cohomology ring $\H^*(Y,\Z_p)=[Y,H\Z_p]_{-*}$ is an $\A_p$-module.}

For $p=2$, the generators of $\A_2$ are Steenrod squares $\Sq^n$.

\cred{
\begin{definition}[Axioms]\label{Sq^i}
For each $i\ge0$, there is a natural transformation
$$\Sq^i:\H^n(-,\Z_2)\to\H^{n+i}(-,\Z_2)$$
such that\\
$\bullet$
$\Sq^0=\text{Id}$\\
$\bullet$
If $i>|x|$, then $\Sq^ix=0$\\
$\bullet$
If $i=|x|$, then $\Sq^ix=x^2$\\
$\bullet$
(Cartan formula)
$\Sq^n(xy)=\sum_{i+j=n}\Sq^i(x)\Sq^j(y)$\\
$\bullet$
(Adem relation)
If $a<2b$, then
$$\Sq^a\Sq^b=\sum_{c=0}^{[\frac{a}{2}]}\binom{b-c-1}{a-2c}\Sq^{a+b-c}\Sq^c$$
\end{definition}
}

 The subalgebra $\A_2(1)$ of $\A_2$ generated by $\Sq^1$ and $\Sq^2$ looks like Figure \ref{fig:A_2(1)}.

\begin{figure}[!h]
\begin{center}
\begin{tikzpicture}[scale=0.5]
\node[below] at (0,0) {$\Sq^0=1$};
\node[right] at (0,1) {$\Sq^1$};
\node[left] at (0,2) {$\Sq^2$};
\node[left] at (0,3) {$\Sq^1\Sq^2$};
\node[right] at (1,3) {$\Sq^2\Sq^1$};
\node[right] at (1,4) {$\Sq^1\Sq^2\Sq^1$};
\node[left] at (1,5) {$\Sq^2\Sq^1\Sq^2$};
\node[right] at (1,6) {$\Sq^2\Sq^1\Sq^2\Sq^1=\Sq^1\Sq^2\Sq^1\Sq^2$};

\draw[fill] (0,0) circle(.1);
\draw[fill] (0,1) circle(.1);
\draw (0,0) -- (0,1);
\draw[fill] (0,2) circle(.1);
\draw (0,0) to [out=150,in=150] (0,2);
\draw[fill] (0,3) circle(.1);
\draw (0,2) -- (0,3);
\draw[fill] (1,3) circle(.1);
\draw[fill] (1,4) circle(.1);
\draw (1,3) -- (1,4);
\draw[fill] (1,5) circle(.1);
\draw[fill] (1,6) circle(.1);
\draw (1,5) -- (1,6);
\draw (0,1) to [out=30,in=150] (1,3);
\draw (0,2) to [out=30,in=150] (1,4);
\draw (0,3) to [out=30,in=150] (1,5);
\draw (1,4) to [out=30,in=30] (1,6);
\end{tikzpicture}
\end{center}
\caption{$\A_2(1)$}
\label{fig:A_2(1)}
\end{figure}

Each dot stands for a $\Z_2$, all relations are from Adem relations \eqref{Adem}.

For odd primes $p$, the generators of $\A_p$ are
the Bockstein homomorphism $\beta_{(p,p)}$ and Steenrod powers $P_p^n$.

Ext functor: Let $R=\A_p$ or $\A_2(1)$.
$\Ext_R^{s,t}$ is the internal degree $t$ part of the $s$-th derived functor of $\Hom_R^*$.

In general, we can find a projective $R$-resolution $P_{\bullet}$ of $L$ to compute $\Ext_R^i(L,\Z_p)$, $\Ext_R^i(L,\Z_p)=\H^i(\Hom_R(P_{\bullet},\Z_p))$ (the $i$-th cohomology of the chain complex $\Hom_R(P_{\bullet},\Z_p)$).

In Adams chart, the horizontal axis is degree $t-s$ and the vertical axis is degree $s$. The differential $d_r^{s,t}:E_r^{s,t}\to E_r^{s+r,t+r-1}$ is an arrow starting at the bidegree $(t-s,s)$ with direction $(-1,r)$.
$E_{r+1}^{s,t}:=\frac{\text{Ker}d_r^{s,t}}{\text{Im}d_r^{s-r,t-r+1}}$ for $r\ge2$. There exists $N$ such that $E_{N+k}=E_N$ for $k>0$, denote $E_{\infty}:=E_N$.

We explain how to read the result from the Adams chart: In the $E_{\infty}$ page, one dot indicates a $\Z_p$, an vertical line connecting $n$ dots indicates a $\Z_{p^n}$, when $n=\infty$, the line indicates a $\Z$. 

In the $H=\tO$ cases, $M\tO$ is the wedge sum of suspensions of the Eilenberg-MacLane spectrum $H\Z_2$, $\H^*(M\tO,\Z_2)$ is the direct sum of suspensions of the Steenrod algebra $\A_2$.

$\H^*(M\tO\wedge X_+,\Z_2)=\H^*(M\tO,\Z_2)\otimes\H^*(X,\Z_2)$ is also the direct sum of suspensions of the Steenrod algebra $\A_2$. 
We have used the K\"unneth formula \eqref{kunneth}.
Let $L=\H^*(M\tO\wedge X_+,\Z_2)$, then $P_0=L$, $P_s=0$ for $s>0$ gives a projective $\A_2$-resolution of $L$.

Since 
\bea
\Ext_{\A_p}^{s,t}(\Sigma^r\A_p,\Z_p)=\left\{\begin{array}{ll}\Hom_{\A_p}^t(\Sigma^r\A_p,\Z_p)=\Z_p&\text{ if }t=r, s=0\\ 0 &\text{ else}\end{array}\right.,
\eea 
all dots are concentrated in $s=0$ in the Adams chart of $\Ext_{\A_2}^{s,t}(\H^*(M\tO\wedge X_+,\Z_2),\Z_2)$, there are no differentials, $E_2=E_{\infty}$,
$\Omega_d^{\tO}(X)$ is a $\Z_2$-vector space.

In the $H=\SO$ cases, the localization of $M\SO$ at the prime 2 is 
\bea
M\SO_{(2)}=H\Z_{(2)}\vee\Sigma^4H\Z_{(2)}\vee\Sigma^5H\Z_2\vee\cdots
\eea
where $H\Z$ is the Eilenberg-MacLane spectrum and $\H^*(H\Z,\Z_2)=\A_2/\A_2\Sq^1$.

\bea
\cdots\To\Sigma^3\A_2\To\Sigma^2\A_2\To\Sigma\A_2\To\A_2\To\A_2/\A_2 \Sq^1
\eea
is an $\A_2$-resolution (denoted by $P_{\bullet}$) where the differentials $d_1$ are induced by $\Sq^1$.

When $X$ is a point, the Adams chart of $\Ext_{\A_2}^{s,t}(\H^*(M\SO,\Z_2),\Z_2)$ is shown in Figure \ref{fig:Ext_{A_2}^{s,t}(H^*(MSO,Z_2),Z_2)}. For general $X$, $P_{\bullet}\otimes\H^*(X,\Z_2)$ is a projective $\A_2$-resolution of $\H^*(H\Z,\Z_2)\otimes\H^*(X,\Z_2)$ (since $P_{\bullet}$ is actually a free $\A_2$-resolution), the differentials $d_1$ are induced by $\Sq^1$.

\begin{figure}[!h]
\begin{center}
\begin{tikzpicture}
\node at (0,-1) {0};
\node at (1,-1) {1};
\node at (2,-1) {2};
\node at (3,-1) {3};
\node at (4,-1) {4};
\node at (5,-1) {5};

\node at (6,-1) {$t-s$};
\node at (-1,0) {0};
\node at (-1,1) {1};
\node at (-1,2) {2};
\node at (-1,3) {3};
\node at (-1,4) {4};
\node at (-1,5) {5};

\node at (-1,6) {$s$};

\draw[->] (-0.5,-0.5) -- (-0.5,6);
\draw[->] (-0.5,-0.5) -- (6,-0.5);

\draw (0,0) -- (0,5);

\draw (4,0) -- (4,5);
\draw[fill] (5,0) circle(0.05);

\end{tikzpicture}
\end{center}
\caption{Adams chart of $\Ext_{\A_2}^{s,t}(\H^*(M\SO,\Z_2),\Z_2)$}
\label{fig:Ext_{A_2}^{s,t}(H^*(MSO,Z_2),Z_2)}
\end{figure}

The localization of $M\SO$ at the prime 3 is the wedge sum of suspensions of the Brown-Peterson spectrum $BP$ ($M\SO_{(3)}=BP\vee\Sigma^8BP\vee\cdots$) and $\H^*(BP,\Z_3)=\A_3/(\beta_{(3,3)})$ where $(\beta_{(3,3)})$ is the two-sided ideal generated by $\beta_{(3,3)}$.
\bea
\cdots\To\Sigma^2\A_3\oplus\Sigma^6\A_3\oplus\cdots\To\Sigma\A_3\oplus\Sigma^5\A_3\oplus\cdots\To\A_3\To\A_3/(\beta_{(3,3)})
\eea
is an $\A_3$-resolution of $\A_3/(\beta_{(3,3)})$ (denoted by $P_{\bullet}'$) where the differentials $d_1$ are induced by $\beta_{(3,3)}$.

When $X$ is a point, the Adams chart of $\Ext_{\A_3}^{s,t}(\H^*(M\SO,\Z_3),\Z_3)$ is shown in Figure \ref{fig:Ext_{A_3}^{s,t}(H^*(MSO,Z_3),Z_3)}. For general $X$, $P_{\bullet}'\otimes\H^*(X,\Z_3)$ is a projective $\A_3$-resolution of $\H^*(BP,\Z_3)\otimes\H^*(X,\Z_3)$ (since $P_{\bullet}'$ is actually a free $\A_3$-resolution), the differentials $d_1$ are induced by $\beta_{(3,3)}$.

\begin{figure}[!h]
\begin{center}
\begin{tikzpicture}
\node at (0,-1) {0};
\node at (1,-1) {1};
\node at (2,-1) {2};
\node at (3,-1) {3};
\node at (4,-1) {4};
\node at (5,-1) {5};

\node at (6,-1) {$t-s$};
\node at (-1,0) {0};
\node at (-1,1) {1};
\node at (-1,2) {2};
\node at (-1,3) {3};
\node at (-1,4) {4};
\node at (-1,5) {5};

\node at (-1,6) {$s$};

\draw[->] (-0.5,-0.5) -- (-0.5,6);
\draw[->] (-0.5,-0.5) -- (6,-0.5);

\draw (0,0) -- (0,5);

\draw (4,1) -- (4,5);

\end{tikzpicture}
\end{center}
\caption{Adams chart of $\Ext_{\A_3}^{s,t}(\H^*(M\SO,\Z_3),\Z_3)$}
\label{fig:Ext_{A_3}^{s,t}(H^*(MSO,Z_3),Z_3)}
\end{figure}

There may be differentials $d_n$ corresponding to the Bockstein homomorphism $\beta_{(p,p^n)}$ \cite{may1981bockstein} for both $p=2$ and $p=3$.
See \ref{Bockstein} for the definition of Bockstein homomorphisms.
Since $M\SO_{(3)}=M\Spin_{(3)}$, $\Omega_d^{\SO}(X)_3^{\wedge}=\Omega_d^{\Spin}(X)_3^{\wedge}$.

In the $H=\Spin/\Pin^{\pm}$ cases, since the mod 2 cohomology of the Thom spectrum $M\Spin$ is
\bea
\H^*(M\Spin,\Z_2)=\A_2\otimes_{\A_2(1)}\{\Z_2\oplus M\}
\eea
where $M$ is a graded $\A_2(1)$-module with the degree $i$ homogeneous part $M_i=0$ for $i<8$.

\cred{
\begin{theorem}[Change of rings/Frobenius reciprocity]
$$\Ext_{\A_2}^{s,t}(\A_2\otimes_{\A_2(1)}L,\Z_2)\cong\Ext_{\A_2(1)}^{s,t}(L,\Z_2)$$
\end{theorem}
}

When we compute $\Omega_d^H(X)_2^{\wedge}$,
we are reduced to compute $\Ext_{\A_2(1)}^{s,t}(L,\Z_2)$ for $t-s<8$, where $L$ is some $\A_2(1)$-module (our cases are some mod 2 cohomology $\H^*(-,\Z_2)$).

Example 1:
$L=\A_2(1)$, 
\cred{
\bea
\Ext_{\A_2(1)}^{s,t}(\A_2(1),\Z_2)=\left\{\begin{array}{ll}\Hom_{\A_2(1)}(\A_2(1),\Z_2)=\Z_2&\text{ if }t=s=0\\ 0 &\text{ else}\end{array}\right.
\eea 
}

Example 2: $L=\Z_2$, the $\A_2(1)$-resolution of $L$ is
\bea
\cdots\to\Sigma^3\A_2(1)\oplus\Sigma^7\A_2(1)\to\Sigma^2\A_2(1)\oplus\Sigma^4\A_2(1)\to\Sigma\A_2(1)\oplus\Sigma^2\A_2(1)\to\A_2(1)\to\Z_2.
\eea

The Adams chart looks like Figure \ref{fig:Ext_{A_2(1)}^{s,t}(Z_2,Z_2)}.

\begin{figure}[!h]
\begin{center}
\begin{tikzpicture}[scale=0.5]
\node at (0,-1) {0};
\node at (1,-1) {1};
\node at (2,-1) {2};
\node at (3,-1) {3};
\node at (4,-1) {4};
\node at (5,-1) {5};
\node at (6,-1) {6};
\node at (7,-1) {7};
\node at (8,-1) {8};
\node at (9,-1) {9};
\node at (10,-1) {10};

\node at (11,-1) {$t-s$};
\node at (-1,0) {0};
\node at (-1,1) {1};
\node at (-1,2) {2};
\node at (-1,3) {3};
\node at (-1,4) {4};
\node at (-1,5) {5};
\node at (-1,6) {6};
\node at (-1,7) {7};
\node at (-1,8) {8};
\node at (-1,9) {9};
\node at (-1,10) {10};

\node at (-1,11) {$s$};

\draw[->] (-0.5,-0.5) -- (-0.5,11);
\draw[->] (-0.5,-0.5) -- (11,-0.5);

\draw (0,0) -- (0,10);
\draw (0,0) -- (2,2);

\draw (4,3) -- (4,10);

\draw (8,4) -- (8,10);
\draw (8,4) -- (10,6);

\draw[->][dashed,color=red] (1,1) -- (0,3);

\draw[->][dashed,color=red] (1,1) -- (0,4);

\draw[->][dashed,color=red] (1,1) -- (0,5);

\end{tikzpicture}
\end{center}
\caption{Adams chart of $\Ext_{\A_2(1)}^{s,t}(\Z_2,\Z_2)$. The dashed arrows indicate the possible differentials.}
\label{fig:Ext_{A_2(1)}^{s,t}(Z_2,Z_2)}
\end{figure}
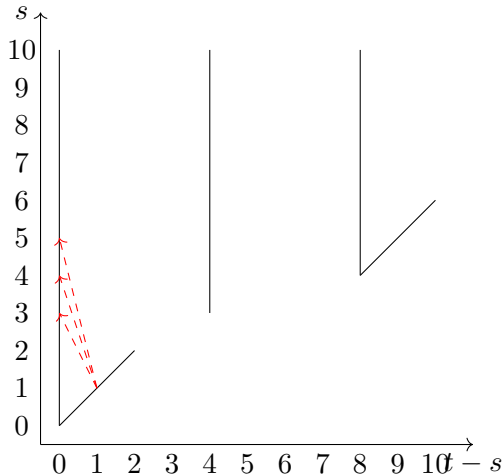

The only possible differentials are $d_r(h_1)=h_0^{r+1}$ where $h_0\in\Ext_{\A_2(1)}^{1,1}(\Z_2\Z_2)$, $h_1\in\Ext_{\A_2(1)}^{1,2}(\Z_2\Z_2)$. If there were such a differential $d_r$ for $r\ge2$, then since $h_0h_1=0$, $0=d_r(h_0h_1)=h_0^{r+2}$ which is not true. Hence $E_2=E_{\infty}$.

This is in fact real Bott periodicity \cred{($\pi_{i+8}ko=\pi_iko$, $\H^*(ko,\Z_2)=\A_2\otimes_{\A_2(1)}\Z_2$).}

Our computation is based on the following fact:
\begin{lemma}\label{principle}
Given a short exact sequence of $\A_2(1)$-modules
\bea
0\to L_1\to L_2\to L_3\to0,
\eea
then for any $t$, there is a long exact sequence
\bea
&&\cdots\to\Ext_{\A_2(1)}^{s,t}(L_3,\Z_2)\to\Ext_{\A_2(1)}^{s,t}(L_2,\Z_2)\to\Ext_{\A_2(1)}^{s,t}(L_1,\Z_2)\\\notag
&&\stackrel{d_1}{\to}\Ext_{\A_2(1)}^{s+1,t}(L_3,\Z_2)\to\Ext_{\A_2(1)}^{s+1,t}(L_2,\Z_2)\to\cdots
\eea
\end{lemma}

After using this fact repeatedly, we obtain the $E_2$ page.

Example 3: 
\bea
\xymatrix{\bullet\ar[r]&\bullet\\&\bullet\ar@{-}@/^/[u]^{\Sq^2}\ar[r]&\bullet}
\eea
is a short exact sequence where the left dot is $L_1$, the middle part is $L_2$, the right dot is $L_3$.

The Adams chart looks like Figure \ref{fig:Ext_{A_2(1)}^{s,t}(L_2,Z_2)}.

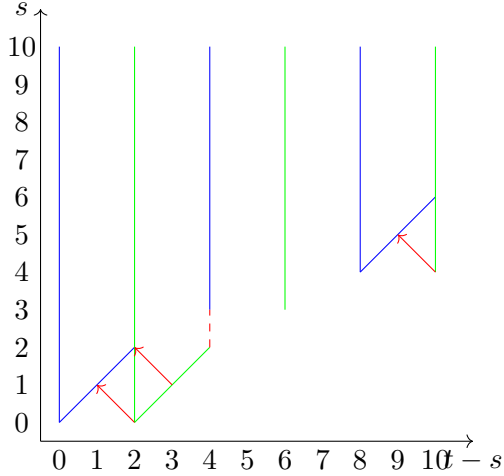
\begin{figure}[!h]
\begin{center}
\begin{tikzpicture}[scale=0.5]
\node at (0,-1) {0};
\node at (1,-1) {1};
\node at (2,-1) {2};
\node at (3,-1) {3};
\node at (4,-1) {4};
\node at (5,-1) {5};
\node at (6,-1) {6};
\node at (7,-1) {7};
\node at (8,-1) {8};
\node at (9,-1) {9};
\node at (10,-1) {10};

\node at (11,-1) {$t-s$};
\node at (-1,0) {0};
\node at (-1,1) {1};
\node at (-1,2) {2};
\node at (-1,3) {3};
\node at (-1,4) {4};
\node at (-1,5) {5};
\node at (-1,6) {6};
\node at (-1,7) {7};
\node at (-1,8) {8};
\node at (-1,9) {9};
\node at (-1,10) {10};

\node at (-1,11) {$s$};

\draw[->] (-0.5,-0.5) -- (-0.5,11);
\draw[->] (-0.5,-0.5) -- (11,-0.5);

\draw[color=blue] (0,0) -- (0,10);
\draw[color=blue] (0,0) -- (2,2);

\draw[color=blue] (4,3) -- (4,10);

\draw[color=blue] (8,4) -- (8,10);
\draw[color=blue] (8,4) -- (10,6);

\draw[color=green] (2,0) -- (2,10);
\draw[color=green] (2,0) -- (4,2);

\draw[color=green] (6,3) -- (6,10);

\draw[color=green] (10,4) -- (10,10);

\draw[->][color=red] (2,0) -- (1,1);
\draw[->][color=red] (3,1) -- (2,2);
\draw[dashed,color=red] (4,2) -- (4,3);

\draw[->][color=red] (10,4) -- (9,5);

\end{tikzpicture}
\end{center}
\caption{Adams chart of $\Ext_{\A_2(1)}^{s,t}(L_2,\Z_2)$. The arrows indicate the differential $d_1$, the dashed line indicates the extension.}
\label{fig:Ext_{A_2(1)}^{s,t}(L_2,Z_2)}
\end{figure}

Example 4: 
\bea
\xymatrix{\bullet\ar[r]&\bullet\\&\bullet\ar@{-}[u]^{\Sq^1}\ar[r]&\bullet}
\eea
is a short exact sequence where the left dot is $L_1'$, the middle part is $L_2'$, the right dot is $L_3'$.

The Adams chart looks like Figure \ref{fig:Ext_{A_2(1)}^{s,t}(L_2',Z_2)}.

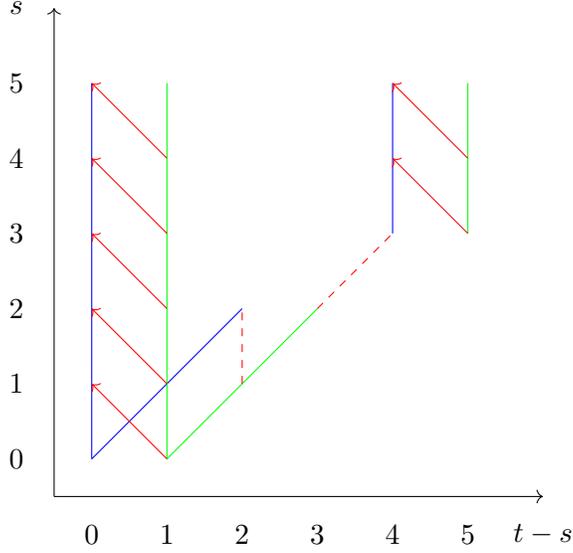
\begin{figure}[!h]
\begin{center}
\begin{tikzpicture}
\node at (0,-1) {0};
\node at (1,-1) {1};
\node at (2,-1) {2};
\node at (3,-1) {3};
\node at (4,-1) {4};
\node at (5,-1) {5};

\node at (6,-1) {$t-s$};
\node at (-1,0) {0};
\node at (-1,1) {1};
\node at (-1,2) {2};
\node at (-1,3) {3};
\node at (-1,4) {4};
\node at (-1,5) {5};

\node at (-1,6) {$s$};

\draw[->] (-0.5,-0.5) -- (-0.5,6);
\draw[->] (-0.5,-0.5) -- (6,-0.5);

\draw[color=blue] (0,0) -- (0,5);
\draw[color=blue] (0,0) -- (2,2);

\draw[color=blue] (4,3) -- (4,5);

\draw[color=green] (1,0) -- (1,5);
\draw[color=green] (1,0) -- (3,2);

\draw[color=green] (5,3) -- (5,5);

\draw[->][color=red] (1,0) -- (0,1);
\draw[->][color=red] (1,1) -- (0,2);
\draw[->][color=red] (1,2) -- (0,3);
\draw[->][color=red] (1,3) -- (0,4);
\draw[->][color=red] (1,4) -- (0,5);
\draw[->][color=red] (5,3) -- (4,4);
\draw[->][color=red] (5,4) -- (4,5);
\draw[dashed,color=red] (2,1) -- (2,2);
\draw[dashed,color=red] (3,2) -- (4,3);

\end{tikzpicture}
\end{center}
\caption{Adams chart of $\Ext_{\A_2(1)}^{s,t}(L_2',\Z_2)$. The arrows indicate the differential $d_1$, the dashed line indicates the extension.}
\label{fig:Ext_{A_2(1)}^{s,t}(L_2',Z_2)}
\end{figure}

\subsubsection{Serre spectral sequence}
Given a fibration $F\to E\to B$, the Serre spectral sequence is the following:
\bea
E_2^{p,q}=\H^p(B,H^q(F,\Z))\Rightarrow\H^{p+q}(E,\Z)
\eea
This can be used in computing the integral cohomology group of the total space of a nontrivial fibration.

There is also a homology version:
\bea
E^2_{p,q}=\H_p(B,H_q(F,\Z))\Rightarrow\H_{p+q}(E,\Z)
\eea

\subsubsection{Atiyah-Hirzebruch spectral sequence}
The Atiyah-Hirzebruch spectral sequence can be viewed as a generalization of the Serre spectral sequence. Given a fibration $F\to E\to B$, the Atiyah-Hirzebruch spectral sequence is the following:
\bea
E^2_{p,q}=\H_p(B,h_q(F,\Z))\Rightarrow h_{p+q}(E,\Z)
\eea
where $h_*$ is an extraordinary homology theory. For example, $h_*$ can be the bordism theory $\Omega_*^H$. In particular, if the fiber $F$ is a point, then the Atiyah-Hirzebruch spectral sequence is of the form:
\bea
\H_p(X,\Omega_q^H)\Rightarrow\Omega_{p+q}^H(X)
\eea

\subsection{Characteristic classes}

\subsubsection{Introduction to characteristic classes}
Characteristic classes are cohomology classes of the base space of a vector bundle.
Stiefel-Whitney classes are defined for real vector bundles, Chern classes are defined for complex vector bundles, Pontryagin classes are defined for real vector bundles. 
All characteristic classes are natural with respect to bundle maps. Characteristic classes of a principal bundle are defined to be the characteristic classes of the associated vector bundle of the principal bundle.

Given a real vector bundle $V\to M$ and a complex vector bundle $E\to M$, the $i$-th Stiefel-Whitney class of $V$ is
$w_i(V)\in\H^i(M,\Z_2)$, the $i$-th Chern class of $E$ is $c_i(E)\in\H^{2i}(M,\Z)$, the $i$-th Pontryagin class of $V$ is $p_i(V)\in\H^{4i}(M,\Z)$.

Pontryagin classes are closely related to Chern classes via complexification:

\bea
p_i(V)=(-1)^ic_{2i}(V\otimes_{\R}\C)\in\H^{4i}(M,\Z)
\eea
where $V\otimes_{\R}\C\to M$ is the complexification of the real vector bundle $V\to M$.

The relation between Pontryagin classes and Stiefel-Whitney classes is 
\bea
p_i(V)=w_{2i}(V)^2\mod2.
\eea

For a manifold $M$, the integrals over $M$ of characteristic classes of a vector bundle over $M$ (the pairing of the characteristic classes with the fundamental class of $M$) are called characteristic numbers.

Let $E_n$ be the universal $n$-bundle over $\B \tO(n)$, the colimit of $E_n-n$ is a virtual bundle $E$ (of dimension 0) over $\B \tO$, the pullback of $E$ along the map $g:M\to \B \tO$ given by the $\tO$-structure on $M$ is just $TM-d$ where $M$ is a 
$d$-manifold and $TM$ is the tangent bundle of $M$. By the naturality of characteristic classes, the pullback of the characteristic classes of $E$ is the characteristic classes of $TM$.

Chern-Simons form:
By Chern-Weil theory, Chern classes (and Pontryagin classes) can also be defined as a closed differential form (in de Rham cohomology). By Poincar\'e Lemma, they are exact locally:
\bea
c_n=\text{d}\text{CS}_{2n-1}
\eea
where d is the exterior differential operator, $\text{CS}_{2n-1}$ is called the Chern-Simons $2n-1$-form.

Whitney sum formula:
Let $w(V)=1+w_1(V)+w_2(V)+\cdots\in\H^*(M,\Z_2)$ be the total Stiefel-Whitney class,
$c(E)=1+c_1(E)+c_2(E)+\cdots\in\H^*(M,\Z)$ be the total Chern class,
$p(V)=1+p_1(V)+p_2(V)+\cdots\in\H^*(M,\Z)$ be the total Pontryagin class,
then 
\bea\label{SWsum}
w(V\oplus V')=w(V)w(V'),
\eea
\bea
c(E\oplus E')=c(E)c(E'),
\eea
\bea
2p(V\oplus V')=2p(V)p(V').
\eea
That is, the total Stiefel-Whitney class and the total Chern class are multiplicative with respect to Whitney sum of vector bundles, the total Pontryagin class is multiplicative modulo 2-torsion with respect to Whitney sum of vector bundles.

\subsubsection{Wu formulas}

The total Stiefel-Whitney class $w=1+w_1+w_2+\cdots$ is related to the
total Wu class $u=1+u_1+u_2+\cdots$ through the total Steenrod square:
\begin{align}
 w=\Sq(u),\ \ \ \Sq=1+\Sq^1+\Sq^2+ \cdots .
\end{align}
Therefore, 
$w_n=\sum_{i=0}^n \Sq^i (u_{n-i})$.
The Steenrod squares satisfy:
\begin{align}
\Sq^i(x_j) &=0, \  i>j, \ \ 
\Sq^j(x_j) =x_jx_j,  \ \  \Sq^0=1,
\end{align}
for any $x_j\in \H^j(M^d;\Z_2)$.
Thus
\begin{align}
u_n=w_n+\sum_{i=1, 2i\leq n} \Sq^i (u_{n-i}).
\end{align}
This allows us to compute $u_n$ iteratively, using Wu formula
\begin{align}
\label{WuF}
\Sq^i(w_j) &=0, \ \ i>j, \ \ \ \ \
\Sq^i(w_i) =w_iw_i, 
\\
 \Sq^i(w_j) &= w_iw_j+\sum_{k=1}^i \binom{j-i-1+k}{k}
w_{i-k} w_{j+k},\ \ i<j ,
\nonumber 
\end{align}
and the Steenrod relation 
\begin{align}\label{steenrel}
	\Sq^n(xy)=\sum_{i=0}^n \Sq^i(x)\Sq^{n-i}(y).
\end{align}
We find
\begin{align}
u_0&=1, 
\ \ \ \ \
u_1=w_1, 
\ \ \	 \ \
u_2=w_1^2+w_2, 
	\nonumber\\
u_3&=w_1w_2, 
\ \ \ \ \
u_4=w_1^4+w_2^2+w_1w_3+w_4, 
	\\
u_5&=w_1^3w_2+w_1w_2^2+w_1^2w_3+w_1w_4.
	\nonumber
\end{align}

On the tangent bundle of $M^d$, the corresponding Wu class and the
Steenrod square satisfy 
\begin{align}
\label{SqWu}
\Sq^{d-j}(x_j)=u_{d-j} x_j,  \text{ for any } x_j \in \H^j(M^d;\Z_2) .
\end{align}
This is also called Wu formula.

\subsection{Bockstein homomorphisms}\label{Bockstein}

In general, given a chain complex $C_{\bullet}$ and a short exact sequence of abelian groups:
\bea
0\to A'\to A\to A''\to 0,
\eea
we have a short exact sequence of cochain complexes:
\bea
0\to\Hom(C_{\bullet},A')\to\Hom(C_{\bullet},A)\to\Hom(C_{\bullet},A'')\to0.
\eea
Hence we obtain a long exact sequence of cohomology groups:
\bea
\cdots\to\H^n(C_{\bullet},A')\to\H^n(C_{\bullet},A)\to\H^n(C_{\bullet},A'')\stackrel{\partial}{\to}\H^{n+1}(C_{\bullet},A')\to\cdots,
\eea
the connecting homomorphism $\partial$ is called Bockstein homomorphism.

For example,
$\beta_{(n,m)}:\H^*(-,\Z_{m})\to\H^{*+1}(-,\Z_{n})$ is the Bockstein homomorphism associated to the extension $\Z_n\stackrel{\cdot m}{\to}\Z_{nm}\to\Z_m$.
%$\beta_n:\H^*(-,\Z_n)\to\H^{*+1}(-,\Z)$ is the Bockstein homomorphism associated to the extension $\Z\stackrel{\cdot n}{\to}\Z\to\Z_n$.

Let $\rho_{(nm,m)}:\H^*(-,\Z_{nm})\to\H^*(-,\Z_m)$ be the mod $m$ reduction map, then $\beta_{(n,m)}\rho_{(nm,m)}=0$ by the long exact sequence.
In particular, $\beta_{(2,2)}\rho_{(4,2)}=0$.

Relations between the Bockstein homomorphisms:
If we have a chain complex $C_{\bullet}$ and a commutative diagram of abelian groups with exact rows:
\bea
\xymatrix{0\ar[r]&C'\ar[r]\ar[d]&C\ar[r]\ar[d]&C''\ar[r]\ar[d]&0\\0\ar[r]&A'\ar[r]&A\ar[r]&A''\ar[r]&0},
\eea
then we have a commutative diagram of cochain complexes with exact rows:
\bea
\xymatrix{0\ar[r]&\Hom(C_{\bullet},C')\ar[r]\ar[d]&\Hom(C_{\bullet},C)\ar[r]\ar[d]&\Hom(C_{\bullet},C'')\ar[r]\ar[d]&0\\0\ar[r]&\Hom(C_{\bullet},A')\ar[r]&\Hom(C_{\bullet},A)\ar[r]&\Hom(C_{\bullet},A'')\ar[r]&0},
\eea

By the naturality of the connecting homomorphism \cite[Theorem 6.13]{rotman2008introduction}, we have a commutative diagram of abelian groups with exact rows:
\bea\label{naturality}
\xymatrix{\cdots\ar[r]&\H^n(C_{\bullet},C')\ar[r]\ar[d]&\H^n(C_{\bullet},C)\ar[r]\ar[d]&\H^n(C_{\bullet},C'')\ar[r]^{\partial}\ar[d]&\H^{n+1}(C_{\bullet},C')\ar[r]\ar[d]&\cdots\\\cdots\ar[r]&\H^n(C_{\bullet},A')\ar[r]&\H^n(C_{\bullet},A)\ar[r]&\H^n(C_{\bullet},A'')\ar[r]^{\partial'}&\H^{n+1}(C_{\bullet},A')\ar[r]&\cdots}
\eea

There are commutative diagrams:
\bea
\xymatrix{\Z_n\ar[r]^{\cdot m}\ar@{=}[d]&\Z_{nm}\ar[r]^{\mod m}\ar[d]^{\cdot k}&\Z_m\ar[d]^{\cdot k}\\
\Z_n\ar[r]^{\cdot km}&\Z_{knm}\ar[r]^{\mod km}&\Z_{km}}
\eea

\bea
\xymatrix{\Z_{kn}\ar[r]^{\cdot m}\ar[d]_{\mod n}&\Z_{knm}\ar[r]^{\mod m}\ar[d]_{\mod nm}&\Z_m\ar@{=}[d]\\
\Z_n\ar[r]^{\cdot m}&\Z_{nm}\ar[r]^{\mod m}&\Z_{km}}
\eea

%\bea
%\xymatrix{\Z\ar[r]^{\cdot m}\ar[d]_{\mod n}&\Z\ar[r]^{\mod m}\ar[d]_{\mod nm}&\Z_m\ar@{=}[d]\\
%\Z_n\ar[r]^{\cdot m}&\Z_{nm}\ar[r]^{\mod m}&\Z_{km}}
%\eea

By \eqref{naturality}, we have the following commutative diagrams:
\bea
\xymatrix{\H^*(-,\Z_m)\ar[r]^{\beta_{(n,m)}}\ar[d]^{\cdot k}&\H^{*+1}(-,\Z_n)\ar@{=}[d]\\
\H^*(-,\Z_{km})\ar[r]^{\beta_{(n,km)}}&\H^{*+1}(-,\Z_n)}
\eea

\bea
\xymatrix{\H^*(-,\Z_m)\ar[r]^{\beta_{(kn,m)}}\ar@{=}[d]&\H^{*+1}(-,\Z_{kn})\ar[d]^{\mod n}\\
\H^*(-,\Z_m)\ar[r]^{\beta_{(n,m)}}&\H^{*+1}(-,\Z_n)}
\eea

%\bea
%\xymatrix{\H^*(-,\Z_m)\ar[r]^{\beta_m}\ar@{=}[d]&\H^{*+1}(-,\Z)\ar[d]^{\mod n}\\
%\H^*(-,\Z_m)\ar[r]^{\beta_{(n,m)}}&\H^{*+1}(-,\Z_n)}
%\eea

Hence we have 

\bea\label{bsrel1}
\beta_{(n,m)}=\beta_{(n,km)}\cdot k,
\eea

and
\bea\label{bsrel2}
\rho_{(kn,n)}\beta_{(kn,m)}=\beta_{(n,m)}.
\eea

By definition,
\bea
\beta_{(2,2^n)}=\frac{1}{2^n}\delta\mod2
\eea
where $\delta$ is the coboundary map.

Moreover, $\Sq^1=\beta_{(2,2)}$.

By \eqref{bsrel2}, $\beta_{(2,4)}=\rho_{(4,2)}\beta_{(4,4)}$, thus 
$\beta_{(2,2)}\beta_{(2,4)}=\beta_{(2,2)}\rho_{(4,2)}\beta_{(4,4)}=0$.

Similarly, $\beta_{(2,8)}=\rho_{(4,2)}\beta_{(4,8)}$, thus 
$\beta_{(2,2)}\beta_{(2,8)}=\beta_{(2,2)}\rho_{(4,2)}\beta_{(4,8)}=0$, etc.

Combining this with the Adem relation $\Sq^1\Sq^1=0$, we obtain the important formula:
\bea\label{sq1beta}
\Sq^1\beta_{(2,2^n)}=0
\eea

\subsection{Useful fomulas}\label{useful}

Adem relations:
\bea\label{Adem}
\Sq^a\Sq^b=\sum_{j=0}^{[a/2]}\binom{b-1-j}{a-2j}\Sq^{a+b-j}\Sq^j
\eea
for $0<a<2b$.
In particular, we have $\Sq^1\Sq^1=0$, $\Sq^1\Sq^2\Sq^1=\Sq^2\Sq^2$.

Recall that 
\bea
\H^*(\B \Z_2,\Z_2)=\Z_2[a]
\eea
\cred{where $a$ is the generator of $\H^1(\B\Z_2,\Z_2)$.}
\bea
\H^*(\B ^2\Z_2,\Z_2)=\Z_2[x_2,x_3,x_5,x_9,\dots]
\eea
\cred{where $x_2$ is the generator of $\H^2(\B^2\Z_2,\Z_2)$, $x_3=\Sq^1x_2$, $x_5=\Sq^2x_3$, $x_9=\Sq^4x_5$, etc.}
\bea
\H^*(\B \PSU(2),\Z_2)=\Z_2[w_2',w_3']
\eea
\cred{where $w_i'$ is the $i$-th Stiefel-Whitney class of the universal $\PSU(2)=\SO(3)$ bundle.}

Combining \eqref{SqWu} and \eqref{steenrel}, we have \cred{the following useful formulas in the presentation of cobordsim invariants:}

\bea
a^2=\Sq^1a&=&w_1a\text{ in 2d}\\
x_3=\Sq^1x_2&=&w_1x_2 \text{ in 3d}\\
w_3'=\Sq^1w_2'&=&w_1w_2' \text{ in 3d}\\
\Sq^1(ax_2)&=&a^2x_2+ax_3=w_1ax_2 \text{ in 4d}\\
\Sq^2(ax_2)&=&ax_2^2+a^2x_3=(w_2+w_1^2)ax_2 \text{ in 5d}\\
x_5=\Sq^2x_3&=&(w_2+w_1^2)x_3 \text{ in 5d}\\
w_2'w_3'=\Sq^2(w_3')&=&(w_2+w_1^2)w_3' \text{ in 5d}\\
\Sq^1(w_2x_2)&=&(w_1w_2+w_3)x_2+w_2x_3=w_1w_2x_2\Rightarrow w_3x_2=w_2x_3\text{ in 5d}\\
\Sq^1(w_1^2x_2)&=&w_1^2x_3=w_1^3x_2\text{ in 5d}\\
\Sq^3x_2&=&w_1w_2x_2=0\text{ in 5d}\\
\Sq^1(w_2w_2')&=&(w_1w_2+w_3)w_2'+w_2w_3'=w_1w_2w_2'\Rightarrow w_3w_2'=w_2w_3'\text{ in 5d}\\
\Sq^1(w_1^2w_2')&=&w_1^2w_3'=w_1^3w_2'\text{ in 5d}\\
\Sq^3w_2'&=&w_1w_2w_2'=0\text{ in 5d}\\
\Sq^1(x_2^2)&=&w_1x_2^2=2x_2x_3=0\text{ in 5d}\\
\Sq^1(w_2'^2)&=&w_1w_2'^2=2w_2'w_3'=0\text{ in 5d}\\
\Sq^1(w_2'x_2)&=&w_3'x_2+w_2'x_3=w_1w_2'x_2\text{ in 5d}
\eea
where $w_i$ is the $i$-th Stiefel-Whitney class of the tangent bundle of $M$, all cohomology classes are pulled back to $M$ \cred{along the maps given in the definition of cobordism groups}.

\section{Warm-Up Examples}

\subsection{Perturbative chiral anomalies in even $d$d 
and associated Chern-Simons $(d+1)$-form theories --- SO- and Spin- Cobordism groups of BU(1)}

\label{sec:p-anom-ex-U1}

\subsubsection{Perturbative bosonic/fermionic anomaly in an even $d$d and U(1) SPTs in an odd $(d+1)$d}

We start from a warming-up example familiar to most physicists and quantum field theorists: the perturbative anomalies that can be captured by
a 1-loop calculation via Feynman-Dyson diagrams involved with a U(1) group. 
Of course, our discussion on the U(1) group can be generalized to any compact semi-simple Lie group such as SU(N), although we focus mostly on U(1) in this section.
We will consider a Dirac fermion theory in any even dimensional spacetime, denoted as $d$d with $d$ as the even integer (say $d=2,4,6,8,10, \dots$). 
The Dirac fermion $\Psi$ (or a complex Dirac spinor) is in a $2^{[d/2]}$-dimensional spinor representation of Spin(1,$d-1$) (or Spin($d$)  in the Euclidean signature, 
 where $\Spin(d)/\Z_2^F=\SO(d)$, with the continuous spacetime rotational symmetry SO$(d)$ and 
the fermion parity $\Z_2^F$ symmetry acts on any fermion $\Psi \to -\Psi$).
The Dirac fermion $\Psi$ can be coupled to non-dynamical U(1) or dynamical U(1) gauge fields, as \ref{m1} and \ref{m2} below respectively. 

\begin{enumerate} [label=\textcolor{blue}{Model (\arabic*)}:, ref={Model (\arabic*)}]
\item  \label{m1}
The U(1) is treated as a U(1) global internal symmetry for the 't Hooft anomaly.
The so-called path integral or partition function ${\bf Z}$ of this Dirac fermion theory is defined as a functional integral (here in Minkowski signature):
\be \label{eq:Dirac-Z}
{\bf Z}[A] :=\int [D \bar \Psi] [D \Psi]\exp\big(+ \ii S_{M,\text{Dirac}}\big)
\equiv
\int [D \bar \Psi] [D \Psi]\exp\big(+ \ii \int_{M^d} \dd^d x ( \bar \Psi (\ii \slashed{D}_A) \Psi \big),
\ee
where $\slashed{D}_A$ is the Dirac operator endorsed with the Feynman slash notation, defined as:
\bea
\slashed{D}_A :=\gamma^\mu  D_{\mu} =\gamma^\mu (\prt_{\mu} - \ii g A_{\mu}).
\eea
The $g$ is the coupling constant for the non-dynamical 1-form U(1) gauge field $A:=A_\mu \dd x^\mu$. 
The  $\gamma^\mu$ with $\mu=0,1,\dots, d-1$ are so-called 
 gamma matrices, satisfying the Clifford algebra $\rm{C}\ell_{1, d-1} (\mathbb{R})$ under the anti-commutator constraint:
 \bea
 \{ \gamma^\mu, \gamma^\nu \} := \gamma^\mu \gamma^\nu + \gamma^\nu \gamma^\mu = 2 \eta^{\mu \nu} \mathbb{I}_{2^{[d/2]}}
 \eea
 The  standard Dirac matrices correspond to $d = 2^[d/2] = 4$.
 The $\eta^{\mu \nu}$ is the Minkowski metric 
 $$
 \eta^{\mu \nu}:=\rm{diag}(+,-,-,\dots,-),
$$
with one $+$ sign and $(d-1)$ $-$ sign along the diagonal.
The hermitian chiral matrix $\gamma^{\text{Chiral}} \equiv \gamma^\text{FIVE}$ can be defined for even $d$ dimensions 
\bea
\gamma^{\text{Chiral}}  \equiv \gamma^\text{FIVE}:=  \ii^{d/2-1}    \gamma^0 \gamma^1 \dots \gamma^{d-1}.
\eea

\item  \label{m2}
The U(1) is treated as a U(1) gauge group for the dynamical gauge anomaly. 
The so-called path integral or partition function ${\bf Z}$ of this Dirac fermion coupled to a dynamicsl U(1) gauge field 
theory is defined as a functional integral (here in Minkowski signature):
\bea
{\bf Z} &:=&\int [D \bar \Psi] [D \Psi][DA]\exp\big(+ \ii S_{M,\text{Dirac-U(1) gauge theory}}\big) \nonumber\\
&\equiv&
\int [D \bar \Psi] [D \Psi][DA]\exp\big(+ \ii \int_{M^d}  \dd^d x (\bar \Psi (\ii \slashed{D}_A) \Psi -\frac{1}{4} F_{\mu\nu} F^{\mu\nu})\big).
\eea
Here the dynamical 1-form U(1) gauge field $A$ is integrated over in the path integral measure $\int [DA]$ as a dynamical gauge variable.  For the 
quantum electrodynamics (QED) as a Dirac-U(1) gauge theory, it is commonly defined as 
$\slashed{D}_A :=\gamma^\mu  D_{\mu} =\gamma^\mu (\prt_{\mu} + \ii e A_{\mu})$ where $e$ is the electric charge constant.

\end{enumerate} 

For the spacetime index $\mu=0,1, \dots, d-1$,\\
the left-moving current $J^{\mu,{\text{L}}}$ is defined as:
\bea
J^{\mu,{\text{L}}} := \bar \Psi \gamma^\mu (\frac{1-\gamma^{\text{Chiral}}}{2}) \Psi .
\eea
The right-moving current $J^{\mu,{\text{R}}}$ is defined as:
\bea
J^{\mu,{\text{R}}}  := \bar \Psi \gamma^\mu (\frac{1+\gamma^{\text{Chiral}}}{2}) \Psi .
\eea
The vector current $J^{\mu}\equiv J^{\mu,{\rm V}}$ is defined as:
\bea
J^{\mu,{\text{V}}}
 := J^{\mu,{\text{L}}} + J^{\mu,{\text{R}}} \equiv \bar \Psi \gamma^\mu  \Psi .
\eea
The axial chiral current $J^{\mu}\equiv J^{\mu,{\rm A}} \equiv J^{\mu}\equiv J^{\mu,{\rm Chiral}}$ is defined as:
\bea
J^{\mu,{\text{A}}}
 := J^{\mu,{\text{L}}} - J^{\mu,{\text{R}}}\equiv \bar \Psi \gamma^\mu \gamma^{\text{Chiral}} \Psi .
\eea
We also define the left and right-handed Weyl fermions, $\Psi_{\text{L}}$ and $\Psi_{\text{R}}$, projected from the Dirac fermion via:
\bea
\Psi_{\text{L}} &:=&(\frac{1-\gamma^{\text{Chiral}}}{2}) \Psi,\\
\Psi_{\text{R}} &:=&(\frac{1+\gamma^{\text{Chiral}}}{2}) \Psi.
\eea
In the classical theory (without doing the path integral), the classical Dirac theory has both the continuous vector symmetry U(1)$_{\text{V}}$ and
the continuous axial symmetry (or the so-called chiral symmetry)  U(1)$_{\text{A}}$, given by the following symmetry transformation: 
\bea
\U(1)_{\text{V}}:&& \Psi_{} \to \exp(\ii \alpha_{\text{V}} )\Psi_{}  \\
&& \Psi_{\text{L}} \to \exp(\ii \alpha_{\text{V}} )\Psi_{\text{L}} \nn\\
&& \Psi_{\text{R}} \to \exp(\ii \alpha_{\text{V}} )\Psi_{\text{R}}. \nn\\
\U(1)_{\text{A}}:&& 
 \Psi_{} \to \exp(\ii \alpha_{\text{A}} \gamma^{\text{Chiral}} )\Psi_{} \\
&& \Psi_{\text{L}} \to \exp(\ii \alpha_{\text{A}} )\Psi_{\text{L}} \nn\\
&& \Psi_{\text{R}} \to \exp(-\ii \alpha_{\text{A}} )\Psi_{\text{R}}.\nn 
\eea
Under the Noether theorem, the corresponding continuous currents are 
$J^{\mu,{\text{V}}}$ and $J^{\mu,{\text{A}}}$ respect to the U(1)$_{\text{V}}$ and U(1)$_{\text{A}}$ symmetry respectively.
In a classical theory, both U(1)$_{\text{V}}$ and U(1)$_{\text{A}}$ symmetries are global symmetries.

However, in the quantum theory, we need do the path integral to get the partition function ${\bf Z}[A]$ for a quantum theory in \eqn{eq:Dirac-Z}.
Now under the continuous axial  (or chiral) symmetry transformation labeled by a U(1) parameter
$\alpha_{\rm A} \in [0, 2 \pi)$, the partition function ${\bf Z}[A]$ shifts to
\begin{multline}
{\bf Z}[A] \to 
\int [D \bar \Psi] [D \Psi]\exp\big(+ \ii \int_{M^d} \dd^d x ( \bar \Psi (\ii \slashed{D}_A) \Psi +\\ \alpha_{\rm A} \Big( \prt_\mu J^{\mu,{\text{Chiral}}}+
\frac{2 g^{d/2}}{({d/2})! (4 \pi)^{{d/2}}}  
\epsilon^{\mu_1\mu_2 \dots \mu_{d}}
F_{\mu_1\mu_2} \dots F_{\mu_{d-1}\mu_{d}}
\Big)  ) \big),
\end{multline}
(In terms of quantum electrodynamics notation, people set the $g=-e$.)
This means the axial (chiral) current is \emph{not} conserved $ \prt_\mu J^{\mu,{\text{Chiral}}} \neq 0$:
If the classical gauge field $A$ has a nontrivial background for $F \wedge \dots F$ term in \ref{m1} where $F= \dd A$. 
The above calculation can be done based on the Fujikawa's path integral method\cite{Fujikawa2004cxSuzukibook}.

The non-conservation of the axial (chiral) current has the form:
\bea \label{eq:dJ5}
 \prt_\mu J^{\mu,{\text{Chiral}}} = -
\frac{2 g^{d/2}}{({d/2})! (4 \pi)^{{d/2}}}  
\epsilon^{\mu_1\mu_2 \dots \mu_{d}}
F_{\mu_1\mu_2} \dots F_{\mu_{d-1}\mu_{d}}.
\eea
In the differential form in terms of the top form paired with the fundamental class of the spacetime manifold, we can rewrite the above formula \eqn{eq:dJ5} as:
\bea  \label{eq:dJ5-diff}
 (\dd \star J^{{\text{Chiral}}} ) \propto -
g^{d/2}  
 (F \wedge \dots  \wedge F).
\eea
The above formula is for the 't Hooft anomaly associated with the probed background Abelian gauge fields (here U(1)).

If we instead consider the background non-Abelian gauge fields (like SU(n)),
then the $(F \wedge \dots  \wedge F)$ with $F= \dd A$ is replaced to
a non-abelian field strength $\mathbf{F} = \dd\mathbf{A}+\mathbf{A}\wedge\mathbf{A}$;
while $(F \wedge \dots  \wedge F)$ is replaced by ${\rm Tr}  \left( \mathbf{F}^{d/2} \right)$:
\bea
\dd \omega_{d-1}={\rm Tr}  \left( \mathbf{F}^{d/2} \right).
\eea
The $\omega_{d-1}$  is the Chern-Simons $(d-1)$-form \cite{Chern1974ft} as the secondary characteristic classes: 
\bea
 \omega_{1} &\propto &{\rm Tr} [ \mathbf{A} ]\\
 \omega_{3} &\propto &{\rm Tr} \left[ \mathbf{F}\wedge\mathbf{A}-\frac{1}{3}\mathbf{A}\wedge\mathbf{A}\wedge\mathbf{A}\right] \nn \\
 \omega_{5} &\propto &{\rm Tr} \left[ \mathbf{F}\wedge\mathbf{F}\wedge\mathbf{A}-\frac{1}{2}\mathbf{F}\wedge\mathbf{A}\wedge\mathbf{A}\wedge\mathbf{A} +\frac{1}{10}\mathbf{A}\wedge\mathbf{A}\wedge\mathbf{A}\wedge\mathbf{A}\wedge\mathbf{A} \right] \nn\\
 && \dots \nn
\eea

Other than Fujikawa's path integral method\cite{Fujikawa2004cxSuzukibook}, we can also capture the perturbative anomaly via a 1-loop Feynman-Dyson diagram calculation.
The vertex term means $g \bar{\Psi} \gamma^\mu A_{\mu} \Psi := g \bar{\Psi} \slashed{A} \Psi $

\begin{figure}[!h]
\centering
(1) \includegraphics[scale=1.1]{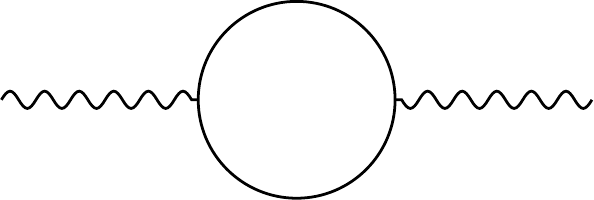}
(2) \includegraphics[scale=.9]{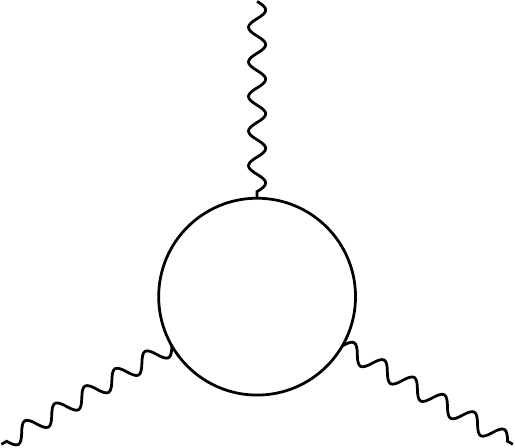}\\
(3) \includegraphics[scale=.9]{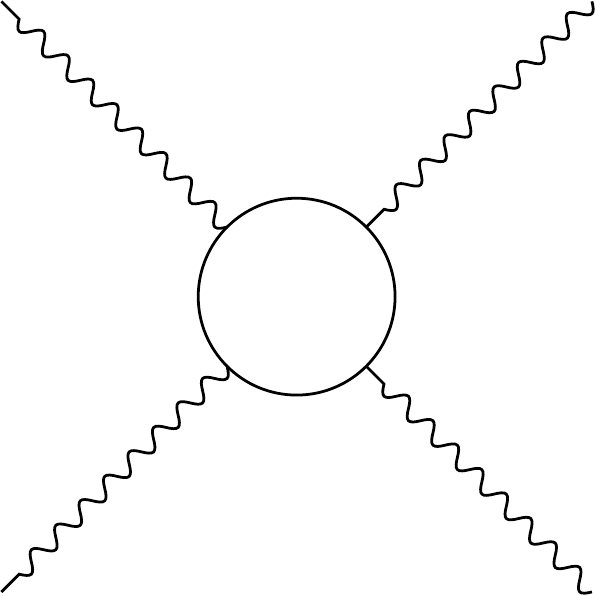}
(4) \includegraphics[scale=1.1]{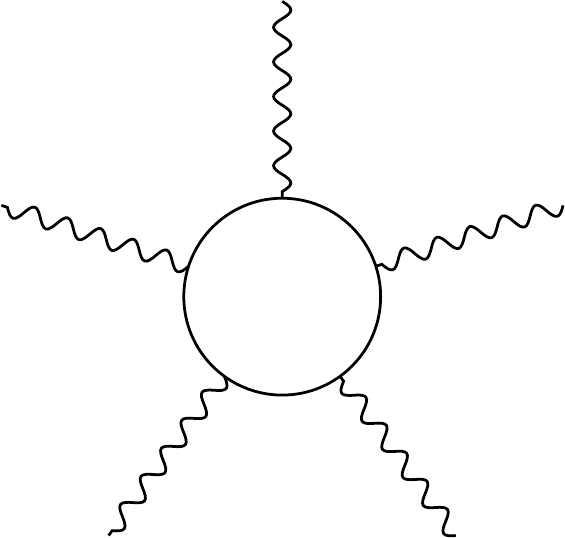}\\
(5)  \includegraphics[scale=1.1]{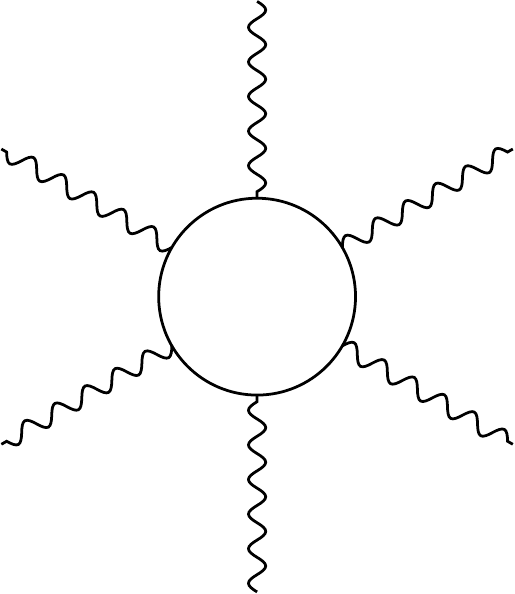}
\caption{Perturbative anomalies with an integer $\mathbb{Z}$ class in an even $d$-dimensional spacetime. Here the anomalies are captured by a
1-loop Feynman diagram with a number $(\frac{d}{2}+1)$ amount of vertex terms $g \bar{\Psi} \slashed{A} \Psi$ 
and one vertex term among the total $(\frac{d}{2}+1)$ vertex terms can be associated with the $J^{{\text{Chiral}}}$ current.
The chiral fermion runs on the solid-line loop (---).
The wavy line (\protect\middlewave{.42cm}) represents the propagator (Green's function) of 1-form vector boson gauge field. 
(The gauge field is a probed background gauge field for the 't Hooft anomaly.)
In the subfigures, we show 1-loop Feynman diagrams for (1) 2d anomaly, (2) 4d anomaly, (3) 6d anomaly,  (4) 8d anomaly, and (5) 10d anomaly, etc., 
of chiral fermions coupled to U(1) background probed gauge fields. A physical explanation of $\mathbb{Z}$ class is given in
\Sec{sec:3d-b-Z} for 2d bosonic anomaly and \Sec{sec:3d-f-Z}  for 2d fermionic anomaly.
Similarly, the analysis can be generalized to any even $d$d by writing down a one-higher dimensional Chern-Simons theory given by Chern-Simons $(d+1)$d form.
}
\end{figure}

\clearpage

How do we obtain an integer $\Z$ class for perturbative anomalies in an even $d$-dimensional spacetime?
Say from the formulas of \eqn{eq:dJ5} and \eqn{eq:dJ5-diff}?
The answer is that we can consider the
modified axial symmetry transformation $\U(1)_{\text{A},k}$ labeled by a integer charge 
$k \in \Z$ class, such that the $\Psi_{\text{L}}$ and $\Psi_{\text{R}}$ transformed differently,
\bea
\U(1)_{\text{A},k}:
&& \Psi_{\text{L}} \to \exp(\ii \alpha_{\text{A},k} )\Psi_{\text{L}}, \\
&& \Psi_{\text{R}} \to \exp(-\ii k \alpha_{\text{A},k} )\Psi_{\text{R}}.\nn \\
&&  k \in \Z. \nn
\eea
The chiral symmetry transformation is labeled by a U(1) parameter $\alpha_{\text{A},k} \in [0, 2 \pi)$.
Below we give
a physical explanation of a $\Z$ class  in
\Sec{sec:3d-b-Z} for 2d bosonic anomaly, and also of a $\Z$ class \Sec{sec:3d-f-Z}  for 2d fermionic anomaly.
Below our derivation is in a similar spirit of 2d anomaly and 3d Chern-Simons theory \cite{Lu2012dt1205.3156, 1405.7689},
but ours is generalizable to arbitary dimensions analogs to \cite{1405.7689}.

In general, we suggest that the for generic even $d$d U(1) bosonic or fermionic anomalies (on non-spin or spin manifolds respectively) 
can be captured by a
partition function depending on the probed U(1) background gauge field $A$:
\bea \label{eq:Z=AFd}
\bZ[A] = \exp[ \ii  \frac{ c \cdot k}{(2 \pi)^{{d}/{2}}} \int A \wedge (F)^{{d}/{2}} ], \;\;\; k \in\Z
\eea
where the precise normalization $c$ depends on the dimensions $d$, and non-spin or spin manifolds,
with the integer $k \in \Z$ class.

\subsubsection{2d anomaly and 3d bosonic-U(1) SPTs: 
Integer $\Z$ class $\in \TP_3(\SO\times\U(1))= \Z^2$} \label{sec:3d-b-Z}

Below we explicitly  derive the analogous \eqn{eq:Z=AFd} for $d=2$ bosonic anomaly (on non-spin manifolds).
The physics idea is that we write down the internal field theory with dynamical gauge fields $a$ coupled to background non-dynamical 
gauge fields $A$, and integrate out those
internal degrees of freedom to get a response theory depending on $A$.

The symmetric bilinear form $K$ matrix Chern-Simons theory with a U(1)$^2$ gauge group 
of internal dynamical $a$ gauge field, and the charge $q$ vector coupling to the background U(1) gauge fields $A$ are the following:
$$K:=K_{IJ}=
\begin{pmatrix}
0 & 1\\
1 & 2k
\end{pmatrix}_{IJ},$$ 
and $q^T=(1,1)$.
In other words, the path integral, written by dynamical gauge fields $a$ and background fields $A$, is
\bea
\bZ[A]=\int [Da] \exp[ \ii  
\big(
\frac{1}{4 \pi} 
\begin{pmatrix}
0 & 1\\
1 & 2k
\end{pmatrix}_{IJ} \int a_I \wedge \dd a_J
+
  \frac{1}{2 \pi} q^T_I  \int A \wedge \dd a_I \vert_{ q^T=(1,1)}
  \big)], 
\eea
Under $\GL(2,\Z)$ or $\SL(2,\Z)$ redefinition of gauge fields, the $K$ can be changed to
$K=\begin{pmatrix}
0 & 1\\
1 & 0
\end{pmatrix},$
and $q^T=(1,k)$.
In other words, the path integral is
\bea \label{eq:Z=AFd-b}
\bZ [A]= \int [Da] \exp[ \ii  
\big(
\frac{1}{4 \pi} 
\begin{pmatrix}
0 & 1\\
1 & 0
\end{pmatrix}_{IJ} \int a_I \wedge \dd a_J
+
  \frac{1}{2 \pi} q^T_I  \int A \wedge \dd a_I \vert_{ q^T=(1,k)}
  \big)], 
\eea

If we integrate out the dynamical internal gauge field $a$ (``emergent'' from the gapped matter field) of SPTs, we obtain the partition function of probed background field
\bea
\bZ [A]=\exp[ \ii  \frac{2k}{4 \pi} \int A \wedge \dd A], \;\;\; k \in\Z
\eea
This Chern-Simons field theory characterizes the low energy physics of a quantum Hall state and its response function.
So the effective bulk quantized Hall conductance is labeled by $2k$ in $2\Z$, as
$$
\sigma_{xy} = \frac{q K^{-1}q}{2 \pi} (\frac{e^2}{\hbar})=\frac{2k}{2 \pi} (\frac{e^2}{\hbar}) ={2k} (\frac{e^2}{h}).
$$
The boundary theory has a $\Z$ class of perturbative Adler-Bell-Jackiw type of U(1)-axial-background gauge anomaly. 
(Due to the L and R chiral fermion carrying imbalanced U(1) charges, in 
$K=\begin{pmatrix}
0 & 1\\
1 & 0
\end{pmatrix},$ 
and $q=(1,k)$.)

The above physics derivation coincides with the mathematical cobordism group calculation,
matching one of the integer $\Z$ class $\in \TP_3(\SO\times\U(1))= \Z^2$ shown later in our Theorem
\ref{thm:TPi(SOU(1)}. 
The reason we require the (co)bordism group of $\SO\times\U(1)$ is due to that the bosonic system has a continuous spacetime SO$(d)$ symmetry (in the 
$d$d Euclidean signature), while the boson has an internal U(1) symmetry.

Similarly, the above analysis can be generalized to any even $d$d by writing down a one-higher dimensional bosonic 
(on manifolds with SO$(d+1)$ or non-spin structures) 
Chern-Simons theory given by a certain Chern-Simons $(d+1)$d form.

\subsubsection{2d anomaly and 3d fermionic-U(1) SPTs:
Integer $\Z$ class %$\in \TP_3(\Spin\times\U(1))= \Z^2$
$\in \TP_3(\frac{\Spin\times\U(1)}{\Z_2})= \Z^2$}
\label{sec:3d-f-Z}

Similar to \Sec{sec:3d-b-Z},
for a fermionic theory, we should have a SPT invariant of U(1) background gauge field,
\bea \label{eq:Z=AFd-f}
\bZ[A]=\exp[ \ii  \frac{k}{4 \pi} \int A \wedge \dd A], \;\;\; k \in\Z.
\eea
This may be obtained from integrating out the fermionic SPTs
$K=\begin{pmatrix}
1 & 0\\
0 & -1
\end{pmatrix},$
and $q^T=(1,-k+1)$ or simply $q^T=(0,-k)$.
In other words, the path integral is
\bea
\bZ[A]= \int [Da] \exp[ \ii  
\big(
\frac{1}{4 \pi} 
\begin{pmatrix}
1 & 0\\
0 & -1
\end{pmatrix}_{IJ} \int a_I \wedge \dd a_J
+
  \frac{1}{2 \pi} q^T_I  \int A \wedge \dd a_I \vert_{ q^T=(1, -k+1)  }
  \big)], 
\eea
This Chern-Simons field theory characterizes the low energy physics of another fermionic quantum Hall state and its response function.
So the effective bulk quantized Hall conductance is labeled by $k$ in $\Z$, as
$$
\sigma_{xy} = \frac{q K^{-1}q}{2 \pi} (\frac{e^2}{\hbar})=\frac{k}{2 \pi} (\frac{e^2}{\hbar}) ={k} (\frac{e^2}{h}).
$$

%Compare to arXiv: 1205.3156 \cite{Lu2012dt1205.3156}.

The above physics derivation coincides with the mathematical cobordism group calculation,
matching one of the integer $\Z$ class $\in \TP_3(\frac{\Spin\times\U(1)}{\Z_2})=\TP_3( {\Spin^c})= \Z^2$ shown later in our Theorem
\ref{thm:TPi(Spinc)}.
%\ref{thm:TPi(SpinU(1)}.
%

The reason we require the (co)bordism group of $(\frac{\Spin\times\U(1)}{\Z_2}) \equiv {\Spin^c} $ is due to that this 
fermionic system has a continuous spacetime Spin$(d)$ symmetry (in the 
$d$d Euclidean signature) under, the extension of SO$(d)$ via $1 \to \Z_2^F \to \Spin(d) \to \SO(d) \to 1$,
  while the fermion has an internal U(1) $\supset \Z_2^F$ containing the fermion parity symmetry.
 A common normal subgroup
${\Z_2^F}$ is mod out due to the fact that rotating a fermion by $2 \pi$ in the
spacetime (i.e., the spin statistics) gives rise to the same fermion parity
minus sign for the fermion field $\Psi \to -\Psi$.

Similarly, the above analysis can be generalized to any even $d$d by writing down a one-higher dimensional fermionic 
(on manifolds with Spin$(d+1)$ thus spin structures)
Chern-Simons theory given by a certain Chern-Simons $(d+1)$d form.

%\clearpage

\subsubsection{$\Omega_d^{\SO}(\B\U(1))$}

Since the computation involves no odd torsion, we can use the Adams spectral sequence 
\bea
E_2^{s,t}=\Ext_{\A_2}^{s,t}(\H^*(M\SO\wedge\B\U(1)_+,\Z_2),\Z_2)\Rightarrow\pi_{t-s}(M\SO\wedge\B\U(1)_+)_2^{\wedge}=\Omega^{\SO}_{t-s}(\B\U(1)).
\eea
The mod 2 cohomology of Thom spectrum $M\SO$ is
\bea
\H^*(M\SO,\Z_2)=\A_2/\A_2 \Sq^1\oplus\Sigma^4\A_2/\A_2 \Sq^1\oplus\Sigma^5\A_2\oplus\cdots.
\eea
\bea
\cdots\To\Sigma^3\A_2\To\Sigma^2\A_2\To\Sigma\A_2\To\A_2\To\A_2/\A_2 \Sq^1
\eea
is an $\A_2$-resolution where the differentials $d_1$ are induced by $\Sq^1$.

We also have
\bea
\H^*(\B\U(1),\Z_2)=\Z_2[c_1]
\eea
where $c_1$ is the first Chern class of the universal $\U(1)$ bundle.

The $E_2$ page is shown in Figure \ref{fig:Omega_*^{SO}(BU(1))}.

\begin{figure}[!h]
\begin{center}
\begin{tikzpicture}
\node at (0,-1) {0};
\node at (1,-1) {1};
\node at (2,-1) {2};
\node at (3,-1) {3};
\node at (4,-1) {4};
\node at (5,-1) {5};
\node at (6,-1) {6};
\node at (7,-1) {7};
\node at (8,-1) {$t-s$};
\node at (-1,0) {0};
\node at (-1,1) {1};
\node at (-1,2) {2};
\node at (-1,3) {3};
\node at (-1,4) {4};
\node at (-1,5) {5};
\node at (-1,6) {$s$};

\draw[->] (-0.5,-0.5) -- (-0.5,6);
\draw[->] (-0.5,-0.5) -- (8,-0.5);

\draw (0,0) -- (0,5);

\draw (2,0) -- (2,5);
\draw (4,0) -- (4,5);
\draw (4.1,0) -- (4.1,5);

\draw[fill] (5,0) circle(0.05);
\draw (6,0) -- (6,5);
\draw (6.1,0) -- (6.1,5);
\draw[fill] (7,0) circle(0.05);
\end{tikzpicture}
\end{center}
\caption{$\Omega_*^{\SO}(\B\U(1))$}
\label{fig:Omega_*^{SO}(BU(1))}
\end{figure}

Hence we have the following theorem
\begin{theorem}
\begin{table}[!h]
\centering
\begin{tabular}{c c}
\hline
$i$ & $\Omega^{\SO}_i(\B\U(1))$\\
\hline
0& $\Z$\\
1& $0$\\
2& $\Z$\\
3 & $0$\\
4 & $\Z^2$\\ 
5 & $\Z_2$\\
6 & $\Z^2$\\
7 & $\Z_2$\\
\hline
\end{tabular}
\end{table}
\end{theorem}
The bordism invariant of $\Omega_{2}^{\SO}(\B\U(1))$ is $c_1$.

The bordism invariants of $\Omega_{4}^{\SO}(\B\U(1))$ are $\sigma,c_1^2$.

Here $\sigma$ is the signature of a 4-manifold.

The bordism invariant of $\Omega^{\SO}_5(\B\U(1))$ is $w_2w_3$.

The bordism invariants of $\Omega_{6}^{\SO}(\B\U(1))$ are $\sigma c_1,c_1^3$.

Here $\sigma c_1=\sigma(\text{PD}(c_1))$ where $\text{PD}(c_1)$ is the submanifold of the 6-manifold which represents the Poinca\'e dual of $c_1$. 

The bordism invariant of $\Omega^{\SO}_7(\B\U(1))$ is $c_1w_2w_3$.

\begin{theorem}
\begin{table}[!h]
\centering
\begin{tabular}{c c}
\hline
$i$ & $\TP_i(\SO\times\U(1))$\\
\hline
0& $0$\\
1& $\Z$\\
2& $0$\\
3 & $\Z^2$\\
4 & $0$\\ 
5 & $\Z^2\times\Z_2$\\
6 & $0$\\ 
\hline
\end{tabular}
\caption{
Note that one of the $\Z$ classes in 
 $\TP_3(\SO\times\U(1))=\Z^2$
is given by
\eqn{eq:Z=AFd-b}. 
Similarly, one of the $\Z$ classes in 
 $\TP_{d+1}(\SO\times\U(1))$ for an even $d$ is given by
 \eqn{eq:Z=AFd}.
}
\end{table}
\label{thm:TPi(SOU(1)}
\end{theorem}

The 1d topological term is the Chern-Simons 1-form $\text{CS}_1^{(\U(1))}$ of the $\U(1)$ bundle.

The 3d topological terms are $\frac{1}{3}\text{CS}_3^{(TM)}$ and $\text{CS}_1^{(\U(1))}c_1$ where $\text{CS}_3^{(TM)}$ is the Chern-Simons 3-form of the tangent bundle.

The 5d topological terms are $c_1\frac{1}{3}\text{CS}_1^{(TM)}$, $\text{CS}_1^{(\U(1))}c_1^2$ and $w_2w_3$.

\subsubsection{$\Omega_d^{\Spin}(\B\U(1))$}

Since the computation involves no odd torsion, we can use the Adams spectral sequence 
\bea
E_2^{s,t}=\Ext_{\A_2}^{s,t}(\H^*(M\Spin\wedge\B\U(1)_+,\Z_2),\Z_2)\Rightarrow\pi_{t-s}(M\Spin\wedge\B\U(1)_+)_2^{\wedge}=\Omega^{\Spin}_{t-s}(\B\U(1)).
\eea
The mod 2 cohomology of Thom spectrum $M\Spin$ is
\bea
\H^*(M\Spin,\Z_2)=\A_2\otimes_{\A_2(1)}\{\Z_2\oplus M\}
\eea
where $M$ is a graded $\A_2(1)$-module with the degree $i$ homogeneous part $M_i=0$ for $i<8$. Here $\A_2(1)$ stands for the subalgebra of $\A_2$ generated by $\Sq^1$
and $\Sq^2$.
For $t-s<8$, we can identify the $E_2$-page with 
$$\Ext_{\A_2(1)}^{s,t}(\H^*(\B\U(1),\Z_2),\Z_2).$$

The $\A_2(1)$-module structure of $\H^*(\B\U(1),\Z_2)$ is shown in Figure \ref{fig:H^*(BU(1),Z_2)}.

\begin{figure}[!h]
\begin{center}
\begin{tikzpicture}[scale=0.5]
\node[below] at (0,0) {1};

\draw[fill] (0,0) circle(0.1);

\node[left] at (0,2) {$c_1$};

\draw[fill] (0,2) circle(0.1);
\node[left] at (0,4) {$c_1^2$};
\draw[fill] (0,4) circle(0.1);

\draw (0,2) to [out=150,in=150] (0,4);
\node[left] at (0,6) {$c_1^3$};
\draw[fill] (0,6) circle(0.1);
\node[left] at (0,8) {$c_1^4$};
\draw[fill] (0,8) circle(0.1);

\draw (0,6) to [out=150,in=150] (0,8);

\end{tikzpicture}
\end{center}
\caption{The $\A_2(1)$-module structure of $\H^*(\B\U(1),\Z_2)$.}
\label{fig:H^*(BU(1),Z_2)}
\end{figure}

The $E_2$ page is shown in Figure \ref{fig:Omega_*^{Spin}(BU(1))}.

\begin{figure}[!h]
\begin{center}
\begin{tikzpicture}
\node at (0,-1) {0};
\node at (1,-1) {1};
\node at (2,-1) {2};
\node at (3,-1) {3};
\node at (4,-1) {4};
\node at (5,-1) {5};
\node at (6,-1) {6};
\node at (7,-1) {7};
\node at (8,-1) {$t-s$};
\node at (-1,0) {0};
\node at (-1,1) {1};
\node at (-1,2) {2};
\node at (-1,3) {3};
\node at (-1,4) {4};
\node at (-1,5) {5};
\node at (-1,6) {$s$};

\draw[->] (-0.5,-0.5) -- (-0.5,6);
\draw[->] (-0.5,-0.5) -- (8,-0.5);

\draw (0,0) -- (0,5);
\draw (0,0) -- (2,2);
\draw (2.1,0) -- (2.1,5);

\draw (4,3) -- (4,5);
\draw (4.1,1) -- (4.1,5);

\draw (6,2) -- (6,5);
\draw (6.1,0) -- (6.1,5);
\end{tikzpicture}
\end{center}
\caption{$\Omega_*^{\Spin}(\B\U(1))$}
\label{fig:Omega_*^{Spin}(BU(1))}
\end{figure}

Hence we have the following theorem
\begin{theorem}
\begin{table}[!h]
\centering
\begin{tabular}{c c}
\hline
$i$ & $\Omega^{\Spin}_i(\B\U(1))$\\
\hline
0& $\Z$\\
1& $\Z_2$\\
2& $\Z\times\Z_2$\\
3 & $0$\\
4 & $\Z^2$\\ 
5 & $0$\\
6 & $\Z^2$\\ 
7 & $0$\\
\hline
\end{tabular}
\end{table}
\end{theorem}
The bordism invariant of $\Omega_1^{\Spin}(\B\U(1))$ is $\tilde{\eta}$.

Here $\tilde{\eta}$ is the ``mod 2 index'' of the 1d Dirac operator (\#zero eigenvalues mod 2, no contribution from spectral asymmetry).

The bordism invariants of $\Omega_2^{\Spin}(\B\U(1))$ are $c_1$ and $\text{Arf}$ (the Arf invariant).

The bordism invariants of $\Omega_4^{\Spin}(\B\U(1))$ are $\frac{\sigma}{16}$ and $\frac{c_1^2}{2}$.

Here $c_1^2$ is divided by 2 since $c_1^2=\Sq^2c_1=(w_2(TM)+w_1(TM)^2)c_1=0\mod2$ on Spin 4-manifolds.

The bordism invariants of $\Omega_6^{\Spin}(\B\U(1))$ are $c_1^3$ and \cred{$\frac{c_1(\sigma-F\cdot F)}{8}$}.

\cred{
Here $\frac{c_1(\sigma-F\cdot F)}{8}$ is defined to be $\frac{1}{8}(\sigma(\text{PD}(c_1))-F\cdot F)$ where $\text{PD}(c_1)$ is the submanifold of a Spin 6-manifold which represents the Poincar\'e dual of $c_1$, $\sigma(\text{PD}(c_1))$ is the signature of $\text{PD}(c_1)$, and $F$ is a characteristic surface of $\text{PD}(c_1)$. By Rokhlin's theorem, $\sigma(\text{PD}(c_1))-F\cdot F$ is a multiple of 8 and $\frac{1}{8}(\sigma(\text{PD}(c_1))-F\cdot F)=\text{Arf}(\text{PD}(c_1),F)\mod2$. See \cite{Saveliev}'s Lecture 10 for more details.
}

\begin{theorem}
\begin{table}[!h]
\centering
\begin{tabular}{c c}
\hline
$i$ & $\TP_i(\Spin\times\U(1))$\\
\hline
0& $0$\\
1& $\Z\times\Z_2$\\
2& $\Z_2$\\
3 & $\Z^2$\\
4 & $0$\\ 
5 & $\Z^2$\\
6 & $0$\\ 
\hline
\end{tabular}
\end{table}
\end{theorem}

The 1d topological terms are $\text{CS}_1^{(\U(1))}$ and $\tilde{\eta}$.

The 2d topological term is $\text{Arf}$.

The 3d topological terms are $\frac{1}{48}\text{CS}_3^{(TM)}$ and $\frac{1}{2}\text{CS}_1^{(\U(1))}c_1$.

The 5d topological terms are $\text{CS}_1^{(\U(1))}c_1^2$ and $\mu(\text{PD}(c_1))$.

Here $\mu(\text{PD}(c_1))$ is the Rokhlin invariant (see \cite{Saveliev}'s Lecture 11) of $\text{PD}(c_1)$ where $\text{PD}(c_1)$ is the submanifold of a Spin 5-manifold which represents the Poincar\'e dual of $c_1$

\subsubsection{$\Omega_d^{(\frac{\Spin\times \U(1)}{\Z_2})}=\Omega_d^{\Spin^c}$}

Since the computation involves no odd torsion, we can use the Adams spectral sequence 
\bea
E_2^{s,t}=\Ext_{\A_2}^{s,t}(\H^*(MT\Spin^c,\Z_2),\Z_2)\Rightarrow\pi_{t-s}(MT\Spin^c)_2^{\wedge}=\Omega^{\Spin^c}_{t-s}.
\eea
By 1801.07530, we have $MT\Spin^c=M\Spin\wedge\Sigma^{-2}M\U(1)$.

The mod 2 cohomology of Thom spectrum $M\Spin$ is
\bea
\H^*(M\Spin,\Z_2)=\A_2\otimes_{\A_2(1)}\{\Z_2\oplus M\}
\eea
where $M$ is a graded $\A_2(1)$-module with the degree $i$ homogeneous part $M_i=0$ for $i<8$. Here $\A_2(1)$ stands for the subalgebra of $\A_2$ generated by $\Sq^1$
and $\Sq^2$.
For $t-s<8$, we can identify the $E_2$-page with 
$$\Ext_{\A_2(1)}^{s,t}(\H^{*+2}(M\U(1),\Z_2),\Z_2).$$

By Thom's isomorphism, $\H^{*+2}(M\U(1),\Z_2)=\Z_2[c_1]U$ where $U$ is the Thom class of the universal $\U(1)$ bundle and $c_1$ is the first Chern class of the universal $\U(1)$ bundle.

The $\A_2(1)$-module structure of $\H^{*+2}(M\U(1),\Z_2)$ is shown in Figure \ref{fig:H^*+2(MU(1),Z_2)}.

\begin{figure}[!h]
\begin{center}
\begin{tikzpicture}[scale=0.5]

\node[left] at (0,2) {$U$};

\draw[fill] (0,2) circle(0.1);
\node[left] at (0,4) {$c_1U$};
\draw[fill] (0,4) circle(0.1);

\draw (0,2) to [out=150,in=150] (0,4);
\node[left] at (0,6) {$c_1^2U$};
\draw[fill] (0,6) circle(0.1);
\node[left] at (0,8) {$c_1^3U$};
\draw[fill] (0,8) circle(0.1);

\draw (0,6) to [out=150,in=150] (0,8);

\end{tikzpicture}
\end{center}
\caption{The $\A_2(1)$-module structure of $\H^{*+2}(M\U(1),\Z_2)$.}
\label{fig:H^*+2(MU(1),Z_2)}
\end{figure}

The $E_2$ page is shown in Figure \ref{fig:Omega_*^{Spin^c}}.

\begin{figure}[!h]
\begin{center}
\begin{tikzpicture}
\node at (0,-1) {0};
\node at (1,-1) {1};
\node at (2,-1) {2};
\node at (3,-1) {3};
\node at (4,-1) {4};
\node at (5,-1) {5};
\node at (6,-1) {6};
\node at (7,-1) {7};
\node at (8,-1) {$t-s$};
\node at (-1,0) {0};
\node at (-1,1) {1};
\node at (-1,2) {2};
\node at (-1,3) {3};
\node at (-1,4) {4};
\node at (-1,5) {5};
\node at (-1,6) {$s$};

\draw[->] (-0.5,-0.5) -- (-0.5,6);
\draw[->] (-0.5,-0.5) -- (8,-0.5);

\draw (0,0) -- (0,5);

\draw (2,1) -- (2,5);

\draw (4,2) -- (4,5);
\draw (4.1,0) -- (4.1,5);

\draw (6,3) -- (6,5);
\draw (6.1,1) -- (6.1,5);
\end{tikzpicture}
\end{center}
\caption{$\Omega_*^{\Spin^c}$}
\label{fig:Omega_*^{Spin^c}}
\end{figure}

Hence we have the following theorem
\begin{theorem}
\begin{table}[!h]
\centering
\begin{tabular}{c c}
\hline
$i$ & $\Omega^{\Spin^c}_i$\\
\hline
0& $\Z$\\
1& $0$\\
2& $\Z$\\
3 & $0$\\
4 & $\Z^2$\\ 
5 & $0$\\
6 & $\Z^2$\\ 
7 & $0$\\
\hline
\end{tabular}
\end{table}
\end{theorem}

The bordism invariants of $\Omega_2^{\Spin^c}$ is $\frac{c_1}{2}$.

Here $c_1$ is divided by 2 since $c_1\mod2=w_2(TM)$ while $w_2(TM)=0$ on $\Spin^c$ 2-manifolds.

The bordism invariants of $\Omega_4^{\Spin^c}$ are $c_1^2$ and $\frac{\sigma-F\cdot F}{8}$.

Here $F$ is a characteristic surface of the $\Spin^c$ 4-manifold $M$. By Rokhlin's theorem, $\sigma-F\cdot F$ is a multiple of 8 and $\frac{1}{8}(\sigma-F\cdot F)=\text{Arf}(M,F)\mod2$. See \cite{Saveliev}'s Lecture 10 for more details.

The bordism invariants of $\Omega_6^{\Spin^c}$ are $\frac{c_1^3}{2}$ and $c_1\frac{\sigma}{16}$.

Here $c_1\frac{\sigma}{16}=\frac{\sigma}{16}(\text{PD}(c_1))$ where $\text{PD}(c_1)$ is the submanifold of the $\Spin^c$ 6-manifold which represents the Poincar\'e dual of $c_1$. Note that $\text{PD}(c_1)$ is Spin.

\begin{theorem}
\begin{table}[!h]
\centering
\begin{tabular}{c c}
\hline
$i$ & $\TP_i(\Spin^c)$\\
\hline
0& $0$\\
1& $\Z$\\
2& $0$\\
3 & $\Z^2$\\
4 & $0$\\ 
5 & $\Z^2$\\
6 & $0$\\ 
\hline
\end{tabular}
\caption{
Note that one of the $\Z$ classes in 
 $\TP_3(\Spin^c)=\Z^2$
is given by
\eqn{eq:Z=AFd-f}. 
Similarly, one of the $\Z$ classes in 
 $\TP_{d+1}(\Spin^c)$ for an even $d$ is given by
 \eqn{eq:Z=AFd}.
}
\end{table}
\label{thm:TPi(Spinc)}
%\label{thm:TPi(SpinU(1)}
\end{theorem}

The 1d topological term is $\frac{1}{2}\text{CS}_1^{(\U(1))}$.

The 3d topological terms are $\text{CS}_1^{(\U(1))}c_1$ and $\mu$.

Here $\mu$ is the Rokhlin invariant (see \cite{Saveliev}'s Lecture 11 ) of the $\Spin^c$ 3-manifold.

The 5d topological terms are $\frac{1}{2}\text{CS}_1^{(\U(1))}c_1^2$ and $c_1\frac{1}{48}\text{CS}_1^{(TM)}$.

\subsection{Non-perturbative global anomalies: Witten's SU(2) Anomaly and A New SU(2) Anomaly in 4d and 5d}

\label{sec:np-anom-ex-SU2}

We  now provide another warm-up example, 
a $(d+1)$-th cobordism group calculation associated with  $d$d \emph{non-perturbative global} anomalies
for the SU(2) anomaly of Witten \cite{Witten:1982fp} and
the new SU(2) anomaly \cite{Wang:2018qoyWWW}.
 
Here these SU(2) anomalies will be interpreted as the 't Hooft anomalies of the internal SU(2) global symmetries in
the QFT, whose fermion multiplets are only in half-integer isospin (say, 1/2, 3/2, 5/2, $\dots$)-representation of SU(2);
while whose bosons are only in an integer isospin (say, 0, 1, 2, $\dots$)-representation of SU(2).
 
Similar to the Spin$^c$  $\equiv (\frac{\Spin\times\U(1)}{\Z_2})$ of \Sec{sec:3d-f-Z},
in the following subsections, we will study the 
(co)bordism group of $(\frac{\Spin\times\SU(2)}{\Z_2})$.

Here the  fermionic system has a continuous spacetime Spin$(d)$ symmetry (in the 
$d$d Euclidean signature) under, the extension of SO$(d)$ via $1 \to \Z_2^F \to \Spin(d) \to \SO(d) \to 1$,
  while the fermion has an internal SU(2) $\supset \Z_2^F$ containing the fermion parity symmetry at SU(2)'s center.
 A common normal subgroup ${\Z_2^F}$ is mod out due to the fact that rotating a fermion by $2 \pi$ in the
spacetime (i.e., the spin statistics) gives rise to the same fermion parity minus sign for the fermion field $\Psi \to -\Psi$.
Thus we need to study the $(\frac{\Spin\times\SU(2)}{\Z_2})$-structure.

We will see that
$\TP_5({\frac{\Spin \times \SU(2)}{\Z_2^F}})=(\Z_2)^2$.
These 5d bordism invariants generates $(\Z_2)^2$, they correspond to the old SU(2)  \cite{Witten:1982fp} and the new SU(2) anomalies \cite{Wang:2018qoyWWW} in 4d, 
shown in Table \ref{table:TPdSpinSU(2)}.

We will see that
$\TP_6({\frac{\Spin \times \SU(2)}{\Z_2^F}})=(\Z_2)^2$.
These 6d bordism invariants generates $(\Z_2)^2$, they correspond to the old SU(2)  and the new SU(2) anomalies in 5d \cite{Wang:2018qoyWWW},
shown in Table \ref{table:TPdSpinSU(2)}.

\subsubsection{
$\Omega_d^{{\frac{\Spin \times \SU(2)}{\Z_2^F}}} =$
$\Omega_d^{\frac{\Spin \times \Spin(3)}{\Z_2^F}}$
}

Let $H=\frac{\Spin \times \Spin(3)}{\Z_2^F}$, we have a homotopy pullback square
\bea
\xymatrix{
\B H\ar[r]\ar[d]&\B\SO(3)\ar[d]^{w_2'}\\
\B\SO\ar[r]^{w_2(TM)}&\B^2\Z_2}
\eea

There is a homotopy equivalence
$f:\B\SO\times \B\SO(3)\xrightarrow{\sim}\B\SO\times \B\SO(3)$ by $(V,W)\mapsto(V-W+3,W)$.
Note that $f^*(w_2)=w_2(V-W)=w_2(V)+w_1(V)w_1(W)+w_2(W)=w_2(TM)+w_2'$.
Then we have the following homotopy pullback
\bea
\xymatrix{
\B H\ar[r]^-{\sim}\ar[d]&\B\Spin\times \B\SO(3)\ar[d]&\\
\B\SO\times \B\SO(3)\ar[r]^{f}\ar[d]_{(V,W)\mapsto V}\ar@/_1pc/[rr]_{w_2(TM)+w_2'}&\B\SO\times \B\SO(3)\ar[r]^-{w_2+0}\ar[ld]^{(V,W)\mapsto V+W-3}&\B^2\Z_2\\
\B\SO&&}
\eea
This implies that $\B H\sim \B\Spin\times \B\SO(3)$.

$MTH=\text{Thom}(\B H;-V)$, where $V$ is the induced virtual bundle (of dimension $0$) by the map $\B H\to \B\tO$.

We can identify $\B H\to \B\tO$ with
$\B\Spin\times \B\SO(3)\xrightarrow{V-V_3+3}\B\SO\hookrightarrow \B\tO$.

The spectrum $MTH$ is homotopy equivalent to $\text{Thom}(\B\Spin\times \B\SO(3);-(V-V_3+3))$, which is $M\Spin\wedge\Sigma^{-3}M\SO(3)$.

For $t-s<8$,
$$\Ext_{\A_2(1)}^{s,t}(\H^{*+3}(M\SO(3),\Z_2),\Z_2)\Rightarrow\Omega_{t-s}^{\frac{\Spin \times \Spin(3)}{\Z_2^F}}.$$

By Thom's isomorphism, $\H^{*+3}(M\SO(3),\Z_2)=\Z_2[w_2',w_3']U$ where $w_i'$ is the Stiefel-Whitney class of the universal $\SO(3)$ bundle and $U$ is the Thom class of the universal $\SO(3)$ bundle.

Since $w_1(TM)=w_1'=0$, and $w_2(TM)=w_2'$ by the gauge bundle constraint, we have $w_3(TM)=w_3'$. Below we use $w_i$ to denote both $w_i(TM)$ and $w_i'$ for $i\le3$.

The $\A_2(1)$-module structure of $\H^{*+3}(M\SO(3),\Z_2)$ and the $E_2$ page are shown in Figure \ref{fig:H^{*+3}(MSO(3),Z_2)}, \ref{fig:Omega_*^{frac{Spin times Spin(3)}{Z_2^F}}}.

\begin{figure}[!h]
\begin{center}
\begin{tikzpicture}[scale=0.5]

\node[below] at (0,0) {$U$};
\node[right] at (0,2) {$w_2U$};
\node[right] at (0,3) {$w_3U$};
\node[right] at (0,4) {$w_2^2U$};
\node[left] at (0,5) {$w_2w_3U$};
\node[left] at (0,6) {$w_3^2U$};
\node[right] at (1,6) {$w_3^2U+w_2^3U$};
\node[right] at (1,7) {$w_2^2w_3U$};
\node[left] at (1,8) {$w_3^2w_2U$};
\node[above] at (1,9) {$w_3^3U$};

\draw[fill] (0,0) circle(.1);
\draw[fill] (0,2) circle(.1);
\draw (0,0) to [out=150,in=150] (0,2);
\draw[fill] (0,3) circle(.1);
\draw (0,2) -- (0,3);
\draw[fill] (0,4) circle(.1);
\draw[fill] (0,5) circle(.1);
\draw[fill] (0,6) circle(.1);
\draw (0,5) -- (0,6);
\draw[fill] (1,6) circle(.1);
\draw (0,4) to [out=30,in=150] (1,6);
\draw[fill] (1,7) circle(.1);
\draw (1,6) -- (1,7);
\draw (0,5) to [out=30,in=150] (1,7);
\draw[fill] (1,8) circle(.1);
\draw (0,6) to [out=30,in=150] (1,8);
\draw[fill] (1,9) circle(.1);
\draw (1,8) -- (1,9);
\draw (1,7) to [out=30,in=30] (1,9);
\end{tikzpicture}
\end{center}
\caption{The $\A_2(1)$-module structure of $\H^{*+3}(M\SO(3),\Z_2)$}
\label{fig:H^{*+3}(MSO(3),Z_2)}
\end{figure}

\begin{figure}[!h]
\begin{center}
\begin{tikzpicture}
\node at (0,-1) {0};
\node at (1,-1) {1};
\node at (2,-1) {2};
\node at (3,-1) {3};
\node at (4,-1) {4};
\node at (5,-1) {5};
\node at (6,-1) {6};
\node at (7,-1) {$t-s$};
\node at (-1,0) {0};
\node at (-1,1) {1};
\node at (-1,2) {2};
\node at (-1,3) {3};
\node at (-1,4) {4};
\node at (-1,5) {5};
\node at (-1,6) {$s$};

\draw[->] (-0.5,-0.5) -- (-0.5,6);
\draw[->] (-0.5,-0.5) -- (7,-0.5);

\draw (0,0) -- (0,5);
\draw (4,1) -- (4,5);
\draw (4,1) -- (6,3);
\draw (4.1,0) -- (4.1,5);
\draw (5,0) -- (6,1);

\end{tikzpicture}
\end{center}
\caption{$\Omega_*^{\frac{\Spin \times \Spin(3)}{\Z_2^F}}$}
\label{fig:Omega_*^{frac{Spin times Spin(3)}{Z_2^F}}}
\end{figure}

\begin{table}[!h]
\centering

\begin{tabular}{c c c }
\hline
\multicolumn{3}{c}{Bordism group}\\
\hline
$d$ & 
$\Omega^{\frac{\Spin \times \SU(2)}{\Z_2^F}}_d$
& bordism invariants \\
\hline
0& $\Z$\\
1& $0$\\
2& $0$\\
3 & $0$\\
4 & $\Z^2$ & ($\sigma,N_0$) \\
5 & $\Z_2^2$ & ($w_2 w_3$, Arf $\cup w_3$) \\
6 & $\Z_2^2$  & ($w_2w_3\cup\tilde{\eta},\frac{\sigma(\text{PD}(w_2))}{16}\mod2)$) \\
\hline
\end{tabular}

\caption{Bordism group.
Here $\sigma$ is the signature of 4-manifolds, $\text{PD}(w_2)$ is the submanifold of a 6-manifold which represents the Poincar\'e dual of $w_2$. Note that $\text{PD}(w_2)$ is Spin.
The $N_0$ is the number of the zero modes of the Dirac operator in 4d. It is defined in \cite{2017arXiv171111587GPW}. On oriented 4-manifolds, $N_0=N_0'=N_+-N_-$ where $N_0'$ is also defined in \cite{2017arXiv171111587GPW}, and $N_{\pm}$ are the numbers of zero modes of the Dirac operator with given chirality. 
}
\end{table}

\begin{table}[!h]
\centering
\begin{tabular}{c c c }
\hline
\multicolumn{3}{c}{Cobordism group}\\
\hline
$d$ & $\TP_d({\frac{\Spin \times \SU(2)}{\Z_2^F}})$
& topological terms \\
\hline
0& $0$\\
1& $0$\\
2& $0$\\
3 & $\Z^2$ & ($\frac{1}{3}\text{CS}_3^{(TM)},X$)\\
4 & $0$\\ 
5 & $\Z_2^2$ & ($w_2 w_3$, Arf $\cup w_3$) \\
6 & $\Z_2^2$ &($w_2w_3\cup\tilde{\eta},\frac{\sigma(\text{PD}(w_2))}{16}\mod2$)\\
\hline
\end{tabular}
\label{table:TPdSpinSU(2)}
\caption{SPT states in $d$-dim spacetime.
$\TP_5({\frac{\Spin \times \SU(2)}{\Z_2^F}})=(\Z_2)^2$.
whose 5d bordism invariants correspond to the old SU(2)  \cite{Witten:1982fp} and the new SU(2) anomalies \cite{Wang:2018qoyWWW} in 4d.
A related cobordism group in one higher dimension,
$\TP_6({\frac{\Spin \times \SU(2)}{\Z_2^F}})=(\Z_2)^2$.
whose 6d bordism invariants correspond to the old SU(2) and the new SU(2) anomalies in 5d \cite{Wang:2018qoyWWW}. Since $\Omega_3^{\Spin \times_{\Z_2} \SU(2)}=0$, for any $\Spin \times_{\Z_2} \SU(2)$ 3-manifold $M^3$, $M^3$ is the boundary of a $\Spin \times_{\Z_2} \SU(2)$ 4-manifold $M^4$, then we define the 3d topological term $X$ on $M^3$ to be $N_0$ of $M^4$. Note that Dirac operator can be defined for manifolds with boundary. If the bulk manifold has Dirac operator which has a mass $m$ being gapped, then the boundary manifold can have Dirac operator no mass $m=0$ being gapless.
}
\end{table}

\section{Higher Group Cobordisms and Non-trivial Fibrations}
\label{sec:higher-G-cobordism}
In this section, we use the Serre spectral sequence method explored in appendix of \cite{Kapustin2013uxa1309.4721} and the Adams spectral sequence method to derive the 5d topological terms for the higher group cobordism $\Omega_5^{\mathbb{G}}$ with non-trivial fibration where $\mathbb{G}$ is defined as follows.

If $G_a$ is a group, $G_b$ is an abelian group, then it is well-known that $\B G_b$ is a group. Consider the group extension
\bea
1\to \B G_b\to\mathbb{G}\to G_a\to1,
\eea 
we have a fibration
\bea
\xymatrix{
\B ^2G_b  \ar[r] &\B \mathbb{G}\ar[d]\\
         &\B G_a}
\eea
which is classified by the Postnikov class $\beta\in\H^3(\B G_a,G_b)$.

\subsection{$(\B G_a,\B ^2G_b):(\B \tO,\B ^2\Z_2)$}\label{(BG_a,B^2G_b):(BO,B^2Z_2)}
We consider the simplest case: $G_a=\tO$ and $G_b=\Z_2$. Note that there is also a group action $\alpha:G_a\to\text{Aut}G_b$ in \cite{Kapustin2013uxa1309.4721}, since $\text{Aut}\Z_2$ is trivial, so $\alpha$ is trivial in this special case. 

For the fibration
\bea
\xymatrix{
\B ^2\Z_2  \ar[r] &\B \mathbb{G}\ar[d]\\
         &\B \tO,}
\eea
there is a Serre spectral sequence
\bea
\H^p(\B \tO,\H^q(\B ^2\Z_2,\mathbb{Z}))\Rightarrow 
\H^{p+q}(\B\mathbb{G},\Z)
\eea
where $\H^p(\B \tO,\H^q(\B ^2\Z_2,\mathbb{Z}))$ actually should be the $\alpha$-equivariant cohomology, but since $\alpha$ is trivial, $\H^p(\B \tO,\H^q(\B ^2\Z_2,\mathbb{Z}))$ is the ordinary cohomology.

Note that $\H^n(\B ^2\Z_2,\Z)$ is computed in Appendix C of \cite{Clement}.
\bea
\H^n(\B ^2\Z_2,\Z)=\left\{\begin{array}{lllllll}\Z&n=0\\
0&n=1\\
0&n=2\\
\Z_2&n=3\\
0&n=4\\
\Z_4&n=5\\
\Z_2&n=6\end{array}\right.
\eea

The $E_2$ page of the Serre spectral sequence is %the $\alpha$-equivariant cohomology 
$\H^p(\B \tO,\H^q(\B ^2\Z_2,\mathbb{Z}))$. The shape of the relevant piece is shown in Figure \ref{fig:SSS for (BO,B^2Z_2)}.
\begin{figure}[!h]
\center
\begin{sseq}[grid=none,labelstep=1,entrysize=1.5cm]{0...7}{0...6}
\ssdrop{\Z}
\ssmoveto 1 0 
\ssdrop{0}
\ssmoveto 2 0
\ssdrop{\Z_2}
\ssmoveto 3 0
\ssdrop{\Z_2}
\ssmoveto 4 0
\ssdrop{\Z\times\Z_2^2}
\ssmoveto 5 0
\ssdrop{\Z_2^2}
\ssmoveto 6 0
\ssdrop{\Z_2^5}
\ssmoveto 7 0
\ssdrop{\Z_2^6}
\ssmoveto 0 1
\ssdrop{0}
\ssmoveto 1 1
\ssdrop{0}
\ssmoveto 2 1
\ssdrop{0}
\ssmoveto 3 1
\ssdrop{0}
\ssmoveto 4 1
\ssdrop{0}
\ssmoveto 5 1
\ssdrop{0}
\ssmoveto 6 1
\ssdrop{0}
\ssmoveto 7 1
\ssdrop{0}
\ssmoveto 0 2
\ssdrop{0}
\ssmoveto 1 2
\ssdrop{0}
\ssmoveto 2 2
\ssdrop{0}
\ssmoveto 3 2
\ssdrop{0}
\ssmoveto 4 2
\ssdrop{0}
\ssmoveto 5 2
\ssdrop{0}
\ssmoveto 6 2
\ssdrop{0}
\ssmoveto 7 2
\ssdrop{0}
\ssmoveto 0 3
\ssdrop{\Z_2}
\ssmoveto 1 3
\ssdrop{\Z_2}
\ssmoveto 2 3
\ssdrop{\Z_2^2}
\ssmoveto 3 3
\ssdrop{\Z_2^3}
\ssmoveto 4 3
\ssdrop{\Z_2^5}
\ssmoveto 5 3
\ssdrop{\Z_2^7}
\ssmoveto 6 3
\ssdrop{\Z_2^{11}}
\ssmoveto 7 3
\ssdrop{\Z_2^{15}}
\ssmoveto 0 4
\ssdrop{0}
\ssmoveto 1 4
\ssdrop{0}
\ssmoveto 2 4
\ssdrop{0}
\ssmoveto 3 4
\ssdrop{0}
\ssmoveto 4 4
\ssdrop{0}
\ssmoveto 5 4
\ssdrop{0}
\ssmoveto 6 4
\ssdrop{0}
\ssmoveto 7 4
\ssdrop{0}
\ssmoveto 0 5
\ssdrop{\Z_4}
\ssmoveto 1 5
\ssdrop{\Z_2}
\ssmoveto 2 5
\ssdrop{\Z_2^2}
\ssmoveto 3 5
\ssdrop{\Z_2^3}
\ssmoveto 4 5
\ssdrop{\Z_4\times\Z_2^4}
\ssmoveto 5 5
\ssdrop{\Z_2^7}
\ssmoveto 6 5
\ssdrop{\Z_2^{11}}
\ssmoveto 7 5
\ssdrop{\Z_2^{15}}
\ssmoveto 0 6
\ssdrop{\Z_2}
\ssmoveto 1 6
\ssdrop{\Z_2}
\ssmoveto 2 6
\ssdrop{\Z_2^2}
\ssmoveto 3 6
\ssdrop{\Z_2^3}
\ssmoveto 4 6
\ssdrop{\Z_2^5}
\ssmoveto 5 6
\ssdrop{\Z_2^7}
\ssmoveto 6 6
\ssdrop{\Z_2^{11}}
\ssmoveto 7 6
\ssdrop{\Z_2^{15}}
\ssmoveto 0 3
\ssarrow[color=red] 4 {-3}
\ssmoveto 1 3
\ssarrow[color=red] 4 {-3}
\ssmoveto 2 3
\ssarrow[color=red] 4 {-3}
\ssmoveto 3 3
\ssarrow[color=red] 4 {-3}
\ssmoveto 0 5
\ssarrow[color=red] 3 {-2}

\end{sseq}
\center
\caption{Serre spectral sequence for $(\B \tO,\B ^2\Z_2)$. Here the row $q=0$ is the result of \cite[Theorem 1.6]{Brown1982}. The row $q=3$ is the result that $\H^*(\B\tO,\Z_2)=\Z_2[w_1,w_2,\dots]$ where $w_i$ is the Stiefel-Whitney class of the virtual bundle (of dimension 0) over $\B\tO$. The rows $q=0$ and $q=3$ are related by the universal coefficient theorem \eqref{uctcohomology2}. The row $q=5$ is resulting from the row $q=0$ by the universal coefficient theorem \eqref{uctcohomology2}. }
\label{fig:SSS for (BO,B^2Z_2)}
\end{figure}

Note that $p$ labels the columns and $q$ labels the rows.

The bottom row is $\H^p(\B \tO,\mathbb{Z})$.

The universal coefficient theorem \eqref{uctcohomology1} tells us that $\H^3(\B ^2\Z_2,\Z)=\H^2(\B ^2\Z_2,\R/\Z)=\Hom(\H_2(\B ^2\Z_2,\Z),\R/\Z)=\Hom(\pi_2(\B ^2\Z_2),\R/\Z)=\Hom(\Z_2,\R/\Z)=\hat \Z_2$, so the $q=3$ row is $\H^p(\B \tO,\hat \Z_2)$. 

It is also known that $\H^5(\B ^2\Z_2,\mathbb{Z})=\H^4(\B ^2\Z_2,\R/\Z)$ is the group of quadratic functions $q:\Z_2\to\R/\Z$ \cite{EM}. The isomorphism is discussed in detail in \cite{2013arXiv1308.2926K}. 

The first possibly non-zero differential is on the $E_3$ page:
\bea
\H^0(\B \tO,\H^5(\B ^2\Z_2,\Z))\to \H^3(\B \tO,\hat \Z_2).
\eea

Following the appendix of \cite{Kapustin2013uxa1309.4721}, this map sends a
quadratic form $q:\Z_2\to\R/\Z$ to $\langle \beta, - \rangle_q$, where the bracket denotes the bilinear pairing $\langle x, y \rangle_q = q(x+y)-q(x)-q(y)$.

The next possibly non-zero differentials are on the $E_4$ page:
\bea
\H^j(\B \tO,\hat \Z_2)\to \H^{j+3}(\B \tO,\R/\Z)\to \H^{j+4}(\B \tO,\Z).
\eea
The first map is contraction with $\beta$.
The second map comes from the long exact sequence
\bea
\cdots\To \H^n(\B \tO,\R)\To \H^n(\B \tO,\R/\Z)\To \H^{n+1}(\B \tO,\Z)\To \H^{n+1}(\B \tO,\R)\To\cdots.
\eea
If $\H^n(\B \tO,\R)=\H^{n+1}(\B \tO,\R)=0$, then $\H^n(\B \tO,\R/\Z)=\H^{n+1}(\B \tO,\Z)$.
Since $\H^n(\B \tO,\R)=\H^n(\B \tO,\Z)\otimes\R$ and $\H^n(\B \tO,\Z)$ is finite if $n$ is not divisible by 4, $\H^n(\B \tO,\R)=0$ if $n$ is not divisible by 4, thus $\H^n(\B \tO,\R/\Z)=\H^{n+1}(\B \tO,\Z)$ for $n=1,2\mod4$.

The last relevant possibly non-zero differential is on the $E_6$ page:
\bea
\H^0(\B \tO,\H^5(\B ^2\Z_2,\Z))\to \H^6(\B \tO,\Z).
\eea

Following the appendix of \cite{Kapustin2013uxa1309.4721}, this differential is actually zero.

So the only possible differentials in Figure \ref{fig:SSS for (BO,B^2Z_2)} below degree 5 are $d_3$ from $(0,5)$ to $(3,3)$ and $d_4$ from the third row to the zeroth row.

By the Universal Coefficient Theorem \eqref{uctcohomology2}, 
\bea
\H^n(\B \mathbb{G},\Z_2)=\H^n(\B \mathbb{G},\Z)\otimes\Z_2\oplus\text{Tor}(\H^{n+1}(\B \mathbb{G},\Z),\Z_2).
\eea

The Madsen-Tillmann spectrum $MT\mathbb{G}=\text{Thom}(\B \mathbb{G};-V)$ where $V$ is the induced virtual bundle over $\B \mathbb{G}$ (of dimension $0$) from $\B \mathbb{G}\to \B \tO$.

By Thom isomorphism \eqref{thomiso}, $\H^*(MT\mathbb{G},\Z_2)=\H^*(\B \mathbb{G},\Z_2)U$ where $U$ is the Thom class of $-V$ with $\Sq^iU=\bar{w}_iU$ where $\bar{w}_i$ is the Stiefel-Whitney class of $-V$ such that $(1+\bar{w}_1+\bar{w}_2+\cdots)(1+w_1+w_2+\cdots)=1$ where $w_i$ is the Stiefel-Whitney class of $V$, i.e., $\bar{w}_1=w_1$, $\bar{w}_2=w_2+w_1^2$, etc. Here the $U$ on the right means the cup product with $U$.

We have the Adams spectral sequence
\bea
\text{Ext}_{\mathcal{A}_2}^{s,t}(\H^*(MT\mathbb{G},\Z_2),\Z_2)\Rightarrow\pi_{t-s}(MT\mathbb{G})=\Omega_{t-s}^{\mathbb{G}}
\eea
where $\mathcal{A}_2$ is the mod 2 Steenrod algebra. The last equality is Pontryagin-Thom isomorphism.

The $\mathcal{A}_2$-module structure of $\H^*(MT\mathbb{G},\Z_2)$ below degree $5$ is shown in Figure \ref{fig:H^*(MTG,Z_2)} where we intentionally omit terms that don't involve the cohomology classes of $\B ^2\Z_2$.

\begin{figure}[!h]
\begin{center}
\begin{tikzpicture}[scale=0.3]
\draw[fill] (0,0) circle(.1);
\draw[fill] (0,1) circle(.1);
\draw (0,0) -- (0,1);
\draw[fill] (0,2) circle(.1);
\draw (0,0) to [out=150,in=150] (0,2);
\draw[fill] (0,3) circle(.1);
\draw[fill] (1,3) circle(.1);
\draw (0,2) -- (0,3);
\draw (0,1) to [out=30,in=150] (1,3);

\node [below] at (0,0) {\tiny{$x_2U$}};
\node [right] at (0,1) {\tiny{$x_3U+x_2w_1U$}};
\node [left] at (0,2) {\tiny{$x_2^2U+x_2(w_2+w_1^2)U+x_3w_1U$}};
\node [left] at (0,3) {\tiny{$x_2^2w_1U+x_3(w_2+w_1^2)U+x_2(w_3+w_1^3)U$}};
\node [right] at (1,3) {\tiny{$x_5U+x_3(w_2+w_1^2)U+x_2^2w_1U+x_2w_1w_2U$}};

\draw[fill] (20,1) circle(.1);
\draw[fill] (20,2) circle(.1);
\draw (20,1) -- (20,2);
\draw[fill] (20,3) circle(.1);
\draw (20,1) to [out=150,in=150] (20,3);

\node [below] at (20,1) {\tiny{$x_3U$}};
\node [right] at (20,2) {\tiny{$x_3w_1U$}};
\node [above] at (20,3) {\tiny{$x_5U+x_3(w_2+w_1^2)U$}};

\draw[fill] (-16,-5) circle(.1);
\draw[fill] (-16,-4) circle(.1);
\draw (-16,-5) -- (-16,-4);

\node [below] at (-16,-5) {\tiny{$x_2w_2U$}};
\node [above] at (-16,-4) {\tiny{$x_3w_2U+x_2w_3U$}};

\draw[fill] (-7,-5) circle(.1);
\draw[fill] (-7,-4) circle(.1);
\draw (-7,-5) -- (-7,-4);

\node [below] at (-7,-5) {\tiny{$x_2w_1^2U$}};
\node [above] at (-7,-4) {\tiny{$x_3w_1^2U+x_2w_1^3U$}};

\draw[fill] (2,-4) circle(.1);

\node [below] at (2,-4) {\tiny{$x_3w_2U(x_2w_3U)$}};

\draw[fill] (11,-4) circle(.1);

\node [below] at (11,-4) {\tiny{$x_3w_1^2U(x_2w_1^3U)$}};

\draw[fill] (20,-4) circle(.1);

\node [below] at (20,-4) {\tiny{$x_2x_3U$}};
\end{tikzpicture}
\end{center}

\caption{The $\mathcal{A}_2$-module structure of $\H^*(MT\mathbb{G},\Z_2)$ below degree $5$.
Here $x_2$ is the generator of $\H^2(\B^2\Z_2,\Z_2)$, $x_3=\Sq^1x_2$, and $x_5=\Sq^2x_3$.
}
\label{fig:H^*(MTG,Z_2)}
\end{figure}

Note that the position $(0,3)$ in Figure \ref{fig:SSS for (BO,B^2Z_2)} contributes to both $\H^2(\B ^2\Z_2,\Z_2)$ which is generated by $x_2$ and $\H^3(\B ^2\Z_2,\Z_2)$ which is generated by $x_3$, $\Sq^1x_2=x_3$. The position $(2,3)$ corresponds to $xw_1^2,xw_2$, the position $(3,3)$ corresponds to $xw_1^3,xw_1w_2,xw_3$ for both $x=x_2$ and $x=x_3$.

Since $\langle \beta,\beta\rangle_q=-2q(\beta)$, $4q(\beta)=q(2\beta)=0$, there are 2 among the 4 choices of $q(\beta)$ such that $q\to\langle\beta,-\rangle_q$ maps to the dual linear function of $\beta$, if we identify $\hat\Z_2$ with $\Z_2$, then the nonzero element in the image of $q\to\langle\beta,-\rangle_q$ is just $\beta$. So
$\text{Im}d_3^{(0,5)}$ is spanned by $x\beta$.

The differential $d_4^{(2,3)}:\H^2(\B \tO,\hat\Z_2)\to \H^5(\B \tO,\R/\Z)=\H^6(\B\tO,\Z)$ is defined by $$d_4^{(2,3)}(\gamma)(v_0,\dots,v_5)=(\gamma(v_0,\dots,v_2))(\beta(v_2,\dots,v_5)).$$
Let $\beta=a_1w_1^3+a_2w_1w_2+a_3w_3$, if we identify $\hat\Z_2$ with $\Z_2$, 
then $d_4^{(2,3)}(\gamma)=\gamma\cup\beta$ which is in $\H^5(\B\tO,\Z_2)$,
while $\H^5(\B \tO,\R/\Z)=\H^6(\B\tO,\Z)$. 
Let $\gamma=b_1w_1^2+b_2w_2$, then $\gamma\cup\beta=a_1b_1w_1^5+(a_1b_2+a_2b_1)w_1^3w_2+a_2b_2w_1w_2^2+a_3b_1w_1^2w_3+a_3b_2w_2w_3$.

The differential $d_4^{(3,3)}:\H^3(\B \tO,\hat\Z_2)\to \H^6(\B \tO,\R/\Z)=\H^7(\B\tO,\Z)$ is defined by $$d_4^{(3,3)}(\zeta)(v_0,\dots,v_6)=(\zeta(v_0,\dots,v_3))(\beta(v_3,\dots,v_6)).$$
Let $\beta=a_1w_1^3+a_2w_1w_2+a_3w_3$, if we identify $\hat\Z_2$ with $\Z_2$, 
then $d_4^{(3,3)}(\zeta)=\zeta\cup\beta$ which is in $\H^6(\B\tO,\Z_2)$,
while $\H^6(\B \tO,\R/\Z)=\H^7(\B\tO,\Z)$. 
Let $\zeta=c_1w_1^3+c_2w_1w_2+c_3w_3$, then $\zeta\cup\beta=a_1c_1w_1^6+a_2c_2w_1^2w_2^2+a_3c_3w_3^2+(a_1c_2+c_1a_2)w_1^4w_2+(a_1c_3+c_1a_3)w_1^3w_3+(a_2c_3+c_2a_3)w_1w_2w_3$.

Note that by the Universal Coefficient Theorem \eqref{uctcohomology2}, 
\bea
\H^n(\B \tO,\Z_2)=\H^n(\B \tO,\Z)\otimes\Z_2\oplus\text{Tor}(\H^{n+1}(\B \tO,\Z),\Z_2).
\eea

By \cite[Theorem 1.6]{Brown1982}:
\bea
\H^*(\B\tO,\Z)\otimes\Z_2=\Z_2[w_2^2,\Sq^1(w_1),\Sq^1(w_2),\Sq^1(w_1w_2),\Sq^1(w_4),\Sq^1(w_1w_4),\Sq^1(w_2w_4),\dots]
\eea
where we list the generators below degree 7.
Among the linear combinations of $w_1^5,w_1^3w_2,w_1w_2^2,w_1^2w_3,w_2w_3$, only $w_1^3w_2+w_1^2w_3$ is in $\H^5(\B \tO,\Z)\otimes\Z_2$
While among the linear combinations of $w_1^6,w_1^2w_2^2,w_3^2,w_1^4w_2,w_1^3w_3,w_1w_2w_3$, only $w_1^2w_2^2$, $w_1^3w_3$, $w_1^6$, and $w_1^2w_2^2+w_3^2$ are in $\H^6(\B \tO,\Z)\otimes\Z_2$. 

So we claim that 
$\text{Ker}d_4^{(2,3)}$ is spanned by $x\gamma$ where $\gamma=b_1w_1^2+b_2w_2$ with 
$a_1b_1=0$, $a_1b_2+a_2b_1=a_3b_1$, $a_2b_2=0$, $a_3b_2=0$.
While $\text{Ker}d_4^{(3,3)}$ is spanned by $x\zeta$ where $\zeta=c_1w_1^3+c_2w_1w_2+c_3w_3$ with
$a_1c_2+c_1a_2=0$, $a_2c_3+c_2a_3=0$.

In the following cases, we only consider the topological terms involving the cohomology classes of $\B ^2\Z_2$.

\textbf{Case 1:} $\beta=0$, there is no differential in Figure \ref{fig:SSS for (BO,B^2Z_2)}, the 5d topological terms are $x_3w_2$ (or $x_2w_3$), $x_3w_1^2$ (or $x_2w_1^3$) and $x_2x_3$ (see Figure \ref{fig:H^*(MTG,Z_2)}). 
Note that $x_3w_2=x_2w_3$, $x_3w_1^2=x_2w_1^3$ and $x_5=x_3(w_1^2+w_2)$ by Wu formula \eqref{SqWu}.
In this case $\B \mathbb{G}=\B \tO\times \B ^2\Z_2$, $MT\mathbb{G}=M\tO\wedge(\B ^2\Z_2)_+$, $\pi_d MT\mathbb{G}=\Omega_d^{\tO}(\B ^2\Z_2)$. This case will be discussed later in another way.

\textbf{Case 2:} $\beta=w_1^3$, $a_1=1$, $a_2=a_3=0$. $\text{Ker}d_4^{(2,3)}$ is spanned by $x\gamma$ where $\gamma=b_1w_1^2+b_2w_2$ with $b_1=b_2=0$.
 $\text{Ker}d_4^{(3,3)}$ is spanned by $x\zeta$ where $\zeta=c_1w_1^3+c_2w_1w_2+c_3w_3$ with $c_2=0$.
 At the position (3,3), $x_2w_1^3$ is killed in the $E_3$ page, $x_2w_3$ survives to the $E_{\infty}$ page, so the $5d$ topological terms are $x_2w_3$ and $x_2x_3$.

\textbf{Case 3:} $\beta=w_1w_2$, $a_1=0$, $a_2=1$, $a_3=0$. $\text{Ker}d_4^{(2,3)}$ is spanned by $x\gamma$ where $\gamma=b_1w_1^2+b_2w_2$ with $b_1=b_2=0$. $\text{Ker}d_4^{(3,3)}$ is spanned by $x\zeta$ where $\zeta=c_1w_1^3+c_2w_1w_2+c_3w_3$ with $c_1=c_3=0$.
 At the position (3,3),
$x_2w_3$ and $x_2w_1^3$ are killed in the $E_4$ page, so the 5d topological term is $x_2x_3$.

\textbf{Case 4:} $\beta=w_3$, $a_1=a_2=0$, $a_3=1$. $\text{Ker}d_4^{(2,3)}$ is spanned by $x\gamma$ where $\gamma=b_1w_1^2+b_2w_2$ with $b_1=b_2=0$. $\text{Ker}d_4^{(3,3)}$ is spanned by $x\zeta$ where $\zeta=c_1w_1^3+c_2w_1w_2+c_3w_3$ with $c_2=0$.
 At the position (3,3),
 $x_2w_3$ is killed in the $E_3$ page, $x_2w_1^3$ survives to the $E_{\infty}$ page, so the 5d topological terms are $x_2w_1^3$ and $x_2x_3$.

\textbf{Case 5:} $\beta=w_1^3+w_1w_2$, $a_1=a_2=1$, $a_3=0$.
$\text{Ker}d_4^{(2,3)}$ is spanned by $x\gamma$ where $\gamma=b_1w_1^2+b_2w_2$ with $b_1=b_2=0$. $\text{Ker}d_4^{(3,3)}$ is spanned by $x\zeta$ where $\zeta=c_1w_1^3+c_2w_1w_2+c_3w_3$ with $c_1+c_2=0$, $c_3=0$.
 At the position (3,3),
 $x_2(w_1^3+w_1w_2)$ is killed in the $E_3$ page, $x_2w_3$ is killed in the $E_4$ page, but since $x_2w_1w_2=\Sq^3x_2=0$ by Wu formula \eqref{SqWu}, so the 5d topological term is $x_2x_3$.

\textbf{Case 6:} $\beta=w_1w_2+w_3$, $a_1=0$, $a_2=a_3=1$.
$\text{Ker}d_4^{(2,3)}$ is spanned by $x\gamma$ where $\gamma=b_1w_1^2+b_2w_2$ with $b_2=0$. $\text{Ker}d_4^{(3,3)}$ is spanned by $x\zeta$ where $\zeta=c_1w_1^3+c_2w_1w_2+c_3w_3$ with $c_1=0$, $c_2+c_3=0$.
 At the position (3,3),
 $x_2(w_1w_2+w_3)$ is killed in the $E_3$ page, $x_2w_1^3$ is killed in the $E_4$ page, but since $x_2w_1w_2=\Sq^3x_2=0$ by Wu formula \eqref{SqWu}, so the 5d topological terms is $x_2x_3$.

\textbf{Case 7:} $\beta=w_1^3+w_3$, $a_1=1$, $a_2=0$, $a_3=1$.
$\text{Ker}d_4^{(2,3)}$ is spanned by $x\gamma$ where $\gamma=b_1w_1^2+b_2w_2$ with $b_1=b_2=0$. $\text{Ker}d_4^{(3,3)}$ is spanned by $x\zeta$ where $\zeta=c_1w_1^3+c_2w_1w_2+c_3w_3$ with $c_2=0$.
 At the position (3,3), 
 $x_2(w_1^3+w_3)$ is killed in the $E_3$ page, so the 5d topological terms are $x_2w_1^3=x_2w_3$ and $x_2x_3$.

\textbf{Case 8:} $\beta=w_1^3+w_1w_2+w_3$, $a_1=a_2=a_3=1$.
$\text{Ker}d_4^{(2,3)}$ is spanned by $x\gamma$ where $\gamma=b_1w_1^2+b_2w_2$ with $b_1=b_2=0$. $\text{Ker}d_4^{(3,3)}$ is spanned by $x\zeta$ where $\zeta=c_1w_1^3+c_2w_1w_2+c_3w_3$ with $c_1+c_2=0$, $c_2+c_3=0$.
 At the position (3,3), 
 $x_2(w_1^3+w_1w_2+w_3)$ is killed in the $E_3$ page, but since $x_2w_1w_2=\Sq^3x_2=0$ by Wu formula \eqref{SqWu}, so the 5d topological term are $x_2w_1^3=x_2w_3$ and $x_2x_3$.

\section{$\tO/\SO/\Spin/\Pin^{\pm}$ bordism groups of classifying spaces}

\label{sec:cobor}

In this section, we compute the $\tO/\SO/\Spin/\Pin^{\pm}$ bordism groups of the classifying space of the group $\mathbb{G}=G_a\times \B G_b$: $\B \mathbb{G}=\B G_a\times \B ^2G_b$.
Here $\B G_b$ is a group since $G_b$ is abelian.

%\section{Difference between a previous cobordism theory and this work}
We briefly comment the difference between a previous cobordism theory \cite{2017arXiv171111587GPW} and this work:
In all Adams charts of the computation in \cite{2017arXiv171111587GPW}, there are no nonzero differentials, 
while in this paper we encounter nonzero differentials $d_n$ due to the $(p,p^n)$-Bockstein homomorphisms in the computation involving $\B ^2\Z_{p^n}$ and $\B \Z_{p^n}$.

\subsection{Introduction}

For $H=\tO/\SO/\Spin/\Pin^{\pm}$ and the group $H\times\mathbb{G}$, define 
\bea
MT(H\times\mathbb{G}):=\text{Thom}(\B (H\times\mathbb{G});-V)
\eea
where $V$ is the induced virtual bundle over $\B (H\times\mathbb{G})$ by the composition $\B (H\times\mathbb{G})\to \B H\to \B \tO$
where the first map is the projection, the second map is the natural homomorphism.

By the Pontryagin-Thom isomorphism \eqref{ponthom} and the property of Thom space \eqref{thomsum}, $\Omega^H_d(\B \mathbb{G})=\pi_d(MTH\wedge \B \mathbb{G}_+)=\pi_d(MT(H\times\mathbb{G}))$.
Hence we can define 
\bea
\Omega_d^{H\times\mathbb{G}}:=\pi_d(MT(H\times\mathbb{G}))=\Omega^H_d(\B \mathbb{G}).
\eea
\bea
\TP_n(H\times\mathbb{G}):=[MT(H\times\mathbb{G}),\Sigma^{n+1}I\Z]
\eea
Here $X_+$ is the disjoint union of $X$ and a point. $MT\tO=M\tO$, $MT\SO=M\SO$, $MT\Spin=M\Spin$, $MT\Pin^+=M\Pin^-$, $MT\Pin^-=M\Pin^+$.
$\pi_d(\mathcal{B})$ is the $d$-th stable homotopy group of the spectrum $\mathcal{B}$. 

$[\mathcal{B},\Sigma^{n+1}I\Z]$ stands for the homotopy classes of maps from spectrum $\mathcal{B}$ to the $(n+1)$-th suspension of spectrum $I\Z$. The Anderson dual $I\Z$ is a spectrum that is the fiber of $I\C\to I\C^{\times}$
where $I\C(I\C^{\times})$ is the Brown-Comenetz dual spectrum defined by 
\bea
[X,I\C]=\Hom(\pi_0X,\C),
\eea
\bea
[X,I\C^{\times}]=\Hom(\pi_0X,\C^{\times}).
\eea

By the work of Freed-Hopkins \cite{Freed2016}, there is a 1:1 correspondence
\bea
\left\{\begin{array}{ccc}\text{deformation classes of reflection positive}\\\text{invertible }n\text{-dimensional extended topological}\\\text{field theories with symmetry group }H_n\times\mathbb{G}\end{array}\right\}\cong[MT(H\times\mathbb{G}),\Sigma^{n+1}I\Z]_{\text{tors}}.
\eea

There is an exact sequence
\bea
0\to\Ext^1(\pi_n\mathcal{B},\Z)\to[\mathcal{B},\Sigma^{n+1}I\Z]\to\Hom(\pi_{n+1}\mathcal{B},\Z)\to0
\eea
for any spectrum $\mathcal{B}$, especially for $MT(H\times\mathbb{G})$.
The torsion part $[MT(H\times\mathbb{G}),\Sigma^{n+1}I\Z]_{\text{tors}}$ is $\Ext^1((\pi_nMT(H\times\mathbb{G}))_{\text{tors}},\Z)=\Hom((\pi_nMT(H\times\mathbb{G}))_{\text{tors}},\U(1))$.

\bea
\H^*(\B \Z_2,\Z_2)=\Z_2[a]
\eea 
where $|a|=1$.

\begin{theorem}[\textbf{Serre}, Ref.~\cite{tamanoi}]
\bea
\H^*(\B ^2\Z_2,\Z_2)=\Z_2[\Sq^Ix_2|I\text{ admissible, }ex(I)<2]=\Z_2[\Sq^{2^{i-1}}\cdots \Sq^2\Sq^1x_2|i\ge0]
\eea
where $x_2$ is the generator of $\H^2(\B ^2\Z_2,\Z_2)$.
\end{theorem}
Denote $\Sq^{2^{i-1}}\cdots \Sq^2\Sq^1x_2=x_{2^i+1}$.

Here $\Sq^I=\Sq^{i_1}\Sq^{i_2}\cdots$ and $I=(i_1,i_2,\dots)$ is admissible if $i_s\ge2i_{s+1}$ for $s\ge1$, $ex(I)=\sum_{s\ge1}(i_s-2i_{s+1})$.

\begin{theorem}[Ref.~\cite{tamanoi}]
\bea
\H^*(\B \Z_3,\Z_3)=F_{\Z_3}[a',b']=\Lambda_{\Z_3}(a')\otimes\Z_3[b']
\eea 
where $|a'|=1$ and $b'=\beta_{(3,3)} a'$.

Here $\beta_{(3,3)}$ is the Bockstein homomorphism in $\A_3$.

$$\H^*(\B ^2\Z_3,\Z_3)=F_{\Z_3}[x_2',\beta_{(3,3)} x_2',Q_ix_2',\beta_{(3,3)} Q_ix_2',i\ge1]=\Z_3[x_2',\beta_{(3,3)} Q_ix_2',i\ge1]\otimes\Lambda_{\Z_3}(\beta_{(3,3)} x_2',Q_ix_2',i\ge1)$$ 

where $|x_2'|=2$ and $Q_i$ is defined inductively by $Q_0=\beta_{(3,3)}$, $Q_i=P^{3^{i-1}}Q_{i-1}-Q_{i-1}P^{3^{i-1}}$ for $i\ge1$. 
Let $x_3'=\beta_{(3,3)} x_2'$, $x_{2\cdot 3^i+1}'=Q_ix_2'$, $x_{2\cdot 3^i+2}'=\beta_{(3,3)} Q_ix_2'$ for $i\ge1$. 

Here $P^n$ is the $n$-th Steenrod power in $\A_3$.

\end{theorem}

\bea
\H^*(\B \PSU(2),\Z_2)=\Z_2[w_2',w_3'].\\
\H^*(\B \PSU(3),\Z_2)=\Z_2[c_2,c_3].
\eea

Here $w_i'$ is the $i$-th Stiefel-Whitney class $w_i(\PSU(2))$ of the universal principal $\PSU(2)$-bundle over $\B \PSU(2)$.
Let $p_i'$ be the $i$-th Pontryagin class $p_i(\PSU(2))$ of the universal principal $\PSU(2)$-bundle over $\B \PSU(2)$, then $p_1'(\mod2)=w_2'^2$.

$c_i$ is the $i$-th Chern class $c_i(\PSU(3))$ of the universal principal $\PSU(3)$-bundle over $\B \PSU(3)$.

Since $\frac{\SU(3)\times \U(1)}{\Z_3}=\U(3)$, $\PSU(3)=\PU(3)$.

\begin{theorem}[Ref.~\cite{kms}]

\bea
\H^*(\B \PSU(3),\Z_3)=F_{\Z_3}[z_2,z_3,z_7,z_8,z_{12}]/J
\eea
where $|z_i|=i$, $J=(z_2z_3,z_2z_7,z_2z_8+z_3z_7)$ is the ideal generated by $z_2z_3,z_2z_7,z_2z_8+z_3z_7$ and $z_3=\beta_{(3,3)} z_2$, $z_7=P^1z_3$, $z_8=\beta_{(3,3)} z_7$.
Note that $c_2(\mod3)=z_2^2$, $c_3(\mod3)=z_2^3$.

\end{theorem}

In the following subsections, all bordism invariants are the pullback of cohomology classes along classifying maps $f:M\to X$ and $g:M\to \B H$.

\subsection{Point}

\subsubsection{$\Omega^{\tO}_d$}
Since the computation involves no odd torsion, we can use the Adams spectral sequence 
\bea
E_2^{s,t}=\Ext_{\A_2}^{s,t}(\H^*(M\tO,\Z_2),\Z_2)\Rightarrow\pi_{t-s}(M\tO)_2^{\wedge}=\Omega^{\tO}_{t-s}.
\eea

Here $\pi_{t-s}(M\tO)_2^{\wedge}$ is the 2-completion of the group $\pi_{t-s}(M\tO)$.

The mod 2 cohomology of Thom spectrum $M\tO$ is
\bea
\H^*(M\tO,\Z_2)=\A_2\otimes\Omega^*
\eea
where $\Omega=\Z_2[y_2,y_4,y_5,y_6,y_8,\dots]$ is the unoriented bordism ring, $\Omega^*$ is the $\Z_2$-linear dual of $\Omega$.

On the other hand, $\H^*(M\tO,\Z_2)=\Z_2[w_1,w_2,w_3,\dots]U$ where $U$ is the Thom class of the virtual bundle (of dimension 0) over $\B \tO$ which is the colimit of $E_n-n$ and $E_n$ is the universal $n$-bundle over $\B \tO(n)$, $w_i$ is the $i$-th Stiefel-Whitney class of the virtual bundle (of dimension 0) over $\B \tO$. Note that the pullback of the virtual bundle (of dimension 0) over $\B \tO$ along the map $g:M\to \B \tO$ is just $TM-d$ where $M$ is a $d$-dimensional manifold and $TM$ is the tangent bundle of $M$, $g$ is given by the $\tO$-structure on $M$. We will not distinguish $w_i$ and $w_i(TM)$.

Here $y_i$ are manifold generators, for example, $y_2=\RP^2$, $y_4=\RP^4$, $y_5$ is Wu manifold $\SU(3)/\SO(3)$.
By Thom's result \cite{thom1954quelques}, two manifolds are unorientedly bordant if and only if they have identical sets of Stiefel-Whitney characteristic numbers. The nonvanishing Stiefel-Whitney numbers of $y_2=\RP^2$ are $w_2$ and $w_1^2$, the nonvanishing Stiefel-Whitney numbers of $y_2^2=\RP^2\times\RP^2$ are $w_2^2$ and $w_4$, 
the nonvanishing Stiefel-Whitney numbers of $y_4=\RP^4$ are $w_1^4$ and $w_4$, the only nonvanishing Stiefel-Whitney number of Wu manifold $\SU(3)/\SO(3)$ is $w_2w_3$.

So $y_2^*=w_1^2$ or $w_2$, $(y_2^2)^*=w_2^2$, $y_4^*=w_1^4$, $y_5^*=w_2w_3$, etc, where $y_i^*$ is the $\Z_2$-linear dual of 
$y_i\in\Omega$.

Below we choose $y_2^*=w_1^2$ by default, this is reasonable since $\Sq^2(x_{d-2})=(w_2+w_1^2)x_{d-2}$ on $d$-manifold by Wu formula \eqref{SqWu}.  

Hence we have the following theorem
\begin{theorem}
\begin{table}[!h]
\centering
\begin{tabular}{c c}
\hline
$i$ & $\Omega^{\tO}_i$\\
\hline
0& $\Z_2$\\
1& $0$\\
2& $\Z_2$\\
3 & $0$\\
4 & $\Z_2^2$\\ 
5 & $\Z_2$\\
\hline
\end{tabular}
\end{table}
\end{theorem}

The bordism invariant of $\Omega_2^{\tO}$ is $w_1^2$.

The bordism invariants of $\Omega_4^{\tO}$ are $w_1^4,w_2^2$.

The bordism invariant of $\Omega_5^{\tO}$ is $w_2w_3$.

\begin{theorem}
\begin{table}[!h]
\centering
\begin{tabular}{c c}
\hline
$i$ & $\TP_i(\tO)$\\
\hline
0& $\Z_2$\\
1& $0$\\
2& $\Z_2$\\
3 & $0$\\
4 & $\Z_2^2$\\ 
5 & $\Z_2$\\
\hline
\end{tabular}
\end{table}
\end{theorem}

The 2d topological term is $w_1^2$.

The 4d topological terms are $w_1^4,w_2^2$.

The 5d topological term is $w_2w_3$.

\subsubsection{$\Omega^{\SO}_d$}

Since the computation involves no odd torsion, we can use the Adams spectral sequence 
\bea
E_2^{s,t}=\Ext_{\A_2}^{s,t}(\H^*(M\SO,\Z_2),\Z_2)\Rightarrow\pi_{t-s}(M\SO)_2^{\wedge}=\Omega^{\SO}_{t-s}.
\eea
The mod 2 cohomology of Thom spectrum $M\SO$ is
\bea
\H^*(M\SO,\Z_2)=\A_2/\A_2 \Sq^1\oplus\Sigma^4\A_2/\A_2 \Sq^1\oplus\Sigma^5\A_2\oplus\cdots.
\eea
\bea
\cdots\To\Sigma^3\A_2\To\Sigma^2\A_2\To\Sigma\A_2\To\A_2\To\A_2/\A_2 \Sq^1
\eea
is an $\A_2$-resolution where the differentials $d_1$ are induced by $\Sq^1$.

The $E_2$ page is shown in Figure \ref{fig:Omega_*^{SO}}.

\begin{figure}[!h]
\begin{center}
\begin{tikzpicture}
\node at (0,-1) {0};
\node at (1,-1) {1};
\node at (2,-1) {2};
\node at (3,-1) {3};
\node at (4,-1) {4};
\node at (5,-1) {5};
\node at (6,-1) {$t-s$};
\node at (-1,0) {0};
\node at (-1,1) {1};
\node at (-1,2) {2};
\node at (-1,3) {3};
\node at (-1,4) {4};
\node at (-1,5) {5};
\node at (-1,6) {$s$};

\draw[->] (-0.5,-0.5) -- (-0.5,6);
\draw[->] (-0.5,-0.5) -- (6,-0.5);

\draw (0,0) -- (0,5);

\draw (4,0) -- (4,5);

\draw[fill] (5,0) circle(0.05);

\end{tikzpicture}
\end{center}
\caption{$\Omega_*^{\SO}$}
\label{fig:Omega_*^{SO}}
\end{figure}

Hence we have the following theorem
\begin{theorem}
\begin{table}[!h]
\centering
\begin{tabular}{c c}
\hline
$i$ & $\Omega^{\SO}_i$\\
\hline
0& $\Z$\\
1& $0$\\
2& $0$\\
3 & $0$\\
4 & $\Z$\\ 
5 & $\Z_2$\\
\hline
\end{tabular}
\end{table}
\end{theorem}

The bordism invariant of $\Omega_{4}^{\SO}$ is $\sigma$.

Here $\sigma$ is the signature of a 4-manifold.

The bordism invariant of $\Omega^{\SO}_5$ is $w_2w_3$.

\begin{theorem}
\begin{table}[!h]
\centering
\begin{tabular}{c c}
\hline
$i$ & $\TP_i(\SO)$\\
\hline
0& $0$\\
1& $0$\\
2& $0$\\
3 & $\Z$\\
4 & $0$\\ 
5 & $\Z_2$\\
\hline
\end{tabular}
\end{table}
\end{theorem}

Since $\sigma=\frac{p_1(TM)}{3}$, $p_1(TM)=\text{d}\text{CS}_3^{(TM)}$, the 3d topological term is $\frac{1}{3}\text{CS}_3^{(TM)}$.

The 5d topological term is $w_2w_3$.

\subsubsection{$\Omega^{\Spin}_d$}
Since the computation involves no odd torsion, we can use the Adams spectral sequence 
\bea
E_2^{s,t}=\Ext_{\A_2}^{s,t}(\H^*(M\Spin,\Z_2),\Z_2)\Rightarrow\pi_{t-s}(M\Spin)_2^{\wedge}=\Omega^{\Spin}_{t-s}.
\eea
The mod 2 cohomology of Thom spectrum $M\Spin$ is
\bea
\H^*(M\Spin,\Z_2)=\A_2\otimes_{\A_2(1)}\{\Z_2\oplus M\}
\eea
where $M$ is a graded $\A_2(1)$-module with the degree $i$ homogeneous part $M_i=0$ for $i<8$. Here $\A_2(1)$ stands for the subalgebra of $\A_2$ generated by $\Sq^1$
and $\Sq^2$.
For $t-s<8$, we can identify the $E_2$-page with 
$$\Ext_{\A_2(1)}^{s,t}(\Z_2,\Z_2).$$

The $E_2$ page is shown in Figure \ref{fig:Omega_*^{Spin}}.

\begin{figure}[!h]
\begin{center}
\begin{tikzpicture}
\node at (0,-1) {0};
\node at (1,-1) {1};
\node at (2,-1) {2};
\node at (3,-1) {3};
\node at (4,-1) {4};
\node at (5,-1) {5};
\node at (6,-1) {$t-s$};
\node at (-1,0) {0};
\node at (-1,1) {1};
\node at (-1,2) {2};
\node at (-1,3) {3};
\node at (-1,4) {4};
\node at (-1,5) {5};
\node at (-1,6) {$s$};

\draw[->] (-0.5,-0.5) -- (-0.5,6);
\draw[->] (-0.5,-0.5) -- (6,-0.5);

\draw (0,0) -- (0,5);
\draw (0,0) -- (2,2);

\draw (4,3) -- (4,5);

\end{tikzpicture}
\end{center}
\caption{$\Omega_*^{\Spin}$}
\label{fig:Omega_*^{Spin}}
\end{figure}

Hence we have the following theorem
\begin{theorem}
\begin{table}[!h]
\centering
\begin{tabular}{c c}
\hline
$i$ & $\Omega^{\Spin}_i$\\
\hline
0& $\Z$\\
1& $\Z_2$\\
2& $\Z_2$\\
3 & $0$\\
4 & $\Z$\\ 
5 & $0$\\
\hline
\end{tabular}
\end{table}
\end{theorem}
The bordism invariant of $\Omega_1^{\Spin}$ is $\tilde{\eta}$.

Here $\tilde{\eta}$ is the ``mod 2 index'' of the 1d Dirac operator (\#zero eigenvalues mod 2, no contribution from spectral asymmetry).

The bordism invariant of $\Omega_2^{\Spin}$ is $\text{Arf}$ (the Arf invariant).

The bordism invariant of $\Omega_4^{\Spin}$ is $\frac{\sigma}{16}$.

\begin{theorem}
\begin{table}[!h]
\centering
\begin{tabular}{c c}
\hline
$i$ & $\TP_i(\Spin)$\\
\hline
0& $0$\\
1& $\Z_2$\\
2& $\Z_2$\\
3 & $\Z$\\
4 & $0$\\ 
5 & $0$\\
\hline
\end{tabular}
\end{table}
\end{theorem}

The 1d topological term is $\tilde{\eta}$.

The 2d topological term is $\text{Arf}$.

The 3d topological term is $\frac{1}{48}\text{CS}_3^{(TM)}$.

\subsubsection{$\Omega^{\Pin^+}_d$}

Since the computation involves no odd torsion, we can use the Adams spectral sequence 
\bea
E_2^{s,t}=\Ext_{\A_2}^{s,t}(\H^*(M\Pin^-,\Z_2),\Z_2)\Rightarrow\pi_{t-s}(M\Pin^-)_2^{\wedge}=\Omega^{\Pin^+}_{t-s}.
\eea
$M\Pin^-=MT\Pin^+\sim M\Spin\wedge S^1\wedge MT\tO(1)$. 

For $t-s<8$, we can identify the $E_2$-page with 
$$\Ext_{\A_2(1)}^{s,t}(\H^{*-1}(MT\tO(1),\Z_2),\Z_2).$$

{By Thom's isomorphism, 
\bea
\H^{*-1}(MT\tO(1),\Z_2)=\Z_2[w_1]U
\eea
where $U$ is the Thom class of the virtual bundle $-E_1$ over $\B \tO(1)$, $E_1$ is the universal 1-bundle over $\B \tO(1)$ and $w_1$ is the 1st Stiefel-Whitney class of $E_1$ over $\B \tO(1)$.
}
The $\A_2(1)$-module structure of $\H^{*-1}(MT\tO(1),\Z_2)$ and the $E_2$ page are shown in Figure \ref{fig:H^{*-1}(MTO(1),Z_2)}, \ref{fig:Omega_*^{Pin^+}}.

\begin{figure}[!h]
\begin{center}
\begin{tikzpicture}[scale=0.5]

\node[below] at (0,0) {$U$};

\draw[fill] (0,0) circle(.1);
\draw[fill] (0,1) circle(.1);
\draw (0,0) -- (0,1);
\draw[fill] (0,2) circle(.1);
\draw (0,0) to [out=150,in=150] (0,2);
\draw[fill] (0,3) circle(.1);
\draw (0,2) -- (0,3);
\draw[fill] (0,4) circle(.1);
\draw[fill] (0,5) circle(.1);
\draw (0,4) -- (0,5);
\draw (0,3) to [out=150,in=150] (0,5);
\draw[fill] (0,6) circle(.1);
\draw (0,4) to [out=30,in=30] (0,6);

\end{tikzpicture}
\end{center}
\caption{The $\A_2(1)$-module structure of $\H^{*-1}(MT\tO(1),\Z_2)$}
\label{fig:H^{*-1}(MTO(1),Z_2)}
\end{figure}

\begin{figure}[!h]
\begin{center}
\begin{tikzpicture}
\node at (0,-1) {0};
\node at (1,-1) {1};
\node at (2,-1) {2};
\node at (3,-1) {3};
\node at (4,-1) {4};
\node at (5,-1) {5};
\node at (6,-1) {$t-s$};
\node at (-1,0) {0};
\node at (-1,1) {1};
\node at (-1,2) {2};
\node at (-1,3) {3};
\node at (-1,4) {4};
\node at (-1,5) {5};
\node at (-1,6) {$s$};

\draw[->] (-0.5,-0.5) -- (-0.5,6);
\draw[->] (-0.5,-0.5) -- (6,-0.5);

\draw[fill] (0,0) circle(0.05);
\draw (2,1) -- (4,3);
\draw (4,3) -- (4,0);

\end{tikzpicture}
\end{center}
\caption{$\Omega_*^{\Pin^+}$}
\label{fig:Omega_*^{Pin^+}}
\end{figure}

Hence we have the following theorem
\begin{theorem}
\begin{table}[!h]
\centering
\begin{tabular}{c c}
\hline
$i$ & $\Omega^{\Pin^+}_i$\\
\hline
0& $\Z_2$\\
1& $0$\\
2& $\Z_2$\\
3 & $\Z_2$\\
4 & $\Z_{16}$\\ 
5 & $0$\\
\hline
\end{tabular}
\end{table}
\end{theorem}

The bordism invariant of $\Omega^{\Pin^+}_2$ is $w_1\cup\tilde{\eta}$.

The bordism invariant of $\Omega^{\Pin^+}_3$ is $w_1\cup\text{Arf}$.

The bordism invariant of $\Omega^{\Pin^+}_4$ is $\eta$.

Here $\eta$ is the usual Atiyah-Patodi-Singer eta-invariant of the 4d Dirac operator (=``\#zero eigenvalues + spectral asymmetry'').

\begin{theorem}
\begin{table}[!h]
\centering
\begin{tabular}{c c}
\hline
$i$ & $\TP_i(\Pin^+)$\\
\hline
0& $\Z_2$\\
1& $0$\\
2& $\Z_2$\\
3 & $\Z_2$\\
4 & $\Z_{16}$\\ 
5 & $0$\\
\hline
\end{tabular}
\end{table}
\end{theorem}

The 2d topological term is $w_1\cup\tilde{\eta}$.

The 3d topological term is $w_1\cup\text{Arf}$.

The 4d topological term is $\eta$.

\subsubsection{$\Omega^{\Pin^-}_d$}

Since the computation involves no odd torsion, we can use the Adams spectral sequence 
\bea
E_2^{s,t}=\Ext_{\A_2}^{s,t}(\H^*(M\Pin^+,\Z_2),\Z_2)\Rightarrow\pi_{t-s}(M\Pin^+)_2^{\wedge}=\Omega^{\Pin^-}_{t-s}.
\eea
$M\Pin^+=MT\Pin^-\sim M\Spin\wedge S^{-1}\wedge M\tO(1)$.

For $t-s<8$, we can identify the $E_2$-page with 
$$\Ext_{\A_2(1)}^{s,t}(\H^{*+1}(M\tO(1),\Z_2),\Z_2).$$

{By Thom's isomorphism, 
\bea
\H^{*+1}(M\tO(1),\Z_2)=\Z_2[w_1]U
\eea
where $U$ is the Thom class of the universal 1-bundle $E_1$ over $\B \tO(1)$ and $w_1$ is the 1st Stiefel-Whitney class of $E_1$ over $\B \tO(1)$.
}
The $\A_2(1)$-module structure of $\H^{*+1}(M\tO(1),\Z_2)$ and the $E_2$ page are shown in Figure \ref{fig:H^{*+1}(MO(1),Z_2)}, \ref{fig:Omega_*^{Pin^-}}.

\begin{figure}[!h]
\begin{center}
\begin{tikzpicture}[scale=0.5]

\node[below] at (0,0) {$U$};

\draw[fill] (0,0) circle(.1);
\draw[fill] (0,1) circle(.1);
\draw (0,0) -- (0,1);
\draw[fill] (0,2) circle(.1);
\draw[fill] (0,3) circle(.1);
\draw (0,2) -- (0,3);
\draw (0,1) to [out=150,in=150] (0,3);
\draw[fill] (0,4) circle(.1);
\draw (0,2) to [out=30,in=30] (0,4);
\draw[fill] (0,5) circle(.1);
\draw (0,4) -- (0,5);

\end{tikzpicture}
\end{center}
\caption{The $\A_2(1)$-module structure of $\H^{*+1}(M\tO(1),\Z_2)$}
\label{fig:H^{*+1}(MO(1),Z_2)}
\end{figure}

\begin{figure}[!h]
\begin{center}
\begin{tikzpicture}
\node at (0,-1) {0};
\node at (1,-1) {1};
\node at (2,-1) {2};
\node at (3,-1) {3};
\node at (4,-1) {4};
\node at (5,-1) {5};
\node at (6,-1) {$t-s$};
\node at (-1,0) {0};
\node at (-1,1) {1};
\node at (-1,2) {2};
\node at (-1,3) {3};
\node at (-1,4) {4};
\node at (-1,5) {5};
\node at (-1,6) {$s$};

\draw[->] (-0.5,-0.5) -- (-0.5,6);
\draw[->] (-0.5,-0.5) -- (6,-0.5);

\draw (0,0) -- (2,2);
\draw (2,2) -- (2,0);

\end{tikzpicture}
\end{center}
\caption{$\Omega_*^{\Pin^-}$}
\label{fig:Omega_*^{Pin^-}}
\end{figure}

Hence we have the following theorem
\begin{theorem}
\begin{table}[!h]
\centering
\begin{tabular}{c c}
\hline
$i$ & $\Omega^{\Pin^-}_i$\\
\hline
0& $\Z_2$\\
1& $\Z_2$\\
2& $\Z_8$\\
3 & $0$\\
4 & $0$\\ 
5 & $0$\\
\hline
\end{tabular}
\end{table}
\end{theorem}

The bordism invariant of $\Omega^{\Pin^-}_1$ is $\tilde{\eta}$.

The bordism invariant of $\Omega^{\Pin^-}_2$ is $\text{ABK}$ (the Arf-Brown-Kervaire invariant).

\begin{theorem}
\begin{table}[!h]
\centering
\begin{tabular}{c c}
\hline
$i$ & $\TP_i(\Pin^-)$\\
\hline
0& $\Z_2$\\
1& $\Z_2$\\
2& $\Z_8$\\
3 & $0$\\
4 & $0$\\ 
5 & $0$\\
\hline
\end{tabular}
\end{table}
\end{theorem}

The 1d topological term is $\tilde{\eta}$.

The 2d topological term is $\text{ABK}$.

\subsection{Atiyah-Hirzebruch spectral sequence}

If $H=\tO/\SO/\Spin/\Pin^{\pm}$, by the Atiyah-Hirzebruch spectral sequence, we have
\bea
\H_p(\B \mathbb{G},\Omega_q^H)\Rightarrow\Omega_{p+q}^H(\B \mathbb{G}).
\eea

If $H=\tO/\Pin^{\pm}$,
since $\Omega_d^H$ are finite, $\Omega_d^{H\times\mathbb{G}}=\Omega_d^H(\B \mathbb{G})$ are also finite, so
$\TP_d(H\times\mathbb{G})=\Omega_d^{H\times\mathbb{G}}$ for $H=\tO/\Pin^{\pm}$.

If $H=\SO/\Spin$, 
\bea
\Omega_q^{\SO}=\left\{\begin{array}{lllllll}\Z&q=0\\0&q=1\\0&q=2\\0&q=3\\\Z&q=4\\\Z_2&q=5\\0&q=6\end{array}\right..
\eea

\bea
\Omega_q^{\Spin}=\left\{\begin{array}{lllllll}\Z&q=0\\\Z_2&q=1\\\Z_2&q=2\\0&q=3\\\Z&q=4\\0&q=5\\0&q=6\end{array}\right..
\eea

If $\H_p(\B \mathbb{G},\Z)$ are finite for $p>0$, then $\Omega_6^H(\B \mathbb{G})$ is finite and
$\TP_5(H\times\mathbb{G})=\Omega_5^H(\B \mathbb{G})$ for $H=\SO/\Spin$.

If $\mathbb{G}=\PSU(2)=\SO(3)$, since 
$\H_2(\B \SO(3),\Z)$ and $\H_6(\B \SO(3),\Z)$ are finite, $\Omega_6^H(\B \mathbb{G})$ is also finite and
$\TP_5(H\times\mathbb{G})=\Omega_5^H(\B \mathbb{G})$ for $H=\SO/\Spin$.

If $\mathbb{G}=\PSU(3)$, then $\H_6(\B \PSU(3),\Z)$ contains a $\Z$ while $\H_2(\B \PSU(3),\Z)$ does not, so $\Omega_6^H(\B \mathbb{G})$ contains a $\Z$ and 
$\TP_5(H\times\mathbb{G})=\Omega_5^H(\B \mathbb{G})\times\Z$ for $H=\SO/\Spin$.

\subsection{$\B ^2G_b:\B ^2\Z_2,\B ^2\Z_3$}

\subsubsection{$\Omega^{\tO}_d(\B ^2\Z_2)$}
\label{sec:OB2Z2}

Since the computation involves no odd torsion, we can use the Adams spectral sequence 
\bea
E_2^{s,t}=\Ext_{\A_2}^{s,t}(\H^*(M\tO\wedge (\B ^2\Z_2)_+,\Z_2),\Z_2)\Rightarrow\pi_{t-s}(M\tO\wedge (\B ^2\Z_2)_+)_2^{\wedge}=\Omega^{\tO}_{t-s}(\B ^2\Z_2).
\eea

\cred{
\begin{theorem}[Thom]
$\bullet$ $\pi_*M\tO=\Omega=\Z_2[y_2,y_4,y_5,y_6,y_8,\dots]$.\\
$\bullet$ $\H^*(M\tO,\Z_2)=\A_2\otimes\Omega^*$ where $\Omega^*$ is the $\Z_2$-linear dual of $\Omega$.
\end{theorem}
$y_2=\RP^2$, $y_4=\RP^4$, $y_5$ is the Wu manifold $\W=\SU(3)/\SO(3)$, $\dots$\\
$y_2^*=w_2(TM)$ or $w_1(TM)^2$, $y_2^*=w_2(TM)^2$, $y_4^*=w_1(TM)^4$, $y_5^*=w_2(TM)w_3(TM)$, $\dots$
\begin{theorem}[Serre]
$$\H^*(\B^2\Z_2,\Z_2)=\Z_2[x_2,x_3,x_5,x_9,\dots]$$
where $x_3=\Sq^1x_2$, $x_5=\Sq^2x_3$, $x_9=\Sq^4x_5$, $\dots$
\end{theorem}
By K\"unneth theorem,
\bea
&&\H^*(M\tO\wedge(\B^2\Z_2)_+,\Z_2)\nn\\
&=&\H^*(M\tO,\Z_2)\otimes\H^*(\B ^2\Z_2,\Z_2)\nn\\
&=&\A_2\otimes\Z_2[y_2,y_4,y_5,y_6,y_8,\dots]^*\otimes\Z_2[x_2,x_3,x_5,x_9,\dots] \nn\\
&=&\A_2\oplus2\Sigma^2\A_2\oplus\Sigma^3\A_2\oplus4\Sigma^4\A_2\oplus4\Sigma^5\A_2\oplus\cdots
\eea
}

{Here $\Sigma^n\A_2$ is the $n$-th iterated shift of the graded algebra $\A_2$.}

\cred{Since 
\bea
\Ext_{\A_2}^{s,t}(\Sigma^r\A_2,\Z_2)=\left\{\begin{array}{ll}\Hom_{\A_2}^t(\Sigma^r\A_2,\Z_2)=\Z_2&\text{ if }t=r, s=0\\ 0 &\text{ else}\end{array}\right.,
\eea 
}
we have the following theorem
%Hence we have the following theorem (see \ref{sec:Adams} for detail)
\begin{theorem}
\begin{table}[!h]
\centering
\begin{tabular}{c c}
\hline
$i$ & $\Omega^{\tO}_i(\B ^2\Z_2)$\\
\hline
0& $\Z_2$\\
1& $0$\\
2& $\Z_2^2$\\
3 & $\Z_2$\\
4 & $\Z_2^4$\\ 
5 & $\Z_2^4$\\
\hline
\end{tabular}
\end{table}
\end{theorem}

The bordism invariants of $\Omega_2^{\tO}(\B ^2\Z_2)$ are $x_2,w_1^2$.

The bordism invariant of $\Omega_3^{\tO}(\B ^2\Z_2)$ is $x_3=w_1x_2$.

The bordism invariants of $\Omega_4^{\tO}(\B ^2\Z_2)$ are $x_2^2,w_1^4,w_1^2x_2,w_2^2$.

The bordism invariants of $\Omega^{\tO}_5(\B ^2\Z_2)$ are $x_2x_3,x_5,w_1^2x_3,w_2w_3.$

\cred{Here $w_i$ is the $i$-th Stiefel-Whitney class of the tangent bundle of $M$, $x_2$ is the generator of $\H^2(\B^2\Z_2,\Z_2)$, $x_3=\Sq^1x_2$, $x_5=\Sq^2\Sq^1x_2$, since there is a map $f:M\to\B^2\Z_2$ in the definition of cobordism group, we identify $x_2$ with $f^*(x_2)=f$. By Wu formula, $\Sq^2x_{d-2}=(w_2+w_1^2)x_{d-2}$ on $d$-manifolds.
}

\begin{theorem}
\begin{table}[!h]
\centering
\begin{tabular}{c c}
\hline
$i$ & $\TP_i(\tO\times \B \Z_2)$\\
\hline
0& $\Z_2$\\
1& $0$\\
2& $\Z_2^2$\\
3 & $\Z_2$\\
4 & $\Z_2^4$\\ 
5 & $\Z_2^4$\\
\hline
\end{tabular}
\end{table}
\end{theorem}

The 2d topological terms are $x_2,w_1^2$.

The 3d topological term is $x_3=w_1x_2$.

The 4d topological terms are $x_2^2,w_1^4,w_1^2x_2,w_2^2$.

The 5d topological terms are $x_2x_3,x_5,w_1^2x_3,w_2w_3.$

\subsubsection{$\Omega^{\SO}_d(\B ^2\Z_2)$}

\label{sec:SOB2Z2}

Since the computation involves no odd torsion, we can use the Adams spectral sequence 
\bea
E_2^{s,t}=\Ext_{\A_2}^{s,t}(\H^*(M\SO\wedge (\B ^2\Z_2)_+,\Z_2),\Z_2)\Rightarrow\pi_{t-s}(M\SO\wedge (\B ^2\Z_2)_+)_2^{\wedge}=\Omega^{\SO}_{t-s}(\B ^2\Z_2).
\eea

\cred{
Since 
$\H^*(\B ^2\Z_2,\Z_2)=\Z_2[x_2,x_3,x_5,x_9,\dots]$ where 
$x_2$ is the generator of $\H^2(\B^2\Z_2,\Z_2)$, $x_3=\Sq^1x_2$, $x_5=\Sq^2\Sq^1x_2$, $x_9=\Sq^4\Sq^2\Sq^1x_2$, etc, 
$\Sq^1x_2=x_3$, $\Sq^1x_3=0$, $\Sq^1(x_2^2)=0$, $\Sq^1(x_2x_3)=\Sq^1(x_5)=x_3^2$.
We have used \eqref{steenrel} and the Adem relations \eqref{Adem}.
}

\cred{
We shift Figure \ref{fig:Ext_{A_2}^{s,t}(H^*(MSO,Z_2),Z_2)} the same times as the dimension of $\H^*(\B ^2\Z_2,\Z_2)$ at each degree as a $\Z_2$-vector space. We obtain the $E_1$ page for $\Omega_*^{\SO}(\B^2\Z_2)$, the differentials $d_1$ are induced by $\Sq^1$, as shown in Figure \ref{fig:E_1pageOmega_*^{SO}(B^2Z_2)}.
 }

\begin{figure}[!h]
\begin{center}
\begin{tikzpicture}
\node at (0,-1) {0};
\node at (1,-1) {1};
\node at (2,-1) {2};
\node at (3,-1) {3};
\node at (4,-1) {4};
\node at (5,-1) {5};
\node at (6,-1) {6};
\node at (7,-1) {$t-s$};
\node at (-1,0) {0};
\node at (-1,1) {1};
\node at (-1,2) {2};
\node at (-1,3) {3};
\node at (-1,4) {4};
\node at (-1,5) {5};

\node at (-1,6) {$s$};

\draw[->] (-0.5,-0.5) -- (-0.5,6);
\draw[->] (-0.5,-0.5) -- (7,-0.5);

\draw (0,0) -- (0,5);

\draw (4,0) -- (4,5);
\draw[fill] (5,0) circle(0.05);

\draw (2,0) -- (2,5);
\draw (6,0) -- (6,5);

\draw (3,0) -- (3,5);

\draw (4.1,0) -- (4.1,5);

\draw (4.9,0) -- (4.9,5);

\draw (5.1,0) -- (5.1,5);

\draw (5.9,0) -- (5.9,5);

\draw[->][color=red] (3,0) -- (2,1);

\draw[->][color=red] (3,1) -- (2,2);

\draw[->][color=red] (3,2) -- (2,3);

\draw[->][color=red] (3,3) -- (2,4);

\draw[->][color=red] (3,4) -- (2,5);

\draw[->][color=red] (5.9,0) -- (5.1,1);

\draw[->][color=red] (5.9,1) -- (5.1,2);

\draw[->][color=red] (5.9,2) -- (5.1,3);

\draw[->][color=red] (5.9,3) -- (5.1,4);

\draw[->][color=red] (5.9,4) -- (5.1,5);

\end{tikzpicture}
\end{center}
\caption{$E_1$ page for $\Omega_*^{\SO}(\B^2\Z_2)$}
\label{fig:E_1pageOmega_*^{SO}(B^2Z_2)}
\end{figure}

\cred{
There is a differential $d_2$ corresponding to the Bockstein homomorphism  $\beta_{(2,4)}:\H^*(-,\Z_4)\to\H^{*+1}(-,\Z_2)$ associated to $0\to\Z_2\to\Z_8\to\Z_4\to0$ \cite{may1981bockstein}. See \ref{Bockstein} for the definition of Bockstein homomorphisms.
}
\cred{
Note that 
\bea\label{x_2x_3+x_5}
\beta_{(2,4)}\cP_2(x_2)&=&\frac{1}{4}\delta\cP_2(x_2)\mod2\nn\\
&=&\frac{1}{4}\delta(x_2\cup x_2+x_2\hcup{1}\delta x_2)\nn\\
&=&\frac{1}{4}(\delta x_2\cup x_2+x_2\cup\delta x_2+\delta(x_2\hcup{1}\delta x_2))\nn\\
&=&\frac{1}{4}(2x_2\cup\delta x_2+\delta x_2\hcup{1}\delta x_2)\nn\\
&=&x_2\cup(\frac{1}{2}\delta x_2)+(\frac{1}{2}\delta x_2)\hcup{1}(\frac{1}{2}\delta x_2)\nn\\
&=&x_2\Sq^1x_2+\Sq^1x_2\hcup{1}\Sq^1x_2\nn\\
&=&x_2\Sq^1x_2+\Sq^2\Sq^1x_2\nn\\
&=&x_2x_3+x_5
\eea
We have used $\beta_{(2,4)}=\frac{1}{4}\delta\mod2$, the Steenrod's formula \eqref{eq:Steenrod's}, $\Sq^1=\beta_{(2,2)}=\frac{1}{2}\delta\mod2$, and the definition $\Sq^kx_n=x_n\hcup{n-k}x_n$.
}

So there is a differential such that
$d_2(x_2x_3+x_5)=x_2^2h_0^2$.

\cred{Then take the differentials $d_2$ into account, we obtain the $E_2$ page for $\Omega_*^{\SO}(\B^2\Z_2)$, as shown in Figure \ref{fig:Omega_*^{SO}(B^2Z_2)}.
}
 
%The $E_2$ page is shown in Figure \ref{fig:Omega_*^{SO}(B^2Z_2)}.

\begin{figure}[!h]
\begin{center}
\begin{tikzpicture}
\node at (0,-1) {0};
\node at (1,-1) {1};
\node at (2,-1) {2};
\node at (3,-1) {3};
\node at (4,-1) {4};
\node at (5,-1) {5};
\node at (6,-1) {$t-s$};
\node at (-1,0) {0};
\node at (-1,1) {1};
\node at (-1,2) {2};
\node at (-1,3) {3};
\node at (-1,4) {4};
\node at (-1,5) {5};
\node at (-1,6) {$s$};

\draw[->] (-0.5,-0.5) -- (-0.5,6);
\draw[->] (-0.5,-0.5) -- (6,-0.5);

\draw (0,0) -- (0,5);

\draw[fill] (2,0) circle(0.05);
\draw (4,0) -- (4,5);
\draw (4.1,0) -- (4.1,5);

\draw[fill] (4.9,0) circle(0.05);
\draw[fill] (5.1,0) circle(0.05);
\draw (5,0) -- (5,5);

\draw[color=red][->] (5,0) -- (4.1,2);
\draw[color=red][->] (5,1) -- (4.1,3);
\draw[color=red][->] (5,2) -- (4.1,4);
\draw[color=red][->] (5,3) -- (4.1,5);

\end{tikzpicture}
\end{center}
\caption{$\Omega_*^{\SO}(\B ^2\Z_2)$}
\label{fig:Omega_*^{SO}(B^2Z_2)}
\end{figure}

Hence we have the following theorem %(see \ref{sec:Adams} for detail)
\begin{theorem}
\begin{table}[!h]
\centering
\begin{tabular}{c c}
\hline
$i$ & $\Omega^{\SO}_i(\B ^2\Z_2)$\\
\hline
0& $\Z$\\
1& $0$\\
2& $\Z_2$\\
3 & $0$\\
4 & $\Z\times\Z_4$\\ 
5 & $\Z_2^2$\\
\hline
\end{tabular}
\end{table}
\end{theorem}

The bordism invariant of $\Omega_2^{\SO}(\B ^2\Z_2)$ is $x_2$.

The bordism invariants of $\Omega_{4}^{\SO}(\B ^2\Z_2)$ are $\sigma$ and $\mathcal{P}_2(x_2)$.

The bordism invariants of $\Omega^{\SO}_5(\B ^2\Z_2)$ are $x_5=x_2x_3$ and $w_2w_3$.

\cred{
Here $\mathcal{P}_2(x_2)$ is the Pontryagin square of $x_2$.
$\sigma$ is the signature of a 4-manifold $M$.
 $x_2x_3+x_5=\frac{1}{2}\tilde{w}_1\cP_2(x_2)$ \cite{Arun2017} where $\tilde{w}_1$ is the twisted first Stiefel-Whitney class of the tangent bundle, in particular, $w_1=0$ implies $\tilde{w}_1=0$, so $x_2x_3=x_5$ on oriented 5-manifolds.
}

\begin{theorem}
\begin{table}[!h]
\centering
\begin{tabular}{c c}
\hline
$i$ & $\TP_i(\SO\times \B \Z_2)$\\
\hline
0& $0$\\
1& $0$\\
2& $\Z_2$\\
3 & $\Z$\\
4 & $\Z_4$\\ 
5 & $\Z_2^2$\\
\hline
\end{tabular}
\end{table}
\end{theorem}

The 2d topological term is $x_2$.

The 3d topological term is $\frac{1}{3}\text{CS}_3^{(TM)}$.

The 4d topological term is $\mathcal{P}_2(x_2)$.

The 5d topological terms are $x_5=x_2x_3$ and $w_2w_3$.

\subsubsection{$\Omega^{\Spin}_d(\B ^2\Z_2)$}
Since the computation involves no odd torsion, we can use the Adams spectral sequence 
\bea
E_2^{s,t}=\Ext_{\A_2}^{s,t}(\H^*(M\Spin\wedge (\B ^2\Z_2)_+,\Z_2),\Z_2)\Rightarrow\pi_{t-s}(M\Spin\wedge (\B ^2\Z_2)_+)_2^{\wedge}=\Omega^{\Spin}_{t-s}(\B ^2\Z_2).
\eea

For $t-s<8$, we can identify the $E_2$-page with 
$$\Ext_{\A_2(1)}^{s,t}(\H^*(\B ^2\Z_2,\Z_2),\Z_2).$$

$\H^*(\B ^2\Z_2,\Z_2)=\Z_2[x_2,x_3,x_5,x_9,\dots]$ where
 $x_2$ is the generator of $\H^2(\B^2\Z_2,\Z_2)$, $x_3=\Sq^1x_2$, $x_5=\Sq^2\Sq^1x_2$, $x_9=\Sq^4\Sq^2\Sq^1x_2$, etc,
 $\Sq^1x_2=x_3$, $\Sq^2x_2=x_2^2$, $\Sq^1x_3=0$, $\Sq^2x_3=x_5$, $\Sq^1(x_2^2)=0$, $\Sq^1(x_2x_3)=x_3^2$, $\Sq^1x_5=\Sq^2x_2^2=x_3^2$, $\Sq^2x_5=0$. $\Sq^2(x_2x_3)=x_2^2x_3+x_2x_5$.
 We have used \eqref{steenrel} and the Adem relations \eqref{Adem}. 

\cred{
There is a differential $d_2$ corresponding to the Bockstein homomorphism  $\beta_{(2,4)}:\H^*(-,\Z_4)\to\H^{*+1}(-,\Z_2)$ associated to $0\to\Z_2\to\Z_8\to\Z_4\to0$ \cite{may1981bockstein}. See \ref{Bockstein} for the definition of Bockstein homomorphisms.
}

By \eqref{x_2x_3+x_5}, there is a differential such that
$d_2(x_2x_3+x_5)=x_2^2h_0^2$. 

%The $\A_2(1)$-module structure of $\H^*(\B ^2\Z_2,\Z_2)$ and the $E_2$ page are shown in Figure \ref{fig:H^*(B^2Z_2,Z_2)}, \ref{fig:Omega_*^{Spin}(B^2Z_2)}.

\begin{figure}[!h]
\begin{center}
\begin{tikzpicture}[scale=0.5]
\node[below] at (0,0) {$1$};

\draw[fill] (0,0) circle(.1);

\node[below] at (0,2) {$x_2$};

\draw[fill] (0,2) circle(.1);
\draw[fill] (0,3) circle(.1);
\draw[fill] (0,4) circle(.1);
\draw (0,2) -- (0,3);
\draw (0,2) to [out=150,in=150] (0,4);
\draw[fill] (1,5) circle(.1);
\draw[fill] (1,6) circle(.1);
\draw (1,5) -- (1,6);
\draw (0,3) to [out=30,in=150] (1,5);
\draw (0,4) to [out=30,in=150] (1,6);
\draw[fill] (2,5) circle(.1);

\node[below] at (2,5) {$x_2x_3$};

\draw (2,5) -- (1,6);
\draw[fill] (2,7) circle(.1);
\draw (2,5) to [out=30,in=30] (2,7);
\draw[fill] (2,8) circle(.1);
\draw (2,7) -- (2,8);
\draw[fill] (2,10) circle(.1);
\draw (2,8) to [out=150,in=150] (2,10);
\end{tikzpicture}
\end{center}
\caption{The $\A_2(1)$-module structure of $\H^*(\B ^2\Z_2,\Z_2)$}
\label{fig:H^*(B^2Z_2,Z_2)}
\end{figure}

The $\A_2(1)$-module structure of $\H^*(\B ^2\Z_2,\Z_2)$ is shown in Figure \ref{fig:H^*(B^2Z_2,Z_2)}.
The dot at the bottom is a $\Z_2$ which has been discussed before. Now we consider the part above the bottom dot. We will use Lemma \ref{principle} several times. Two steps are shown in Figure \ref{fig:Spin-B2Z2-first-step}, \ref{fig:Spin-B2Z2-second-step}.

\begin{figure}[!h]
\begin{center}
\begin{tikzpicture}[scale=0.5]

\node at (-4,0) {$L$};

\draw[fill=red,color=red] (-4,6) circle(.1);

\draw[fill=red,color=red] (-4,10) circle(.1);

\draw[->] (-3,0) -- (-1,0);

\draw[color=red] (1,6) circle(.3);

\draw[color=red] (2,10) circle(.3);

\node at (1,0) {$M$}; 

\node[below] at (0,2) {$x_2$};

\draw[fill] (0,2) circle(.1);
\draw[fill] (0,3) circle(.1);
\draw[fill] (0,4) circle(.1);
\draw (0,2) -- (0,3);
\draw (0,2) to [out=150,in=150] (0,4);
\draw[fill] (1,5) circle(.1);
\draw[fill] (1,6) circle(.1);
\draw (1,5) -- (1,6);
\draw (0,3) to [out=30,in=150] (1,5);
\draw (0,4) to [out=30,in=150] (1,6);
\draw[fill] (2,5) circle(.1);

\node[below] at (2,5) {$x_2x_3$};

\draw (2,5) -- (1,6);
\draw[fill] (2,7) circle(.1);
\draw (2,5) to [out=30,in=30] (2,7);
\draw[fill] (2,8) circle(.1);
\draw (2,7) -- (2,8);
\draw[fill] (2,10) circle(.1);
\draw (2,8) to [out=150,in=150] (2,10);

\draw[->] (3,0) -- (5,0);

\node at (6,0) {$N$};

\node[below] at (6,2) {$x_2$};

\draw[fill=blue,color=blue] (6,2) circle(.1);
\draw[fill=blue,color=blue] (6,3) circle(.1);
\draw[fill=blue,color=blue] (6,4) circle(.1);
\draw[color=blue] (6,2) -- (6,3);
\draw[color=blue] (6,2) to [out=150,in=150] (6,4);
\draw[fill=blue,color=blue] (7,5) circle(.1);

\draw[color=blue] (6,3) to [out=30,in=150] (7,5);

\draw[fill=blue,color=blue] (8,5) circle(.1);

\node[below] at (8,5) {$x_2x_3$};

\draw[fill=blue,color=blue] (8,7) circle(.1);
\draw[color=blue] (8,5) to [out=30,in=30] (8,7);
\draw[fill=blue,color=blue] (8,8) circle(.1);
\draw[color=blue] (8,7) -- (8,8);

\end{tikzpicture}
\end{center}
\caption{First step to get the $E_2$ page of $\Omega_*^{\Spin}(\B^2\Z_2)$.}
\label{fig:Spin-B2Z2-first-step}
\end{figure}

\begin{figure}[!h]
\begin{center}
\begin{tikzpicture}[scale=0.5]

\node at (2,0) {$P$};

\draw[->] (3,0) -- (5,0);

\draw[fill=green,color=green] (2,4) circle(.1);
\draw[fill=green,color=green] (2,5) circle(.1);
\draw[fill=green,color=green] (2,8) circle(.1);

\node at (7,0) {$N$};

\draw[color=green] (6,4) circle(.3);
\draw[color=green] (7,5) circle(.3);
\draw[color=green] (8,8) circle(.3);

\node[below] at (6,2) {$x_2$};

\draw[fill=blue,color=blue] (6,2) circle(.1);
\draw[fill=blue,color=blue] (6,3) circle(.1);
\draw[fill=blue,color=blue] (6,4) circle(.1);
\draw[color=blue] (6,2) -- (6,3);
\draw[color=blue] (6,2) to [out=150,in=150] (6,4);
\draw[fill=blue,color=blue] (7,5) circle(.1);

\draw[color=blue] (6,3) to [out=30,in=150] (7,5);

\draw[fill=blue,color=blue] (8,5) circle(.1);

\node[below] at (8,5) {$x_2x_3$};

\draw[fill=blue,color=blue] (8,7) circle(.1);
\draw[color=blue] (8,5) to [out=30,in=30] (8,7);
\draw[fill=blue,color=blue] (8,8) circle(.1);
\draw[color=blue] (8,7) -- (8,8);

\node at (12,0) {$Q$};

\draw[->] (9,0) -- (11,0);

\node[below] at (12,2) {$x_2$};

\draw[fill=orange,color=orange] (12,2) circle(.1);
\draw[fill=orange,color=orange] (12,3) circle(.1);

\draw[color=orange] (12,2) -- (12,3);

\draw[fill=orange,color=orange] (14,5) circle(.1);

\node[below] at (14,5) {$x_2x_3$};

\draw[fill=orange,color=orange] (14,7) circle(.1);
\draw[color=orange] (14,5) to [out=30,in=30] (14,7);

\end{tikzpicture}
\end{center}
\caption{Second step to get the $E_2$ page of $\Omega_*^{\Spin}(\B^2\Z_2)$.}
\label{fig:Spin-B2Z2-second-step}
\end{figure}

We will proceed in the reversed order.

First, we apply Lemma \ref{principle} to the short exact sequence of $\A_2(1)$-modules:
$0\to P\to N\to Q\to 0$ in
the second step (as shown in Figure \ref{fig:Spin-B2Z2-second-step}), 
the Adams chart of $\Ext_{\A_2(1)}^{s,t}(N,\Z_2)$ is shown in Figure \ref{fig:Ext_{A_2(1)}^{s,t}(N,Z_2)}.

\begin{figure}[!h]
\begin{center}
\begin{tikzpicture}
\node at (0,-1) {0};
\node at (1,-1) {1};
\node at (2,-1) {2};
\node at (3,-1) {3};
\node at (4,-1) {4};
\node at (5,-1) {5};
\node at (6,-1) {6};

\node at (7,-1) {$t-s$};
\node at (-1,0) {0};
\node at (-1,1) {1};
\node at (-1,2) {2};
\node at (-1,3) {3};
\node at (-1,4) {4};
\node at (-1,5) {5};

\node at (-1,6) {$s$};

\draw[->] (-0.5,-0.5) -- (-0.5,6);
\draw[->] (-0.5,-0.5) -- (7,-0.5);

\draw[color=orange] (2,0) -- (4,2);

\draw[color=orange] (4,1) -- (4,2);

\draw[color=orange] (4,1) -- (6,3);
\draw[color=orange] (5,0) -- (5,5);
\draw[color=green] (4.1,0) -- (4.1,5);

\draw[color=green] (4.1,0) -- (6.1,2);
\draw[color=green] (5.1,0) -- (5.1,5);
\draw[color=green] (5.1,0) -- (6.1,1);

\draw[->][color=purple] (4.1,0) -- (3,1);
\draw[->][color=purple] (5.1,0) -- (4,1);
\draw[->][color=purple] (5.1,1) -- (4,2);
\draw[->][color=purple] (6.1,1) -- (5,2);

\end{tikzpicture}
\end{center}
\caption{Adams chart of $\Ext_{\A_2(1)}^{s,t}(N,\Z_2)$. The arrows indicate the differential $d_1$.}
\label{fig:Ext_{A_2(1)}^{s,t}(N,Z_2)}
\end{figure}

Next, we apply Lemma \ref{principle} to the short exact sequence of $\A_2(1)$-modules:
$0\to L\to M\to N\to 0$ in
the first step (as shown in Figure \ref{fig:Spin-B2Z2-first-step}), 
the Adams chart of $\Ext_{\A_2(1)}^{s,t}(M,\Z_2)$ is shown in Figure \ref{fig:Ext_{A_2(1)}^{s,t}(M,Z_2)}.

\begin{figure}[!h]
\begin{center}
\begin{tikzpicture}
\node at (0,-1) {0};
\node at (1,-1) {1};
\node at (2,-1) {2};
\node at (3,-1) {3};
\node at (4,-1) {4};
\node at (5,-1) {5};
\node at (6,-1) {6};

\node at (7,-1) {$t-s$};
\node at (-1,0) {0};
\node at (-1,1) {1};
\node at (-1,2) {2};
\node at (-1,3) {3};
\node at (-1,4) {4};
\node at (-1,5) {5};

\node at (-1,6) {$s$};

\draw[->] (-0.5,-0.5) -- (-0.5,6);
\draw[->] (-0.5,-0.5) -- (7,-0.5);

\draw[fill=blue,color=blue] (2,0) circle(0.05);
\draw[color=blue] (4,1) -- (4,5);

\draw[color=blue] (5,0) -- (5,5);

\draw[color=blue] (5.1,1) -- (5.1,5);
\draw[color=blue] (5.1,1) -- (6.1,2);

\draw[color=red] (6,0) -- (6,5);

\draw[->][color=purple] (6,0) -- (5.1,1);
\draw[->][color=purple] (6,1) -- (5.1,2);
\draw[->][color=purple] (6,2) -- (5.1,3);
\draw[->][color=purple] (6,3) -- (5.1,4);
\draw[->][color=purple] (6,4) -- (5.1,5);

\end{tikzpicture}
\end{center}
\caption{Adams chart of $\Ext_{\A_2(1)}^{s,t}(M,\Z_2)$. The arrows indicate the differential $d_1$.}
\label{fig:Ext_{A_2(1)}^{s,t}(M,Z_2)}
\end{figure}

Then take the differentials $d_2$ into account, we obtain the $E_2$ page for $\Omega_*^{\Spin}(\B^2\Z_2)$, as shown in Figure \ref{fig:Omega_*^{Spin}(B^2Z_2)}.

\begin{figure}[!h]
\begin{center}
\begin{tikzpicture}
\node at (0,-1) {0};
\node at (1,-1) {1};
\node at (2,-1) {2};
\node at (3,-1) {3};
\node at (4,-1) {4};
\node at (5,-1) {5};
\node at (6,-1) {$t-s$};
\node at (-1,0) {0};
\node at (-1,1) {1};
\node at (-1,2) {2};
\node at (-1,3) {3};
\node at (-1,4) {4};
\node at (-1,5) {5};
\node at (-1,6) {$s$};

\draw[->] (-0.5,-0.5) -- (-0.5,6);
\draw[->] (-0.5,-0.5) -- (6,-0.5);

\draw (0,0) -- (0,5);
\draw (0,0) -- (2,2);
\draw[fill] (2,0) circle(0.05);
\draw (4,3) -- (4,5);
\draw (4.1,1) -- (4.1,5);
\draw (5,0) -- (5,5);

\draw[color=red][->] (5,0) -- (4.1,2);
\draw[color=red][->] (5,1) -- (4.1,3);
\draw[color=red][->] (5,2) -- (4.1,4);
\draw[color=red][->] (5,3) -- (4.1,5);

\end{tikzpicture}
\end{center}
\caption{$\Omega_*^{\Spin}(\B ^2\Z_2)$}
\label{fig:Omega_*^{Spin}(B^2Z_2)}
\end{figure}

Hence we have the following theorem %(see \ref{sec:Adams} for detail)
\begin{theorem}
\begin{table}[!h]
\centering
\begin{tabular}{c c}
\hline
$i$ & $\Omega^{\Spin}_i(\B ^2\Z_2)$\\
\hline
0& $\Z$\\
1& $\Z_2$\\
2& $\Z_2^2$\\
3 & $0$\\
4 & $\Z\times\Z_2$\\ 
5 & $0$\\
\hline
\end{tabular}
\end{table}
\end{theorem}

The bordism invariants of $\Omega_{2}^{\Spin}(\B ^2\Z_2)$ are $x_2$ and Arf.

The bordism invariants of $\Omega_{4}^{\Spin}(\B ^2\Z_2)$ are $\frac{\sigma}{16}$ and $\frac{\mathcal{P}_2(x_2)}{2}$.

\cred{
By Wu formula, $x_2^2=\Sq^2(x_2)=(w_2(TM)+w_1(TM)^2)x_2=0$ on Spin 4-manifolds, $x_5=\Sq^2(x_3)=(w_2(TM)+w_1(TM)^2)x_3=0$ on Spin 5-manifolds, $\mathcal{P}_2(x_2)=x_2^2=0\mod2$ on Spin 4-manifolds.
}

\cred{
Here Arf is the Arf invariant.
 $\sigma$ is the signature of Spin 4-manifold, it is a multiple of 16 by Rokhlin's theorem.
}

\begin{theorem}
\begin{table}[!h]
\centering
\begin{tabular}{c c}
\hline
$i$ & $\TP_i(\Spin\times \B \Z_2)$\\
\hline
0& $0$\\
1& $\Z_2$\\
2& $\Z_2^2$\\
3 & $\Z$\\
4 & $\Z_2$\\ 
5 & $0$\\
\hline
\end{tabular}
\end{table}
\end{theorem}

The 2d topological terms are $x_2$ and Arf.

The 3d topological term is $\frac{1}{48}\text{CS}_3^{(TM)}$.

The 4d topological term is $\frac{\mathcal{P}_2(x_2)}{2}$.

\subsubsection{$\Omega^{\Pin^+}_d(\B ^2\Z_2)$}

Since the computation involves no odd torsion, we can use the Adams spectral sequence 
\bea
E_2^{s,t}=\Ext_{\A_2}^{s,t}(\H^*(M\Pin^-\wedge (\B ^2\Z_2)_+,\Z_2),\Z_2)\Rightarrow\pi_{t-s}(M\Pin^-\wedge (\B ^2\Z_2)_+)_2^{\wedge}=\Omega^{\Pin^+}_{t-s}(\B ^2\Z_2).
\eea
$M\Pin^-=MT\Pin^+\sim M\Spin\wedge S^1\wedge MT\tO(1)$.

For $t-s<8$, we can identify the $E_2$-page with 
$$\Ext_{\A_2(1)}^{s,t}(\H^{*-1}(MT\tO(1),\Z_2)\otimes\H^*(\B ^2\Z_2,\Z_2),\Z_2).$$

The $\A_2(1)$-module structure of $\H^{*-1}(MT\tO(1),\Z_2)\otimes\H^*(\B ^2\Z_2,\Z_2)$ and the $E_2$ page are shown in Figure \ref{fig:H^{*-1}(MTO(1),Z_2)otimesH^*(B^2Z_2,Z_2)}, \ref{fig:Omega_*^{Pin^+}(B^2Z_2)}.

\begin{figure}[!h]
\begin{center}
\begin{tikzpicture}[scale=0.5]

\node[below] at (0,0) {$U$};

\draw[fill] (0,0) circle(.1);
\draw[fill] (0,1) circle(.1);
\draw (0,0) -- (0,1);
\draw[fill] (0,2) circle(.1);
\draw (0,0) to [out=150,in=150] (0,2);
\draw[fill] (0,3) circle(.1);
\draw (0,2) -- (0,3);
\draw[fill] (0,4) circle(.1);
\draw[fill] (0,5) circle(.1);
\draw (0,4) -- (0,5);
\draw (0,3) to [out=150,in=150] (0,5);
\draw[fill] (0,6) circle(.1);
\draw (0,4) to [out=30,in=30] (0,6);

\node[below] at (10,0) {$1$};

\draw[fill] (10,0) circle(.1);

\node[below] at (10,2) {$x_2$};

\draw[fill] (10,2) circle(.1);
\draw[fill] (10,3) circle(.1);
\draw[fill] (10,4) circle(.1);
\draw (10,2) -- (10,3);
\draw (10,2) to [out=150,in=150] (10,4);
\draw[fill] (11,5) circle(.1);
\draw[fill] (11,6) circle(.1);
\draw (11,5) -- (11,6);
\draw (10,3) to [out=30,in=150] (11,5);
\draw (10,4) to [out=30,in=150] (11,6);
\draw[fill] (12,5) circle(.1);

\node[below] at (12,5) {$x_2x_3$};

\draw (12,5) -- (11,6);
\draw[fill] (12,7) circle(.1);
\draw (12,5) to [out=30,in=30] (12,7);
\draw[fill] (12,8) circle(.1);
\draw (12,7) -- (12,8);
\draw[fill] (12,10) circle(.1);
\draw (12,8) to [out=150,in=150] (12,10);

\node at (5,5) {$\bigotimes$};
\node at (-2,-10) {$=$};

\node[below] at (0,-15) {$U$};

\draw[fill] (0,-15) circle(.1);
\draw[fill] (0,-14) circle(.1);
\draw (0,-15) -- (0,-14);
\draw[fill] (0,-13) circle(.1);
\draw (0,-15) to [out=150,in=150] (0,-13);
\draw[fill] (0,-12) circle(.1);
\draw (0,-13) -- (0,-12);
\draw[fill] (0,-11) circle(.1);
\draw[fill] (0,-10) circle(.1);
\draw (0,-11) -- (0,-10);
\draw (0,-12) to [out=150,in=150] (0,-10);
\draw[fill] (0,-9) circle(.1);
\draw (0,-11) to [out=30,in=30] (0,-9);

\node[below] at (2,-13) {$x_2U$};

\draw[fill] (2,-13) circle(.1);
\draw[fill] (2,-12) circle(.1);
\draw (2,-13) -- (2,-12);
\draw[fill] (2,-11) circle(.1);
\draw (2,-13) to [out=150,in=150] (2,-11);
\draw[fill] (3,-10) circle(.1);
\draw (2,-12) to [out=30,in=150] (3,-10);
\draw[fill] (2,-10) circle(.1);
\draw (2,-11) -- (2,-10);
\draw[fill] (3,-9) circle(.1);
\draw (3,-10) -- (3,-9);
\draw (2,-11) to [out=30,in=150] (3,-9);
\draw[fill] (3,-8) circle(.1);
\draw (2,-10) to [out=30,in=150] (3,-8);
\draw[fill] (3,-7) circle(.1);
\draw (3,-8) -- (3,-7);
\draw (3,-9) to [out=30,in=30] (3,-7); 
 
\node[below] at (5,-12) {$w_1x_2U$};

\draw[fill] (5,-12) circle(.1);
\draw[fill] (5,-11) circle(.1);
\draw[fill] (5,-10) circle(.1);
\draw (5,-12) -- (5,-11);
\draw (5,-12) to [out=150,in=150] (5,-10);
\draw[fill] (6,-9) circle(.1);
\draw[fill] (6,-8) circle(.1);
\draw (6,-9) -- (6,-8);
\draw (5,-11) to [out=30,in=150] (6,-9);
\draw (5,-10) to [out=30,in=150] (6,-8);
\draw[fill] (7,-9) circle(.1);
\draw (7,-9) -- (6,-8);
\draw[fill] (7,-7) circle(.1);
\draw (7,-9) to [out=30,in=30] (7,-7);
\draw[fill] (7,-6) circle(.1);
\draw (7,-7) -- (7,-6);
\draw[fill] (7,-4) circle(.1);
\draw (7,-6) to [out=150,in=150] (7,-4);

\node[below] at (4,-11) {$x_2^2U$};

\draw[fill] (4,-11) circle(.1);
\draw (4,-11) -- (5,-10);
\draw[fill] (4,-9) circle(.1);
\draw (4,-11) to [out=150,in=150] (4,-9);
\draw[fill] (4,-8) circle(.1);
\draw (4,-9) -- (4,-8);
\draw[fill] (4,-7) circle(.1);
\draw[fill] (4,-6) circle(.1);
\draw (4,-7) -- (4,-6);
\draw (4,-8) to [out=150,in=150] (4,-6);
\draw[fill] (4,-5) circle(.1);
\draw (4,-7) to [out=30,in=30] (4,-5);

\node[below] at (8,-10) {$w_1^2x_3U$};

\draw[fill] (8,-10) circle(.1);
\draw[fill] (8,-9) circle(.1);
\draw (8,-10) -- (8,-9);
\draw[fill] (8,-8) circle(.1);
\draw (8,-10) to [out=150,in=150] (8,-8);
\draw[fill] (9,-7) circle(.1);
\draw (8,-9) to [out=30,in=150] (9,-7);
\draw[fill] (8,-7) circle(.1);
\draw (8,-8) -- (8,-7);
\draw[fill] (9,-6) circle(.1);
\draw (9,-7) -- (9,-6);
\draw (8,-8) to [out=30,in=150] (9,-6);
\draw[fill] (9,-5) circle(.1);
\draw (8,-7) to [out=30,in=150] (9,-5);
\draw[fill] (9,-4) circle(.1);
\draw (9,-5) -- (9,-4);
\draw (9,-6) to [out=30,in=30] (9,-4);

\node[below] at (10,-10) {$x_2x_3U$};

\draw[fill] (10,-10) circle(.1);
\draw[fill] (10,-9) circle(.1);
\draw (10,-10) -- (10,-9);
\draw[fill] (10,-8) circle(.1);
\draw (10,-10) to [out=150,in=150] (10,-8);
\draw[fill] (11,-7) circle(.1);
\draw (10,-9) to [out=30,in=150] (11,-7);
\draw[fill] (10,-7) circle(.1);
\draw (10,-8) -- (10,-7);
\draw[fill] (11,-6) circle(.1);
\draw (11,-7) -- (11,-6);
\draw (10,-8) to [out=30,in=150] (11,-6);
\draw[fill] (11,-5) circle(.1);
\draw (10,-7) to [out=30,in=150] (11,-5);
\draw[fill] (11,-4) circle(.1);
\draw (11,-5) -- (11,-4);
\draw (11,-6) to [out=30,in=30] (11,-4); 

\end{tikzpicture}
\end{center}
\caption{The $\A_2(1)$-module structure of $\H^{*-1}(MT\tO(1),\Z_2)\otimes\H^*(\B ^2\Z_2,\Z_2)$}
\label{fig:H^{*-1}(MTO(1),Z_2)otimesH^*(B^2Z_2,Z_2)}
\end{figure}

\begin{figure}[!h]
\begin{center}
\begin{tikzpicture}
\node at (0,-1) {0};
\node at (1,-1) {1};
\node at (2,-1) {2};
\node at (3,-1) {3};
\node at (4,-1) {4};
\node at (5,-1) {5};
\node at (6,-1) {$t-s$};
\node at (-1,0) {0};
\node at (-1,1) {1};
\node at (-1,2) {2};
\node at (-1,3) {3};
\node at (-1,4) {4};
\node at (-1,5) {5};
\node at (-1,6) {$s$};

\draw[->] (-0.5,-0.5) -- (-0.5,6);
\draw[->] (-0.5,-0.5) -- (6,-0.5);

\draw[fill] (0,0) circle(0.05);
\draw (2,1) -- (4,3);
\draw (4,3) -- (4,0);
\draw[fill] (2,0) circle(0.05);
\draw (3,0) -- (3.9,1);
\draw (3.9,1) -- (3.9,0);
\draw[fill] (4.9,0) circle(0.05);
\draw[fill] (5.1,0) circle(0.05);

\end{tikzpicture}
\end{center}
\caption{$\Omega_*^{\Pin^+}(\B ^2\Z_2)$}
\label{fig:Omega_*^{Pin^+}(B^2Z_2)}
\end{figure}

Hence we have the following theorem %(see \ref{sec:Adams} for detail)
\begin{theorem}
\begin{table}[!h]
\centering
\begin{tabular}{c c}
\hline
$i$ & $\Omega^{\Pin^+}_i(\B ^2\Z_2)$\\
\hline
0& $\Z_2$\\
1& $0$\\
2& $\Z_2^2$\\
3 & $\Z_2^2$\\
4 & $\Z_4\times\Z_{16}$\\ 
5 & $\Z_2^2$\\
\hline
\end{tabular}
\end{table}
\end{theorem}

The bordism invariants of $\Omega^{\Pin^+}_2(\B ^2\Z_2)$ are $x_2$ and $w_1\tilde{\eta}$.

The bordism invariants of $\Omega^{\Pin^+}_3(\B ^2\Z_2)$ are $w_1x_2=x_3$ and $w_1\text{Arf}$.

The bordism invariants of $\Omega^{\Pin^+}_4(\B ^2\Z_2)$ are $q_s(x_2)$ and $\eta$.

The bordism invariants of $\Omega^{\Pin^+}_5(\B ^2\Z_2)$ are $x_2x_3$ and $w_1^2x_3{(=x_5)}$.

Here $\tilde{\eta}$ is the ``mod 2
index'' of the 1d Dirac operator (\#zero eigenvalues mod 2, no
contribution from spectral asymmetry).

 $x_3=\Sq^1x_2=w_1x_2$ on 3-manifolds by Wu formula.

$q_s$ is explained in the footnotes of Table \ref{4d bordism groups}.

 $\eta$ is the usual Atiyah-Patodi-Singer eta-invariant of the 4d Dirac operator (=``\#zero eigenvalues + spectral asymmetry'').

 $x_5=\Sq^2x_3=(w_2+w_1^2)x_3=w_1^2x_3$ on Pin$^+$ 5-manifolds by Wu formula.

\begin{theorem}
\begin{table}[!h]
\centering
\begin{tabular}{c c}
\hline
$i$ & $\TP_i(\Pin^+\times \B \Z_2)$\\
\hline
0& $\Z_2$\\
1& $0$\\
2& $\Z_2^2$\\
3 & $\Z_2^2$\\
4 & $\Z_4\times\Z_{16}$\\ 
5 & $\Z_2^2$\\
\hline
\end{tabular}
\end{table}
\end{theorem}

The 2d topological terms are $x_2$ and $w_1\tilde{\eta}$.

The 3d topological terms are $w_1x_2=x_3$ and $w_1\text{Arf}$.

The 4d topological terms are $q_s(x_2)$ and $\eta$.

The 5d topological terms are $x_2x_3$ and $w_1^2x_3{(=x_5)}$.

\subsubsection{$\Omega^{\Pin^-}_d(\B ^2\Z_2)$}

Since the computation involves no odd torsion, we can use the Adams spectral sequence 
\bea
E_2^{s,t}=\Ext_{\A_2}^{s,t}(\H^*(M\Pin^+\wedge (\B ^2\Z_2)_+,\Z_2),\Z_2)\Rightarrow\pi_{t-s}(M\Pin^+\wedge (\B ^2\Z_2)_+)_2^{\wedge}=\Omega^{\Pin^-}_{t-s}(\B ^2\Z_2).
\eea
$M\Pin^+=MT\Pin^-\sim M\Spin\wedge S^{-1}\wedge M\tO(1)$.

For $t-s<8$, we can identify the $E_2$-page with 
$$\Ext_{\A_2(1)}^{s,t}(\H^{*+1}(M\tO(1),\Z_2)\otimes\H^*(\B ^2\Z_2,\Z_2),\Z_2).$$

The $\A_2(1)$-module structure of $\H^{*+1}(M\tO(1),\Z_2)\otimes\H^*(\B ^2\Z_2,\Z_2)$ and the $E_2$ page are shown in Figure \ref{fig:H^{*+1}(MO(1),Z_2)otimesH^*(B^2Z_2,Z_2)}, \ref{fig:Omega_*^{Pin^-}(B^2Z_2)}.

\begin{figure}[!h]
\begin{center}
\begin{tikzpicture}[scale=0.5]

\node[below] at (0,0) {$U$};

\draw[fill] (0,0) circle(.1);
\draw[fill] (0,1) circle(.1);
\draw (0,0) -- (0,1);
\draw[fill] (0,2) circle(.1);
\draw[fill] (0,3) circle(.1);
\draw (0,2) -- (0,3);
\draw (0,1) to [out=150,in=150] (0,3);
\draw[fill] (0,4) circle(.1);
\draw (0,2) to [out=30,in=30] (0,4);
\draw[fill] (0,5) circle(.1);
\draw (0,4) -- (0,5);

\node[below] at (10,0) {$1$};

\draw[fill] (10,0) circle(.1);

\node[below] at (10,2) {$x_2$};

\draw[fill] (10,2) circle(.1);
\draw[fill] (10,3) circle(.1);
\draw[fill] (10,4) circle(.1);
\draw (10,2) -- (10,3);
\draw (10,2) to [out=150,in=150] (10,4);
\draw[fill] (11,5) circle(.1);
\draw[fill] (11,6) circle(.1);
\draw (11,5) -- (11,6);
\draw (10,3) to [out=30,in=150] (11,5);
\draw (10,4) to [out=30,in=150] (11,6);
\draw[fill] (12,5) circle(.1);

\node[below] at (12,5) {$x_2x_3$};

\draw (12,5) -- (11,6);
\draw[fill] (12,7) circle(.1);
\draw (12,5) to [out=30,in=30] (12,7);
\draw[fill] (12,8) circle(.1);
\draw (12,7) -- (12,8);
\draw[fill] (12,10) circle(.1);
\draw (12,8) to [out=150,in=150] (12,10);

\node at (5,5) {$\bigotimes$};
\node at (-2,-10) {$=$};

\node[below] at (0,-15) {$U$};

\draw[fill] (0,-15) circle(.1);
\draw[fill] (0,-14) circle(.1);
\draw (0,-15) -- (0,-14);
\draw[fill] (0,-13) circle(.1);
\draw[fill] (0,-12) circle(.1);
\draw (0,-13) -- (0,-12);
\draw (0,-14) to [out=150,in=150] (0,-12);
\draw[fill] (0,-11) circle(.1);
\draw (0,-13) to [out=30,in=30] (0,-11);
\draw[fill] (0,-10) circle(.1);
\draw (0,-11) -- (0,-10);

\node[below] at (2,-13) {$x_2U$};

\draw[fill] (2,-13) circle(.1);
\draw[fill] (2,-12) circle(.1);
\draw[fill] (2,-11) circle(.1);
\draw (2,-13) -- (2,-12);
\draw (2,-13) to [out=150,in=150] (2,-11);
\draw[fill] (3,-10) circle(.1);
\draw[fill] (3,-9) circle(.1);
\draw (3,-10) -- (3,-9);
\draw (2,-12) to [out=30,in=150] (3,-10);
\draw (2,-11) to [out=30,in=150] (3,-9);
\draw[fill] (2,-10) circle(.1);
\draw (2,-11) -- (2,-10);
\draw[fill] (2,-9) circle(.1);
\draw[fill] (2,-8) circle(.1);
\draw (2,-10) to [out=30,in=30] (2,-8);
\draw (2,-9) -- (2,-8);
\draw[fill] (2,-7) circle(.1);
\draw (2,-9) to [out=150,in=150] (2,-7);

\node[below] at (4,-12) {$x_3U$};
 
\draw[fill] (4,-12) circle(.1);
\draw[fill] (4,-11) circle(.1);
\draw (4,-12) -- (4,-11);
\draw[fill] (4,-10) circle(.1);
\draw (4,-12) to [out=150,in=150] (4,-10);
\draw[fill] (5,-9) circle(.1);
\draw (4,-11) to [out=30,in=150] (5,-9);
\draw[fill] (4,-9) circle(.1);
\draw (4,-10) -- (4,-9);
\draw[fill] (5,-8) circle(.1);
\draw (5,-9) -- (5,-8);
\draw (4,-10) to [out=30,in=150] (5,-8);
\draw[fill] (5,-7) circle(.1);
\draw (4,-9) to [out=30,in=150] (5,-7);
\draw[fill] (5,-6) circle(.1);
\draw (5,-7) -- (5,-6);
\draw (5,-8) to [out=30,in=30] (5,-6); 
 
\node[below] at (6,-11) {$w_1^2x_2U$}; 
 
\draw[fill] (6,-11) circle(.1);
\draw[fill] (6,-10) circle(.1);
\draw (6,-11) -- (6,-10);
\draw[fill] (6,-9) circle(.1);
\draw (6,-11) to [out=150,in=150] (6,-9);
\draw[fill] (7,-8) circle(.1);
\draw (6,-10) to [out=30,in=150] (7,-8);
\draw[fill] (6,-8) circle(.1);
\draw (6,-9) -- (6,-8);
\draw[fill] (7,-7) circle(.1);
\draw (7,-8) -- (7,-7);
\draw (6,-9) to [out=30,in=150] (7,-7);
\draw[fill] (7,-6) circle(.1);
\draw (6,-8) to [out=30,in=150] (7,-6);
\draw[fill] (7,-5) circle(.1);
\draw (7,-6) -- (7,-5);
\draw (7,-7) to [out=30,in=30] (7,-5); 

\node[below] at (8,-10) {$x_2x_3U$};

\draw[fill] (8,-10) circle(.1);
\draw[fill] (8,-9) circle(.1);
\draw (8,-10) -- (8,-9);
\draw[fill] (8,-8) circle(.1);
\draw (8,-10) to [out=150,in=150] (8,-8);
\draw[fill] (9,-7) circle(.1);
\draw (8,-9) to [out=30,in=150] (9,-7);
\draw[fill] (8,-7) circle(.1);
\draw (8,-8) -- (8,-7);
\draw[fill] (9,-6) circle(.1);
\draw (9,-7) -- (9,-6);
\draw (8,-8) to [out=30,in=150] (9,-6);
\draw[fill] (9,-5) circle(.1);
\draw (8,-7) to [out=30,in=150] (9,-5);
\draw[fill] (9,-4) circle(.1);
\draw (9,-5) -- (9,-4);
\draw (9,-6) to [out=30,in=30] (9,-4);

\end{tikzpicture}
\end{center}
\caption{The $\A_2(1)$-module structure of $\H^{*+1}(M\tO(1),\Z_2)\otimes\H^*(\B ^2\Z_2,\Z_2)$}
\label{fig:H^{*+1}(MO(1),Z_2)otimesH^*(B^2Z_2,Z_2)}
\end{figure}

\begin{figure}[!h]
\begin{center}
\begin{tikzpicture}
\node at (0,-1) {0};
\node at (1,-1) {1};
\node at (2,-1) {2};
\node at (3,-1) {3};
\node at (4,-1) {4};
\node at (5,-1) {5};
\node at (6,-1) {$t-s$};
\node at (-1,0) {0};
\node at (-1,1) {1};
\node at (-1,2) {2};
\node at (-1,3) {3};
\node at (-1,4) {4};
\node at (-1,5) {5};
\node at (-1,6) {$s$};

\draw[->] (-0.5,-0.5) -- (-0.5,6);
\draw[->] (-0.5,-0.5) -- (6,-0.5);

\draw (0,0) -- (2,2);
\draw (2,2) -- (2,0);
\draw[fill] (2.1,0) circle(0.05);
\draw[fill] (3,0) circle(0.05);
\draw[fill] (4,0) circle(0.05);
\draw[fill] (5,0) circle(0.05);

\end{tikzpicture}
\end{center}
\caption{$\Omega_*^{\Pin^-}(\B ^2\Z_2)$}
\label{fig:Omega_*^{Pin^-}(B^2Z_2)}
\end{figure}

Hence we have the following theorem %(see \ref{sec:Adams} for detail)
\begin{theorem}
\begin{table}[!h]
\centering
\begin{tabular}{c c}
\hline
$i$ & $\Omega^{\Pin^-}_i(\B ^2\Z_2)$\\
\hline
0& $\Z_2$\\
1& $\Z_2$\\
2& $\Z_2\times\Z_8$\\
3 & $\Z_2$\\
4 & $\Z_2$\\ 
5 & $\Z_2$\\
\hline
\end{tabular}
\end{table}
\end{theorem}

The bordism invariants of $\Omega^{\Pin^-}_2(\B ^2\Z_2)$ are $x_2$ and ABK.

The bordism invariant of $\Omega^{\Pin^-}_3(\B ^2\Z_2)$ is $w_1x_2=x_3$.

The bordism invariant of $\Omega^{\Pin^-}_4(\B ^2\Z_2)$ is $w_1^2x_2$.

The bordism invariant of $\Omega^{\Pin^-}_5(\B ^2\Z_2)$ is $x_2x_3$.

\cred{
Here ABK is the Arf-Brown-Kervaire invariant.
 $x_3=\Sq^1x_2=w_1x_2$ on 3-manifolds by Wu formula.
}

\begin{theorem}
\begin{table}[!h]
\centering
\begin{tabular}{c c}
\hline
$i$ & $\TP_i(\Pin^-\times \B \Z_2)$\\
\hline
0& $\Z_2$\\
1& $\Z_2$\\
2& $\Z_2\times\Z_8$\\
3 & $\Z_2$\\
4 & $\Z_2$\\ 
5 & $\Z_2$\\
\hline
\end{tabular}
\end{table}
\end{theorem}

The 2d topological terms are $x_2$ and ABK.

The 3d topological term is $w_1x_2=x_3$.

The 4d topological term is $w_1^2x_2$.

The 5d topological term is $x_2x_3$.

\subsubsection{$\Omega_d^{\tO}(\B ^2\Z_3)$}
\bea
\Ext_{\A_2}^{s,t}(\H^*(M\tO\wedge (\B ^2\Z_3)_+,\Z_2),\Z_2)\Rightarrow\Omega_{t-s}^{\tO}(\B ^2\Z_3)_2^{\wedge}.
\eea
\bea
\Ext_{\A_3}^{s,t}(\H^*(M\tO\wedge (\B ^2\Z_3)_+,\Z_3),\Z_3)\Rightarrow\Omega_{t-s}^{\tO}(\B ^2\Z_3)_3^{\wedge}.
\eea
{Since $M\tO$ is the wedge sum of suspensions of the Eilenberg-MacLane spectrum $H\Z_2$,
$\H^*(M\tO,\Z_3)=0$, thus $\Omega_d^{\tO}(\B ^2\Z_3)_3^{\wedge}=0$.
}

Since $\H^*(\B ^2\Z_3,\Z_2)=\Z_2$, we have $\Omega_d^{\tO}(\B ^2\Z_3)_2^{\wedge}=\Omega_d^{\tO}$.

Hence $\Omega_d^{\tO}(\B ^2\Z_3)=\Omega_d^{\tO}$.

\begin{theorem}
\begin{table}[!h]
\centering
\begin{tabular}{c c}
\hline
$i$ & $\Omega^{\tO}_i(\B ^2\Z_3)$\\
\hline
0& $\Z_2$\\
1& $0$\\
2& $\Z_2$\\
3 & $0$\\
4 & $\Z_2^2$\\ 
5 & $\Z_2$\\
\hline
\end{tabular}
\end{table}
\end{theorem}

The bordism invariant of $\Omega^{\tO}_2(\B ^2\Z_3)$ is $w_1^2$.

The bordism invariants of $\Omega^{\tO}_4(\B ^2\Z_3)$ are $w_1^4,w_2^2$.

The bordism invariant of $\Omega^{\tO}_5(\B ^2\Z_3)$ is $w_2w_3$.

\begin{theorem}
\begin{table}[!h]
\centering
\begin{tabular}{c c}
\hline
$i$ & $\TP_i(\tO\times \B \Z_3)$\\
\hline
0& $\Z_2$\\
1& $0$\\
2& $\Z_2$\\
3 & $0$\\
4 & $\Z_2^2$\\ 
5 & $\Z_2$\\
\hline
\end{tabular}
\end{table}
\end{theorem}

The 2d topological term is $w_1^2$.

The 4d topological terms are $w_1^4,w_2^2$.

The 5d topological term is $w_2w_3$.

\subsubsection{$\Omega_d^{\SO}(\B ^2\Z_3)$}
\bea
\Ext_{\A_2}^{s,t}(\H^*(M\SO\wedge (\B ^2\Z_3)_+,\Z_2),\Z_2)\Rightarrow\Omega_{t-s}^{\SO}(\B ^2\Z_3)_2^{\wedge}.
\eea
Since $\H^*(\B ^2\Z_3,\Z_2)=\Z_2$, we have $\Omega_d^{\SO}(\B ^2\Z_3)_2^{\wedge}=\Omega_d^{\SO}$.
\bea
\Ext_{\A_3}^{s,t}(\H^*(M\SO\wedge (\B ^2\Z_3)_+,\Z_3),\Z_3)\Rightarrow\Omega_{t-s}^{\SO}(\B ^2\Z_3)_3^{\wedge}.
\eea

The dual of $\A_3=\H^*(H\Z_3,\Z_3)$ is 
\bea
\A_{3*}=\H_*(H\Z_3,\Z_3)=\Lambda_{\Z_3}(\tau_0,\tau_1,\dots)\otimes\Z_3[\xi_1,\xi_2,\dots]
\eea
where $\tau_i=(P^{3^{i-1}}\cdots P^3P^1\beta_{(3,3)})^*$ and $\xi_i=(P^{3^{i-1}}\cdots P^3P^1)^*$. Let $C=\Z_3[\xi_1,\xi_2,\dots]\subseteq\A_{3*}$, then 
\bea
\H_*(M\SO,\Z_3)=C\otimes\Z_3[z_1',z_2',\dots]
\eea
where $|z_k'|=4k$ for $k\ne\frac{3^t-1}{2}$.
\bea
\H^*(M\SO,\Z_3)=(\Z_3[z_1',z_2',\dots])^*\otimes C^*=C^*\oplus\Sigma^8C^*\oplus\cdots
\eea
where $C^*=\A_3/(\beta_{(3,3)})$ and $(\beta_{(3,3)})$ is the two-sided ideal of $\A_3$ generated by $\beta_{(3,3)}$.
\bea
\cdots\To\Sigma^2\A_3\oplus\Sigma^6\A_3\oplus\cdots\To\Sigma\A_3\oplus\Sigma^5\A_3\oplus\cdots\To\A_3\To\A_3/(\beta_{(3,3)})
\eea
is an $\A_3$-resolution of $\A_3/(\beta_{(3,3)})$ where the differentials $d_1$ are induced by $\beta_{(3,3)}$..

\bea
\H^*(\B ^2\Z_3,\Z_3)=\Z_3[x_2',x_8',\dots]\otimes\Lambda_{\Z_3}(x_3',x_7',\dots)
\eea

$\beta_{(3,3)} x_2'=x_3'$, $\beta_{(3,3)} x_2'^2=2x_2'x_3'$.

The $E_2$ page is shown in Figure \ref{fig:Omega_*^{SO}(B^2Z_3)_3}.

\begin{figure}[!h]
\begin{center}
\begin{tikzpicture}
\node at (0,-1) {0};
\node at (1,-1) {1};
\node at (2,-1) {2};
\node at (3,-1) {3};
\node at (4,-1) {4};
\node at (5,-1) {5};
\node at (6,-1) {$t-s$};
\node at (-1,0) {0};
\node at (-1,1) {1};
\node at (-1,2) {2};
\node at (-1,3) {3};
\node at (-1,4) {4};
\node at (-1,5) {5};
\node at (-1,6) {$s$};

\draw[->] (-0.5,-0.5) -- (-0.5,6);
\draw[->] (-0.5,-0.5) -- (6,-0.5);

\draw (0,0) -- (0,5);
\draw (4,1) -- (4,5);

\draw[fill] (2,0) circle(0.05);
\draw[fill] (4,0) circle(0.05);

\end{tikzpicture}
\end{center}
\caption{$\Omega_*^{\SO}(\B ^2\Z_3)_3^{\wedge}$}
\label{fig:Omega_*^{SO}(B^2Z_3)_3}
\end{figure}

Hence we have the following 
\begin{theorem}
\begin{table}[!h]
\centering
\begin{tabular}{c c}
\hline
$i$ & $\Omega^{\SO}_i(\B ^2\Z_3)$\\
\hline
0& $\Z$\\
1& $0$\\
2& $\Z_3$\\
3 & $0$\\
4 & $\Z\times\Z_3$\\ 
5 & $\Z_2$\\
\hline
\end{tabular}
\end{table}
\end{theorem}

The bordism invariant of $\Omega^{\SO}_2(\B ^2\Z_3)$ is $x_2'$.

The bordism invariants of $\Omega^{\SO}_{4}(\B ^2\Z_3)$ are $\sigma$ and $x_2'^2$.

The bordism invariant of $\Omega^{\SO}_5(\B ^2\Z_3)$ is $w_2w_3$.

\begin{theorem}
\begin{table}[!h]
\centering
\begin{tabular}{c c}
\hline
$i$ & $\TP_i(\SO\times \B \Z_3)$\\
\hline
0& $0$\\
1& $0$\\
2& $\Z_3$\\
3 & $\Z$\\
4 & $\Z_3$\\ 
5 & $\Z_2$\\
\hline
\end{tabular}
\end{table}
\end{theorem}

The 2d topological term is $x_2'$.

The 3d topological term is $\frac{1}{3}\text{CS}_3^{(TM)}$.

The 4d topological term is $x_2'^2$.

The 5d topological term is $w_2w_3$.

\subsubsection{$\Omega_d^{\Spin}(\B ^2\Z_3)$}
\bea
\Ext_{\A_2}^{s,t}(\H^*(M\Spin\wedge (\B ^2\Z_3)_+,\Z_2),\Z_2)\Rightarrow\Omega_{t-s}^{\Spin}(\B ^2\Z_3)_2^{\wedge}.
\eea
Since $\H^*(\B ^2\Z_3,\Z_2)=\Z_2$, we have $\Omega_d^{\Spin}(\B ^2\Z_3)_2^{\wedge}=\Omega_d^{\Spin}$.
\bea
\Ext_{\A_3}^{s,t}(\H^*(M\Spin\wedge (\B ^2\Z_3)_+,\Z_3),\Z_3)\Rightarrow\Omega_{t-s}^{\Spin}(\B ^2\Z_3)_3^{\wedge}.
\eea
{Since there is a short exact sequence of groups
\bea
1\To\Z_2\To \Spin\To \SO\To1,
\eea
we have a fibration 
\bea
\xymatrix{
\B \Z_2  \ar[r] &\B \Spin\ar[d]\\
        &\B \SO}
\eea
Take the localization at prime 3, we have a homotopy equivalence $\B \Spin_{(3)}\sim \B \SO_{(3)}$ since the localization of $\B \Z_2$ at 3 is trivial. Take the Thom spectra, we have a homotopy equivalence $M\Spin_{(3)}\sim M\SO_{(3)}$. Hence
\bea
\H^*(M\Spin,\Z_3)=\H^*(M\SO,\Z_3).
\eea
}
We have the following
\begin{theorem}
\begin{table}[!h]
\centering
\begin{tabular}{c c}
\hline
$i$ & $\Omega^{\Spin}_i(\B ^2\Z_3)$\\
\hline
0& $\Z$\\
1& $\Z_2$\\
2& $\Z_2\times\Z_3$\\
3 & $0$\\
4 & $\Z\times\Z_3$\\ 
5 & $0$\\
\hline
\end{tabular}
\end{table}
\end{theorem}

The bordism invariants of $\Omega^{\Spin}_{2}(\B ^2\Z_3)$ are Arf and $x_2'$.

The bordism invariants of $\Omega^{\Spin}_{4}(\B ^2\Z_3)$ are $\frac{\sigma}{16}$ and $x_2'^2$.

\begin{theorem}
\begin{table}[!h]
\centering
\begin{tabular}{c c}
\hline
$i$ & $\TP_i(\Spin\times \B \Z_3)$\\
\hline
0& $0$\\
1& $\Z_2$\\
2& $\Z_2\times\Z_3$\\
3 & $\Z$\\
4 & $\Z_3$\\ 
5 & $0$\\
\hline
\end{tabular}
\end{table}
\end{theorem}

The 2d topological terms are Arf and $x_2'$.

The 3d topological term is $\frac{1}{48}\text{CS}_3^{(TM)}$.

The 4d topological term is $x_2'^2$.

\subsubsection{$\Omega_d^{\Pin^+}(\B ^2\Z_3)$}
\bea
\Ext_{\A_2}^{s,t}(\H^*(M\Pin^-\wedge (\B ^2\Z_3)_+,\Z_2),\Z_2)\Rightarrow\Omega_{t-s}^{\Pin^+}(\B ^2\Z_3)_2^{\wedge}.
\eea
\bea
\Ext_{\A_3}^{s,t}(\H^*(M\Pin^+\wedge (\B ^2\Z_3)_+,\Z_3),\Z_3)\Rightarrow\Omega_{t-s}^{\Pin^-}(\B ^2\Z_3)_3^{\wedge}.
\eea
{Since $MT\Pin^+=M\Pin^-\sim M\Spin\wedge S^1\wedge MT\tO(1)$ and $\H^*(MT\tO(1),\Z_3)=0$,
we have $\H^*(M\Pin^-,\Z_3)=0$, thus $\Omega_d^{\Pin^+}(\B ^2\Z_3)_3^{\wedge}=0$.
}

Since $\H^*(\B ^2\Z_3,\Z_2)=\Z_2$, we have $\Omega_d^{\Pin^+}(\B ^2\Z_3)_2^{\wedge}=\Omega_d^{\Pin^+}$.

Hence $\Omega_d^{\Pin^+}(\B ^2\Z_3)=\Omega_d^{\Pin^+}$.

\begin{theorem}
\begin{table}[!h]
\centering
\begin{tabular}{c c}
\hline
$i$ & $\Omega^{\Pin^+}_i(\B ^2\Z_3)$\\
\hline
0& $\Z_2$\\
1& $0$\\
2& $\Z_2$\\
3 & $\Z_2$\\
4 & $\Z_{16}$\\ 
5 & $0$\\
\hline
\end{tabular}
\end{table}
\end{theorem}

The bordism invariant of $\Omega^{\Pin^+}_2(\B ^2\Z_3)$ is $w_1\tilde{\eta}$.

The bordism invariant of $\Omega^{\Pin^+}_3(\B ^2\Z_3)$ is $w_1\text{Arf}$.

The bordism invariant of $\Omega^{\Pin^+}_4(\B ^2\Z_3)$ is $\eta$.

\begin{theorem}
\begin{table}[!h]
\centering
\begin{tabular}{c c}
\hline
$i$ & $\TP_i(\Pin^+\times \B \Z_3)$\\
\hline
0& $\Z_2$\\
1& $0$\\
2& $\Z_2$\\
3 & $\Z_2$\\
4 & $\Z_{16}$\\ 
5 & $0$\\
\hline
\end{tabular}
\end{table}
\end{theorem}

The 2d topological term is $w_1\tilde{\eta}$.

The 3d topological term is $w_1\text{Arf}$.

The 4d topological term is $\eta$.

\subsubsection{$\Omega_d^{\Pin^-}(\B ^2\Z_3)$}
\bea
\Ext_{\A_2}^{s,t}(\H^*(M\Pin^+\wedge (\B ^2\Z_3)_+,\Z_2),\Z_2)\Rightarrow\Omega_{t-s}^{\Pin^-}(\B ^2\Z_3)_2^{\wedge}.
\eea
\bea
\Ext_{\A_3}^{s,t}(\H^*(M\Pin^+\wedge (\B ^2\Z_3)_+,\Z_3),\Z_3)\Rightarrow\Omega_{t-s}^{\Pin^-}(\B ^2\Z_3)_3^{\wedge}.
\eea
{Since $MT\Pin^-=M\Pin^+\sim M\Spin\wedge S^{-1}\wedge M\tO(1)$ and $\H^*(M\tO(1),\Z_3)=0$,
we have $\H^*(M\Pin^+,\Z_3)=0$, thus $\Omega_d^{\Pin^-}(\B ^2\Z_3)_3^{\wedge}=0$.
}
Since $\H^*(\B ^2\Z_3,\Z_2)=\Z_2$, we have $\Omega_d^{\Pin^-}(\B ^2\Z_3)_2^{\wedge}=\Omega_d^{\Pin^-}$.

Hence $\Omega_d^{\Pin^-}(\B ^2\Z_3)=\Omega_d^{\Pin^-}$.

\begin{theorem}
\begin{table}[!h]
\centering
\begin{tabular}{c c}
\hline
$i$ & $\Omega^{\Pin^-}_i(\B ^2\Z_3)$\\
\hline
0& $\Z_2$\\
1& $\Z_2$\\
2& $\Z_8$\\
3 & $0$\\
4 & $0$\\ 
5 & $0$\\
\hline
\end{tabular}
\end{table}
\end{theorem}

The bordism invariant of $\Omega^{\Pin^-}_2(\B ^2\Z_3)$ is ABK.

\begin{theorem}
\begin{table}[!h]
\centering
\begin{tabular}{c c}
\hline
$i$ & $\TP_i(\Pin^-\times \B \Z_3)$\\
\hline
0& $\Z_2$\\
1& $\Z_2$\\
2& $\Z_8$\\
3 & $0$\\
4 & $0$\\ 
5 & $0$\\
\hline
\end{tabular}
\end{table}
\end{theorem}

The 2d topological term is ABK.

\subsection{$\B G_a:\B \PSU(2),\B \PSU(3)$}

\label{sec:BSUn}

\subsubsection{$\Omega^{\tO}_d(\B \PSU(2))$}

\bea
\H^*(M\tO,\Z_2)\otimes \H^*(\B \PSU(2),\Z_2)=\A_2\oplus2\Sigma^2\A_2\oplus\Sigma^3\A_2\oplus4\Sigma^4\A_2\oplus3\Sigma^5\A_2\oplus\cdots.
\eea
\bea
\Ext_{\A_2}^{s,t}(\H^*(M\tO\wedge (\B \PSU(2))_+,\Z_2),\Z_2)\Rightarrow\Omega_{t-s}^{\tO}(\B \PSU(2))_2^{\wedge}
\eea
\begin{theorem}
\begin{table}[!h]
\centering
\begin{tabular}{c c}
\hline
$i$ & $\Omega^{\tO}_i(\B \PSU(2))$\\
\hline
0& $\Z_2$\\
1& $0$\\
2& $\Z_2^2$\\
3 & $\Z_2$\\
4 & $\Z_2^4$\\ 
5 & $\Z_2^3$\\
\hline
\end{tabular}
\end{table}
\end{theorem}

The bordism invariants of $\Omega_2^{\tO}(\B \PSU(2))$ are $w_2',w_1^2$.

The bordism invariant of $\Omega_3^{\tO}(\B \PSU(2))$ is $w_3'=w_1w_2'$.

The bordism invariants of $\Omega_4^{\tO}(\B \PSU(2))$ are $w_2'^2,w_1^4,w_1^2w_2',w_2^2$.

The bordism invariants of $\Omega_5^{\tO}(\B \PSU(2))$ are $w_2w_3,w_1^2w_3',w_2'w_3'$.

\begin{theorem}
\begin{table}[!h]
\centering
\begin{tabular}{c c}
\hline
$i$ & $\TP_i(\tO\times \PSU(2))$\\
\hline
0& $\Z_2$\\
1& $0$\\
2& $\Z_2^2$\\
3 & $\Z_2$\\
4 & $\Z_2^4$\\ 
5 & $\Z_2^3$\\
\hline
\end{tabular}
\end{table}
\end{theorem}

The 2d topological terms are $w_2',w_1^2$.

The 3d topological term is $w_3'=w_1w_2'$.

The 4d topological terms are $w_2'^2,w_1^4,w_1^2w_2',w_2^2$.

The 5d topological terms are $w_2w_3,w_1^2w_3',w_2'w_3'$.

\subsubsection{$\Omega^{\SO}_d(\B \PSU(2))$}
\bea
\Ext_{\A_2}^{s,t}(\H^*(M\SO\wedge (\B \PSU(2))_+,\Z_2),\Z_2)\Rightarrow\Omega_{t-s}^{\SO}(\B \PSU(2))_2^{\wedge}.
\eea
The $E_2$ page is shown in Figure \ref{fig:Omega_*^{SO}(BPSU(2))_2}.

\begin{figure}[!h]
\begin{center}
\begin{tikzpicture}
\node at (0,-1) {0};
\node at (1,-1) {1};
\node at (2,-1) {2};
\node at (3,-1) {3};
\node at (4,-1) {4};
\node at (5,-1) {5};
\node at (6,-1) {$t-s$};
\node at (-1,0) {0};
\node at (-1,1) {1};
\node at (-1,2) {2};
\node at (-1,3) {3};
\node at (-1,4) {4};
\node at (-1,5) {5};
\node at (-1,6) {$s$};

\draw[->] (-0.5,-0.5) -- (-0.5,6);
\draw[->] (-0.5,-0.5) -- (6,-0.5);

\draw (0,0) -- (0,5);
\draw (4,0) -- (4,5);
\draw (4.1,0) -- (4.1,5);

\draw[fill] (2,0) circle(0.05);
\draw[fill] (4.9,0) circle(0.05);
\draw[fill] (5.1,0) circle(0.05);

\end{tikzpicture}
\end{center}
\caption{$\Omega_*^{\SO}(\B \PSU(2))_2^{\wedge}$}
\label{fig:Omega_*^{SO}(BPSU(2))_2}
\end{figure}

\begin{theorem}
\begin{table}[!h]
\centering
\begin{tabular}{c c}
\hline
$i$ & $\Omega^{\SO}_i(\B \PSU(2))$\\
\hline
0& $\Z$\\
1& $0$\\
2& $\Z_2$\\
3 & $0$\\
4 & $\Z^2$\\ 
5 & $\Z_2^2$\\
\hline
\end{tabular}
\end{table}
\end{theorem}

The bordism invariant of $\Omega_2^{\SO}(\B \PSU(2))$ is $w_2'$.

The bordism invariants of $\Omega_4^{\SO}(\B \PSU(2))$ are $\sigma,p_1'$.

The bordism invariants of $\Omega_5^{\SO}(\B \PSU(2))$ are $w_2w_3,w_2'w_3'$.

{The manifold generators of $\Omega_4^{\SO}(\B \PSU(2))$ are $(\CP^2,3)$ and $(\CP^2,L_{\C}+1)$ where $n$ is the trivial real $n$-plane bundle and $L_{\C}$ is the tautological complex line bundle over $\CP^2$.
Note that the principal $\SO(3)$-bundle $P$ associated to $L_{\C}+1$ is the induce bundle $P'\times_{\SO(2)}\SO(3)$ from $P'$
\bea
\xymatrix{S^1=\SO(2)\ar[r]&S^5\ar[d]\\
&\CP^2}
\eea
by the group homomorphism $\phi:\SO(2)\to \SO(3)$ which is the inclusion map, that means $P=P'\times_{\SO(2)}\SO(3)=(P'\times \SO(3))/\SO(2)$ which is the quotient of $P'\times \SO(3)$ by the right $\SO(2)$ action 
\bea
(p,g)h=(ph,\phi(h^{-1})g).
\eea
\bea
p_1(L_{\C}+1)=p_1(L_{\C})=-c_2(L_{\C}\otimes_{\R}\C)=-c_2(L_{\C}\oplus\bar{L}_{\C})=-c_1(L_{\C})c_1(\bar{L}_{\C})=c_1(L_{\C})^2.
\eea
So 
\bea
\int_{\CP^2}p_1(L_{\C}+1)=1.
\eea}

\begin{theorem}
\begin{table}[!h]
\centering
\begin{tabular}{c c}
\hline
$i$ & $\TP_i(\SO\times \PSU(2))$\\
\hline
0& $0$\\
1& $0$\\
2& $\Z_2$\\
3 & $\Z^2$\\
4 & $0$\\ 
5 & $\Z_2^2$\\
\hline
\end{tabular}
\end{table}
\end{theorem}

The 2d topological term is $w_2'$.

Since $p_1'=\text{d}\text{CS}_3^{(\SO(3))}$, the 3d topological terms are $\frac{1}{3}\text{CS}_3^{(TM)}$ and $\text{CS}_3^{(\SO(3))}$.

The 5d topological terms are $w_2w_3,w_2'w_3'$.

\subsubsection{$\Omega^{\Spin}_d(\B \PSU(2))$}
\bea
\Ext_{\A_2}^{s,t}(\H^*(M\Spin\wedge (\B \PSU(2))_+,\Z_2),\Z_2)\Rightarrow\Omega_{t-s}^{\Spin}(\B \PSU(2))_2^{\wedge}.
\eea
For $t-s<8$,
\bea
\Ext_{\A_2(1)}^{s,t}(\H^*(\B \PSU(2),\Z_2),\Z_2)\Rightarrow\Omega_{t-s}^{\Spin}(\B \PSU(2))_2^{\wedge}.
\eea

The $\A_2(1)$-module structure of $\H^*(\B \PSU(2),\Z_2)$ and the $E_2$ page are shown in Figure \ref{fig:H^*(BPSU(2),Z_2)}, \ref{fig:Omega_*^{Spin}(BPSU(2))_2}.

\begin{figure}[!h]
\begin{center}
\begin{tikzpicture}[scale=0.5]
\node[below] at (0,0) {$1$};

\draw[fill] (0,0) circle(.1);
\draw[fill] (0,2) circle(.1);

\node[below] at (0,2) {$w_2'$};

\draw[fill] (0,3) circle(.1);
\draw[fill] (0,4) circle(.1);
\draw (0,2) -- (0,3);
\draw (0,2) to [out=150,in=150] (0,4);
\draw[fill] (1,5) circle(.1);
\draw[fill] (1,6) circle(.1);
\draw (1,5) -- (1,6);
\draw (0,3) to [out=30,in=150] (1,5);
\draw (0,4) to [out=30,in=150] (1,6);

\end{tikzpicture}
\end{center}
\caption{The $\A_2(1)$-module structure of $\H^*(\B \PSU(2),\Z_2)$}
\label{fig:H^*(BPSU(2),Z_2)}
\end{figure}

\begin{figure}[!h]
\begin{center}
\begin{tikzpicture}
\node at (0,-1) {0};
\node at (1,-1) {1};
\node at (2,-1) {2};
\node at (3,-1) {3};
\node at (4,-1) {4};
\node at (5,-1) {5};
\node at (6,-1) {$t-s$};
\node at (-1,0) {0};
\node at (-1,1) {1};
\node at (-1,2) {2};
\node at (-1,3) {3};
\node at (-1,4) {4};
\node at (-1,5) {5};
\node at (-1,6) {$s$};

\draw[->] (-0.5,-0.5) -- (-0.5,6);
\draw[->] (-0.5,-0.5) -- (6,-0.5);

\draw (0,0) -- (0,5);
\draw (0,0) -- (2,2);
\draw (4,3) -- (4,5);
\draw (4.1,1) -- (4.1,5);

\draw[fill] (2,0) circle(0.05);

\end{tikzpicture}
\end{center}
\caption{$\Omega_*^{\Spin}(\B \PSU(2))_2^{\wedge}$}
\label{fig:Omega_*^{Spin}(BPSU(2))_2}
\end{figure}

\begin{theorem}
\begin{table}[!h]
\centering
\begin{tabular}{c c}
\hline
$i$ & $\Omega^{\Spin}_i(\B \PSU(2))$\\
\hline
0& $\Z$\\
1& $\Z_2$\\
2& $\Z_2^2$\\
3 & $0$\\
4 & $\Z^2$\\ 
5 & $0$\\
\hline
\end{tabular}
\end{table}
\end{theorem}

The bordism invariants of $\Omega_2^{\Spin}(\B \PSU(2))$ are $w_2'$ and Arf.

By Wu formula \eqref{SqWu}, $w_2'^2=\Sq^2(w_2')=(w_2(TM)+w_1(TM)^2)w_2'=0$ on Spin 4-manifolds, $p_1'=w_2'^2=0\mod2$ on Spin 4-manifolds. 

The bordism invariants of $\Omega_4^{\Spin}(\B \PSU(2))$ are $\frac{\sigma}{16}$ and $\frac{p_1'}{2}$.

\begin{theorem}
\begin{table}[!h]
\centering
\begin{tabular}{c c}
\hline
$i$ & $\TP_i(\Spin\times \PSU(2))$\\
\hline
0& $0$\\
1& $\Z_2$\\
2& $\Z_2^2$\\
3 & $\Z^2$\\
4 & $0$\\ 
5 & $0$\\
\hline
\end{tabular}
\end{table}
\end{theorem}

The 2d topological terms are $w_2'$ and Arf.

The 3d topological terms are $\frac{1}{48}\text{CS}_3^{(TM)}$ and $\frac{1}{2}\text{CS}_3^{(\SO(3))}$.

\subsubsection{$\Omega^{\Pin^+}_d(\B \PSU(2))$}
\bea
\Ext_{\A_2}^{s,t}(\H^*(MT\Pin^+\wedge (\B \PSU(2))_+,\Z_2),\Z_2)\Rightarrow\Omega_{t-s}^{\Pin^+}(\B \PSU(2))_2^{\wedge}.
\eea
$MT\Pin^+=M\Spin\wedge S^1\wedge MT\tO(1)$.

For $t-s<8$,
\bea
\Ext_{\A_2(1)}^{s,t}(\H^{*-1}(MT\tO(1),\Z_2)\otimes\H^*(\B \PSU(2),\Z_2),\Z_2)\Rightarrow\Omega_{t-s}^{\Pin^+}(\B \PSU(2))_2^{\wedge}.
\eea

The $\A_2(1)$-module structure of $\H^{*-1}(MT\tO(1),\Z_2)\otimes\H^*(\B \PSU(2),\Z_2)$ and the $E_2$ page are shown in Figure \ref{fig:H^{*-1}(MTO(1),Z_2)otimesH^*(BPSU(2),Z_2)}, \ref{fig:Omega_*^{Pin^+}(BPSU(2))_2}.

\begin{figure}[!h]
\begin{center}
\begin{tikzpicture}[scale=0.5]

\node[below] at (0,0) {$U$};

\draw[fill] (0,0) circle(.1);
\draw[fill] (0,1) circle(.1);
\draw (0,0) -- (0,1);
\draw[fill] (0,2) circle(.1);
\draw (0,0) to [out=150,in=150] (0,2);
\draw[fill] (0,3) circle(.1);
\draw (0,2) -- (0,3);
\draw[fill] (0,4) circle(.1);
\draw[fill] (0,5) circle(.1);
\draw (0,4) -- (0,5);
\draw (0,3) to [out=150,in=150] (0,5);
\draw[fill] (0,6) circle(.1);
\draw (0,4) to [out=30,in=30] (0,6);

\node[below] at (10,0) {$1$};

\draw[fill] (10,0) circle(.1);
\draw[fill] (10,2) circle(.1);

\node[below] at (10,2) {$w_2'$};

\draw[fill] (10,3) circle(.1);
\draw[fill] (10,4) circle(.1);
\draw (10,2) -- (10,3);
\draw (10,2) to [out=150,in=150] (10,4);
\draw[fill] (11,5) circle(.1);
\draw[fill] (11,6) circle(.1);
\draw (11,5) -- (11,6);
\draw (10,3) to [out=30,in=150] (11,5);
\draw (10,4) to [out=30,in=150] (11,6);

\node at (5,5) {$\bigotimes$};
\node at (-2,-10) {$=$};

\node[below] at (0,-15) {$U$};

\draw[fill] (0,-15) circle(.1);
\draw[fill] (0,-14) circle(.1);
\draw (0,-15) -- (0,-14);
\draw[fill] (0,-13) circle(.1);
\draw (0,-15) to [out=150,in=150] (0,-13);
\draw[fill] (0,-12) circle(.1);
\draw (0,-13) -- (0,-12);
\draw[fill] (0,-11) circle(.1);
\draw[fill] (0,-10) circle(.1);
\draw (0,-11) -- (0,-10);
\draw (0,-12) to [out=150,in=150] (0,-10);
\draw[fill] (0,-9) circle(.1);
\draw (0,-11) to [out=30,in=30] (0,-9);

\node[below] at (2,-13) {$w_2'U$};

\draw[fill] (2,-13) circle(.1);
\draw[fill] (2,-12) circle(.1);
\draw (2,-13) -- (2,-12);
\draw[fill] (2,-11) circle(.1);
\draw (2,-13) to [out=150,in=150] (2,-11);
\draw[fill] (3,-10) circle(.1);
\draw (2,-12) to [out=30,in=150] (3,-10);
\draw[fill] (2,-10) circle(.1);
\draw (2,-11) -- (2,-10);
\draw[fill] (3,-9) circle(.1);
\draw (3,-10) -- (3,-9);
\draw (2,-11) to [out=30,in=150] (3,-9);
\draw[fill] (3,-8) circle(.1);
\draw (2,-10) to [out=30,in=150] (3,-8);
\draw[fill] (3,-7) circle(.1);
\draw (3,-8) -- (3,-7);
\draw (3,-9) to [out=30,in=30] (3,-7); 
 
\node[below] at (5,-12) {$w_1w_2'U$};

\draw[fill] (5,-12) circle(.1);
\draw[fill] (5,-11) circle(.1);
\draw[fill] (5,-10) circle(.1);
\draw (5,-12) -- (5,-11);
\draw (5,-12) to [out=150,in=150] (5,-10);
\draw[fill] (6,-9) circle(.1);
\draw[fill] (6,-8) circle(.1);
\draw (6,-9) -- (6,-8);
\draw (5,-11) to [out=30,in=150] (6,-9);
\draw (5,-10) to [out=30,in=150] (6,-8);

\node[below] at (4,-11) {$w_2'^2U$};

\draw[fill] (4,-11) circle(.1);
\draw (4,-11) -- (5,-10);
\draw[fill] (4,-9) circle(.1);
\draw (4,-11) to [out=150,in=150] (4,-9);
\draw[fill] (4,-8) circle(.1);
\draw (4,-9) -- (4,-8);
\draw[fill] (4,-7) circle(.1);
\draw[fill] (4,-6) circle(.1);
\draw (4,-7) -- (4,-6);
\draw (4,-8) to [out=150,in=150] (4,-6);
\draw[fill] (4,-5) circle(.1);
\draw (4,-7) to [out=30,in=30] (4,-5);

\node[below] at (8,-10) {$w_1^2w_3'U$};

\draw[fill] (8,-10) circle(.1);
\draw[fill] (8,-9) circle(.1);
\draw (8,-10) -- (8,-9);
\draw[fill] (8,-8) circle(.1);
\draw (8,-10) to [out=150,in=150] (8,-8);
\draw[fill] (9,-7) circle(.1);
\draw (8,-9) to [out=30,in=150] (9,-7);
\draw[fill] (8,-7) circle(.1);
\draw (8,-8) -- (8,-7);
\draw[fill] (9,-6) circle(.1);
\draw (9,-7) -- (9,-6);
\draw (8,-8) to [out=30,in=150] (9,-6);
\draw[fill] (9,-5) circle(.1);
\draw (8,-7) to [out=30,in=150] (9,-5);
\draw[fill] (9,-4) circle(.1);
\draw (9,-5) -- (9,-4);
\draw (9,-6) to [out=30,in=30] (9,-4);

\end{tikzpicture}
\end{center}
\caption{The $\A_2(1)$-module structure of $\H^{*-1}(MT\tO(1),\Z_2)\otimes\H^*(\B \PSU(2),\Z_2)$}
\label{fig:H^{*-1}(MTO(1),Z_2)otimesH^*(BPSU(2),Z_2)}
\end{figure}

\begin{figure}[!h]
\begin{center}
\begin{tikzpicture}
\node at (0,-1) {0};
\node at (1,-1) {1};
\node at (2,-1) {2};
\node at (3,-1) {3};
\node at (4,-1) {4};
\node at (5,-1) {5};
\node at (6,-1) {$t-s$};
\node at (-1,0) {0};
\node at (-1,1) {1};
\node at (-1,2) {2};
\node at (-1,3) {3};
\node at (-1,4) {4};
\node at (-1,5) {5};
\node at (-1,6) {$s$};

\draw[->] (-0.5,-0.5) -- (-0.5,6);
\draw[->] (-0.5,-0.5) -- (6,-0.5);

\draw[fill] (0,0) circle(0.05);
\draw[fill] (2,0) circle(0.05);
\draw (2,1) -- (4,3);
\draw (4,3) -- (4,0);
\draw (3,0) -- (3.9,1);
\draw (3.9,1) -- (3.9,0);
\draw[fill] (5,0) circle(0.05);

\end{tikzpicture}
\end{center}
\caption{$\Omega_*^{\Pin^+}(\B \PSU(2))_2^{\wedge}$}
\label{fig:Omega_*^{Pin^+}(BPSU(2))_2}
\end{figure}

\begin{theorem}
\begin{table}[!h]
\centering
\begin{tabular}{c c}
\hline
$i$ & $\Omega^{\Pin^+}_i(\B \PSU(2))$\\
\hline
0& $\Z_2$\\
1& $0$\\
2& $\Z_2^2$\\
3 & $\Z_2^2$\\
4 & $\Z_4\times\Z_{16}$\\ 
5 & $\Z_2$\\
\hline
\end{tabular}
\end{table}
\end{theorem}

The bordism invariants of $\Omega_2^{\Pin^+}(\B \PSU(2))$ are $w_2'$ and $w_1\tilde{\eta}$.

The bordism invariants of $\Omega_3^{\Pin^+}(\B \PSU(2))$ are $w_1w_2'=w_3'$ and $w_1\text{Arf}$.

The bordism invariants of $\Omega_4^{\Pin^+}(\B \PSU(2))$ are $q_s(w_2')$ (this invariant has another form, see the footnotes of Table \ref{4d bordism groups}) and $\eta$.

The bordism invariant of $\Omega_5^{\Pin^+}(\B \PSU(2))$ is $w_1^2w_3'{(=w_2'w_3')}$.

\begin{theorem}
\begin{table}[!h]
\centering
\begin{tabular}{c c}
\hline
$i$ & $\TP_i(\Pin^+\times \PSU(2))$\\
\hline
0& $\Z_2$\\
1& $0$\\
2& $\Z_2^2$\\
3 & $\Z_2^2$\\
4 & $\Z_4\times\Z_{16}$\\ 
5 & $\Z_2$\\ 
\hline
\end{tabular}
\end{table}
\end{theorem}

The 2d topological terms are $w_2'$ and $w_1\tilde{\eta}$.

The 3d topological terms are $w_1w_2'=w_3'$ and $w_1\text{Arf}$.

The 4d topological terms are $q_s(w_2')$ and $\eta$.

The 5d topological term is $w_1^2w_3'{(=w_2'w_3')}$.

\subsubsection{$\Omega^{\Pin^-}_d(\B \PSU(2))$}
\bea
\Ext_{\A_2}^{s,t}(\H^*(MT\Pin^-\wedge (\B \PSU(2))_+,\Z_2),\Z_2)\Rightarrow\Omega_{t-s}^{\Pin^-}(\B \PSU(2))_2^{\wedge}.
\eea
$MT\Pin^-=M\Spin\wedge S^{-1}\wedge M\tO(1)$.

For $t-s<8$,
\bea
\Ext_{\A_2(1)}^{s,t}(\H^{*+1}(M\tO(1),\Z_2)\otimes\H^*(\B \PSU(2),\Z_2),\Z_2)\Rightarrow\Omega_{t-s}^{\Pin^-}(\B \PSU(2))_2^{\wedge}.
\eea

The $\A_2(1)$-module structure of $\H^{*+1}(M\tO(1),\Z_2)\otimes\H^*(\B \PSU(2),\Z_2)$ and the $E_2$ page are shown in Figure \ref{fig:H^{*+1}(MO(1),Z_2)otimesH^*(BPSU(2),Z_2)}, \ref{fig:Omega_*^{Pin^-}(BPSU(2))_2}.

\begin{figure}[!h]
\begin{center}
\begin{tikzpicture}[scale=0.5]
\node[below] at (0,0) {$U$};

\draw[fill] (0,0) circle(.1);
\draw[fill] (0,1) circle(.1);
\draw (0,0) -- (0,1);
\draw[fill] (0,2) circle(.1);
\draw[fill] (0,3) circle(.1);
\draw (0,2) -- (0,3);
\draw (0,1) to [out=150,in=150] (0,3);
\draw[fill] (0,4) circle(.1);
\draw (0,2) to [out=30,in=30] (0,4);
\draw[fill] (0,5) circle(.1);
\draw (0,4) -- (0,5);

\node[below] at (10,0) {$1$};

\draw[fill] (10,0) circle(.1);
\draw[fill] (10,2) circle(.1);

\node[below] at (10,2) {$w_2'$};

\draw[fill] (10,3) circle(.1);
\draw[fill] (10,4) circle(.1);
\draw (10,2) -- (10,3);
\draw (10,2) to [out=150,in=150] (10,4);
\draw[fill] (11,5) circle(.1);
\draw[fill] (11,6) circle(.1);
\draw (11,5) -- (11,6);
\draw (10,3) to [out=30,in=150] (11,5);
\draw (10,4) to [out=30,in=150] (11,6);

\node at (5,5) {$\bigotimes$};
\node at (-2,-10) {$=$};

\node[below] at (0,-15) {$U$};

\draw[fill] (0,-15) circle(.1);
\draw[fill] (0,-14) circle(.1);
\draw (0,-15) -- (0,-14);
\draw[fill] (0,-13) circle(.1);
\draw[fill] (0,-12) circle(.1);
\draw (0,-13) -- (0,-12);
\draw (0,-14) to [out=150,in=150] (0,-12);
\draw[fill] (0,-11) circle(.1);
\draw (0,-13) to [out=30,in=30] (0,-11);
\draw[fill] (0,-10) circle(.1);
\draw (0,-11) -- (0,-10);

\node[below] at (2,-13) {$w_2'U$};

\draw[fill] (2,-13) circle(.1);
\draw[fill] (2,-12) circle(.1);
\draw[fill] (2,-11) circle(.1);
\draw (2,-13) -- (2,-12);
\draw (2,-13) to [out=150,in=150] (2,-11);
\draw[fill] (3,-10) circle(.1);
\draw[fill] (3,-9) circle(.1);
\draw (3,-10) -- (3,-9);
\draw (2,-12) to [out=30,in=150] (3,-10);
\draw (2,-11) to [out=30,in=150] (3,-9);
\draw[fill] (2,-10) circle(.1);
\draw (2,-11) -- (2,-10);
\draw[fill] (2,-9) circle(.1);
\draw[fill] (2,-8) circle(.1);
\draw (2,-10) to [out=30,in=30] (2,-8);
\draw (2,-9) -- (2,-8);
\draw[fill] (2,-7) circle(.1);
\draw (2,-9) to [out=150,in=150] (2,-7);

\node[below] at (4,-12) {$w_3'U$};
 
\draw[fill] (4,-12) circle(.1);
\draw[fill] (4,-11) circle(.1);
\draw (4,-12) -- (4,-11);
\draw[fill] (4,-10) circle(.1);
\draw (4,-12) to [out=150,in=150] (4,-10);
\draw[fill] (5,-9) circle(.1);
\draw (4,-11) to [out=30,in=150] (5,-9);
\draw[fill] (4,-9) circle(.1);
\draw (4,-10) -- (4,-9);
\draw[fill] (5,-8) circle(.1);
\draw (5,-9) -- (5,-8);
\draw (4,-10) to [out=30,in=150] (5,-8);
\draw[fill] (5,-7) circle(.1);
\draw (4,-9) to [out=30,in=150] (5,-7);
\draw[fill] (5,-6) circle(.1);
\draw (5,-7) -- (5,-6);
\draw (5,-8) to [out=30,in=30] (5,-6); 

\node[below] at (6,-11) {$w_1^2w_2'U$};
 
\draw[fill] (6,-11) circle(.1);
\draw[fill] (6,-10) circle(.1);
\draw (6,-11) -- (6,-10);
\draw[fill] (6,-9) circle(.1);
\draw (6,-11) to [out=150,in=150] (6,-9);
\draw[fill] (7,-8) circle(.1);
\draw (6,-10) to [out=30,in=150] (7,-8);
\draw[fill] (6,-8) circle(.1);
\draw (6,-9) -- (6,-8);
\draw[fill] (7,-7) circle(.1);
\draw (7,-8) -- (7,-7);
\draw (6,-9) to [out=30,in=150] (7,-7);
\draw[fill] (7,-6) circle(.1);
\draw (6,-8) to [out=30,in=150] (7,-6);
\draw[fill] (7,-5) circle(.1);
\draw (7,-6) -- (7,-5);
\draw (7,-7) to [out=30,in=30] (7,-5);

\end{tikzpicture}
\end{center}
\caption{The $\A_2(1)$-module structure of $\H^{*+1}(M\tO(1),\Z_2)\otimes\H^*(\B \PSU(2),\Z_2)$}
\label{fig:H^{*+1}(MO(1),Z_2)otimesH^*(BPSU(2),Z_2)}
\end{figure}

\begin{figure}[!h]
\begin{center}
\begin{tikzpicture}
\node at (0,-1) {0};
\node at (1,-1) {1};
\node at (2,-1) {2};
\node at (3,-1) {3};
\node at (4,-1) {4};
\node at (5,-1) {5};
\node at (6,-1) {$t-s$};
\node at (-1,0) {0};
\node at (-1,1) {1};
\node at (-1,2) {2};
\node at (-1,3) {3};
\node at (-1,4) {4};
\node at (-1,5) {5};
\node at (-1,6) {$s$};

\draw[->] (-0.5,-0.5) -- (-0.5,6);
\draw[->] (-0.5,-0.5) -- (6,-0.5);

\draw (0,0) -- (2,2);
\draw (2,2) -- (2,0);
\draw[fill] (2.1,0) circle(0.05);
\draw[fill] (3,0) circle(0.05);
\draw[fill] (4,0) circle(0.05);

\end{tikzpicture}
\end{center}
\caption{$\Omega_*^{\Pin^-}(\B \PSU(2))_2^{\wedge}$}
\label{fig:Omega_*^{Pin^-}(BPSU(2))_2}
\end{figure}

\begin{theorem}
\begin{table}[!h]
\centering
\begin{tabular}{c c}
\hline
$i$ & $\Omega^{\Pin^-}_i(\B \PSU(2))$\\
\hline
0& $\Z_2$\\
1& $\Z_2$\\
2& $\Z_2\times\Z_8$\\
3 & $\Z_2$\\
4 & $\Z_2$\\
5 & $0$\\
\hline
\end{tabular}
\end{table}
\end{theorem}

The bordism invariants of $\Omega^{\Pin^-}_2(\B \PSU(2))$ are $w_2'$ and ABK.

The bordism invariant of $\Omega^{\Pin^-}_3(\B \PSU(2))$ is $w_1w_2'=w_3'$.

The bordism invariant of $\Omega^{\Pin^-}_4(\B \PSU(2))$ is $w_1^2w_2'$.

\begin{theorem}
\begin{table}[!h]
\centering
\begin{tabular}{c c}
\hline
$i$ & $\TP_i(\Pin^-\times \PSU(2))$\\
\hline
0& $\Z_2$\\
1& $\Z_2$\\
2& $\Z_2\times\Z_8$\\
3 & $\Z_2$\\
4 & $\Z_2$\\
5 & $0$\\
\hline
\end{tabular}
\end{table}
\end{theorem}

The 2d topological terms are $w_2'$ and ABK.

The 3d topological term is $w_1w_2'=w_3'$.

The 4d topological term is $w_1^2w_2'$.

\subsubsection{$\Omega^{\tO}_d(\B \PSU(3))$}
\bea
\Ext_{\A_3}^{s,t}(\H^*(M\tO,\Z_3)\otimes\H^*(\B \PSU(3),\Z_3),\Z_3)\Rightarrow\Omega^{\tO}_{t-s}(\B \PSU(3))_3^{\wedge}.
\eea
Since $\H^*(M\tO,\Z_3)=0$, $\Omega^{\tO}_d(\B \PSU(3))_3^{\wedge}=0$.
\bea
\Ext_{\A_2}^{s,t}(\H^*(M\tO,\Z_2)\otimes\H^*(\B \PSU(3),\Z_2),\Z_2)\Rightarrow\Omega^{\tO}_{t-s}(\B \PSU(3))_2^{\wedge}.
\eea
\bea
\H^*(M\tO,\Z_2)\otimes\H^*(\B \PSU(3),\Z_2)=\A_2\oplus\Sigma^2\A_2\oplus3\Sigma^4\A_2\oplus\Sigma^5\A_2\oplus\cdots.
\eea
\begin{theorem}
\begin{table}[!h]
\centering
\begin{tabular}{c c}
\hline
$i$ & $\Omega^{\tO}_i(\B \PSU(3))$\\
\hline
0& $\Z_2$\\
1& $0$\\
2& $\Z_2$\\
3 & $0$\\
4 & $\Z_2^3$\\
5 & $\Z_2$\\
\hline
\end{tabular}
\end{table}
\end{theorem}

The bordism invariant of $\Omega_2^{\tO}(\B \PSU(3))$ is $w_1^2$.

The bordism invariants of $\Omega_4^{\tO}(\B \PSU(3))$ are $w_1^4,w_2^2,c_2(\mod2)$.

The bordism invariant of $\Omega_5^{\tO}(\B \PSU(3))$ is $w_2w_3$.

\begin{theorem}
\begin{table}[!h]
\centering
\begin{tabular}{c c}
\hline
$i$ & $\TP_i(\tO\times \PSU(3))$\\
\hline
0& $\Z_2$\\
1& $0$\\
2& $\Z_2$\\
3 & $0$\\
4 & $\Z_2^3$\\
5 & $\Z_2$\\
\hline
\end{tabular}
\end{table}
\end{theorem}

The 2d topological term is $w_1^2$.

The 4d topological terms are $w_1^4,w_2^2,c_2(\mod2)$.

The 5d topological term is $w_2w_3$.

\subsubsection{$\Omega^{\SO}_d(\B \PSU(3))$}
\bea
\Ext_{\A_2}^{s,t}(\H^*(M\SO,\Z_2)\otimes\H^*(\B \PSU(3),\Z_2),\Z_2)\Rightarrow\Omega^{\SO}_{t-s}(\B \PSU(3))_2^{\wedge}.
\eea

\bea
\H^*(\B \PSU(3),\Z_2)=\Z_2[c_2,c_3].
\eea

The $E_2$ page is shown in Figure \ref{fig:Omega_*^{SO}(BPSU(3))_2}.

\begin{figure}[!h]
\begin{center}
\begin{tikzpicture}
\node at (0,-1) {0};
\node at (1,-1) {1};
\node at (2,-1) {2};
\node at (3,-1) {3};
\node at (4,-1) {4};
\node at (5,-1) {5};
\node at (6,-1) {6};
\node at (7,-1) {$t-s$};
\node at (-1,0) {0};
\node at (-1,1) {1};
\node at (-1,2) {2};
\node at (-1,3) {3};
\node at (-1,4) {4};
\node at (-1,5) {5};
\node at (-1,6) {$s$};

\draw[->] (-0.5,-0.5) -- (-0.5,6);
\draw[->] (-0.5,-0.5) -- (7,-0.5);

\draw (0,0) -- (0,5);
\draw (4,0) -- (4,5);
\draw (4.1,0) -- (4.1,5);
\draw[fill] (5,0) circle(0.05);
\draw (6,0) -- (6,5);

\end{tikzpicture}
\end{center}
\caption{$\Omega_*^{\SO}(\B \PSU(3))_2^{\wedge}$}
\label{fig:Omega_*^{SO}(BPSU(3))_2}
\end{figure}
\bea
\Ext_{\A_3}^{s,t}(\H^*(M\SO,\Z_3)\otimes\H^*(\B \PSU(3),\Z_3),\Z_3)\Rightarrow\Omega^{\SO}_{t-s}(\B \PSU(3))_3^{\wedge}.
\eea

\bea
\H^*(\B \PSU(3),\Z_3)=(\Z_3[z_2,z_8,z_{12}]\otimes\Lambda_{\Z_3}(z_3,z_7))/(z_2z_3,z_2z_7,z_2z_8+z_3z_7)
\eea

$\beta_{(3,3)} z_2=z_3$, $\beta_{(3,3)} z_2^2=2z_2z_3=0$, $\beta_{(3,3)} z_2^3=0$.

The $E_2$ page is shown in Figure \ref{fig:Omega_*^{SO}(BPSU(3))_3}.

\begin{figure}[!h]
\begin{center}
\begin{tikzpicture}
\node at (0,-1) {0};
\node at (1,-1) {1};
\node at (2,-1) {2};
\node at (3,-1) {3};
\node at (4,-1) {4};
\node at (5,-1) {5};
\node at (6,-1) {6};
\node at (7,-1) {$t-s$};
\node at (-1,0) {0};
\node at (-1,1) {1};
\node at (-1,2) {2};
\node at (-1,3) {3};
\node at (-1,4) {4};
\node at (-1,5) {5};
\node at (-1,6) {$s$};

\draw[->] (-0.5,-0.5) -- (-0.5,6);
\draw[->] (-0.5,-0.5) -- (7,-0.5);

\draw (0,0) -- (0,5);
\draw (4,1) -- (4,5);
\draw (4.1,0) -- (4.1,5);
\draw[fill] (2,0) circle(0.05);
\draw (6,0) -- (6,5);

\end{tikzpicture}
\end{center}
\caption{$\Omega_*^{\SO}(\B \PSU(3))_3^{\wedge}$}
\label{fig:Omega_*^{SO}(BPSU(3))_3}
\end{figure}

\begin{theorem}
\begin{table}[!h]
\centering
\begin{tabular}{c c}
\hline
$i$ & $\Omega^{\SO}_i(\B \PSU(3))$\\
\hline
0& $\Z$\\
1& $0$\\
2& $\Z_3$\\
3 & $0$\\
4 & $\Z^2$\\
5 & $\Z_2$\\
6 & $\Z$\\
\hline
\end{tabular}
\end{table}
\end{theorem}

The bordism invariant of $\Omega_2^{\SO}(\B \PSU(3))$ is $z_2$.

The bordism invariants of $\Omega_4^{\SO}(\B \PSU(3))$ are $\sigma,c_2$.

The bordism invariant of $\Omega_5^{\SO}(\B \PSU(3))$ is $w_2w_3$.

The bordism invariant of $\Omega_6^{\SO}(\B \PSU(3))$ is $c_3$.

{The manifold generators of $\Omega_4^{\SO}(\B \PSU(3))$ are $(\CP^2,\CP^2\times \PSU(3))$ and $(S^4,H)$ where $H$ is the induced bundle from the Hopf fibration $H'$ 
\bea
\xymatrix{S^3=\SU(2)\ar[r] &S^7\ar[d]\\
&S^4}
\eea
by the group homomorphism $\rho:\SU(2)\to \PSU(3)$ which is the composition of the inclusion map $\SU(2)\to \SU(3)$ and the quotient map $\SU(3)\to \PSU(3)$, that means $H=H'\times_{\SU(2)}\PSU(3)=(H'\times \PSU(3))/\SU(2)$ which is the quotient of $H'\times \PSU(3)$ by the right $\SU(2)$ action 
\bea
(p,g)h=(ph,\rho(h^{-1})g).
\eea}

\begin{theorem}
\begin{table}[!h]
\centering
\begin{tabular}{c c}
\hline
$i$ & $\TP_i(\SO\times \PSU(3))$\\
\hline
0& $0$\\
1& $0$\\
2& $\Z_3$\\
3 & $\Z^2$\\
4 & $0$\\
5 & $\Z\times\Z_2$\\
\hline
\end{tabular}
\end{table}
\end{theorem}

The 2d topological term is $z_2$.

Since $c_2=\text{d}\text{CS}_3^{(\PSU(3))}$, the 3d topological terms are $\frac{1}{3}\text{CS}_3^{(TM)}$ and $\text{CS}_3^{(\PSU(3))}$.

Since $c_3=\text{d}\text{CS}_5^{(\PSU(3))}$, the 5d topological term are $\text{CS}_5^{(\PSU(3))}$ and $w_2w_3$.

\subsubsection{$\Omega^{\Spin}_d(\B \PSU(3))$}

For $t-s<8$,
\bea
\Ext_{\A_2(1)}^{s,t}(\H^*(\B \PSU(3),\Z_2),\Z_2)\Rightarrow\Omega^{\Spin}_{t-s}(\B \PSU(3))_2^{\wedge}.
\eea

The $E_2$ page is shown in Figure \ref{fig:Omega_*^{Spin}(BPSU(3))_2}.

\begin{figure}[!h]
\begin{center}
\begin{tikzpicture}
\node at (0,-1) {0};
\node at (1,-1) {1};
\node at (2,-1) {2};
\node at (3,-1) {3};
\node at (4,-1) {4};
\node at (5,-1) {5};
\node at (6,-1) {6};
\node at (7,-1) {$t-s$};
\node at (-1,0) {0};
\node at (-1,1) {1};
\node at (-1,2) {2};
\node at (-1,3) {3};
\node at (-1,4) {4};
\node at (-1,5) {5};
\node at (-1,6) {$s$};

\draw[->] (-0.5,-0.5) -- (-0.5,6);
\draw[->] (-0.5,-0.5) -- (7,-0.5);

\draw (0,0) -- (0,5);
\draw (0,0) -- (2,2);
\draw (4,3) -- (4,5);
\draw (4.1,0) -- (4.1,5);
\draw (6,1) -- (6,5);

\end{tikzpicture}
\end{center}
\caption{$\Omega_*^{\Spin}(\B \PSU(3))_2^{\wedge}$}
\label{fig:Omega_*^{Spin}(BPSU(3))_2}
\end{figure}
\bea
\Ext_{\A_3}^{s,t}(\H^*(M\Spin,\Z_3)\otimes\H^*(\B \PSU(3),\Z_3),\Z_3)\Rightarrow\Omega^{\Spin}_{t-s}(\B \PSU(3))_3^{\wedge}.
\eea

Since $\H^*(M\Spin,\Z_3)=\H^*(M\SO,\Z_3)$, the $E_2$ page is shown in Figure \ref{fig:Omega_*^{Spin}(BPSU(3))_3}.
\begin{figure}[!h]
\begin{center}
\begin{tikzpicture}
\node at (0,-1) {0};
\node at (1,-1) {1};
\node at (2,-1) {2};
\node at (3,-1) {3};
\node at (4,-1) {4};
\node at (5,-1) {5};
\node at (6,-1) {6};
\node at (7,-1) {$t-s$};
\node at (-1,0) {0};
\node at (-1,1) {1};
\node at (-1,2) {2};
\node at (-1,3) {3};
\node at (-1,4) {4};
\node at (-1,5) {5};
\node at (-1,6) {$s$};

\draw[->] (-0.5,-0.5) -- (-0.5,6);
\draw[->] (-0.5,-0.5) -- (7,-0.5);

\draw (0,0) -- (0,5);
\draw (4,1) -- (4,5);
\draw (4.1,0) -- (4.1,5);
\draw[fill] (2,0) circle(0.05);

\draw (6,0) -- (6,5);

\end{tikzpicture}
\end{center}
\caption{$\Omega_*^{\Spin}(\B \PSU(3))_3^{\wedge}$}
\label{fig:Omega_*^{Spin}(BPSU(3))_3}
\end{figure}

\begin{theorem}
\begin{table}[!h]
\centering
\begin{tabular}{c c}
\hline
$i$ & $\Omega^{\Spin}_i(\B \PSU(3))$\\
\hline
0& $\Z$\\
1& $\Z_2$\\
2& $\Z_2\times\Z_3$\\
3 & $0$\\
4 & $\Z^2$\\
5 & $0$\\
6 & $\Z$\\
\hline
\end{tabular}
\end{table}
\end{theorem}

The bordism invariants of $\Omega_2^{\Spin}(\B \PSU(3))$ are Arf and $z_2$.

The bordism invariants of $\Omega_4^{\Spin}(\B \PSU(3))$ are $\frac{\sigma}{16}$ and $c_2$.

By Wu formula \eqref{SqWu}, $c_3=\Sq^2c_2=(w_2(TM)+w_1^2(TM))c_2=0\mod2$ on Spin 6-manifolds.

The bordism invariant of $\Omega_6^{\Spin}(\B \PSU(3))$ is $\frac{c_3}{2}$.

\begin{theorem}
\begin{table}[!h]
\centering
\begin{tabular}{c c}
\hline
$i$ & $\TP_i(\Spin\times \PSU(3))$\\
\hline
0& $0$\\
1& $\Z_2$\\
2& $\Z_2\times\Z_3$\\
3 & $\Z^2$\\
4 & $0$\\
5 & $\Z$\\
\hline
\end{tabular}
\end{table}
\end{theorem}

The 2d topological terms are Arf and $z_2$.

The 3d topological terms are $\frac{1}{48}\text{CS}_3^{(TM)}$ and $\text{CS}_3^{(\PSU(3))}$.

The 5d topological term is $\frac{1}{2}\text{CS}_5^{(\PSU(3))}$.

\subsubsection{$\Omega^{\Pin^+}_d(\B \PSU(3))$}
\bea
\Ext_{\A_3}^{s,t}(\H^*(M\Pin^-,\Z_3)\otimes\H^*(\B \PSU(3),\Z_3),\Z_3)\Rightarrow\Omega^{\Pin^+}_{t-s}(\B \PSU(3))_3^{\wedge}.
\eea

Since $\H^*(M\Pin^-,\Z_3)=\H^*(M\tO,\Z_3)=0$, $\Omega^{\Pin^+}_{t-s}(\B \PSU(3))_3^{\wedge}=0$.

\bea
\Ext_{\A_2}^{s,t}(\H^*(M\Pin^-,\Z_2)\otimes\H^*(\B \PSU(3),\Z_2),\Z_2)\Rightarrow\Omega^{\Pin^+}_{t-s}(\B \PSU(3))_2^{\wedge}.
\eea
For $t-s<8$,
\bea
\Ext_{\A_2(1)}^{s,t}(\H^{*-1}(MT\tO(1),\Z_2)\otimes\H^*(\B \PSU(3),\Z_2),\Z_2)\Rightarrow\Omega^{\Pin^+}_{t-s}(\B \PSU(3))_2^{\wedge}.
\eea

The $\A_2(1)$-module structure of $\H^{*-1}(MT\tO(1),\Z_2)\otimes \H^*(\B \PSU(3),\Z_2)$ and the $E_2$ page are shown in Figure \ref{fig:H^{*-1}(MTO(1),Z_2)otimesH^*(BPSU(3),Z_2)}, \ref{fig:Omega_*^{Pin^+}(BPSU(3))_2}.

\begin{figure}[!h]
\begin{center}
\begin{tikzpicture}[scale=0.5]
\node[below] at (0,0) {$U$};

\draw[fill] (0,0) circle(.1);
\draw (0,0) -- (0,1);
\draw[fill] (0,1) circle(.1);
\draw[fill] (0,2) circle(.1);
\draw (0,0) to [out=150,in=150] (0,2);
\draw (0,2) -- (0,3);
\draw[fill] (0,3) circle(.1);
\draw (0,3) to [out=150,in=150] (0,5);
\draw[fill] (0,4) circle(.1);
\draw[fill] (0,5) circle(.1);
\draw (0,4) -- (0,5);
\draw (0,4) to [out=30,in=30] (0,6);
\draw[fill] (0,6) circle(.1);

\node[below] at (10,0) {$1$};

\draw[fill] (10,0) circle(.1);

\node[below] at (10,4) {$c_2(\mod2)$};

\draw[fill] (10,4) circle(.1);
\draw[fill] (10,6) circle(.1);
\draw (10,4) to [out=150,in=150] (10,6);

\node at (5,5) {$\bigotimes$};

\node at (-2,-10) {$=$};

\node[below] at (0,-15) {$U$};

\draw[fill] (0,-15) circle(.1);
\draw (0,-15) -- (0,-14);
\draw[fill] (0,-14) circle(.1);
\draw[fill] (0,-13) circle(.1);
\draw (0,-15) to [out=150,in=150] (0,-13);
\draw (0,-13) -- (0,-12);
\draw[fill] (0,-12) circle(.1);
\draw (0,-12) to [out=150,in=150] (0,-10);
\draw[fill] (0,-11) circle(.1);
\draw[fill] (0,-10) circle(.1);
\draw (0,-11) -- (0,-10);
\draw (0,-11) to [out=30,in=30] (0,-9);
\draw[fill] (0,-9) circle(.1);

\node[below] at (4,-11) {$c_2(\mod2)U$};

\draw[fill] (4,-11) circle(.1);
\draw[fill] (4,-10) circle(.1);
\draw (4,-11) -- (4,-10);
\draw[fill] (4,-9) circle(.1);
\draw (4,-11) to [out=150,in=150] (4,-9);
\draw[fill] (4,-8) circle(.1);
\draw (4,-9) -- (4,-8);
\draw[fill] (4,-7) circle(.1);
\draw[fill] (4,-6) circle(.1);
\draw (4,-7) -- (4,-6);
\draw (4,-8) to [out=150,in=150] (4,-6);
\draw[fill] (4,-5) circle(.1);
\draw (4,-7) to [out=30,in=30] (4,-5);
\draw[fill] (4,-4) circle(.1);
\draw (4,-5) -- (4,-4);
\draw[fill] (5,-9) circle(.1);
\draw[fill] (5,-8) circle(.1);
\draw (5,-9) -- (5,-8);
\draw (4,-10) to [out=30,in=150] (5,-8);
\draw[fill] (5,-7) circle(.1);
\draw (5,-9) to [out=30,in=30] (5,-7);
\draw[fill] (5,-6) circle(.1);
\draw (5,-7) -- (5,-6);
\draw[fill] (5,-5) circle(.1);
\draw[fill] (5,-4) circle(.1);
\draw (5,-5) -- (5,-4);
\draw[fill] (5,-3) circle(.1);
\draw (5,-6) to [out=30,in=30] (5,-4); 
\draw (5,-5) to [out=150,in=150] (5,-3);

\end{tikzpicture}
\end{center}

\caption{The $\A_2(1)$-module structure of $\H^{*-1}(MT\tO(1),\Z_2)\otimes \H^*(\B \PSU(3),\Z_2)$}
\label{fig:H^{*-1}(MTO(1),Z_2)otimesH^*(BPSU(3),Z_2)}
\end{figure}

\begin{figure}[!h]
\begin{center}
\begin{tikzpicture}
\node at (0,-1) {0};
\node at (1,-1) {1};
\node at (2,-1) {2};
\node at (3,-1) {3};
\node at (4,-1) {4};
\node at (5,-1) {5};
\node at (6,-1) {$t-s$};
\node at (-1,0) {0};
\node at (-1,1) {1};
\node at (-1,2) {2};
\node at (-1,3) {3};
\node at (-1,4) {4};
\node at (-1,5) {5};
\node at (-1,6) {$s$};

\draw[->] (-0.5,-0.5) -- (-0.5,6);
\draw[->] (-0.5,-0.5) -- (6,-0.5);

\draw[fill] (0,0) circle(0.05);
\draw (2,1) -- (4,3);
\draw (4,3) -- (4,0);
\draw[fill] (4.1,0) circle(0.05);

\end{tikzpicture}
\end{center}
\caption{$\Omega_*^{\Pin^+}(\B \PSU(3))_2^{\wedge}$}
\label{fig:Omega_*^{Pin^+}(BPSU(3))_2}
\end{figure}

\begin{theorem}
\begin{table}[!h]
\centering
\begin{tabular}{c c}
\hline
$i$ & $\Omega^{\Pin^+}_i(\B \PSU(3))$\\
\hline
0& $\Z_2$\\
1& $0$\\
2& $\Z_2$\\
3 & $\Z_2$\\
4 & $\Z_2\times\Z_{16}$\\
5 & $0$\\
\hline
\end{tabular}
\end{table}
\end{theorem}

The bordism invariant of $\Omega^{\Pin^+}_2(\B \PSU(3))$ is $w_1\tilde{\eta}$.

The bordism invariant of $\Omega^{\Pin^+}_3(\B \PSU(3))$ is $w_1\text{Arf}$.

The bordism invariants of $\Omega^{\Pin^+}_4(\B \PSU(3))$ are $c_2(\mod2)$ and $\eta$.

\begin{theorem}
\begin{table}[!h]
\centering
\begin{tabular}{c c}
\hline
$i$ & $\TP_i(\Pin^+\times \PSU(3))$\\
\hline
0& $\Z_2$\\
1& $0$\\
2& $\Z_2$\\
3 & $\Z_2$\\
4 & $\Z_2\times\Z_{16}$\\
5 & $0$\\
\hline
\end{tabular}
\end{table}
\end{theorem}

The 2d topological term is $w_1\tilde{\eta}$.

The 3d topological term is $w_1\text{Arf}$.

The 4d topological terms are $c_2(\mod2)$ and $\eta$.

\subsubsection{$\Omega^{\Pin^-}_d(\B \PSU(3))$}

\bea
\Ext_{\A_3}^{s,t}(\H^*(M\Pin^+,\Z_3)\otimes\H^*(\B \PSU(3),\Z_3),\Z_3)\Rightarrow\Omega^{\Pin^-}_{t-s}(\B \PSU(3))_3^{\wedge}.
\eea

Since $\H^*(M\Pin^+,\Z_3)=\H^*(M\tO,\Z_3)=0$, $\Omega^{\Pin^-}_{t-s}(\B \PSU(3))_3^{\wedge}=0$.
\bea
\Ext_{\A_2}^{s,t}(\H^*(M\Pin^+,\Z_2)\otimes\H^*(\B \PSU(3),\Z_2),\Z_2)\Rightarrow\Omega^{\Pin^-}_{t-s}(\B \PSU(3))_2^{\wedge}.
\eea
For $t-s<8$,
\bea
\Ext_{\A_2(1)}^{s,t}(\H^{*+1}(M\tO(1),\Z_2)\otimes\H^*(\B \PSU(3),\Z_2),\Z_2)\Rightarrow\Omega^{\Pin^-}_{t-s}(\B \PSU(3))_2^{\wedge}.
\eea

The $\A_2(1)$-module structure of $\H^{*+1}(M\tO(1),\Z_2)\otimes \H^*(\B \PSU(3),\Z_2)$ and the $E_2$ page are shown in Figure \ref{fig:H^{*+1}(MO(1),Z_2)otimesH^*(BPSU(3),Z_2)}, \ref{fig:Omega_*^{Pin^-}(BPSU(3))_2}.

\begin{figure}[!h]
\begin{center}
\begin{tikzpicture}[scale=0.5]

\node[below] at (0,0) {$U$};

\draw[fill] (0,0) circle(.1);
\draw (0,0) -- (0,1);
\draw[fill] (0,1) circle(.1);
\draw (0,1) to [out=150,in=150] (0,3);
\draw[fill] (0,2) circle(.1);
\draw[fill] (0,3) circle(.1);
\draw (0,2) -- (0,3);
\draw (0,2) to [out=30,in=30] (0,4);
\draw[fill] (0,4) circle(.1);
\draw (0,4) -- (0,5);
\draw[fill] (0,5) circle(.1);

\node[below] at (10,0) {$1$};

\draw[fill] (10,0) circle(.1);

\node[below] at (10,4) {$c_2(\mod2)$};

\draw[fill] (10,4) circle(.1);
\draw[fill] (10,6) circle(.1);
\draw (10,4) to [out=150,in=150] (10,6);

\node at (5,5) {$\bigotimes$};

\node at (-2,-10) {$=$};

\node[below] at (0,-15) {$U$};

\draw[fill] (0,-15) circle(.1);
\draw (0,-15) -- (0,-14);
\draw[fill] (0,-14) circle(.1);
\draw (0,-14) to [out=150,in=150] (0,-12);
\draw[fill] (0,-13) circle(.1);
\draw[fill] (0,-12) circle(.1);
\draw (0,-13) -- (0,-12);
\draw (0,-13) to [out=30,in=30] (0,-11);
\draw[fill] (0,-11) circle(.1);
\draw (0,-11) -- (0,-10);
\draw[fill] (0,-10) circle(.1);

\node[below] at (4,-11) {$c_2(\mod2)U$};

\draw[fill] (4,-11) circle(.1);
\draw[fill] (4,-10) circle(.1);
\draw (4,-11) -- (4,-10);
\draw[fill] (4,-9) circle(.1);
\draw (4,-11) to [out=150,in=150] (4,-9);
\draw[fill] (4,-8) circle(.1);
\draw (4,-9) -- (4,-8);
\draw[fill] (4,-7) circle(.1);
\draw[fill] (4,-6) circle(.1);
\draw (4,-7) -- (4,-6);
\draw (4,-8) to [out=150,in=150] (4,-6);
\draw[fill] (4,-5) circle(.1);
\draw (4,-7) to [out=30,in=30] (4,-5);
\draw[fill] (4,-4) circle(.1);
\draw (4,-5) -- (4,-4);
\draw[fill] (5,-9) circle(.1);
\draw[fill] (5,-8) circle(.1);
\draw (5,-9) -- (5,-8);
\draw (4,-10) to [out=30,in=150] (5,-8);
\draw[fill] (5,-7) circle(.1);
\draw (5,-9) to [out=30,in=30] (5,-7);
\draw[fill] (5,-6) circle(.1);
\draw (5,-7) -- (5,-6);
\draw[fill] (5,-5) circle(.1);
\draw[fill] (5,-4) circle(.1);
\draw (5,-5) -- (5,-4);
\draw[fill] (5,-3) circle(.1);
\draw (5,-6) to [out=30,in=30] (5,-4); 
\draw (5,-5) to [out=150,in=150] (5,-3);

\end{tikzpicture}
\end{center}
\caption{The $\A_2(1)$-module structure of $\H^{*+1}(M\tO(1),\Z_2)\otimes \H^*(\B \PSU(3),\Z_2)$}
\label{fig:H^{*+1}(MO(1),Z_2)otimesH^*(BPSU(3),Z_2)}
\end{figure}

\begin{figure}[!h]
\begin{center}
\begin{tikzpicture}
\node at (0,-1) {0};
\node at (1,-1) {1};
\node at (2,-1) {2};
\node at (3,-1) {3};
\node at (4,-1) {4};
\node at (5,-1) {5};
\node at (6,-1) {$t-s$};
\node at (-1,0) {0};
\node at (-1,1) {1};
\node at (-1,2) {2};
\node at (-1,3) {3};
\node at (-1,4) {4};
\node at (-1,5) {5};
\node at (-1,6) {$s$};

\draw[->] (-0.5,-0.5) -- (-0.5,6);
\draw[->] (-0.5,-0.5) -- (6,-0.5);

\draw (0,0) -- (2,2);
\draw (2,2) -- (2,0);
\draw[fill] (4,0) circle(0.05);

\end{tikzpicture}
\end{center}
\caption{$\Omega_*^{\Pin^-}(\B \PSU(3))_2^{\wedge}$}
\label{fig:Omega_*^{Pin^-}(BPSU(3))_2}
\end{figure}

\begin{theorem}
\begin{table}[!h]
\centering
\begin{tabular}{c c}
\hline
$i$ & $\Omega^{\Pin^-}_i(\B \PSU(3))$\\
\hline
0& $\Z_2$\\
1& $\Z_2$\\
2& $\Z_8$\\
3 & $0$\\
4 & $\Z_2$\\
5 & $0$\\
\hline
\end{tabular}
\end{table}
\end{theorem}

The bordism invariant of $\Omega^{\Pin^-}_2(\B \PSU(3))$ is ABK.

The bordism invariant of $\Omega^{\Pin^-}_4(\B \PSU(3))$ is $c_2(\mod2)$.

\begin{theorem}
\begin{table}[!h]
\centering
\begin{tabular}{c c}
\hline
$i$ & $\TP_i(\Pin^-\times \PSU(3))$\\
\hline
0& $\Z_2$\\
1& $\Z_2$\\
2& $\Z_8$\\
3 & $0$\\
4 & $\Z_2$\\
5 & $0$\\
\hline
\end{tabular}
\end{table}
\end{theorem}

The 2d topological term is ABK.

The 4d topological term is $c_2(\mod2)$.

\subsection{$(\B G_a,\B ^2G_b):(\B \Z_2,\B ^2\Z_2),(\B \Z_3,\B ^2\Z_3)$}

\label{sec:BZnB2Zn}

\subsubsection{$\Omega^{\tO}_d(\B \Z_2\times \B ^2\Z_2)$}

Since the computation involves no odd torsion, we can use the Adams spectral sequence 
\bea
&&E_2^{s,t}=\Ext_{\A_2}^{s,t}(\H^*(M\tO\wedge(\B \Z_2\times \B ^2\Z_2)_+,\Z_2),\Z_2)\notag\\
&\Rightarrow&\pi_{t-s}(M\tO\wedge(\B \Z_2\times \B ^2\Z_2)_+)_2^{\wedge}=\Omega^{\tO}_{t-s}(\B \Z_2\times \B ^2\Z_2).
\eea

\bea
&&\H^*(M\tO,\Z_2)\otimes\H^*(\B \Z_2\times \B ^2\Z_2,\Z_2)=\A_2\otimes\Z_2[y_2,y_4,y_5,y_6,y_8,\dots]^*\otimes\Z_2[a,x_2,x_3,x_5,x_9,\dots]\notag\\
&=&\A_2\oplus\Sigma\A_2\oplus3\Sigma^2\A_2\oplus4\Sigma^3\A_2\oplus8\Sigma^4\A_2\oplus12\Sigma^5\A_2\oplus\cdots
\eea

\begin{theorem}
\begin{table}[!h]
\centering
\begin{tabular}{c c}
\hline
$i$ & $\Omega^{\tO}_i(\B \Z_2\times \B ^2\Z_2)$\\
\hline
0& $\Z_2$\\
1& $\Z_2$\\
2& $\Z_2^3$\\
3 & $\Z_2^4$\\
4 & $\Z_2^8$\\ 
5 & $\Z_2^{12}$\\
\hline
\end{tabular}
\end{table}
\end{theorem}

The bordism invariants of $\Omega^{\tO}_2(\B \Z_2\times \B ^2\Z_2)$ are $a^2,x_2,w_1^2$.

The bordism invariants of $\Omega^{\tO}_3(\B \Z_2\times \B ^2\Z_2)$ are $x_3=w_1x_2,ax_2,aw_1^2,a^3$.

The bordism invariants of $\Omega^{\tO}_4(\B \Z_2\times \B ^2\Z_2)$ are $w_1^4,w_2^2,a^4,a^2x_2,ax_3,x_2^2,w_1^2a^2,w_1^2x_2$.

The bordism invariants of $\Omega^{\tO}_5(\B \Z_2\times \B ^2\Z_2)$ are $$a^5,a^2x_3,a^3x_2,a^3w_1^2,ax_2^2,aw_1^4,ax_2w_1^2,aw_2^2,x_2x_3,w_1^2x_3,x_5,w_2w_3.$$

\begin{theorem}
\begin{table}[!h]
\centering
\begin{tabular}{c c}
\hline
$i$ & $\TP_i(\tO\times\Z_2\times \B \Z_2)$\\
\hline
0& $\Z_2$\\
1& $\Z_2$\\
2& $\Z_2^3$\\
3 & $\Z_2^4$\\
4 & $\Z_2^8$\\ 
5 & $\Z_2^{12}$\\
\hline
\end{tabular}
\end{table}
\end{theorem}

The 2d topological terms are $a^2,x_2,w_1^2$.

The 3d topological terms are $x_3=w_1x_2,ax_2,aw_1^2,a^3$.

The 4d topological terms are $w_1^4,w_2^2,a^4,a^2x_2,ax_3,x_2^2,w_1^2a^2,w_1^2x_2$.

The 5d topological terms are $$a^5,a^2x_3,a^3x_2,a^3w_1^2,ax_2^2,aw_1^4,ax_2w_1^2,aw_2^2,x_2x_3,w_1^2x_3,x_5,w_2w_3.$$

\subsubsection{$\Omega^{\SO}_d(\B \Z_2\times \B ^2\Z_2)$}
Since the computation involves no odd torsion, we can use the Adams spectral sequence 
\bea
&&E_2^{s,t}=\Ext_{\A_2}^{s,t}(\H^*(M\SO\wedge (\B \Z_2\times \B ^2\Z_2)_+,\Z_2),\Z_2)\notag\\
&\Rightarrow&\pi_{t-s}(M\SO\wedge (\B \Z_2\times \B ^2\Z_2)_+)_2^{\wedge}=\Omega^{\SO}_{t-s}(\B \Z_2\times \B ^2\Z_2).
\eea

\cred{
There is a differential $d_2$ corresponding to the Bockstein homomorphism  $\beta_{(2,4)}:\H^*(-,\Z_4)\to\H^{*+1}(-,\Z_2)$ associated to $0\to\Z_2\to\Z_8\to\Z_4\to0$ \cite{may1981bockstein}. See \ref{Bockstein} for the definition of Bockstein homomorphisms.
}

By \eqref{x_2x_3+x_5}, there is a differential such that
$d_2(x_2x_3+x_5)=x_2^2h_0^2$.

The $E_2$ page is shown in Figure \ref{fig:Omega_*^{SO}(BZ_2timesB^2Z_2)}.

\begin{figure}[!h]
\begin{center}
\begin{tikzpicture}
\node at (0,-1) {0};
\node at (1,-1) {1};
\node at (2,-1) {2};
\node at (3,-1) {3};
\node at (4,-1) {4};
\node at (5,-1) {5};
\node at (6,-1) {$t-s$};
\node at (-1,0) {0};
\node at (-1,1) {1};
\node at (-1,2) {2};
\node at (-1,3) {3};
\node at (-1,4) {4};
\node at (-1,5) {5};
\node at (-1,6) {$s$};

\draw[->] (-0.5,-0.5) -- (-0.5,6);
\draw[->] (-0.5,-0.5) -- (6,-0.5);

\draw (0,0) -- (0,5);
\draw[fill] (1,0) circle(0.05);
\draw[fill] (2,0) circle(0.05);
\draw[fill] (2.9,0) circle(0.05);
\draw[fill] (3.1,0) circle(0.05);
\draw[fill] (3.9,0) circle(0.05);
\draw (4,0) -- (4,5);
\draw (4.1,0) -- (4.1,5);
\draw[fill] (4.7,0) circle(0.05);
\draw[fill] (4.8,0) circle(0.05);
\draw[fill] (4.9,0) circle(0.05);
\draw[fill] (5.1,0) circle(0.05);
\draw[fill] (5.2,0) circle(0.05);
\draw[fill] (5.3,0) circle(0.05);
\draw (5,0) -- (5,5);

\draw[color=red][->] (5,0) -- (4.1,2);
\draw[color=red][->] (5,1) -- (4.1,3);
\draw[color=red][->] (5,2) -- (4.1,4);
\draw[color=red][->] (5,3) -- (4.1,5);

\end{tikzpicture}
\end{center}
\caption{$\Omega_*^{\SO}(\B \Z_2\times \B ^2\Z_2)$}
\label{fig:Omega_*^{SO}(BZ_2timesB^2Z_2)}
\end{figure}

\begin{theorem}
\begin{table}[!h]
\centering
\begin{tabular}{c c}
\hline
$i$ & $\Omega^{\SO}_i(\B \Z_2\times \B ^2\Z_2)$\\
\hline
0& $\Z$\\
1& $\Z_2$\\
2& $\Z_2$\\
3 & $\Z_2^2$\\
4 & $\Z\times\Z_2\times\Z_4$\\ 
5 & $\Z_2^6$\\
\hline
\end{tabular}
\end{table}
\end{theorem}

The bordism invariant of $\Omega_{2}^{\SO}(\B \Z_2\times \B ^2\Z_2)$ is $x_2$.

The bordism invariants of $\Omega_{3}^{\SO}(\B \Z_2\times \B ^2\Z_2)$ are $ax_2,a^3$.

The bordism invariants of $\Omega_{4}^{\SO}(\B \Z_2\times \B ^2\Z_2)$ are $\sigma$, $ax_3(=a^2x_2)$ and $\mathcal{P}_2(x_2)$.

The bordism invariants of $\Omega_5^{\SO}(\B \Z_2\times \B ^2\Z_2)$ are $ax_2^2,a^5,x_5,a^3x_2,w_2w_3,aw_2^2$.

\begin{theorem}
\begin{table}[!h]
\centering
\begin{tabular}{c c}
\hline
$i$ & $\TP_i(\SO\times\Z_2\times \B \Z_2)$\\
\hline
0& $0$\\
1& $\Z_2$\\
2& $\Z_2$\\
3 & $\Z\times\Z_2^2$\\
4 & $\Z_2\times\Z_4$\\ 
5 & $\Z_2^6$\\
\hline
\end{tabular}
\end{table}
\end{theorem}

The 2d topological term is $x_2$.

The 3d topological terms are $\frac{1}{3}\text{CS}_3^{(TM)},ax_2,a^3$.

The 4d topological terms are $ax_3(=a^2x_2)$ and $\mathcal{P}_2(x_2)$.

The 5d topological terms are $ax_2^2,a^5,x_5,a^3x_2,w_2w_3,aw_2^2$.

\subsubsection{$\Omega^{\Spin}_d(\B \Z_2\times \B ^2\Z_2)$}

Since the computation involves no odd torsion, we can use the Adams spectral sequence 
\bea
&&E_2^{s,t}=\Ext_{\A_2}^{s,t}(\H^*(M\Spin\wedge (\B \Z_2\times \B ^2\Z_2)_+,\Z_2),\Z_2)\notag\\
&\Rightarrow&\pi_{t-s}(M\Spin\wedge (\B \Z_2\times \B ^2\Z_2)_+)_2^{\wedge}=\Omega^{\Spin}_{t-s}(\B \Z_2\times \B ^2\Z_2).
\eea
For $t-s<8$,
\bea
\Ext_{\A_2(1)}^{s,t}(\H^*(\B \Z_2\times \B ^2\Z_2,\Z_2),\Z_2)\Rightarrow\Omega^{\Spin}_{t-s}(\B \Z_2\times \B ^2\Z_2).
\eea

$\H^*(\B \Z_2\times \B ^2\Z_2,\Z_2)=\Z_2[a,x_2,x_3,x_5,x_9,\dots]$ where $\Sq^1x_2=x_3$, $\Sq^2x_2=x_2^2$, $\Sq^1x_3=0$, $\Sq^2x_3=x_5$, $\Sq^1x_5=\Sq^2x_2^2=x_3^2$, $\Sq^2x_5=0$.

\cred{
There is a differential $d_2$ corresponding to the Bockstein homomorphism  $\beta_{(2,4)}:\H^*(-,\Z_4)\to\H^{*+1}(-,\Z_2)$ associated to $0\to\Z_2\to\Z_8\to\Z_4\to0$ \cite{may1981bockstein}. See \ref{Bockstein} for the definition of Bockstein homomorphisms.
}

By \eqref{x_2x_3+x_5}, there is a differential such that
$d_2(x_2x_3+x_5)=x_2^2h_0^2$.

The $\A_2(1)$-module structure of $\H^*(\B \Z_2\times \B ^2\Z_2,\Z_2)$ and the $E_2$ page are shown in Figure \ref{fig:H^*(BZ_2timesB^2Z_2,Z_2)}, \ref{Omega_*^{Spin}(BZ_2timesB^2Z_2)}.

\begin{figure}[!h]
\begin{center}
\begin{tikzpicture}[scale=0.5]

\node[below] at (0,0) {$1$};

\draw[fill] (0,0) circle(.1);

\node[right] at (0,1) {$a$};

\draw[fill] (0,1) circle(.1);
\draw[fill] (0,2) circle(.1);
\draw (0,1) -- (0,2);
\draw[fill] (0,3) circle(.1);
\draw[fill] (0,4) circle(.1);
\draw (0,3) -- (0,4);
\draw (0,2) to [out=150,in=150] (0,4);
\draw[fill] (0,5) circle(.1);
\draw (0,3) to [out=30,in=30] (0,5);
\draw[fill] (0,6) circle(.1);
\draw (0,5) -- (0,6);

\node[below] at (10,0) {$1$};

\draw[fill] (10,0) circle(.1);

\node[below] at (10,2) {$x_2$};

\draw[fill] (10,2) circle(.1);
\draw[fill] (10,3) circle(.1);
\draw[fill] (10,4) circle(.1);
\draw (10,2) -- (10,3);
\draw (10,2) to [out=150,in=150] (10,4);
\draw[fill] (11,5) circle(.1);
\draw[fill] (11,6) circle(.1);
\draw (11,5) -- (11,6);
\draw (10,3) to [out=30,in=150] (11,5);
\draw (10,4) to [out=30,in=150] (11,6);
\draw[fill] (12,5) circle(.1);

\node[below] at (12,5) {$x_2x_3$};

\draw (12,5) -- (11,6);
\draw[fill] (12,7) circle(.1);
\draw (12,5) to [out=30,in=30] (12,7);
\draw[fill] (12,8) circle(.1);
\draw (12,7) -- (12,8);
\draw[fill] (12,10) circle(.1);
\draw (12,8) to [out=150,in=150] (12,10);

\node at (5,5) {$\bigotimes$};
\node at (-2,-10) {$=$};

\node[below] at (0,-15) {$1$};

\draw[fill] (0,-15) circle(.1);

\node[right] at (0,-14) {$a$};

\draw[fill] (0,-14) circle(.1);
\draw[fill] (0,-13) circle(.1);
\draw (0,-14) -- (0,-13);
\draw[fill] (0,-12) circle(.1);
\draw[fill] (0,-11) circle(.1);
\draw (0,-12) -- (0,-11);
\draw (0,-13) to [out=150,in=150] (0,-11);
\draw[fill] (0,-10) circle(.1);
\draw (0,-12) to [out=30,in=30] (0,-10);
\draw[fill] (0,-9) circle(.1);
\draw (0,-10) -- (0,-9);

\node[below] at (2,-13) {$x_2$};

\draw[fill] (2,-13) circle(.1);
\draw[fill] (2,-12) circle(.1);
\draw[fill] (2,-11) circle(.1);
\draw (2,-13) -- (2,-12);
\draw (2,-13) to [out=150,in=150] (2,-11);
\draw[fill] (3,-10) circle(.1);
\draw[fill] (3,-9) circle(.1);
\draw (3,-10) -- (3,-9);
\draw (2,-12) to [out=30,in=150] (3,-10);
\draw (2,-11) to [out=30,in=150] (3,-9);
\draw[fill] (4,-10) circle(.1);
\draw (4,-10) -- (3,-9);
\draw[fill] (4,-8) circle(.1);
\draw (4,-10) to [out=30,in=30] (4,-8);
\draw[fill] (4,-7) circle(.1);
\draw (4,-8) -- (4,-7);
\draw[fill] (4,-5) circle(.1);
\draw (4,-7) to [out=150,in=150] (4,-5);

\node[below] at (6,-12) {$ax_2$};

\draw[fill] (6,-12) circle(.1);
\draw[fill] (6,-11) circle(.1);
\draw[fill] (6,-10) circle(.1);
\draw (6,-12) -- (6,-11);
\draw (6,-12) to [out=150,in=150] (6,-10);
\draw[fill] (7,-9) circle(.1);
\draw[fill] (7,-8) circle(.1);
\draw (7,-9) -- (7,-8);
\draw (6,-11) to [out=30,in=150] (7,-9);
\draw (6,-10) to [out=30,in=150] (7,-8);
\draw[fill] (6,-9) circle(.1);
\draw (6,-10) -- (6,-9);
\draw[fill] (6,-8) circle(.1);
\draw[fill] (6,-7) circle(.1);
\draw (6,-9) to [out=30,in=30] (6,-7);
\draw (6,-8) -- (6,-7);
\draw[fill] (6,-6) circle(.1);
\draw (6,-8) to [out=150,in=150] (6,-6);

\node[below] at (8,-11) {$ax_3$};
 
\draw[fill] (8,-11) circle(.1);
\draw[fill] (8,-10) circle(.1);
\draw (8,-11) -- (8,-10);
\draw[fill] (8,-9) circle(.1);
\draw (8,-11) to [out=150,in=150] (8,-9);
\draw[fill] (9,-8) circle(.1);
\draw (8,-10) to [out=30,in=150] (9,-8);
\draw[fill] (8,-8) circle(.1);
\draw (8,-9) -- (8,-8);
\draw[fill] (9,-7) circle(.1);
\draw (9,-8) -- (9,-7);
\draw (8,-9) to [out=30,in=150] (9,-7);
\draw[fill] (9,-6) circle(.1);
\draw (8,-8) to [out=30,in=150] (9,-6);
\draw[fill] (9,-5) circle(.1);
\draw (9,-6) -- (9,-5);
\draw (9,-7) to [out=30,in=30] (9,-5); 

\node[below] at (10,-10) {$a^3x_2$};
 
\draw[fill] (10,-10) circle(.1);
\draw[fill] (10,-9) circle(.1);
\draw (10,-10) -- (10,-9);
\draw[fill] (10,-8) circle(.1);
\draw (10,-10) to [out=150,in=150] (10,-8);
\draw[fill] (11,-7) circle(.1);
\draw (10,-9) to [out=30,in=150] (11,-7);
\draw[fill] (10,-7) circle(.1);
\draw (10,-8) -- (10,-7);
\draw[fill] (11,-6) circle(.1);
\draw (11,-7) -- (11,-6);
\draw (10,-8) to [out=30,in=150] (11,-6);
\draw[fill] (11,-5) circle(.1);
\draw (10,-7) to [out=30,in=150] (11,-5);
\draw[fill] (11,-4) circle(.1);
\draw (11,-5) -- (11,-4);
\draw (11,-6) to [out=30,in=30] (11,-4);

\end{tikzpicture}
\end{center}
\caption{The $\A_2(1)$-module structure of $\H^*(\B \Z_2\times \B ^2\Z_2,\Z_2)$}
\label{fig:H^*(BZ_2timesB^2Z_2,Z_2)}
\end{figure}

\begin{figure}[!h]
\begin{center}
\begin{tikzpicture}
\node at (0,-1) {0};
\node at (1,-1) {1};
\node at (2,-1) {2};
\node at (3,-1) {3};
\node at (4,-1) {4};
\node at (5,-1) {5};
\node at (6,-1) {$t-s$};
\node at (-1,0) {0};
\node at (-1,1) {1};
\node at (-1,2) {2};
\node at (-1,3) {3};
\node at (-1,4) {4};
\node at (-1,5) {5};
\node at (-1,6) {$s$};

\draw[->] (-0.5,-0.5) -- (-0.5,6);
\draw[->] (-0.5,-0.5) -- (6,-0.5);

\draw (0,0) -- (0,5);
\draw (0,0) -- (2,2);
\draw (1,0) -- (3,2);
\draw (3,2) -- (3,0);
\draw[fill] (2,0) circle(0.05);
\draw[fill] (3.1,0) circle(0.05);
\draw[fill] (4,0) circle(0.05);
\draw[fill] (5.1,0) circle(0.05);
\draw (4,3) -- (4,5);
\draw (4.1,1) -- (4.1,5);
\draw (5,0) -- (5,5);

\draw[color=red][->] (5,0) -- (4.1,2);
\draw[color=red][->] (5,1) -- (4.1,3);
\draw[color=red][->] (5,2) -- (4.1,4);
\draw[color=red][->] (5,3) -- (4.1,5);

\end{tikzpicture}
\end{center}
\caption{$\Omega_*^{\Spin}(\B \Z_2\times \B ^2\Z_2)$}
\label{Omega_*^{Spin}(BZ_2timesB^2Z_2)}
\end{figure}

\begin{theorem}
\begin{table}[!h]
\centering
\begin{tabular}{c c}
\hline
$i$ & $\Omega^{\Spin}_i(\B \Z_2\times \B ^2\Z_2)$\\
\hline
0& $\Z$\\
1& $\Z_2^2$\\
2& $\Z_2^3$\\
3 & $\Z_2\times\Z_8$\\
4 & $\Z\times\Z_2^2$\\ 
5 & $\Z_2$\\
\hline
\end{tabular}
\end{table}
\end{theorem}

The bordism invariants of $\Omega_{2}^{\Spin}(\B \Z_2\times \B ^2\Z_2)$ are $x_2,\text{Arf},a\tilde{\eta}$.

The bordism invariants of $\Omega_{3}^{\Spin}(\B \Z_2\times \B ^2\Z_2)$ are $ax_2,a\text{ABK}$.

The bordism invariants of $\Omega_{4}^{\Spin}(\B \Z_2\times \B ^2\Z_2)$ are $\frac{\sigma}{16}$, $ax_3(=a^2x_2)$ and $\frac{\mathcal{P}_2(x_2)}{2}$.

The bordism invariant of $\Omega_5^{\Spin}(\B \Z_2\times \B ^2\Z_2)$ is $a^3x_2$.

\begin{theorem}
\begin{table}[!h]
\centering
\begin{tabular}{c c}
\hline
$i$ & $\TP_i(\Spin\times\Z_2\times \B \Z_2)$\\
\hline
0& $0$\\
1& $\Z_2^2$\\
2& $\Z_2^3$\\
3 & $\Z\times\Z_2\times\Z_8$\\
4 & $\Z_2^2$\\ 
5 & $\Z_2$\\
\hline
\end{tabular}
\end{table}
\end{theorem}

The 2d topological terms are $x_2,\text{Arf},a\tilde{\eta}$.

The 3d topological terms are $\frac{1}{48}\text{CS}_3^{(TM)},ax_2,a\text{ABK}$.

The 4d topological terms are $ax_3(=a^2x_2)$ and $\frac{\mathcal{P}_2(x_2)}{2}$.

The 5d topological term is $a^3x_2$.

\subsubsection{$\Omega^{\Pin^+}_d(\B \Z_2\times \B ^2\Z_2)$}

Since the computation involves no odd torsion, we can use the Adams spectral sequence 
\bea
&&E_2^{s,t}=\Ext_{\A_2}^{s,t}(\H^*(M\Pin^-\wedge (\B \Z_2\times \B ^2\Z_2)_+,\Z_2),\Z_2)\notag\\
&\Rightarrow&\pi_{t-s}(M\Pin^-\wedge (\B \Z_2\times \B ^2\Z_2)_+)_2^{\wedge}=\Omega^{\Pin^+}_{t-s}(\B \Z_2\times \B ^2\Z_2).
\eea
$M\Pin^-=MT\Pin^+\sim M\Spin\wedge S^1\wedge MT\tO(1)$.

For $t-s<8$,
\bea
&&\Ext_{\A_2(1)}^{s,t}(\H^{*-1}(MT\tO(1),\Z_2)\otimes\H^*(\B \Z_2\times \B ^2\Z_2,\Z_2),\Z_2)\Rightarrow\Omega^{\Pin^+}_{t-s}(\B \Z_2\times \B ^2\Z_2).
\eea

The $\A_2(1)$-module structure of $\H^{*-1}(MT\tO(1),\Z_2)\otimes\H^*(\B \Z_2\times \B ^2\Z_2,\Z_2)$ and the $E_2$ page are shown in Figure \ref{fig:H^{*-1}(MTO(1),Z_2)otimesH^*(BZ_2timesB^2Z_2,Z_2)}, \ref{fig:Omega_*^{Pin^+}(BZ_2timesB^2Z_2)}.

\begin{figure}[!h]
\begin{center}
\begin{tikzpicture}[scale=0.5]

\node[below] at (0,0) {$U$};

\draw[fill] (0,0) circle(.1);
\draw[fill] (0,1) circle(.1);
\draw (0,0) -- (0,1);
\draw[fill] (0,2) circle(.1);
\draw (0,0) to [out=150,in=150] (0,2);
\draw[fill] (0,3) circle(.1);
\draw (0,2) -- (0,3);
\draw[fill] (0,4) circle(.1);
\draw[fill] (0,5) circle(.1);
\draw (0,4) -- (0,5);
\draw (0,3) to [out=150,in=150] (0,5);
\draw[fill] (0,6) circle(.1);
\draw (0,4) to [out=30,in=30] (0,6);

\node[below] at (4,0) {$1$};

\draw[fill] (4,0) circle(.1);
\draw[fill] (4,1) circle(.1);

\node[right] at (4,1) {$a$};

\draw[fill] (4,2) circle(.1);
\draw (4,1) -- (4,2);
\draw[fill] (4,3) circle(.1);
\draw[fill] (4,4) circle(.1);
\draw (4,3) -- (4,4);
\draw (4,2) to [out=150,in=150] (4,4);
\draw[fill] (4,5) circle(.1);
\draw (4,3) to [out=30,in=30] (4,5);
\draw[fill] (4,6) circle(.1);
\draw (4,5) -- (4,6);

\node[below] at (6,2) {$x_2$};

\draw[fill] (6,2) circle(.1);
\draw[fill] (6,3) circle(.1);
\draw[fill] (6,4) circle(.1);
\draw (6,2) -- (6,3);
\draw (6,2) to [out=150,in=150] (6,4);
\draw[fill] (7,5) circle(.1);
\draw[fill] (7,6) circle(.1);
\draw (7,5) -- (7,6);
\draw (6,3) to [out=30,in=150] (7,5);
\draw (6,4) to [out=30,in=150] (7,6);
\draw[fill] (8,5) circle(.1);

\node[below] at (8,5) {$x_2x_3$};

\draw (8,5) -- (7,6);
\draw[fill] (8,7) circle(.1);
\draw (8,5) to [out=30,in=30] (8,7);
\draw[fill] (8,8) circle(.1);
\draw (8,7) -- (8,8);
\draw[fill] (8,10) circle(.1);
\draw (8,8) to [out=150,in=150] (8,10);

\node[below] at (10,3) {$ax_2$};

\draw[fill] (10,3) circle(.1);
\draw[fill] (10,4) circle(.1);
\draw[fill] (10,5) circle(.1);
\draw (10,3) -- (10,4);
\draw (10,3) to [out=150,in=150] (10,5);
\draw[fill] (11,6) circle(.1);
\draw[fill] (11,7) circle(.1);
\draw (11,6) -- (11,7);
\draw (10,4) to [out=30,in=150] (11,6);
\draw (10,5) to [out=30,in=150] (11,7);
\draw[fill] (10,6) circle(.1);
\draw (10,5) -- (10,6);
\draw[fill] (10,7) circle(.1);
\draw[fill] (10,8) circle(.1);
\draw (10,6) to [out=30,in=30] (10,8);
\draw (10,7) -- (10,8);
\draw[fill] (10,9) circle(.1);
\draw (10,7) to [out=150,in=150] (10,9);

\node[below] at (12,4) {$ax_3$};
 
\draw[fill] (12,4) circle(.1);
\draw[fill] (12,5) circle(.1);
\draw (12,4) -- (12,5);
\draw[fill] (12,6) circle(.1);
\draw (12,4) to [out=150,in=150] (12,6);
\draw[fill] (13,7) circle(.1);
\draw (12,5) to [out=30,in=150] (13,7);
\draw[fill] (12,7) circle(.1);
\draw (12,6) -- (12,7);
\draw[fill] (13,8) circle(.1);
\draw (13,7) -- (13,8);
\draw (12,6) to [out=30,in=150] (13,8);
\draw[fill] (13,9) circle(.1);
\draw (12,7) to [out=30,in=150] (13,9);
\draw[fill] (13,10) circle(.1);
\draw (13,9) -- (13,10);
\draw (13,8) to [out=30,in=30] (13,10); 

\node[below] at (14,5) {$a^3x_2$};
 
\draw[fill] (14,5) circle(.1);
\draw[fill] (14,6) circle(.1);
\draw (14,5) -- (14,6);
\draw[fill] (14,7) circle(.1);
\draw (14,5) to [out=150,in=150] (14,7);
\draw[fill] (15,8) circle(.1);
\draw (14,6) to [out=30,in=150] (15,8);
\draw[fill] (14,8) circle(.1);
\draw (14,7) -- (14,8);
\draw[fill] (15,9) circle(.1);
\draw (15,8) -- (15,9);
\draw (14,7) to [out=30,in=150] (15,9);
\draw[fill] (15,10) circle(.1);
\draw (14,8) to [out=30,in=150] (15,10);
\draw[fill] (15,11) circle(.1);
\draw (15,10) -- (15,11);
\draw (15,9) to [out=30,in=30] (15,11);

\node at (2,5) {$\bigotimes$};
\node at (-2,-10) {$=$};

\node[below] at (0,-15) {$U$};

\draw[fill] (0,-15) circle(.1);
\draw[fill] (0,-14) circle(.1);
\draw (0,-15) -- (0,-14);
\draw[fill] (0,-13) circle(.1);
\draw (0,-15) to [out=150,in=150] (0,-13);
\draw[fill] (0,-12) circle(.1);
\draw (0,-13) -- (0,-12);
\draw[fill] (0,-11) circle(.1);
\draw[fill] (0,-10) circle(.1);
\draw (0,-11) -- (0,-10);
\draw (0,-12) to [out=150,in=150] (0,-10);
\draw[fill] (0,-9) circle(.1);
\draw (0,-11) to [out=30,in=30] (0,-9);

\node[below] at (2,-14) {$aU$};

\draw[fill] (2,-14) circle(.1);
\draw[fill] (2,-13) circle(.1);
\draw (2,-14) -- (2,-13);
\draw[fill] (2,-12) circle(.1);
\draw (2,-14) to [out=150,in=150] (2,-12);
\draw[fill] (3,-11) circle(.1);
\draw (2,-13) to [out=30,in=150] (3,-11);
\draw[fill] (2,-11) circle(.1);
\draw (2,-12) -- (2,-11);
\draw[fill] (3,-10) circle(.1);
\draw (3,-11) -- (3,-10);
\draw (2,-12) to [out=30,in=150] (3,-10);
\draw[fill] (3,-9) circle(.1);
\draw (2,-11) to [out=30,in=150] (3,-9);
\draw[fill] (3,-8) circle(.1);
\draw (3,-9) -- (3,-8);
\draw (3,-10) to [out=30,in=30] (3,-8); 

\node[below] at (4,-13) {$w_1aU$};

\draw[fill] (4,-13) circle(.1);
\draw[fill] (4,-12) circle(.1);
\draw (4,-13) -- (4,-12);
\draw[fill] (4,-11) circle(.1);
\draw[fill] (4,-10) circle(.1);
\draw (4,-11) -- (4,-10);
\draw (4,-12) to [out=150,in=150] (4,-10);
\draw[fill] (4,-9) circle(.1);
\draw (4,-11) to [out=30,in=30] (4,-9);
\draw[fill] (4,-8) circle(.1);
\draw (4,-9) -- (4,-8);

\node[below] at (6,-12) {$a^3U$};

\draw[fill] (6,-12) circle(.1);
\draw[fill] (6,-11) circle(.1);
\draw (6,-12) -- (6,-11);
\draw[fill] (6,-10) circle(.1);
\draw (6,-12) to [out=150,in=150] (6,-10);
\draw[fill] (7,-9) circle(.1);
\draw (6,-11) to [out=30,in=150] (7,-9);
\draw[fill] (6,-9) circle(.1);
\draw (6,-10) -- (6,-9);
\draw[fill] (7,-8) circle(.1);
\draw (7,-9) -- (7,-8);
\draw (6,-10) to [out=30,in=150] (7,-8);
\draw[fill] (7,-7) circle(.1);
\draw (6,-9) to [out=30,in=150] (7,-7);
\draw[fill] (7,-6) circle(.1);
\draw (7,-7) -- (7,-6);
\draw (7,-8) to [out=30,in=30] (7,-6); 

\node[below] at (8,-10) {$w_1^4aU$};

\draw[fill] (8,-10) circle(.1);
\draw[fill] (8,-9) circle(.1);
\draw (8,-10) -- (8,-9);
\draw[fill] (8,-8) circle(.1);
\draw (8,-10) to [out=150,in=150] (8,-8);
\draw[fill] (9,-7) circle(.1);
\draw (8,-9) to [out=30,in=150] (9,-7);
\draw[fill] (8,-7) circle(.1);
\draw (8,-8) -- (8,-7);
\draw[fill] (9,-6) circle(.1);
\draw (9,-7) -- (9,-6);
\draw (8,-8) to [out=30,in=150] (9,-6);
\draw[fill] (9,-5) circle(.1);
\draw (8,-7) to [out=30,in=150] (9,-5);
\draw[fill] (9,-4) circle(.1);
\draw (9,-5) -- (9,-4);
\draw (9,-6) to [out=30,in=30] (9,-4); 

\node[below] at (10,-10) {$a^5U$};

\draw[fill] (10,-10) circle(.1);
\draw[fill] (10,-9) circle(.1);
\draw (10,-10) -- (10,-9);
\draw[fill] (10,-8) circle(.1);
\draw (10,-10) to [out=150,in=150] (10,-8);
\draw[fill] (11,-7) circle(.1);
\draw (10,-9) to [out=30,in=150] (11,-7);
\draw[fill] (10,-7) circle(.1);
\draw (10,-8) -- (10,-7);
\draw[fill] (11,-6) circle(.1);
\draw (11,-7) -- (11,-6);
\draw (10,-8) to [out=30,in=150] (11,-6);
\draw[fill] (11,-5) circle(.1);
\draw (10,-7) to [out=30,in=150] (11,-5);
\draw[fill] (11,-4) circle(.1);
\draw (11,-5) -- (11,-4);
\draw (11,-6) to [out=30,in=30] (11,-4); 

\node[below] at (12,-13) {$x_2U$};

\draw[fill] (12,-13) circle(.1);
\draw[fill] (12,-12) circle(.1);
\draw (12,-13) -- (12,-12);
\draw[fill] (12,-11) circle(.1);
\draw (12,-13) to [out=150,in=150] (12,-11);
\draw[fill] (13,-10) circle(.1);
\draw (12,-12) to [out=30,in=150] (13,-10);
\draw[fill] (12,-10) circle(.1);
\draw (12,-11) -- (12,-10);
\draw[fill] (13,-9) circle(.1);
\draw (13,-10) -- (13,-9);
\draw (12,-11) to [out=30,in=150] (13,-9);
\draw[fill] (13,-8) circle(.1);
\draw (12,-10) to [out=30,in=150] (13,-8);
\draw[fill] (13,-7) circle(.1);
\draw (13,-8) -- (13,-7);
\draw (13,-9) to [out=30,in=30] (13,-7);

\node[below] at (15,-12) {$w_1x_2U$};

\draw[fill] (15,-12) circle(.1);
\draw[fill] (15,-11) circle(.1);
\draw[fill] (15,-10) circle(.1);
\draw (15,-12) -- (15,-11);
\draw (15,-12) to [out=150,in=150] (15,-10);
\draw[fill] (16,-9) circle(.1);
\draw[fill] (16,-8) circle(.1);
\draw (16,-9) -- (16,-8);
\draw (15,-11) to [out=30,in=150] (16,-9);
\draw (15,-10) to [out=30,in=150] (16,-8);
\draw[fill] (17,-9) circle(.1);
\draw (17,-9) -- (16,-8);
\draw[fill] (17,-7) circle(.1);
\draw (17,-9) to [out=30,in=30] (17,-7);
\draw[fill] (17,-6) circle(.1);
\draw (17,-7) -- (17,-6);
\draw[fill] (17,-4) circle(.1);
\draw (17,-6) to [out=150,in=150] (17,-4);

\node[below] at (14,-11) {$x_2^2U$};

\draw[fill] (14,-11) circle(.1);
\draw (14,-11) -- (15,-10);
\draw[fill] (14,-9) circle(.1);
\draw (14,-11) to [out=150,in=150] (14,-9);
\draw[fill] (14,-8) circle(.1);
\draw (14,-9) -- (14,-8);
\draw[fill] (14,-7) circle(.1);
\draw[fill] (14,-6) circle(.1);
\draw (14,-7) -- (14,-6);
\draw (14,-8) to [out=150,in=150] (14,-6);
\draw[fill] (14,-5) circle(.1);
\draw (14,-7) to [out=30,in=30] (14,-5);

\node[below] at (18,-10) {$w_1^2x_3U$};

\draw[fill] (18,-10) circle(.1);
\draw[fill] (18,-9) circle(.1);
\draw (18,-10) -- (18,-9);
\draw[fill] (18,-8) circle(.1);
\draw (18,-10) to [out=150,in=150] (18,-8);
\draw[fill] (19,-7) circle(.1);
\draw (18,-9) to [out=30,in=150] (19,-7);
\draw[fill] (18,-7) circle(.1);
\draw (18,-8) -- (18,-7);
\draw[fill] (19,-6) circle(.1);
\draw (19,-7) -- (19,-6);
\draw (18,-8) to [out=30,in=150] (19,-6);
\draw[fill] (19,-5) circle(.1);
\draw (18,-7) to [out=30,in=150] (19,-5);
\draw[fill] (19,-4) circle(.1);
\draw (19,-5) -- (19,-4);
\draw (19,-6) to [out=30,in=30] (19,-4); 

\node[below] at (20,-10) {$x_2x_3U$};

\draw[fill] (20,-10) circle(.1);
\draw[fill] (20,-9) circle(.1);
\draw (20,-10) -- (20,-9);
\draw[fill] (20,-8) circle(.1);
\draw (20,-10) to [out=150,in=150] (20,-8);
\draw[fill] (21,-7) circle(.1);
\draw (20,-9) to [out=30,in=150] (21,-7);
\draw[fill] (20,-7) circle(.1);
\draw (20,-8) -- (20,-7);
\draw[fill] (21,-6) circle(.1);
\draw (21,-7) -- (21,-6);
\draw (20,-8) to [out=30,in=150] (21,-6);
\draw[fill] (21,-5) circle(.1);
\draw (20,-7) to [out=30,in=150] (21,-5);
\draw[fill] (21,-4) circle(.1);
\draw (21,-5) -- (21,-4);
\draw (21,-6) to [out=30,in=30] (21,-4); 

\node[below] at (22,-12) {$ax_2U$};

\draw[fill] (22,-12) circle(.1);
\draw[fill] (22,-11) circle(.1);
\draw (22,-12) -- (22,-11);
\draw[fill] (22,-10) circle(.1);
\draw (22,-12) to [out=150,in=150] (22,-10);
\draw[fill] (23,-9) circle(.1);
\draw (22,-11) to [out=30,in=150] (23,-9);
\draw[fill] (22,-9) circle(.1);
\draw (22,-10) -- (22,-9);
\draw[fill] (23,-8) circle(.1);
\draw (23,-9) -- (23,-8);
\draw (22,-10) to [out=30,in=150] (23,-8);
\draw[fill] (23,-7) circle(.1);
\draw (22,-9) to [out=30,in=150] (23,-7);
\draw[fill] (23,-6) circle(.1);
\draw (23,-7) -- (23,-6);
\draw (23,-8) to [out=30,in=30] (23,-6); 

\node[below] at (24,-11) {$w_1ax_2U$};

\draw[fill] (24,-11) circle(.1);
\draw[fill] (24,-10) circle(.1);
\draw[fill] (24,-9) circle(.1);
\draw (24,-11) -- (24,-10);
\draw (24,-11) to [out=150,in=150] (24,-9);
\draw[fill] (25,-8) circle(.1);
\draw[fill] (25,-7) circle(.1);
\draw (25,-8) -- (25,-7);
\draw (24,-10) to [out=30,in=150] (25,-8);
\draw (24,-9) to [out=30,in=150] (25,-7);
\draw[fill] (24,-8) circle(.1);
\draw (24,-9) -- (24,-8);
\draw[fill] (24,-7) circle(.1);
\draw[fill] (24,-6) circle(.1);
\draw (24,-8) to [out=30,in=30] (24,-6);
\draw (24,-7) -- (24,-6);
\draw[fill] (24,-5) circle(.1);
\draw (24,-7) to [out=150,in=150] (24,-5);

\node[below] at (26,-10) {$w_1^2ax_2U$};

\draw[fill] (26,-10) circle(.1);
\draw[fill] (26,-9) circle(.1);
\draw (26,-10) -- (26,-9);
\draw[fill] (26,-8) circle(.1);
\draw (26,-10) to [out=150,in=150] (26,-8);
\draw[fill] (27,-7) circle(.1);
\draw (26,-9) to [out=30,in=150] (27,-7);
\draw[fill] (26,-7) circle(.1);
\draw (26,-8) -- (26,-7);
\draw[fill] (27,-6) circle(.1);
\draw (27,-7) -- (27,-6);
\draw (26,-8) to [out=30,in=150] (27,-6);
\draw[fill] (27,-5) circle(.1);
\draw (26,-7) to [out=30,in=150] (27,-5);
\draw[fill] (27,-4) circle(.1);
\draw (27,-5) -- (27,-4);
\draw (27,-6) to [out=30,in=30] (27,-4); 

\node[below] at (28,-11) {$ax_3U$};

\draw[fill] (28,-11) circle(.1);
\draw[fill] (28,-10) circle(.1);
\draw (28,-11) -- (28,-10);
\draw[fill] (28,-9) circle(.1);
\draw (28,-11) to [out=150,in=150] (28,-9);
\draw[fill] (29,-8) circle(.1);
\draw (28,-10) to [out=30,in=150] (29,-8);
\draw[fill] (28,-8) circle(.1);
\draw (28,-9) -- (28,-8);
\draw[fill] (29,-7) circle(.1);
\draw (29,-8) -- (29,-7);
\draw (28,-9) to [out=30,in=150] (29,-7);
\draw[fill] (29,-6) circle(.1);
\draw (28,-8) to [out=30,in=150] (29,-6);
\draw[fill] (29,-5) circle(.1);
\draw (29,-6) -- (29,-5);
\draw (29,-7) to [out=30,in=30] (29,-5); 

\node[below] at (30,-10) {$w_1ax_3U$};

\draw[fill] (30,-10) circle(.1);
\draw[fill] (30,-9) circle(.1);
\draw (30,-10) -- (30,-9);
\draw[fill] (30,-8) circle(.1);
\draw (30,-10) to [out=150,in=150] (30,-8);
\draw[fill] (31,-7) circle(.1);
\draw (30,-9) to [out=30,in=150] (31,-7);
\draw[fill] (30,-7) circle(.1);
\draw (30,-8) -- (30,-7);
\draw[fill] (31,-6) circle(.1);
\draw (31,-7) -- (31,-6);
\draw (30,-8) to [out=30,in=150] (31,-6);
\draw[fill] (31,-5) circle(.1);
\draw (30,-7) to [out=30,in=150] (31,-5);
\draw[fill] (31,-4) circle(.1);
\draw (31,-5) -- (31,-4);
\draw (31,-6) to [out=30,in=30] (31,-4); 

\node[below] at (32,-10) {$a^3x_2U$};

\draw[fill] (32,-10) circle(.1);
\draw[fill] (32,-9) circle(.1);
\draw (32,-10) -- (32,-9);
\draw[fill] (32,-8) circle(.1);
\draw (32,-10) to [out=150,in=150] (32,-8);
\draw[fill] (33,-7) circle(.1);
\draw (32,-9) to [out=30,in=150] (33,-7);
\draw[fill] (32,-7) circle(.1);
\draw (32,-8) -- (32,-7);
\draw[fill] (33,-6) circle(.1);
\draw (33,-7) -- (33,-6);
\draw (32,-8) to [out=30,in=150] (33,-6);
\draw[fill] (33,-5) circle(.1);
\draw (32,-7) to [out=30,in=150] (33,-5);
\draw[fill] (33,-4) circle(.1);
\draw (33,-5) -- (33,-4);
\draw (33,-6) to [out=30,in=30] (33,-4);

\end{tikzpicture}
\end{center}
\caption{The $\A_2(1)$-module structure of $\H^{*-1}(MT\tO(1),\Z_2)\otimes\H^*(\B \Z_2\times \B ^2\Z_2,\Z_2)$}
\label{fig:H^{*-1}(MTO(1),Z_2)otimesH^*(BZ_2timesB^2Z_2,Z_2)}
\end{figure}

\begin{figure}[!h]
\begin{center}
\begin{tikzpicture}
\node at (0,-1) {0};
\node at (1,-1) {1};
\node at (2,-1) {2};
\node at (3,-1) {3};
\node at (4,-1) {4};
\node at (5,-1) {5};
\node at (6,-1) {$t-s$};
\node at (-1,0) {0};
\node at (-1,1) {1};
\node at (-1,2) {2};
\node at (-1,3) {3};
\node at (-1,4) {4};
\node at (-1,5) {5};
\node at (-1,6) {$s$};

\draw[->] (-0.5,-0.5) -- (-0.5,6);
\draw[->] (-0.5,-0.5) -- (6,-0.5);

\draw[fill] (0,0) circle(0.05);
\draw[fill] (1,0) circle(0.05);
\draw (2,1) -- (4,3);
\draw (4,3) -- (4,0);
\draw (2,0) -- (3.9,2);
\draw (3.9,2) -- (3.9,0);

\draw[fill] (1.9,0) circle(0.05);
\draw[fill] (2.8,0) circle(0.05);
\draw[fill] (2.9,0) circle(0.05);
\draw (3,0) -- (3.8,1);
\draw (3.8,1) -- (3.8,0);
\draw[fill] (4.1,0) circle(0.05);
\draw[fill] (4.2,0) circle(0.05);

\draw[fill] (4.7,0) circle(0.05);
\draw[fill] (4.8,0) circle(0.05);
\draw[fill] (4.9,0) circle(0.05);
\draw[fill] (5,0) circle(0.05);
\draw[fill] (5.1,0) circle(0.05);
\draw[fill] (5.2,0) circle(0.05);
\draw[fill] (5.3,0) circle(0.05);

\end{tikzpicture}
\end{center}
\caption{$\Omega_*^{\Pin^+}(\B \Z_2\times \B ^2\Z_2)$}
\label{fig:Omega_*^{Pin^+}(BZ_2timesB^2Z_2)}
\end{figure}

\begin{theorem}
\begin{table}[!h]
\centering
\begin{tabular}{c c}
\hline
$i$ & $\Omega^{\Pin^+}_i(\B \Z_2\times \B ^2\Z_2)$\\
\hline
0& $\Z_2$\\
1& $\Z_2$\\
2& $\Z_2^3$\\
3 & $\Z_2^5$\\
4 & $\Z_2^2\times\Z_4\times\Z_8\times\Z_{16}$\\ 
5 & $\Z_2^7$\\
\hline
\end{tabular}
\end{table}
\end{theorem}

The bordism invariants of $\Omega^{\Pin^+}_2(\B \Z_2\times \B ^2\Z_2)$ are $w_1a=a^2,x_2,w_1\tilde{\eta}$.

The bordism invariants of $\Omega^{\Pin^+}_3(\B \Z_2\times \B ^2\Z_2)$ are $a^3,w_1x_2=x_3,ax_2,w_1a\tilde{\eta},w_1\text{Arf}$.

The bordism invariants of $\Omega^{\Pin^+}_4(\B \Z_2\times \B ^2\Z_2)$ are $ax_3,w_1ax_2{(=a^2x_2+ax_3)},q_s(x_2),w_1a(\text{ABK}),\eta$.

The bordism invariants of $\Omega^{\Pin^+}_5(\B \Z_2\times \B ^2\Z_2)$ are $$w_1^4a,a^5(=w_1^2a^3),w_1^2x_3{(=x_5)},x_2x_3,w_1^2ax_2{(=ax_2^2+a^2x_3)},w_1ax_3{(=a^2x_3)},a^3x_2.$$

\begin{theorem}
\begin{table}[!h]
\centering
\begin{tabular}{c c}
\hline
$i$ & $\TP_i(\Pin^+\times\Z_2\times \B \Z_2)$\\
\hline
0& $\Z_2$\\
1& $\Z_2$\\
2& $\Z_2^3$\\
3 & $\Z_2^5$\\
4 & $\Z_2^2\times\Z_4\times\Z_8\times\Z_{16}$\\ 
5 & $\Z_2^7$\\
\hline
\end{tabular}
\end{table}
\end{theorem}

The 2d topological terms are $w_1a=a^2,x_2,w_1\tilde{\eta}$.

The 3d topological terms are $a^3,w_1x_2=x_3,ax_2,w_1a\tilde{\eta},w_1\text{Arf}$.

The 4d topological terms are $ax_3,w_1ax_2{(=a^2x_2+ax_3)},q_s(x_2),w_1a(\text{ABK}),\eta$.

The 5d topological terms are $$w_1^4a,a^5(=w_1^2a^3),w_1^2x_3{(=x_5)},x_2x_3,w_1^2ax_2{(=ax_2^2+a^2x_3)},w_1ax_3{(=a^2x_3)},a^3x_2.$$

\subsubsection{$\Omega^{\Pin^-}_d(\B \Z_2\times \B ^2\Z_2)$}

Since the computation involves no odd torsion, we can use the Adams spectral sequence 
\bea
&&E_2^{s,t}=\Ext_{\A_2}^{s,t}(\H^*(M\Pin^+\wedge (\B \Z_2\times \B ^2\Z_2)_+,\Z_2),\Z_2)\notag\\
&\Rightarrow&\pi_{t-s}(M\Pin^+\wedge (\B \Z_2\times \B ^2\Z_2)_+)_2^{\wedge}=\Omega^{\Pin^-}_{t-s}(\B \Z_2\times \B ^2\Z_2).
\eea
$M\Pin^+=MT\Pin^-\sim M\Spin\wedge S^{-1}\wedge M\tO(1)$.

For $t-s<8$,
\bea
&&\Ext_{\A_2(1)}^{s,t}(\H^{*+1}(M\tO(1),\Z_2)\otimes\H^*(\B \Z_2\times \B ^2\Z_2,\Z_2),\Z_2)\Rightarrow\Omega^{\Pin^-}_{t-s}(\B \Z_2\times \B ^2\Z_2).
\eea

The $\A_2(1)$-module structure of $\H^{*+1}(M\tO(1),\Z_2)\otimes\H^*(\B \Z_2\times \B ^2\Z_2,\Z_2)$ and the $E_2$ page are shown in Figure \ref{fig:H^{*+1}(MO(1),Z_2)otimesH^*(BZ_2timesB^2Z_2,Z_2)}, \ref{fig:Omega_*^{Pin^-}(BZ_2timesB^2Z_2)}.

\begin{figure}[!h]
\begin{center}
\begin{tikzpicture}[scale=0.5]

\node[below] at (0,0) {$U$};

\draw[fill] (0,0) circle(.1);
\draw[fill] (0,1) circle(.1);
\draw (0,0) -- (0,1);
\draw[fill] (0,2) circle(.1);
\draw[fill] (0,3) circle(.1);
\draw (0,2) -- (0,3);
\draw (0,1) to [out=150,in=150] (0,3);
\draw[fill] (0,4) circle(.1);
\draw (0,2) to [out=30,in=30] (0,4);
\draw[fill] (0,5) circle(.1);
\draw (0,4) -- (0,5);

\node[below] at (4,0) {$1$};

\draw[fill] (4,0) circle(.1);

\node[right] at (4,1) {$a$};

\draw[fill] (4,1) circle(.1);
\draw[fill] (4,2) circle(.1);
\draw (4,1) -- (4,2);
\draw[fill] (4,3) circle(.1);
\draw[fill] (4,4) circle(.1);
\draw (4,3) -- (4,4);
\draw (4,2) to [out=150,in=150] (4,4);
\draw[fill] (4,5) circle(.1);
\draw (4,3) to [out=30,in=30] (4,5);
\draw[fill] (4,6) circle(.1);
\draw (4,5) -- (4,6);

\node[below] at (6,2) {$x_2$};

\draw[fill] (6,2) circle(.1);
\draw[fill] (6,3) circle(.1);
\draw[fill] (6,4) circle(.1);
\draw (6,2) -- (6,3);
\draw (6,2) to [out=150,in=150] (6,4);
\draw[fill] (7,5) circle(.1);
\draw[fill] (7,6) circle(.1);
\draw (7,5) -- (7,6);
\draw (6,3) to [out=30,in=150] (7,5);
\draw (6,4) to [out=30,in=150] (7,6);
\draw[fill] (8,5) circle(.1);

\node[below] at (8,5) {$x_2x_3$};

\draw (8,5) -- (7,6);
\draw[fill] (8,7) circle(.1);
\draw (8,5) to [out=30,in=30] (8,7);
\draw[fill] (8,8) circle(.1);
\draw (8,7) -- (8,8);
\draw[fill] (8,10) circle(.1);
\draw (8,8) to [out=150,in=150] (8,10);

\node[below] at (10,3) {$ax_2$};

\draw[fill] (10,3) circle(.1);
\draw[fill] (10,4) circle(.1);
\draw[fill] (10,5) circle(.1);
\draw (10,3) -- (10,4);
\draw (10,3) to [out=150,in=150] (10,5);
\draw[fill] (11,6) circle(.1);
\draw[fill] (11,7) circle(.1);
\draw (11,6) -- (11,7);
\draw (10,4) to [out=30,in=150] (11,6);
\draw (10,5) to [out=30,in=150] (11,7);
\draw[fill] (10,6) circle(.1);
\draw (10,5) -- (10,6);
\draw[fill] (10,7) circle(.1);
\draw[fill] (10,8) circle(.1);
\draw (10,6) to [out=30,in=30] (10,8);
\draw (10,7) -- (10,8);
\draw[fill] (10,9) circle(.1);
\draw (10,7) to [out=150,in=150] (10,9);
 
\node[below] at (12,4) {$ax_3$}; 
 
\draw[fill] (12,4) circle(.1);
\draw[fill] (12,5) circle(.1);
\draw (12,4) -- (12,5);
\draw[fill] (12,6) circle(.1);
\draw (12,4) to [out=150,in=150] (12,6);
\draw[fill] (13,7) circle(.1);
\draw (12,5) to [out=30,in=150] (13,7);
\draw[fill] (12,7) circle(.1);
\draw (12,6) -- (12,7);
\draw[fill] (13,8) circle(.1);
\draw (13,7) -- (13,8);
\draw (12,6) to [out=30,in=150] (13,8);
\draw[fill] (13,9) circle(.1);
\draw (12,7) to [out=30,in=150] (13,9);
\draw[fill] (13,10) circle(.1);
\draw (13,9) -- (13,10);
\draw (13,8) to [out=30,in=30] (13,10); 

\node[below] at (14,5) {$a^3x_2$};
 
\draw[fill] (14,5) circle(.1);
\draw[fill] (14,6) circle(.1);
\draw (14,5) -- (14,6);
\draw[fill] (14,7) circle(.1);
\draw (14,5) to [out=150,in=150] (14,7);
\draw[fill] (15,8) circle(.1);
\draw (14,6) to [out=30,in=150] (15,8);
\draw[fill] (14,8) circle(.1);
\draw (14,7) -- (14,8);
\draw[fill] (15,9) circle(.1);
\draw (15,8) -- (15,9);
\draw (14,7) to [out=30,in=150] (15,9);
\draw[fill] (15,10) circle(.1);
\draw (14,8) to [out=30,in=150] (15,10);
\draw[fill] (15,11) circle(.1);
\draw (15,10) -- (15,11);
\draw (15,9) to [out=30,in=30] (15,11);

\node at (2,5) {$\bigotimes$};
\node at (-2,-10) {$=$};

\node[below] at (0,-15) {$U$};
 
\draw[fill] (0,-15) circle(.1);
\draw[fill] (0,-14) circle(.1);
\draw (0,-15) -- (0,-14);
\draw[fill] (0,-13) circle(.1);
\draw[fill] (0,-12) circle(.1);
\draw (0,-13) -- (0,-12);
\draw (0,-14) to [out=150,in=150] (0,-12);
\draw[fill] (0,-11) circle(.1);
\draw (0,-13) to [out=30,in=30] (0,-11);
\draw[fill] (0,-10) circle(.1);
\draw (0,-11) -- (0,-10);

\node[below] at (2,-14) {$aU$};
  
 \draw[fill] (2,-14) circle(.1);
\draw[fill] (2,-13) circle(.1);
\draw (2,-14) -- (2,-13);
\draw[fill] (2,-11) circle(.1);
\draw (2,-13) to [out=150,in=150] (2,-11);
\draw[fill] (2,-10) circle(.1);
\draw (2,-11) -- (2,-10);
\draw[fill] (3,-13) circle(.1);

\node[below] at (3,-13) {$a^2U$};

\draw[fill] (3,-12) circle(.1);
\draw (3,-13) -- (3,-12);
\draw (2,-14) to [out=30,in=150] (3,-12);
\draw (3,-12) to [out=150,in=30] (2,-10);
\draw[fill] (3,-11) circle(.1);
\draw (3,-13) to [out=30,in=30] (3,-11);
\draw[fill] (3,-10) circle(.1);
\draw (3,-11) -- (3,-10);
\draw[fill] (3,-9) circle(.1);
\draw[fill] (3,-8) circle(.1);
\draw (3,-9) -- (3,-8);
\draw (3,-10) to [out=150,in=150] (3,-8);
\draw[fill] (3,-7) circle(.1);
\draw (3,-9) to [out=30,in=30] (3,-7); 

\node[below] at (5,-12) {$a^3U$};

\draw[fill] (5,-12) circle(.1);
\draw[fill] (5,-11) circle(.1);
\draw (5,-12) -- (5,-11);
\draw[fill] (5,-10) circle(.1);
\draw (5,-12) to [out=150,in=150] (5,-10);
\draw[fill] (6,-9) circle(.1);
\draw (5,-11) to [out=30,in=150] (6,-9);
\draw[fill] (5,-9) circle(.1);
\draw (5,-10) -- (5,-9);
\draw[fill] (6,-8) circle(.1);
\draw (6,-9) -- (6,-8);
\draw (5,-10) to [out=30,in=150] (6,-8);
\draw[fill] (6,-7) circle(.1);
\draw (5,-9) to [out=30,in=150] (6,-7);
\draw[fill] (6,-6) circle(.1);
\draw (6,-7) -- (6,-6);
\draw (6,-8) to [out=30,in=30] (6,-6); 

\node[below] at (7,-12) {$w_1^2aU$};

\draw[fill] (7,-12) circle(.1);
\draw[fill] (7,-11) circle(.1);
\draw (7,-12) -- (7,-11);
\draw[fill] (7,-10) circle(.1);
\draw (7,-12) to [out=150,in=150] (7,-10);
\draw[fill] (8,-9) circle(.1);
\draw (7,-11) to [out=30,in=150] (8,-9);
\draw[fill] (7,-9) circle(.1);
\draw (7,-10) -- (7,-9);
\draw[fill] (8,-8) circle(.1);
\draw (8,-9) -- (8,-8);
\draw (7,-10) to [out=30,in=150] (8,-8);
\draw[fill] (8,-7) circle(.1);
\draw (7,-9) to [out=30,in=150] (8,-7);
\draw[fill] (8,-6) circle(.1);
\draw (8,-7) -- (8,-6);
\draw (8,-8) to [out=30,in=30] (8,-6); 

\node[below] at (9,-10) {$w_1^2a^3U$};

\draw[fill] (9,-10) circle(.1);
\draw[fill] (9,-9) circle(.1);
\draw (9,-10) -- (9,-9);
\draw[fill] (9,-8) circle(.1);
\draw (9,-10) to [out=150,in=150] (9,-8);
\draw[fill] (10,-7) circle(.1);
\draw (9,-9) to [out=30,in=150] (10,-7);
\draw[fill] (9,-7) circle(.1);
\draw (9,-8) -- (9,-7);
\draw[fill] (10,-6) circle(.1);
\draw (10,-7) -- (10,-6);
\draw (9,-8) to [out=30,in=150] (10,-6);
\draw[fill] (10,-5) circle(.1);
\draw (9,-7) to [out=30,in=150] (10,-5);
\draw[fill] (10,-4) circle(.1);
\draw (10,-5) -- (10,-4);
\draw (10,-6) to [out=30,in=30] (10,-4); 

\node[below] at (11,-13) {$x_2U$};

\draw[fill] (11,-13) circle(.1);
\draw[fill] (11,-12) circle(.1);
\draw[fill] (11,-11) circle(.1);
\draw (11,-13) -- (11,-12);
\draw (11,-13) to [out=150,in=150] (11,-11);
\draw[fill] (12,-10) circle(.1);
\draw[fill] (12,-9) circle(.1);
\draw (12,-10) -- (12,-9);
\draw (11,-12) to [out=30,in=150] (12,-10);
\draw (11,-11) to [out=30,in=150] (12,-9);
\draw[fill] (11,-10) circle(.1);
\draw (11,-11) -- (11,-10);
\draw[fill] (11,-9) circle(.1);
\draw[fill] (11,-8) circle(.1);
\draw (11,-10) to [out=30,in=30] (11,-8);
\draw (11,-9) -- (11,-8);
\draw[fill] (11,-7) circle(.1);
\draw (11,-9) to [out=150,in=150] (11,-7);

\node[below] at (13,-12) {$x_3U$};

\draw[fill] (13,-12) circle(.1);
\draw[fill] (13,-11) circle(.1);
\draw (13,-12) -- (13,-11);
\draw[fill] (13,-10) circle(.1);
\draw (13,-12) to [out=150,in=150] (13,-10);
\draw[fill] (14,-9) circle(.1);
\draw (13,-11) to [out=30,in=150] (14,-9);
\draw[fill] (13,-9) circle(.1);
\draw (13,-10) -- (13,-9);
\draw[fill] (14,-8) circle(.1);
\draw (14,-9) -- (14,-8);
\draw (13,-10) to [out=30,in=150] (14,-8);
\draw[fill] (14,-7) circle(.1);
\draw (13,-9) to [out=30,in=150] (14,-7);
\draw[fill] (14,-6) circle(.1);
\draw (14,-7) -- (14,-6);
\draw (14,-8) to [out=30,in=30] (14,-6); 

\node[below] at (15,-11) {$w_1^2x_2U$};

\draw[fill] (15,-11) circle(.1);
\draw[fill] (15,-10) circle(.1);
\draw (15,-11) -- (15,-10);
\draw[fill] (15,-9) circle(.1);
\draw (15,-11) to [out=150,in=150] (15,-9);
\draw[fill] (16,-8) circle(.1);
\draw (15,-10) to [out=30,in=150] (16,-8);
\draw[fill] (15,-8) circle(.1);
\draw (15,-9) -- (15,-8);
\draw[fill] (16,-7) circle(.1);
\draw (16,-8) -- (16,-7);
\draw (15,-9) to [out=30,in=150] (16,-7);
\draw[fill] (16,-6) circle(.1);
\draw (15,-8) to [out=30,in=150] (16,-6);
\draw[fill] (16,-5) circle(.1);
\draw (16,-6) -- (16,-5);
\draw (16,-7) to [out=30,in=30] (16,-5); 

\node[below] at (17,-10) {$x_2x_3U$};

\draw[fill] (17,-10) circle(.1);
\draw[fill] (17,-9) circle(.1);
\draw (17,-10) -- (17,-9);
\draw[fill] (17,-8) circle(.1);
\draw (17,-10) to [out=150,in=150] (17,-8);
\draw[fill] (18,-7) circle(.1);
\draw (17,-9) to [out=30,in=150] (18,-7);
\draw[fill] (17,-7) circle(.1);
\draw (17,-8) -- (17,-7);
\draw[fill] (18,-6) circle(.1);
\draw (18,-7) -- (18,-6);
\draw (17,-8) to [out=30,in=150] (18,-6);
\draw[fill] (18,-5) circle(.1);
\draw (17,-7) to [out=30,in=150] (18,-5);
\draw[fill] (18,-4) circle(.1);
\draw (18,-5) -- (18,-4);
\draw (18,-6) to [out=30,in=30] (18,-4); 

\node[below] at (19,-12) {$ax_2U$};

\draw[fill] (19,-12) circle(.1);
\draw[fill] (19,-11) circle(.1);
\draw (19,-12) -- (19,-11);
\draw[fill] (19,-10) circle(.1);
\draw (19,-12) to [out=150,in=150] (19,-10);
\draw[fill] (20,-9) circle(.1);
\draw (19,-11) to [out=30,in=150] (20,-9);
\draw[fill] (19,-9) circle(.1);
\draw (19,-10) -- (19,-9);
\draw[fill] (20,-8) circle(.1);
\draw (20,-9) -- (20,-8);
\draw (19,-10) to [out=30,in=150] (20,-8);
\draw[fill] (20,-7) circle(.1);
\draw (19,-9) to [out=30,in=150] (20,-7);
\draw[fill] (20,-6) circle(.1);
\draw (20,-7) -- (20,-6);
\draw (20,-8) to [out=30,in=30] (20,-6); 

\node[below] at (21,-11) {$w_1ax_2U$};

\draw[fill] (21,-11) circle(.1);
\draw[fill] (21,-10) circle(.1);
\draw (21,-11) -- (21,-10);
\draw[fill] (21,-9) circle(.1);
\draw (21,-11) to [out=150,in=150] (21,-9);
\draw[fill] (22,-8) circle(.1);
\draw (21,-10) to [out=30,in=150] (22,-8);
\draw[fill] (21,-8) circle(.1);
\draw (21,-9) -- (21,-8);
\draw[fill] (22,-7) circle(.1);
\draw (22,-8) -- (22,-7);
\draw (21,-9) to [out=30,in=150] (22,-7);
\draw[fill] (22,-6) circle(.1);
\draw (21,-8) to [out=30,in=150] (22,-6);
\draw[fill] (22,-5) circle(.1);
\draw (22,-6) -- (22,-5);
\draw (22,-7) to [out=30,in=30] (22,-5); 

\node[below] at (23,-10) {$w_1^2ax_2U$};

\draw[fill] (23,-10) circle(.1);
\draw[fill] (23,-9) circle(.1);
\draw (23,-10) -- (23,-9);
\draw[fill] (23,-8) circle(.1);
\draw (23,-10) to [out=150,in=150] (23,-8);
\draw[fill] (24,-7) circle(.1);
\draw (23,-9) to [out=30,in=150] (24,-7);
\draw[fill] (23,-7) circle(.1);
\draw (23,-8) -- (23,-7);
\draw[fill] (24,-6) circle(.1);
\draw (24,-7) -- (24,-6);
\draw (23,-8) to [out=30,in=150] (24,-6);
\draw[fill] (24,-5) circle(.1);
\draw (23,-7) to [out=30,in=150] (24,-5);
\draw[fill] (24,-4) circle(.1);
\draw (24,-5) -- (24,-4);
\draw (24,-6) to [out=30,in=30] (24,-4);

\node[below] at (25,-11) {$ax_3U$};

\draw[fill] (25,-11) circle(.1);
\draw[fill] (25,-10) circle(.1);
\draw (25,-11) -- (25,-10);
\draw[fill] (25,-9) circle(.1);
\draw (25,-11) to [out=150,in=150] (25,-9);
\draw[fill] (26,-8) circle(.1);
\draw (25,-10) to [out=30,in=150] (26,-8);
\draw[fill] (25,-8) circle(.1);
\draw (25,-9) -- (25,-8);
\draw[fill] (26,-7) circle(.1);
\draw (26,-8) -- (26,-7);
\draw (25,-9) to [out=30,in=150] (26,-7);
\draw[fill] (26,-6) circle(.1);
\draw (25,-8) to [out=30,in=150] (26,-6);
\draw[fill] (26,-5) circle(.1);
\draw (26,-6) -- (26,-5);
\draw (26,-7) to [out=30,in=30] (26,-5); 

\node[below] at (27,-10) {$w_1ax_3U$};

\draw[fill] (27,-10) circle(.1);
\draw[fill] (27,-9) circle(.1);
\draw (27,-10) -- (27,-9);
\draw[fill] (27,-8) circle(.1);
\draw (27,-10) to [out=150,in=150] (27,-8);
\draw[fill] (28,-7) circle(.1);
\draw (27,-9) to [out=30,in=150] (28,-7);
\draw[fill] (27,-7) circle(.1);
\draw (27,-8) -- (27,-7);
\draw[fill] (28,-6) circle(.1);
\draw (28,-7) -- (28,-6);
\draw (27,-8) to [out=30,in=150] (28,-6);
\draw[fill] (28,-5) circle(.1);
\draw (27,-7) to [out=30,in=150] (28,-5);
\draw[fill] (28,-4) circle(.1);
\draw (28,-5) -- (28,-4);
\draw (28,-6) to [out=30,in=30] (28,-4); 

\node[below] at (29,-10) {$a^3x_2U$};

\draw[fill] (29,-10) circle(.1);
\draw[fill] (29,-9) circle(.1);
\draw (29,-10) -- (29,-9);
\draw[fill] (29,-8) circle(.1);
\draw (29,-10) to [out=150,in=150] (29,-8);
\draw[fill] (30,-7) circle(.1);
\draw (29,-9) to [out=30,in=150] (30,-7);
\draw[fill] (29,-7) circle(.1);
\draw (29,-8) -- (29,-7);
\draw[fill] (30,-6) circle(.1);
\draw (30,-7) -- (30,-6);
\draw (29,-8) to [out=30,in=150] (30,-6);
\draw[fill] (30,-5) circle(.1);
\draw (29,-7) to [out=30,in=150] (30,-5);
\draw[fill] (30,-4) circle(.1);
\draw (30,-5) -- (30,-4);
\draw (30,-6) to [out=30,in=30] (30,-4);

\end{tikzpicture}
\end{center}
\caption{The $\A_2(1)$-module structure of $\H^{*+1}(M\tO(1),\Z_2)\otimes\H^*(\B \Z_2\times \B ^2\Z_2,\Z_2)$}
\label{fig:H^{*+1}(MO(1),Z_2)otimesH^*(BZ_2timesB^2Z_2,Z_2)}
\end{figure}

\begin{figure}[!h]
\begin{center}
\begin{tikzpicture}
\node at (0,-1) {0};
\node at (1,-1) {1};
\node at (2,-1) {2};
\node at (3,-1) {3};
\node at (4,-1) {4};
\node at (5,-1) {5};
\node at (6,-1) {$t-s$};
\node at (-1,0) {0};
\node at (-1,1) {1};
\node at (-1,2) {2};
\node at (-1,3) {3};
\node at (-1,4) {4};
\node at (-1,5) {5};
\node at (-1,6) {$s$};

\draw[->] (-0.5,-0.5) -- (-0.5,6);
\draw[->] (-0.5,-0.5) -- (6,-0.5);

\draw (0,0) -- (2,2);
\draw (2,2) -- (2,0);
\draw (1,0) -- (1.9,1);
\draw (1.9,1) -- (1.9,0);
\draw[fill] (2.1,0) circle(0.05);
\draw[fill] (2.9,0) circle(0.05);
\draw[fill] (3,0) circle(0.05);
\draw[fill] (3.1,0) circle(0.05);
\draw[fill] (3.2,0) circle(0.05);
\draw[fill] (3.9,0) circle(0.05);
\draw[fill] (4,0) circle(0.05);
\draw[fill] (4.1,0) circle(0.05);
\draw[fill] (4.8,0) circle(0.05);
\draw[fill] (4.9,0) circle(0.05);
\draw[fill] (5,0) circle(0.05);
\draw[fill] (5.1,0) circle(0.05);
\draw[fill] (5.2,0) circle(0.05);

\end{tikzpicture}
\end{center}
\caption{$\Omega_*^{\Pin^-}(\B \Z_2\times \B ^2\Z_2)$}
\label{fig:Omega_*^{Pin^-}(BZ_2timesB^2Z_2)}
\end{figure}

\begin{theorem}
\begin{table}[!h]
\centering
\begin{tabular}{c c}
\hline
$i$ & $\Omega^{\Pin^-}_i(\B \Z_2\times \B ^2\Z_2)$\\
\hline
0& $\Z_2$\\
1& $\Z_2^2$\\
2& $\Z_2\times\Z_4\times\Z_8$\\
3 & $\Z_2^4$\\
4 & $\Z_2^3$\\ 
5 & $\Z_2^5$\\
\hline
\end{tabular}
\end{table}
\end{theorem}

The bordism invariants of $\Omega^{\Pin^-}_2(\B \Z_2\times \B ^2\Z_2)$ are $x_2,q(a),\text{ABK}$. ($q(a)$ is explained in the footnotes of Table \ref{2d bordism groups}.)

The bordism invariants of $\Omega^{\Pin^-}_3(\B \Z_2\times \B ^2\Z_2)$ are $a^3,w_1^2a,x_3=w_1x_2,ax_2$.

The bordism invariants of $\Omega^{\Pin^-}_4(\B \Z_2\times \B ^2\Z_2)$ are $w_1^2x_2,w_1ax_2{(=a^2x_2+ax_3)},ax_3$.

The bordism invariants of $\Omega^{\Pin^-}_5(\B \Z_2\times \B ^2\Z_2)$ are $w_1^2a^3,x_2x_3,w_1^2ax_2,w_1ax_3{(=a^2x_3)},a^3x_2$.

\begin{theorem}
\begin{table}[!h]
\centering
\begin{tabular}{c c}
\hline
$i$ & $\TP_i(\Pin^-\times\Z_2\times \B \Z_2)$\\
\hline
0& $\Z_2$\\
1& $\Z_2^2$\\
2& $\Z_2\times\Z_4\times\Z_8$\\
3 & $\Z_2^4$\\
4 & $\Z_2^3$\\ 
5 & $\Z_2^5$\\
\hline
\end{tabular}
\end{table}
\end{theorem}

The 2d topological terms are $x_2,q(a),\text{ABK}$. 

The 3d topological terms are $a^3,w_1^2a,x_3=w_1x_2,ax_2$.

The 4d topological terms are $w_1^2x_2,w_1ax_2{(=a^2x_2+ax_3)},ax_3$.

The 5d topological terms are $w_1^2a^3,x_2x_3,w_1^2ax_2,w_1ax_3{(=a^2x_3)},a^3x_2$.

\subsubsection{$\Omega_d^{\tO}(\B \Z_3\times \B ^2\Z_3)$}
\bea
\Ext_{\A_2}^{s,t}(\H^*(M\tO\wedge (\B \Z_3\times \B ^2\Z_3)_+,\Z_2),\Z_2)\Rightarrow\Omega_{t-s}^{\tO}(\B \Z_3\times \B ^2\Z_3)_2^{\wedge}.
\eea
\bea
\Ext_{\A_3}^{s,t}(\H^*(M\tO\wedge (\B \Z_3\times \B ^2\Z_3)_+,\Z_3),\Z_3)\Rightarrow\Omega_{t-s}^{\tO}(\B \Z_3\times \B ^2\Z_3)_3^{\wedge}.
\eea

Since $\H^*(M\tO,\Z_3)=0$, we have $\Omega_d^{\tO}(\B \Z_3\times \B ^2\Z_3)_3^{\wedge}=0$.

Since $\H^*(\B \Z_3\times \B ^2\Z_3,\Z_2)=\Z_2$, we have $\Omega_d^{\tO}(\B \Z_3\times \B ^2\Z_3)_2^{\wedge}=\Omega_d^{\tO}$.

Hence $\Omega_d^{\tO}(\B \Z_3\times \B ^2\Z_3)=\Omega_d^{\tO}$.

\begin{theorem}
\begin{table}[!h]
\centering
\begin{tabular}{c c}
\hline
$i$ & $\Omega^{\tO}_i(\B \Z_3\times \B ^2\Z_3)$\\
\hline
0& $\Z_2$\\
1& $0$\\
2& $\Z_2$\\
3 & $0$\\
4 & $\Z_2^2$\\ 
5 & $\Z_2$\\
\hline
\end{tabular}
\end{table}
\end{theorem}

The bordism invariant of $\Omega^{\tO}_2(\B \Z_3\times \B ^2\Z_3)$ is $w_1^2$.

The bordism invariants of $\Omega^{\tO}_4(\B \Z_3\times \B ^2\Z_3)$ are $w_1^4,w_2^2$.

The bordism invariant of $\Omega^{\tO}_5(\B \Z_3\times \B ^2\Z_3)$ is $w_2w_3$.

\begin{theorem}
\begin{table}[!h]
\centering
\begin{tabular}{c c}
\hline
$i$ & $\TP_i(\tO\times\Z_3\times \B \Z_3)$\\
\hline
0& $\Z_2$\\
1& $0$\\
2& $\Z_2$\\
3 & $0$\\
4 & $\Z_2^2$\\ 
5 & $\Z_2$\\
\hline
\end{tabular}
\end{table}
\end{theorem}

The 2d topological term is $w_1^2$.

The 4d topological terms are $w_1^4,w_2^2$.

The 5d topological term is $w_2w_3$.

\subsubsection{$\Omega_d^{\SO}(\B \Z_3\times \B ^2\Z_3)$}
\bea
\Ext_{\A_2}^{s,t}(\H^*(M\SO\wedge (\B \Z_3\times \B ^2\Z_3)_+,\Z_2),\Z_2)\Rightarrow\Omega_{t-s}^{\SO}(\B \Z_3\times \B ^2\Z_3)_2^{\wedge}.
\eea
Since $\H^*(\B \Z_3\times \B ^2\Z_3,\Z_2)=\Z_2$, we have $\Omega_d^{\SO}(\B \Z_3\times \B ^2\Z_3)_2^{\wedge}=\Omega_d^{\SO}$.
\bea
\Ext_{\A_3}^{s,t}(\H^*(M\SO\wedge (\B \Z_3\times \B ^2\Z_3)_+,\Z_3),\Z_3)\Rightarrow\Omega_{t-s}^{\SO}(\B \Z_3\times \B ^2\Z_3)_3^{\wedge}.
\eea

\bea
\H^*(\B \Z_3\times \B ^2\Z_3,\Z_3)=\Z_3[b',x_2',x_8',\dots]\otimes\Lambda_{\Z_3}(a',x_3',x_7',\dots)
\eea

$\beta_{(3,3)} a'=b'$, $\beta_{(3,3)} x_2'=x_3'$, $\beta_{(3,3)} x_2'^2=2x_2'x_3'$.

The $E_2$ page is shown in Figure \ref{fig:Omega_*^{SO}(BZ_3timesB^2Z_3)_3}.

\begin{figure}[!h]
\begin{center}
\begin{tikzpicture}
\node at (0,-1) {0};
\node at (1,-1) {1};
\node at (2,-1) {2};
\node at (3,-1) {3};
\node at (4,-1) {4};
\node at (5,-1) {5};
\node at (6,-1) {$t-s$};
\node at (-1,0) {0};
\node at (-1,1) {1};
\node at (-1,2) {2};
\node at (-1,3) {3};
\node at (-1,4) {4};
\node at (-1,5) {5};
\node at (-1,6) {$s$};

\draw[->] (-0.5,-0.5) -- (-0.5,6);
\draw[->] (-0.5,-0.5) -- (6,-0.5);

\draw (0,0) -- (0,5);
\draw (4,1) -- (4,5);
\draw[fill] (1,0) circle(0.05);
\draw[fill] (2,0) circle(0.05);
\draw[fill] (2.9,0) circle(0.05);
\draw[fill] (3.1,0) circle(0.05);
\draw[fill] (3.9,0) circle(0.05);
\draw[fill] (4.1,0) circle(0.05);

\draw (5,0) -- (5,1);
\draw[fill] (4.9,0) circle(0.05);
\draw[fill] (5.1,0) circle(0.05);
\end{tikzpicture}
\end{center}
\caption{$\Omega_*^{\SO}(\B \Z_3\times \B ^2\Z_3)_3^{\wedge}$}
\label{fig:Omega_*^{SO}(BZ_3timesB^2Z_3)_3}
\end{figure}

Hence we have the following 

\begin{theorem}
\begin{table}[!h]
\centering
\begin{tabular}{c c}
\hline
$i$ & $\Omega^{\SO}_i(\B \Z_3\times \B ^2\Z_3)$\\
\hline
0& $\Z$\\
1& $\Z_3$\\
2& $\Z_3$\\
3 & $\Z_3^2$\\
4 & $\Z\times\Z_3^2$\\ 
5 & $\Z_2\times\Z_3^2\times\Z_9$\\
\hline
\end{tabular}
\end{table}
\end{theorem}

The bordism invariant of $\Omega^{\SO}_{2}(\B \Z_3\times \B ^2\Z_3)$ is $x_2'$.

The bordism invariants of $\Omega^{\SO}_{3}(\B \Z_3\times \B ^2\Z_3)$ are $a'b',a'x_2'$.

The bordism invariants of $\Omega^{\SO}_{4}(\B \Z_3\times \B ^2\Z_3)$ are $\sigma$, $a'x_3'(=b'x_2')$ and $x_2'^2$.

The bordism invariants of $\Omega^{\SO}_5(\B \Z_3\times \B ^2\Z_3)$ are $w_2w_3,a'b'x_2',a'x_2'^2,\mathfrak{P}_3(b')$.

Here $\mathfrak{P}_3$ is the Postnikov square.

\begin{theorem}
\begin{table}[!h]
\centering
\begin{tabular}{c c}
\hline
$i$ & $\TP_i(\SO\times\Z_3\times \B \Z_3)$\\
\hline
0& $0$\\
1& $\Z_3$\\
2& $\Z_3$\\
3 & $\Z\times\Z_3^2$\\
4 & $\Z_3^2$\\ 
5 & $\Z_2\times\Z_3^2\times\Z_9$\\
\hline
\end{tabular}
\end{table}
\end{theorem}

The 2d topological term is $x_2'$.

The 3d topological terms are $\frac{1}{3}\text{CS}_3^{(TM)},a'b',a'x_2'$.

The 4d topological terms are $a'x_3'(=b'x_2')$ and $x_2'^2$.

The 5d topological terms are $w_2w_3,a'b'x_2',a'x_2'^2,\mathfrak{P}_3(b')$.

\subsubsection{$\Omega_d^{\Spin}(\B \Z_3\times \B ^2\Z_3)$}
\bea
\Ext_{\A_2}^{s,t}(\H^*(M\Spin\wedge (\B \Z_3\times \B ^2\Z_3)_+,\Z_2),\Z_2)\Rightarrow\Omega_{t-s}^{\Spin}(\B \Z_3\times \B ^2\Z_3)_2^{\wedge}.
\eea
Since $\H^*(\B \Z_3\times \B ^2\Z_3,\Z_2)=\Z_2$, we have $\Omega_d^{\Spin}(\B \Z_3\times \B ^2\Z_3)_2^{\wedge}=\Omega_d^{\Spin}$.
\bea
\Ext_{\A_3}^{s,t}(\H^*(M\Spin\wedge (\B \Z_3\times \B ^2\Z_3)_+,\Z_3),\Z_3)\Rightarrow\Omega_{t-s}^{\Spin}(\B \Z_3\times \B ^2\Z_3)_3^{\wedge}.
\eea
Since $$\H^*(M\Spin,\Z_3)=\H^*(M\SO,\Z_3),$$
we have the following

\begin{theorem}
\begin{table}[!h]
\centering
\begin{tabular}{c c}
\hline
$i$ & $\Omega^{\Spin}_i(\B \Z_3\times \B ^2\Z_3)$\\
\hline
0& $\Z$\\
1& $\Z_2\times\Z_3$\\
2& $\Z_2\times\Z_3$\\
3 & $\Z_3^2$\\
4 & $\Z\times\Z_3^2$\\ 
5 & $\Z_3^2\times\Z_9$\\
\hline
\end{tabular}
\end{table}
\end{theorem}

The bordism invariants of $\Omega^{\Spin}_{2}(\B \Z_3\times \B ^2\Z_3)$ are Arf and $x_2'$.

The bordism invariants of $\Omega^{\Spin}_{3}(\B \Z_3\times \B ^2\Z_3)$ are $a'b',a'x_2'$.

The bordism invariants of $\Omega^{\Spin}_{4}(\B \Z_3\times \B ^2\Z_3)$ are $\frac{\sigma}{16}$, $a'x_3'(=b'x_2')$ and $x_2'^2$.

The bordism invariants of $\Omega^{\Spin}_5(\B \Z_3\times \B ^2\Z_3)$ are $a'b'x_2',a'x_2'^2,\mathfrak{P}_3(b')$.

\begin{theorem}
\begin{table}[!h]
\centering
\begin{tabular}{c c}
\hline
$i$ & $\TP_i(\Spin\times\Z_3\times \B \Z_3)$\\
\hline
0& $0$\\
1& $\Z_2\times\Z_3$\\
2& $\Z_2\times\Z_3$\\
3 & $\Z\times\Z_3^2$\\
4 & $\Z_3^2$\\ 
5 & $\Z_3^2\times\Z_9$\\
\hline
\end{tabular}
\end{table}
\end{theorem}

The 2d topological terms are Arf and $x_2'$.

The 3d topological terms are $\frac{1}{48}\text{CS}_3^{(TM)},a'b',a'x_2'$.

The 4d topological terms are $a'x_3'(=b'x_2')$ and $x_2'^2$.

The 5d topological terms are $a'b'x_2',a'x_2'^2,\mathfrak{P}_3(b')$.

\subsubsection{$\Omega_d^{\Pin^+}(\B \Z_3\times \B ^2\Z_3)$}
\bea
\Ext_{\A_2}^{s,t}(\H^*(M\Pin^-\wedge (\B \Z_3\times \B ^2\Z_3)_+,\Z_2),\Z_2)\Rightarrow\Omega_{t-s}^{\Pin^+}(\B \Z_3\times \B ^2\Z_3)_2^{\wedge}.
\eea
\bea
\Ext_{\A_3}^{s,t}(\H^*(M\Pin^+\wedge (\B \Z_3\times \B ^2\Z_3)_+,\Z_3),\Z_3)\Rightarrow\Omega_{t-s}^{\Pin^-}(\B \Z_3\times \B ^2\Z_3)_3^{\wedge}.
\eea

Since $\H^*(M\Pin^-,\Z_3)=0$, we have $\Omega_d^{\Pin^+}(\B \Z_3\times \B ^2\Z_3)_3^{\wedge}=0$.

Since $\H^*(\B \Z_3\times \B ^2\Z_3,\Z_2)=\Z_2$, we have $\Omega_d^{\Pin^+}(\B \Z_3\times \B ^2\Z_3)_2^{\wedge}=\Omega_d^{\Pin^+}$.

Hence $\Omega_d^{\Pin^+}(\B \Z_3\times \B ^2\Z_3)=\Omega_d^{\Pin^+}$.

\begin{theorem}
\begin{table}[!h]
\centering
\begin{tabular}{c c}
\hline
$i$ & $\Omega^{\Pin^+}_i(\B \Z_3\times \B ^2\Z_3)$\\
\hline
0& $\Z_2$\\
1& $0$\\
2& $\Z_2$\\
3 & $\Z_2$\\
4 & $\Z_{16}$\\ 
5 & $0$\\
\hline
\end{tabular}
\end{table}
\end{theorem}

The bordism invariant of $\Omega^{\Pin^+}_2(\B \Z_3\times \B ^2\Z_3)$ is $w_1\tilde{\eta}$.

The bordism invariant of $\Omega^{\Pin^+}_3(\B \Z_3\times \B ^2\Z_3)$ is $w_1\text{Arf}$.

The bordism invariant of $\Omega^{\Pin^+}_4(\B \Z_3\times \B ^2\Z_3)$ is $\eta$.

\begin{theorem}
\begin{table}[!h]
\centering
\begin{tabular}{c c}
\hline
$i$ & $\TP_i(\Pin^+\times\Z_3\times \B \Z_3)$\\
\hline
0& $\Z_2$\\
1& $0$\\
2& $\Z_2$\\
3 & $\Z_2$\\
4 & $\Z_{16}$\\ 
5 & $0$\\
\hline
\end{tabular}
\end{table}
\end{theorem}

The 2d topological term is $w_1\tilde{\eta}$.

The 3d topological term is $w_1\text{Arf}$.

The 4d topological term is $\eta$.

\subsubsection{$\Omega_d^{\Pin^-}(\B \Z_3\times \B ^2\Z_3)$}
\bea
\Ext_{\A_2}^{s,t}(\H^*(M\Pin^+\wedge (\B \Z_3\times \B ^2\Z_3)_+,\Z_2),\Z_2)\Rightarrow\Omega_{t-s}^{\Pin^-}(\B \Z_3\times \B ^2\Z_3)_2^{\wedge}.
\eea
\bea
\Ext_{\A_3}^{s,t}(\H^*(M\Pin^+\wedge (\B \Z_3\times \B ^2\Z_3)_+,\Z_3),\Z_3)\Rightarrow\Omega_{t-s}^{\Pin^-}(\B \Z_3\times \B ^2\Z_3)_3^{\wedge}.
\eea

Since $\H^*(M\Pin^+,\Z_3)=0$, we have $\Omega_d^{\Pin^-}(\B \Z_3\times \B ^2\Z_3)_3^{\wedge}=0$.

Since $\H^*(\B \Z_3\times \B ^2\Z_3,\Z_2)=\Z_2$, we have $\Omega_d^{\Pin^-}(\B \Z_3\times \B ^2\Z_3)_2^{\wedge}=\Omega_d^{\Pin^-}$.

Hence $\Omega_d^{\Pin^-}(\B \Z_3\times \B ^2\Z_3)=\Omega_d^{\Pin^-}$.

\begin{theorem}
\begin{table}[!h]
\centering
\begin{tabular}{c c}
\hline
$i$ & $\Omega^{\Pin^-}_i(\B \Z_3\times \B ^2\Z_3)$\\
\hline
0& $\Z_2$\\
1& $\Z_2$\\
2& $\Z_8$\\
3 & $0$\\
4 & $0$\\ 
5 & $0$\\
\hline
\end{tabular}
\end{table}
\end{theorem}

The bordism invariant of $\Omega^{\Pin^-}_2(\B \Z_3\times \B ^2\Z_3)$ is ABK.

\begin{theorem}
\begin{table}[!h]
\centering
\begin{tabular}{c c}
\hline
$i$ & $\TP_i(\Pin^-\times\Z_3\times \B \Z_3)$\\
\hline
0& $\Z_2$\\
1& $\Z_2$\\
2& $\Z_8$\\
3 & $0$\\
4 & $0$\\ 
5 & $0$\\
\hline
\end{tabular}
\end{table}
\end{theorem}

The 2d topological term is ABK.

\subsection{$(\B G_a,\B ^2G_b):(\B \PSU(2),\B ^2\Z_2),(\B \PSU(3),\B ^2\Z_3)$}

\label{sec:BPSU(m)B2ZN}

\subsubsection{$\Omega^{\tO}_d(\B \PSU(2)\times \B ^2\Z_2)$}
\bea
\Ext_{\A_2}^{s,t}(\H^*(M\tO,\Z_2)\otimes\H^*(\B \PSU(2)\times \B ^2\Z_2,\Z_2),\Z_2)\Rightarrow\Omega^{\tO}_{t-s}(\B \PSU(2)\times \B ^2\Z_2).
\eea
\bea
\H^*(\B \PSU(2),\Z_2)=\Z_2[w_2',w_3'],
\eea
\bea
\H^*(\B ^2\Z_2,\Z_2)=\Z_2[x_2,x_3,x_5,x_9,\dots],
\eea
\bea
\H^*(M\tO,\Z_2)=\A_2\otimes\Z_2[y_2,y_4,y_5,y_6,y_8,\dots]^*.
\eea
\bea
&&\H^*(M\tO,\Z_2)\otimes\H^*(\B \PSU(2),\Z_2)\otimes\H^*(\B ^2\Z_2,\Z_2)\notag\\
&=&\A_2\oplus3\Sigma^2\A_2\oplus2\Sigma^3\A_2\oplus7\Sigma^4\A_2\oplus8\Sigma^5\A_2\oplus\cdots.
\eea

\begin{theorem}
\begin{table}[!h]
\centering
\begin{tabular}{c c}
\hline
$i$ & $\Omega^{\tO}_i(\B \PSU(2)\times \B ^2\Z_2)$\\
\hline
0& $\Z_2$\\
1& $0$\\
2& $\Z_2^3$\\
3 & $\Z_2^2$\\
4 & $\Z_2^7$\\ 
5 & $\Z_2^8$\\
\hline
\end{tabular}
\end{table}
\end{theorem}

The bordism invariants of $\Omega^{\tO}_2(\B \PSU(2)\times \B ^2\Z_2)$ are $w_2',x_2,w_1^2$.

The bordism invariants of $\Omega^{\tO}_3(\B \PSU(2)\times \B ^2\Z_2)$ are $x_3=w_1x_2,w_3'=w_1w_2'$.

The bordism invariants of $\Omega^{\tO}_4(\B \PSU(2)\times \B ^2\Z_2)$ are $w_1^4,w_2^2,x_2^2,w_2'^2,x_2w_1^2,w_2'w_1^2,w_2'x_2$.

The bordism invariants of $\Omega^{\tO}_5(\B \PSU(2)\times \B ^2\Z_2)$ are $w_2'w_3',x_2w_3',w_1^2w_3',w_2'x_3,x_2x_3,w_1^2x_3,x_5,w_2w_3$.

\begin{theorem}
\begin{table}[!h]
\centering
\begin{tabular}{c c}
\hline
$i$ & $\TP_i(\tO\times \PSU(2)\times \B \Z_2)$\\
\hline
0& $\Z_2$\\
1& $0$\\
2& $\Z_2^3$\\
3 & $\Z_2^2$\\
4 & $\Z_2^7$\\ 
5 & $\Z_2^8$\\
\hline
\end{tabular}
\end{table}
\end{theorem}

The 2d topological terms are $w_2',x_2,w_1^2$.

The 3d topological terms are $x_3=w_1x_2,w_3'=w_1w_2'$.

The 4d topological terms are $w_1^4,w_2^2,x_2^2,w_2'^2,x_2w_1^2,w_2'w_1^2,w_2'x_2$.

The 5d topological terms are $w_2'w_3',x_2w_3',w_1^2w_3',w_2'x_3,x_2x_3,w_1^2x_3,x_5,w_2w_3$.

\subsubsection{$\Omega^{\SO}_d(\B \PSU(2)\times \B ^2\Z_2)$}

\bea
\Ext_{\A_2}^{s,t}(\H^*(M\SO,\Z_2)\otimes\H^*(\B \PSU(2)\times \B ^2\Z_2,\Z_2),\Z_2)\Rightarrow\Omega_{t-s}^{\SO}(\B \PSU(2)\times \B ^2\Z_2).
\eea

\cred{
There is a differential $d_2$ corresponding to the Bockstein homomorphism  $\beta_{(2,4)}:\H^*(-,\Z_4)\to\H^{*+1}(-,\Z_2)$ associated to $0\to\Z_2\to\Z_8\to\Z_4\to0$ \cite{may1981bockstein}. See \ref{Bockstein} for the definition of Bockstein homomorphisms.
}

By \eqref{x_2x_3+x_5}, there is a differential such that
$d_2(x_2x_3+x_5)=x_2^2h_0^2$.

The $E_2$ page is shown in Figure \ref{fig:Omega_*^{SO}(BPSU(2)timesB^2Z_2)}.

\begin{figure}[!h]
\begin{center}
\begin{tikzpicture}
\node at (0,-1) {0};
\node at (1,-1) {1};
\node at (2,-1) {2};
\node at (3,-1) {3};
\node at (4,-1) {4};
\node at (5,-1) {5};
\node at (6,-1) {$t-s$};
\node at (-1,0) {0};
\node at (-1,1) {1};
\node at (-1,2) {2};
\node at (-1,3) {3};
\node at (-1,4) {4};
\node at (-1,5) {5};
\node at (-1,6) {$s$};

\draw[->] (-0.5,-0.5) -- (-0.5,6);
\draw[->] (-0.5,-0.5) -- (6,-0.5);

\draw (0,0) -- (0,5);
\draw (4,0) -- (4,5);
\draw (3.9,0) -- (3.9,5);
\draw (4.1,0) -- (4.1,5);
\draw (5,0) -- (5,5);

\draw[fill] (1.9,0) circle(0.05);
\draw[fill] (2.1,0) circle(0.05);
\draw[fill] (3.8,0) circle(0.05);
\draw[fill] (4.8,0) circle(0.05);
\draw[fill] (5.2,0) circle(0.05);
\draw[fill] (4.9,0) circle(0.05);
\draw[fill] (5.1,0) circle(0.05);

\draw[color=red][->] (5,0) -- (4.1,2);
\draw[color=red][->] (5,1) -- (4.1,3);
\draw[color=red][->] (5,2) -- (4.1,4);
\draw[color=red][->] (5,3) -- (4.1,5);
\end{tikzpicture}
\end{center}
\caption{$\Omega_*^{\SO}(\B \PSU(2)\times \B ^2\Z_2)$}
\label{fig:Omega_*^{SO}(BPSU(2)timesB^2Z_2)}
\end{figure}

\begin{theorem}
\begin{table}[!h]
\centering
\begin{tabular}{c c}
\hline
$i$ & $\Omega^{\SO}_i(\B \PSU(2)\times \B ^2\Z_2)$\\
\hline
0& $\Z$\\
1& $0$\\
2& $\Z_2^2$\\
3 & $0$\\
4 & $\Z^2\times\Z_2\times\Z_4$\\ 
5 & $\Z_2^4$\\
\hline
\end{tabular}
\end{table}
\end{theorem}

The bordism invariants of $\Omega^{\SO}_{2}(\B \PSU(2)\times \B ^2\Z_2)$ are $w_2',x_2$.

The bordism invariants of $\Omega^{\SO}_{4}(\B \PSU(2)\times \B ^2\Z_2)$ are $\sigma$, $p_1'$, $w_2'x_2$ and $\mathcal{P}_2(x_2)$.

The bordism invariants of $\Omega^{\SO}_5(\B \PSU(2)\times \B ^2\Z_2)$ are $w_2'w_3',x_5,w_3'x_2{(=w_2'x_3)},w_2w_3$.

\begin{theorem}
\begin{table}[!h]
\centering
\begin{tabular}{c c}
\hline
$i$ & $\TP_i(\SO\times \PSU(2)\times \B \Z_2)$\\
\hline
0& $0$\\
1& $0$\\
2& $\Z_2^2$\\
3 & $\Z^2$\\
4 & $\Z_2\times\Z_4$\\ 
5 & $\Z_2^4$\\
\hline
\end{tabular}
\end{table}
\end{theorem}

The 2d topological terms are $w_2',x_2$.

The 3d topological terms are $\frac{1}{3}\text{CS}_3^{(TM)}$, $\text{CS}_3^{(\SO(3))}$.

The 4d topological terms are $w_2'x_2$ and $\mathcal{P}_2(x_2)$.

The 5d topological terms are $w_2'w_3',x_5,w_3'x_2{(=w_2'x_3)},w_2w_3$.

\subsubsection{$\Omega^{\Spin}_d(\B \PSU(2)\times \B ^2\Z_2)$}
For $t-s<8$,
\bea
\Ext_{\A_2(1)}^{s,t}(\H^*(\B \PSU(2)\times \B ^2\Z_2,\Z_2),\Z_2)\Rightarrow\Omega_{t-s}^{\Spin}(\B \PSU(2)\times \B ^2\Z_2).
\eea

\cred{
There is a differential $d_2$ corresponding to the Bockstein homomorphism  $\beta_{(2,4)}:\H^*(-,\Z_4)\to\H^{*+1}(-,\Z_2)$ associated to $0\to\Z_2\to\Z_8\to\Z_4\to0$ \cite{may1981bockstein}. See \ref{Bockstein} for the definition of Bockstein homomorphisms.
}

By \eqref{x_2x_3+x_5}, there is a differential such that
$d_2(x_2x_3+x_5)=x_2^2h_0^2$.

The $\A_2(1)$-module structure of $\H^*(\B \PSU(2)\times \B ^2\Z_2,\Z_2)$ and the $E_2$ page is shown in Figure \ref{fig:H^*(BPSU(2)timesB^2Z_2,Z_2)}, \ref{fig:Omega_*^{Spin}(BPSU(2)timesB^2Z_2)}.

\begin{figure}[!h]
\begin{center}
\begin{tikzpicture}[scale=0.5]

\node[below] at (0,0) {$1$};

\draw[fill] (0,0) circle(.1);

\node[below] at (0,2) {$w_2'$};

\draw[fill] (0,2) circle(.1);
\draw[fill] (0,3) circle(.1);
\draw[fill] (0,4) circle(.1);
\draw (0,2) -- (0,3);
\draw (0,2) to [out=150,in=150] (0,4);
\draw[fill] (1,5) circle(.1);
\draw[fill] (1,6) circle(.1);
\draw (1,5) -- (1,6);
\draw (0,3) to [out=30,in=150] (1,5);
\draw (0,4) to [out=30,in=150] (1,6);

\node[below] at (10,0) {$1$};

\draw[fill] (10,0) circle(.1);

\node[below] at (10,2) {$x_2$};

\draw[fill] (10,2) circle(.1);
\draw[fill] (10,3) circle(.1);
\draw[fill] (10,4) circle(.1);
\draw (10,2) -- (10,3);
\draw (10,2) to [out=150,in=150] (10,4);
\draw[fill] (11,5) circle(.1);
\draw[fill] (11,6) circle(.1);
\draw (11,5) -- (11,6);
\draw (10,3) to [out=30,in=150] (11,5);
\draw (10,4) to [out=30,in=150] (11,6);
\draw[fill] (12,5) circle(.1);

\node[below] at (12,5) {$x_2x_3$};

\draw (12,5) -- (11,6);
\draw[fill] (12,7) circle(.1);
\draw (12,5) to [out=30,in=30] (12,7);
\draw[fill] (12,8) circle(.1);
\draw (12,7) -- (12,8);
\draw[fill] (12,10) circle(.1);
\draw (12,8) to [out=150,in=150] (12,10);

\node at (5,5) {$\bigotimes$};
\node at (-2,-10) {$=$};

\node[below] at (0,-15) {$1$};

\draw[fill] (0,-15) circle(.1);

\node[below] at (0,-13) {$w_2'$};

\draw[fill] (0,-13) circle(.1);
\draw[fill] (0,-12) circle(.1);
\draw[fill] (0,-11) circle(.1);
\draw (0,-13) -- (0,-12);
\draw (0,-13) to [out=150,in=150] (0,-11);
\draw[fill] (1,-10) circle(.1);
\draw[fill] (1,-9) circle(.1);
\draw (1,-10) -- (1,-9);
\draw (0,-12) to [out=30,in=150] (1,-10);
\draw (0,-11) to [out=30,in=150] (1,-9);

\node[below] at (2,-13) {$x_2$};

\draw[fill] (2,-13) circle(.1);
\draw[fill] (2,-12) circle(.1);
\draw[fill] (2,-11) circle(.1);
\draw (2,-13) -- (2,-12);
\draw (2,-13) to [out=150,in=150] (2,-11);
\draw[fill] (3,-10) circle(.1);
\draw[fill] (3,-9) circle(.1);
\draw (3,-10) -- (3,-9);
\draw (2,-12) to [out=30,in=150] (3,-10);
\draw (2,-11) to [out=30,in=150] (3,-9);
\draw[fill] (4,-10) circle(.1);

\node[below] at (4,-10) {$x_2x_3$};

\draw (4,-10) -- (3,-9);
\draw[fill] (4,-8) circle(.1);
\draw (4,-10) to [out=30,in=30] (4,-8);
\draw[fill] (4,-7) circle(.1);
\draw (4,-8) -- (4,-7);
\draw[fill] (4,-5) circle(.1);
\draw (4,-7) to [out=150,in=150] (4,-5);

\node[below] at (6,-11) {$w_2'x_2$};

\draw[fill] (6,-11) circle(.1);
\draw[fill] (6,-10) circle(.1);
\draw (6,-11) -- (6,-10);
\draw[fill] (6,-9) circle(.1);
\draw (6,-11) to [out=150,in=150] (6,-9);
\draw[fill] (7,-8) circle(.1);
\draw (6,-10) to [out=30,in=150] (7,-8);
\draw[fill] (6,-8) circle(.1);
\draw (6,-9) -- (6,-8);
\draw[fill] (7,-7) circle(.1);
\draw (7,-8) -- (7,-7);
\draw (6,-9) to [out=30,in=150] (7,-7);
\draw[fill] (7,-6) circle(.1);
\draw (6,-8) to [out=30,in=150] (7,-6);
\draw[fill] (7,-5) circle(.1);
\draw (7,-6) -- (7,-5);
\draw (7,-7) to [out=30,in=30] (7,-5);

\node[below] at (8,-10) {$w_3'x_2$};
 
\draw[fill] (8,-10) circle(.1);
\draw[fill] (8,-9) circle(.1);
\draw (8,-10) -- (8,-9);
\draw[fill] (8,-8) circle(.1);
\draw (8,-10) to [out=150,in=150] (8,-8);
\draw[fill] (9,-7) circle(.1);
\draw (8,-9) to [out=30,in=150] (9,-7);
\draw[fill] (8,-7) circle(.1);
\draw (8,-8) -- (8,-7);
\draw[fill] (9,-6) circle(.1);
\draw (9,-7) -- (9,-6);
\draw (8,-8) to [out=30,in=150] (9,-6);
\draw[fill] (9,-5) circle(.1);
\draw (8,-7) to [out=30,in=150] (9,-5);
\draw[fill] (9,-4) circle(.1);
\draw (9,-5) -- (9,-4);
\draw (9,-6) to [out=30,in=30] (9,-4);

\end{tikzpicture}
\end{center}
\caption{The $\A_2(1)$-module structure of $\H^*(\B \PSU(2)\times \B ^2\Z_2,\Z_2)$}
\label{fig:H^*(BPSU(2)timesB^2Z_2,Z_2)}
\end{figure}

\begin{figure}[!h]
\begin{center}
\begin{tikzpicture}
\node at (0,-1) {0};
\node at (1,-1) {1};
\node at (2,-1) {2};
\node at (3,-1) {3};
\node at (4,-1) {4};
\node at (5,-1) {5};
\node at (6,-1) {$t-s$};
\node at (-1,0) {0};
\node at (-1,1) {1};
\node at (-1,2) {2};
\node at (-1,3) {3};
\node at (-1,4) {4};
\node at (-1,5) {5};
\node at (-1,6) {$s$};

\draw[->] (-0.5,-0.5) -- (-0.5,6);
\draw[->] (-0.5,-0.5) -- (6,-0.5);

\draw (0,0) -- (0,5);
\draw (0,0) -- (2,2);
\draw (4,3) -- (4,5);
\draw (3.9,1) -- (3.9,5);
\draw (4.1,1) -- (4.1,5);
\draw (5,0) -- (5,5);
\draw[fill] (1.9,0) circle(0.05);
\draw[fill] (2.1,0) circle(0.05);
\draw[fill] (4,0) circle(0.05);
\draw[fill] (5.1,0) circle(0.05);
\draw[color=red][->] (5,0) -- (4.1,2);
\draw[color=red][->] (5,1) -- (4.1,3);
\draw[color=red][->] (5,2) -- (4.1,4);
\draw[color=red][->] (5,3) -- (4.1,5);
\end{tikzpicture}
\end{center}
\caption{$\Omega_*^{\Spin}(\B \PSU(2)\times \B ^2\Z_2)$}
\label{fig:Omega_*^{Spin}(BPSU(2)timesB^2Z_2)}
\end{figure}

\begin{theorem}
\begin{table}[!h]
\centering
\begin{tabular}{c c}
\hline
$i$ & $\Omega^{\Spin}_i(\B \PSU(2)\times \B ^2\Z_2)$\\
\hline
0& $\Z$\\
1& $\Z_2$\\
2& $\Z_2^3$\\
3 & $0$\\
4 & $\Z^2\times\Z_2^2$\\ 
5 & $\Z_2$\\
\hline
\end{tabular}
\end{table}
\end{theorem}

The bordism invariants of $\Omega^{\Spin}_{2}(\B \PSU(2)\times \B ^2\Z_2)$ are $w_2',x_2,\text{Arf}$.

The bordism invariants of $\Omega^{\Spin}_{4}(\B \PSU(2)\times \B ^2\Z_2)$ are $\frac{\sigma}{16}$, $\frac{p_1'}{2}$, $w_2'x_2$ and $\frac{\mathcal{P}_2(x_2)}{2}$.

The bordism invariant of $\Omega^{\Spin}_5(\B \PSU(2)\times \B ^2\Z_2)$ is $w_3'x_2{(=w_2'x_3)}$.

\begin{theorem}
\begin{table}[!h]
\centering
\begin{tabular}{c c}
\hline
$i$ & $\TP_i(\Spin\times \PSU(2)\times \B \Z_2)$\\
\hline
0& $0$\\
1& $\Z_2$\\
2& $\Z_2^3$\\
3 & $\Z^2$\\
4 & $\Z_2^2$\\ 
5 & $\Z_2$\\
\hline
\end{tabular}
\end{table}
\end{theorem}

The 2d topological terms are $w_2',x_2,\text{Arf}$.

The 3d topological terms are $\frac{1}{48}\text{CS}_3^{(TM)}$, $\frac{1}{2}\text{CS}_3^{(\SO(3))}$.

The 4d topological terms are $w_2'x_2$ and $\frac{\mathcal{P}_2(x_2)}{2}$.

The 5d topological term is $w_3'x_2{(=w_2'x_3)}$.

\subsubsection{$\Omega^{\Pin^+}_d(\B \PSU(2)\times \B ^2\Z_2)$}
For $t-s<8$,
\bea
&&\Ext_{\A_2(1)}^{s,t}(\H^{*-1}(MT\tO(1),\Z_2)\otimes\H^*(\B \PSU(2)\times \B ^2\Z_2,\Z_2),\Z_2)\notag\\
&\Rightarrow&\Omega_{t-s}^{\Pin^+}(\B \PSU(2)\times \B ^2\Z_2).
\eea

The $\A_2(1)$-module structure of $\H^{*-1}(MT\tO(1),\Z_2)\otimes\H^*(\B \PSU(2)\times \B ^2\Z_2,\Z_2)$ and the $E_2$ page are shown in Figure \ref{fig:H^{*-1}(MTO(1),Z_2)otimesH^*(BPSU(2)timesB^2Z_2,Z_2)}, \ref{fig:Omega_*^{Pin^+}(BPSU(2)timesB^2Z_2)}.

\begin{figure}[!h]
\begin{center}
\begin{tikzpicture}[scale=0.5]

\node[below] at (0,0) {$U$};

\draw[fill] (0,0) circle(.1);
\draw[fill] (0,1) circle(.1);
\draw (0,0) -- (0,1);
\draw[fill] (0,2) circle(.1);
\draw (0,0) to [out=150,in=150] (0,2);
\draw[fill] (0,3) circle(.1);
\draw (0,2) -- (0,3);
\draw[fill] (0,4) circle(.1);
\draw[fill] (0,5) circle(.1);
\draw (0,4) -- (0,5);
\draw (0,3) to [out=150,in=150] (0,5);
\draw[fill] (0,6) circle(.1);
\draw (0,4) to [out=30,in=30] (0,6);

\node[below] at (4,0) {$1$};

\draw[fill] (4,0) circle(.1);

\node[below] at (4,2) {$w_2'$};

\draw[fill] (4,2) circle(.1);
\draw[fill] (4,3) circle(.1);
\draw[fill] (4,4) circle(.1);
\draw (4,2) -- (4,3);
\draw (4,2) to [out=150,in=150] (4,4);
\draw[fill] (5,5) circle(.1);
\draw[fill] (5,6) circle(.1);
\draw (5,5) -- (5,6);
\draw (4,3) to [out=30,in=150] (5,5);
\draw (4,4) to [out=30,in=150] (5,6);

\node[below] at (6,2) {$x_2$};

\draw[fill] (6,2) circle(.1);
\draw[fill] (6,3) circle(.1);
\draw[fill] (6,4) circle(.1);
\draw (6,2) -- (6,3);
\draw (6,2) to [out=150,in=150] (6,4);
\draw[fill] (7,5) circle(.1);
\draw[fill] (7,6) circle(.1);
\draw (7,5) -- (7,6);
\draw (6,3) to [out=30,in=150] (7,5);
\draw (6,4) to [out=30,in=150] (7,6);
\draw[fill] (8,5) circle(.1);

\node[below] at (8,5) {$x_2x_3$};

\draw (8,5) -- (7,6);
\draw[fill] (8,7) circle(.1);
\draw (8,5) to [out=30,in=30] (8,7);
\draw[fill] (8,8) circle(.1);
\draw (8,7) -- (8,8);
\draw[fill] (8,10) circle(.1);
\draw (8,8) to [out=150,in=150] (8,10);

\node[below] at (10,4) {$w_2'x_2$};
 
\draw[fill] (10,4) circle(.1);
\draw[fill] (10,5) circle(.1);
\draw (10,4) -- (10,5);
\draw[fill] (10,6) circle(.1);
\draw (10,4) to [out=150,in=150] (10,6);
\draw[fill] (11,7) circle(.1);
\draw (10,5) to [out=30,in=150] (11,7);
\draw[fill] (10,7) circle(.1);
\draw (10,6) -- (10,7);
\draw[fill] (11,8) circle(.1);
\draw (11,7) -- (11,8);
\draw (10,6) to [out=30,in=150] (11,8);
\draw[fill] (11,9) circle(.1);
\draw (10,7) to [out=30,in=150] (11,9);
\draw[fill] (11,10) circle(.1);
\draw (11,9) -- (11,10);
\draw (11,8) to [out=30,in=30] (11,10); 

\node[below] at (12,5) {$w_3'x_2$};
 
\draw[fill] (12,5) circle(.1);
\draw[fill] (12,6) circle(.1);
\draw (12,5) -- (12,6);
\draw[fill] (12,7) circle(.1);
\draw (12,5) to [out=150,in=150] (12,7);
\draw[fill] (13,8) circle(.1);
\draw (12,6) to [out=30,in=150] (13,8);
\draw[fill] (12,8) circle(.1);
\draw (12,7) -- (12,8);
\draw[fill] (13,9) circle(.1);
\draw (13,8) -- (13,9);
\draw (12,7) to [out=30,in=150] (13,9);
\draw[fill] (13,10) circle(.1);
\draw (12,8) to [out=30,in=150] (13,10);
\draw[fill] (13,11) circle(.1);
\draw (13,10) -- (13,11);
\draw (13,9) to [out=30,in=30] (13,11);

\node at (2,5) {$\bigotimes$};
\node at (-2,-10) {$=$};

\node[below] at (0,-15) {$U$};

\draw[fill] (0,-15) circle(.1);
\draw[fill] (0,-14) circle(.1);
\draw (0,-15) -- (0,-14);
\draw[fill] (0,-13) circle(.1);
\draw (0,-15) to [out=150,in=150] (0,-13);
\draw[fill] (0,-12) circle(.1);
\draw (0,-13) -- (0,-12);
\draw[fill] (0,-11) circle(.1);
\draw[fill] (0,-10) circle(.1);
\draw (0,-11) -- (0,-10);
\draw (0,-12) to [out=150,in=150] (0,-10);
\draw[fill] (0,-9) circle(.1);
\draw (0,-11) to [out=30,in=30] (0,-9);

\node[below] at (2,-13) {$w_2'U$};

\draw[fill] (2,-13) circle(.1);
\draw[fill] (2,-12) circle(.1);
\draw (2,-13) -- (2,-12);
\draw[fill] (2,-11) circle(.1);
\draw (2,-13) to [out=150,in=150] (2,-11);
\draw[fill] (3,-10) circle(.1);
\draw (2,-12) to [out=30,in=150] (3,-10);
\draw[fill] (2,-10) circle(.1);
\draw (2,-11) -- (2,-10);
\draw[fill] (3,-9) circle(.1);
\draw (3,-10) -- (3,-9);
\draw (2,-11) to [out=30,in=150] (3,-9);
\draw[fill] (3,-8) circle(.1);
\draw (2,-10) to [out=30,in=150] (3,-8);
\draw[fill] (3,-7) circle(.1);
\draw (3,-8) -- (3,-7);
\draw (3,-9) to [out=30,in=30] (3,-7); 

\node[below] at (5,-12) {$w_1w_2'U$};

\draw[fill] (5,-12) circle(.1);
\draw[fill] (5,-11) circle(.1);
\draw[fill] (5,-10) circle(.1);
\draw (5,-12) -- (5,-11);
\draw (5,-12) to [out=150,in=150] (5,-10);
\draw[fill] (6,-9) circle(.1);
\draw[fill] (6,-8) circle(.1);
\draw (6,-9) -- (6,-8);
\draw (5,-11) to [out=30,in=150] (6,-9);
\draw (5,-10) to [out=30,in=150] (6,-8);

\node[below] at (4,-11) {$w_2'^2U$};

\draw[fill] (4,-11) circle(.1);
\draw (4,-11) -- (5,-10);
\draw[fill] (4,-9) circle(.1);
\draw (4,-11) to [out=150,in=150] (4,-9);
\draw[fill] (4,-8) circle(.1);
\draw (4,-9) -- (4,-8);
\draw[fill] (4,-7) circle(.1);
\draw[fill] (4,-6) circle(.1);
\draw (4,-7) -- (4,-6);
\draw (4,-8) to [out=150,in=150] (4,-6);
\draw[fill] (4,-5) circle(.1);
\draw (4,-7) to [out=30,in=30] (4,-5);

\node[below] at (8,-10) {$w_1^2w_3'U$};

\draw[fill] (8,-10) circle(.1);
\draw[fill] (8,-9) circle(.1);
\draw (8,-10) -- (8,-9);
\draw[fill] (8,-8) circle(.1);
\draw (8,-10) to [out=150,in=150] (8,-8);
\draw[fill] (9,-7) circle(.1);
\draw (8,-9) to [out=30,in=150] (9,-7);
\draw[fill] (8,-7) circle(.1);
\draw (8,-8) -- (8,-7);
\draw[fill] (9,-6) circle(.1);
\draw (9,-7) -- (9,-6);
\draw (8,-8) to [out=30,in=150] (9,-6);
\draw[fill] (9,-5) circle(.1);
\draw (8,-7) to [out=30,in=150] (9,-5);
\draw[fill] (9,-4) circle(.1);
\draw (9,-5) -- (9,-4);
\draw (9,-6) to [out=30,in=30] (9,-4);

\node[below] at (10,-13) {$x_2U$};

\draw[fill] (10,-13) circle(.1);
\draw[fill] (10,-12) circle(.1);
\draw (10,-13) -- (10,-12);
\draw[fill] (10,-11) circle(.1);
\draw (10,-13) to [out=150,in=150] (10,-11);
\draw[fill] (11,-10) circle(.1);
\draw (10,-12) to [out=30,in=150] (11,-10);
\draw[fill] (10,-10) circle(.1);
\draw (10,-11) -- (10,-10);
\draw[fill] (11,-9) circle(.1);
\draw (11,-10) -- (11,-9);
\draw (10,-11) to [out=30,in=150] (11,-9);
\draw[fill] (11,-8) circle(.1);
\draw (10,-10) to [out=30,in=150] (11,-8);
\draw[fill] (11,-7) circle(.1);
\draw (11,-8) -- (11,-7);
\draw (11,-9) to [out=30,in=30] (11,-7);

\node[below] at (13,-12) {$w_1x_2U$};

\draw[fill] (13,-12) circle(.1);
\draw[fill] (13,-11) circle(.1);
\draw[fill] (13,-10) circle(.1);
\draw (13,-12) -- (13,-11);
\draw (13,-12) to [out=150,in=150] (13,-10);
\draw[fill] (14,-9) circle(.1);
\draw[fill] (14,-8) circle(.1);
\draw (14,-9) -- (14,-8);
\draw (13,-11) to [out=30,in=150] (14,-9);
\draw (13,-10) to [out=30,in=150] (14,-8);
\draw[fill] (15,-9) circle(.1);
\draw (15,-9) -- (14,-8);
\draw[fill] (15,-7) circle(.1);
\draw (15,-9) to [out=30,in=30] (15,-7);
\draw[fill] (15,-6) circle(.1);
\draw (15,-7) -- (15,-6);
\draw[fill] (15,-4) circle(.1);
\draw (15,-6) to [out=150,in=150] (15,-4);

\node[below] at (12,-11) {$x_2^2U$};

\draw[fill] (12,-11) circle(.1);
\draw (12,-11) -- (13,-10);
\draw[fill] (12,-9) circle(.1);
\draw (12,-11) to [out=150,in=150] (12,-9);
\draw[fill] (12,-8) circle(.1);
\draw (12,-9) -- (12,-8);
\draw[fill] (12,-7) circle(.1);
\draw[fill] (12,-6) circle(.1);
\draw (12,-7) -- (12,-6);
\draw (12,-8) to [out=150,in=150] (12,-6);
\draw[fill] (12,-5) circle(.1);
\draw (12,-7) to [out=30,in=30] (12,-5);

\node[below] at (16,-10) {$w_1^2x_3U$};

\draw[fill] (16,-10) circle(.1);
\draw[fill] (16,-9) circle(.1);
\draw (16,-10) -- (16,-9);
\draw[fill] (16,-8) circle(.1);
\draw (16,-10) to [out=150,in=150] (16,-8);
\draw[fill] (17,-7) circle(.1);
\draw (16,-9) to [out=30,in=150] (17,-7);
\draw[fill] (16,-7) circle(.1);
\draw (16,-8) -- (16,-7);
\draw[fill] (17,-6) circle(.1);
\draw (17,-7) -- (17,-6);
\draw (16,-8) to [out=30,in=150] (17,-6);
\draw[fill] (17,-5) circle(.1);
\draw (16,-7) to [out=30,in=150] (17,-5);
\draw[fill] (17,-4) circle(.1);
\draw (17,-5) -- (17,-4);
\draw (17,-6) to [out=30,in=30] (17,-4); 

\node[below] at (18,-10) {$x_2x_3U$};

\draw[fill] (18,-10) circle(.1);
\draw[fill] (18,-9) circle(.1);
\draw (18,-10) -- (18,-9);
\draw[fill] (18,-8) circle(.1);
\draw (18,-10) to [out=150,in=150] (18,-8);
\draw[fill] (19,-7) circle(.1);
\draw (18,-9) to [out=30,in=150] (19,-7);
\draw[fill] (18,-7) circle(.1);
\draw (18,-8) -- (18,-7);
\draw[fill] (19,-6) circle(.1);
\draw (19,-7) -- (19,-6);
\draw (18,-8) to [out=30,in=150] (19,-6);
\draw[fill] (19,-5) circle(.1);
\draw (18,-7) to [out=30,in=150] (19,-5);
\draw[fill] (19,-4) circle(.1);
\draw (19,-5) -- (19,-4);
\draw (19,-6) to [out=30,in=30] (19,-4); 

\node[below] at (20,-11) {$w_2'x_2U$};

\draw[fill] (20,-11) circle(.1);
\draw[fill] (20,-10) circle(.1);
\draw (20,-11) -- (20,-10);
\draw[fill] (20,-9) circle(.1);
\draw (20,-11) to [out=150,in=150] (20,-9);
\draw[fill] (21,-8) circle(.1);
\draw (20,-10) to [out=30,in=150] (21,-8);
\draw[fill] (20,-8) circle(.1);
\draw (20,-9) -- (20,-8);
\draw[fill] (21,-7) circle(.1);
\draw (21,-8) -- (21,-7);
\draw (20,-9) to [out=30,in=150] (21,-7);
\draw[fill] (21,-6) circle(.1);
\draw (20,-8) to [out=30,in=150] (21,-6);
\draw[fill] (21,-5) circle(.1);
\draw (21,-6) -- (21,-5);
\draw (21,-7) to [out=30,in=30] (21,-5); 

\node[below] at (22,-10) {$w_1w_2'x_2U$};

\draw[fill] (22,-10) circle(.1);
\draw[fill] (22,-9) circle(.1);
\draw (22,-10) -- (22,-9);
\draw[fill] (22,-8) circle(.1);
\draw (22,-10) to [out=150,in=150] (22,-8);
\draw[fill] (23,-7) circle(.1);
\draw (22,-9) to [out=30,in=150] (23,-7);
\draw[fill] (22,-7) circle(.1);
\draw (22,-8) -- (22,-7);
\draw[fill] (23,-6) circle(.1);
\draw (23,-7) -- (23,-6);
\draw (22,-8) to [out=30,in=150] (23,-6);
\draw[fill] (23,-5) circle(.1);
\draw (22,-7) to [out=30,in=150] (23,-5);
\draw[fill] (23,-4) circle(.1);
\draw (23,-5) -- (23,-4);
\draw (23,-6) to [out=30,in=30] (23,-4); 

\node[below] at (24,-10) {$w_3'x_2U$};

\draw[fill] (24,-10) circle(.1);
\draw[fill] (24,-9) circle(.1);
\draw (24,-10) -- (24,-9);
\draw[fill] (24,-8) circle(.1);
\draw (24,-10) to [out=150,in=150] (24,-8);
\draw[fill] (25,-7) circle(.1);
\draw (24,-9) to [out=30,in=150] (25,-7);
\draw[fill] (24,-7) circle(.1);
\draw (24,-8) -- (24,-7);
\draw[fill] (25,-6) circle(.1);
\draw (25,-7) -- (25,-6);
\draw (24,-8) to [out=30,in=150] (25,-6);
\draw[fill] (25,-5) circle(.1);
\draw (24,-7) to [out=30,in=150] (25,-5);
\draw[fill] (25,-4) circle(.1);
\draw (25,-5) -- (25,-4);
\draw (25,-6) to [out=30,in=30] (25,-4);

\end{tikzpicture}
\end{center}
\caption{The $\A_2(1)$-module structure of $\H^{*-1}(MT\tO(1),\Z_2)\otimes\H^*(\B \PSU(2)\times \B ^2\Z_2,\Z_2)$}
\label{fig:H^{*-1}(MTO(1),Z_2)otimesH^*(BPSU(2)timesB^2Z_2,Z_2)}
\end{figure}

\begin{figure}[!h]
\begin{center}
\begin{tikzpicture}
\node at (0,-1) {0};
\node at (1,-1) {1};
\node at (2,-1) {2};
\node at (3,-1) {3};
\node at (4,-1) {4};
\node at (5,-1) {5};
\node at (6,-1) {$t-s$};
\node at (-1,0) {0};
\node at (-1,1) {1};
\node at (-1,2) {2};
\node at (-1,3) {3};
\node at (-1,4) {4};
\node at (-1,5) {5};
\node at (-1,6) {$s$};

\draw[->] (-0.5,-0.5) -- (-0.5,6);
\draw[->] (-0.5,-0.5) -- (6,-0.5);

\draw[fill] (0,0) circle(0.05);
\draw[fill] (1.9,0) circle(0.05);
\draw[fill] (2.1,0) circle(0.05);
\draw (2,1) -- (4,3);
\draw (4,3) -- (4,0);
\draw (3,0) -- (3.9,1);
\draw (3.9,1) -- (3.9,0);
\draw (2.9,0) -- (3.8,1);
\draw (3.8,1) -- (3.8,0);
\draw[fill] (4.1,0) circle(0.05);
\draw[fill] (4.8,0) circle(0.05);
\draw[fill] (4.9,0) circle(0.05);
\draw[fill] (5,0) circle(0.05);
\draw[fill] (5.1,0) circle(0.05);
\draw[fill] (5.2,0) circle(0.05);

\end{tikzpicture}
\end{center}
\caption{$\Omega_*^{\Pin^+}(\B \PSU(2)\times \B ^2\Z_2)$}
\label{fig:Omega_*^{Pin^+}(BPSU(2)timesB^2Z_2)}
\end{figure}

\begin{theorem}
\begin{table}[!h]
\centering
\begin{tabular}{c c}
\hline
$i$ & $\Omega^{\Pin^+}_i(\B \PSU(2)\times \B ^2\Z_2)$\\
\hline
0& $\Z_2$\\
1& $0$\\
2& $\Z_2^3$\\
3 & $\Z_2^3$\\
4 & $\Z_4^2\times\Z_{16}\times\Z_2$\\ 
5 & $\Z_2^5$\\
\hline
\end{tabular}
\end{table}
\end{theorem}

The bodism invariants of $\Omega^{\Pin^+}_2(\B \PSU(2)\times \B ^2\Z_2)$ are $w_2',x_2,w_1\tilde{\eta}$.

The bodism invariants of $\Omega^{\Pin^+}_3(\B \PSU(2)\times \B ^2\Z_2)$ are $w_1w_2'=w_3',w_1x_2=x_3,w_1\text{Arf}$.

The bodism invariants of $\Omega^{\Pin^+}_4(\B \PSU(2)\times \B ^2\Z_2)$ are $q_s(w_2'),q_s(x_2),\eta,w_2'x_2$.

The bodism invariants of $\Omega^{\Pin^+}_5(\B \PSU(2)\times \B ^2\Z_2)$ are $$w_1^2w_3'{(=w_2'w_3')},w_1^2x_3{(=x_5)},x_2x_3,w_3'x_2,w_1w_2'x_2{(=w_2'x_3+w_3'x_2)}.$$

\begin{theorem}
\begin{table}[!h]
\centering
\begin{tabular}{c c}
\hline
$i$ & $\TP_i(\Pin^+\times \PSU(2)\times \B \Z_2)$\\
\hline
0& $\Z_2$\\
1& $0$\\
2& $\Z_2^3$\\
3 & $\Z_2^3$\\
4 & $\Z_4^2\times\Z_{16}\times\Z_2$\\ 
5 & $\Z_2^5$\\
\hline
\end{tabular}
\end{table}
\end{theorem}

The 2d topological terms are $w_2',x_2,w_1\tilde{\eta}$.

The 3d topological terms are $w_1w_2'=w_3',w_1x_2=x_3,w_1\text{Arf}$.

The 4d topological terms are $q_s(w_2'),q_s(x_2),\eta,w_2'x_2$.

The 5d topological terms are $$w_1^2w_3'{(=w_2'w_3')},w_1^2x_3{(=x_5)},x_2x_3,w_3'x_2,w_1w_2'x_2{(=w_2'x_3+w_3'x_2)}.$$

\subsubsection{$\Omega^{\Pin^-}_d(\B \PSU(2)\times \B ^2\Z_2)$}
For $t-s<8$,
\bea
\Ext_{\A_2(1)}^{s,t}(\H^{*+1}(M\tO(1),\Z_2)\otimes\H^*(\B \PSU(2)\times \B ^2\Z_2,\Z_2),\Z_2)\Rightarrow\Omega_{t-s}^{\Pin^-}(\B \PSU(2)\times \B ^2\Z_2).
\eea

The $\A_2(1)$-module structure of $\H^{*+1}(M\tO(1).\Z_2)\otimes\H^*(\B \PSU(2)\times \B ^2\Z_2,\Z_2)$ and the $E_2$ page are shown in Figure \ref{fig:H^{*+1}(MO(1),Z_2)otimesH^*(BPSU(2)timesB^2Z_2,Z_2)}, \ref{fig:Omega_*^{Pin^-}(BPSU(2)timesB^2Z_2)}.

\begin{figure}[!h]
\begin{center}
\begin{tikzpicture}[scale=0.5]

\node[below] at (0,0) {$U$};

\draw[fill] (0,0) circle(.1);
\draw[fill] (0,1) circle(.1);
\draw (0,0) -- (0,1);
\draw[fill] (0,2) circle(.1);
\draw[fill] (0,3) circle(.1);
\draw (0,2) -- (0,3);
\draw (0,1) to [out=150,in=150] (0,3);
\draw[fill] (0,4) circle(.1);
\draw (0,2) to [out=30,in=30] (0,4);
\draw[fill] (0,5) circle(.1);
\draw (0,4) -- (0,5);

\node[below] at (4,0) {$1$};

\draw[fill] (4,0) circle(.1);

\node[below] at (4,2) {$w_2'$};

\draw[fill] (4,2) circle(.1);
\draw[fill] (4,3) circle(.1);
\draw[fill] (4,4) circle(.1);
\draw (4,2) -- (4,3);
\draw (4,2) to [out=150,in=150] (4,4);
\draw[fill] (5,5) circle(.1);
\draw[fill] (5,6) circle(.1);
\draw (5,5) -- (5,6);
\draw (4,3) to [out=30,in=150] (5,5);
\draw (4,4) to [out=30,in=150] (5,6);

\node[below] at (6,2) {$x_2$};

\draw[fill] (6,2) circle(.1);
\draw[fill] (6,3) circle(.1);
\draw[fill] (6,4) circle(.1);
\draw (6,2) -- (6,3);
\draw (6,2) to [out=150,in=150] (6,4);
\draw[fill] (7,5) circle(.1);
\draw[fill] (7,6) circle(.1);
\draw (7,5) -- (7,6);
\draw (6,3) to [out=30,in=150] (7,5);
\draw (6,4) to [out=30,in=150] (7,6);
\draw[fill] (8,5) circle(.1);

\node[below] at (8,5) {$x_2x_3$};

\draw (8,5) -- (7,6);
\draw[fill] (8,7) circle(.1);
\draw (8,5) to [out=30,in=30] (8,7);
\draw[fill] (8,8) circle(.1);
\draw (8,7) -- (8,8);
\draw[fill] (8,10) circle(.1);
\draw (8,8) to [out=150,in=150] (8,10);

\node[below] at (10,4) {$w_2'x_2$};
 
\draw[fill] (10,4) circle(.1);
\draw[fill] (10,5) circle(.1);
\draw (10,4) -- (10,5);
\draw[fill] (10,6) circle(.1);
\draw (10,4) to [out=150,in=150] (10,6);
\draw[fill] (11,7) circle(.1);
\draw (10,5) to [out=30,in=150] (11,7);
\draw[fill] (10,7) circle(.1);
\draw (10,6) -- (10,7);
\draw[fill] (11,8) circle(.1);
\draw (11,7) -- (11,8);
\draw (10,6) to [out=30,in=150] (11,8);
\draw[fill] (11,9) circle(.1);
\draw (10,7) to [out=30,in=150] (11,9);
\draw[fill] (11,10) circle(.1);
\draw (11,9) -- (11,10);
\draw (11,8) to [out=30,in=30] (11,10); 

\node[below] at (12,5) {$w_3'x_2$};
 
\draw[fill] (12,5) circle(.1);
\draw[fill] (12,6) circle(.1);
\draw (12,5) -- (12,6);
\draw[fill] (12,7) circle(.1);
\draw (12,5) to [out=150,in=150] (12,7);
\draw[fill] (13,8) circle(.1);
\draw (12,6) to [out=30,in=150] (13,8);
\draw[fill] (12,8) circle(.1);
\draw (12,7) -- (12,8);
\draw[fill] (13,9) circle(.1);
\draw (13,8) -- (13,9);
\draw (12,7) to [out=30,in=150] (13,9);
\draw[fill] (13,10) circle(.1);
\draw (12,8) to [out=30,in=150] (13,10);
\draw[fill] (13,11) circle(.1);
\draw (13,10) -- (13,11);
\draw (13,9) to [out=30,in=30] (13,11);

\node at (2,5) {$\bigotimes$};
\node at (-2,-10) {$=$};

\node[below] at (0,-15) {$U$};

\draw[fill] (0,-15) circle(.1);
\draw[fill] (0,-14) circle(.1);
\draw (0,-15) -- (0,-14);
\draw[fill] (0,-13) circle(.1);
\draw[fill] (0,-12) circle(.1);
\draw (0,-13) -- (0,-12);
\draw (0,-14) to [out=150,in=150] (0,-12);
\draw[fill] (0,-11) circle(.1);
\draw (0,-13) to [out=30,in=30] (0,-11);
\draw[fill] (0,-10) circle(.1);
\draw (0,-11) -- (0,-10);

\node[below] at (2,-13) {$w_2'U$};

\draw[fill] (2,-13) circle(.1);
\draw[fill] (2,-12) circle(.1);
\draw[fill] (2,-11) circle(.1);
\draw (2,-13) -- (2,-12);
\draw (2,-13) to [out=150,in=150] (2,-11);
\draw[fill] (3,-10) circle(.1);
\draw[fill] (3,-9) circle(.1);
\draw (3,-10) -- (3,-9);
\draw (2,-12) to [out=30,in=150] (3,-10);
\draw (2,-11) to [out=30,in=150] (3,-9);
\draw[fill] (2,-10) circle(.1);
\draw (2,-11) -- (2,-10);
\draw[fill] (2,-9) circle(.1);
\draw[fill] (2,-8) circle(.1);
\draw (2,-10) to [out=30,in=30] (2,-8);
\draw (2,-9) -- (2,-8);
\draw[fill] (2,-7) circle(.1);
\draw (2,-9) to [out=150,in=150] (2,-7);

\node[below] at (4,-12) {$w_3'U$};

\draw[fill] (4,-12) circle(.1);
\draw[fill] (4,-11) circle(.1);
\draw (4,-12) -- (4,-11);
\draw[fill] (4,-10) circle(.1);
\draw (4,-12) to [out=150,in=150] (4,-10);
\draw[fill] (5,-9) circle(.1);
\draw (4,-11) to [out=30,in=150] (5,-9);
\draw[fill] (4,-9) circle(.1);
\draw (4,-10) -- (4,-9);
\draw[fill] (5,-8) circle(.1);
\draw (5,-9) -- (5,-8);
\draw (4,-10) to [out=30,in=150] (5,-8);
\draw[fill] (5,-7) circle(.1);
\draw (4,-9) to [out=30,in=150] (5,-7);
\draw[fill] (5,-6) circle(.1);
\draw (5,-7) -- (5,-6);
\draw (5,-8) to [out=30,in=30] (5,-6); 

\node[below] at (6,-11) {$w_1^2w_2'U$};

\draw[fill] (6,-11) circle(.1);
\draw[fill] (6,-10) circle(.1);
\draw (6,-11) -- (6,-10);
\draw[fill] (6,-9) circle(.1);
\draw (6,-11) to [out=150,in=150] (6,-9);
\draw[fill] (7,-8) circle(.1);
\draw (6,-10) to [out=30,in=150] (7,-8);
\draw[fill] (6,-8) circle(.1);
\draw (6,-9) -- (6,-8);
\draw[fill] (7,-7) circle(.1);
\draw (7,-8) -- (7,-7);
\draw (6,-9) to [out=30,in=150] (7,-7);
\draw[fill] (7,-6) circle(.1);
\draw (6,-8) to [out=30,in=150] (7,-6);
\draw[fill] (7,-5) circle(.1);
\draw (7,-6) -- (7,-5);
\draw (7,-7) to [out=30,in=30] (7,-5);

\node[below] at (9,-13) {$x_2U$};

\draw[fill] (9,-13) circle(.1);
\draw[fill] (9,-12) circle(.1);
\draw[fill] (9,-11) circle(.1);
\draw (9,-13) -- (9,-12);
\draw (9,-13) to [out=150,in=150] (9,-11);
\draw[fill] (10,-10) circle(.1);
\draw[fill] (10,-9) circle(.1);
\draw (10,-10) -- (10,-9);
\draw (9,-12) to [out=30,in=150] (10,-10);
\draw (9,-11) to [out=30,in=150] (10,-9);
\draw[fill] (9,-10) circle(.1);
\draw (9,-11) -- (9,-10);
\draw[fill] (9,-9) circle(.1);
\draw[fill] (9,-8) circle(.1);
\draw (9,-10) to [out=30,in=30] (9,-8);
\draw (9,-9) -- (9,-8);
\draw[fill] (9,-7) circle(.1);
\draw (9,-9) to [out=150,in=150] (9,-7);

\node[below] at (11,-12) {$x_3U$};

\draw[fill] (11,-12) circle(.1);
\draw[fill] (11,-11) circle(.1);
\draw (11,-12) -- (11,-11);
\draw[fill] (11,-10) circle(.1);
\draw (11,-12) to [out=150,in=150] (11,-10);
\draw[fill] (12,-9) circle(.1);
\draw (11,-11) to [out=30,in=150] (12,-9);
\draw[fill] (11,-9) circle(.1);
\draw (11,-10) -- (11,-9);
\draw[fill] (12,-8) circle(.1);
\draw (12,-9) -- (12,-8);
\draw (11,-10) to [out=30,in=150] (12,-8);
\draw[fill] (12,-7) circle(.1);
\draw (11,-9) to [out=30,in=150] (12,-7);
\draw[fill] (12,-6) circle(.1);
\draw (12,-7) -- (12,-6);
\draw (12,-8) to [out=30,in=30] (12,-6); 

\node[below] at (13,-11) {$w_1^2x_2U$};

\draw[fill] (13,-11) circle(.1);
\draw[fill] (13,-10) circle(.1);
\draw (13,-11) -- (13,-10);
\draw[fill] (13,-9) circle(.1);
\draw (13,-11) to [out=150,in=150] (13,-9);
\draw[fill] (14,-8) circle(.1);
\draw (13,-10) to [out=30,in=150] (14,-8);
\draw[fill] (13,-8) circle(.1);
\draw (13,-9) -- (13,-8);
\draw[fill] (14,-7) circle(.1);
\draw (14,-8) -- (14,-7);
\draw (13,-9) to [out=30,in=150] (14,-7);
\draw[fill] (14,-6) circle(.1);
\draw (13,-8) to [out=30,in=150] (14,-6);
\draw[fill] (14,-5) circle(.1);
\draw (14,-6) -- (14,-5);
\draw (14,-7) to [out=30,in=30] (14,-5);

\node[below] at (15,-10) {$x_2x_3U$};

\draw[fill] (15,-10) circle(.1);
\draw[fill] (15,-9) circle(.1);
\draw (15,-10) -- (15,-9);
\draw[fill] (15,-8) circle(.1);
\draw (15,-10) to [out=150,in=150] (15,-8);
\draw[fill] (16,-7) circle(.1);
\draw (15,-9) to [out=30,in=150] (16,-7);
\draw[fill] (15,-7) circle(.1);
\draw (15,-8) -- (15,-7);
\draw[fill] (16,-6) circle(.1);
\draw (16,-7) -- (16,-6);
\draw (15,-8) to [out=30,in=150] (16,-6);
\draw[fill] (16,-5) circle(.1);
\draw (15,-7) to [out=30,in=150] (16,-5);
\draw[fill] (16,-4) circle(.1);
\draw (16,-5) -- (16,-4);
\draw (16,-6) to [out=30,in=30] (16,-4);

\node[below] at (20,-11) {$w_2'x_2U$};

\draw[fill] (20,-11) circle(.1);
\draw[fill] (20,-10) circle(.1);
\draw (20,-11) -- (20,-10);
\draw[fill] (20,-9) circle(.1);
\draw (20,-11) to [out=150,in=150] (20,-9);
\draw[fill] (21,-8) circle(.1);
\draw (20,-10) to [out=30,in=150] (21,-8);
\draw[fill] (20,-8) circle(.1);
\draw (20,-9) -- (20,-8);
\draw[fill] (21,-7) circle(.1);
\draw (21,-8) -- (21,-7);
\draw (20,-9) to [out=30,in=150] (21,-7);
\draw[fill] (21,-6) circle(.1);
\draw (20,-8) to [out=30,in=150] (21,-6);
\draw[fill] (21,-5) circle(.1);
\draw (21,-6) -- (21,-5);
\draw (21,-7) to [out=30,in=30] (21,-5); 

\node[below] at (22,-10) {$w_1w_2'x_2U$};

\draw[fill] (22,-10) circle(.1);
\draw[fill] (22,-9) circle(.1);
\draw (22,-10) -- (22,-9);
\draw[fill] (22,-8) circle(.1);
\draw (22,-10) to [out=150,in=150] (22,-8);
\draw[fill] (23,-7) circle(.1);
\draw (22,-9) to [out=30,in=150] (23,-7);
\draw[fill] (22,-7) circle(.1);
\draw (22,-8) -- (22,-7);
\draw[fill] (23,-6) circle(.1);
\draw (23,-7) -- (23,-6);
\draw (22,-8) to [out=30,in=150] (23,-6);
\draw[fill] (23,-5) circle(.1);
\draw (22,-7) to [out=30,in=150] (23,-5);
\draw[fill] (23,-4) circle(.1);
\draw (23,-5) -- (23,-4);
\draw (23,-6) to [out=30,in=30] (23,-4); 

\node[below] at (24,-10) {$w_3'x_2U$};

\draw[fill] (24,-10) circle(.1);
\draw[fill] (24,-9) circle(.1);
\draw (24,-10) -- (24,-9);
\draw[fill] (24,-8) circle(.1);
\draw (24,-10) to [out=150,in=150] (24,-8);
\draw[fill] (25,-7) circle(.1);
\draw (24,-9) to [out=30,in=150] (25,-7);
\draw[fill] (24,-7) circle(.1);
\draw (24,-8) -- (24,-7);
\draw[fill] (25,-6) circle(.1);
\draw (25,-7) -- (25,-6);
\draw (24,-8) to [out=30,in=150] (25,-6);
\draw[fill] (25,-5) circle(.1);
\draw (24,-7) to [out=30,in=150] (25,-5);
\draw[fill] (25,-4) circle(.1);
\draw (25,-5) -- (25,-4);
\draw (25,-6) to [out=30,in=30] (25,-4);

\end{tikzpicture}
\end{center}
\caption{The $\A_2(1)$-module structure of $\H^{*+1}(M\tO(1),\Z_2)\otimes\H^*(\B \PSU(2)\times \B ^2\Z_2,\Z_2)$}
\label{fig:H^{*+1}(MO(1),Z_2)otimesH^*(BPSU(2)timesB^2Z_2,Z_2)}
\end{figure}

\begin{figure}[!h]
\begin{center}
\begin{tikzpicture}
\node at (0,-1) {0};
\node at (1,-1) {1};
\node at (2,-1) {2};
\node at (3,-1) {3};
\node at (4,-1) {4};
\node at (5,-1) {5};
\node at (6,-1) {$t-s$};
\node at (-1,0) {0};
\node at (-1,1) {1};
\node at (-1,2) {2};
\node at (-1,3) {3};
\node at (-1,4) {4};
\node at (-1,5) {5};
\node at (-1,6) {$s$};

\draw[->] (-0.5,-0.5) -- (-0.5,6);
\draw[->] (-0.5,-0.5) -- (6,-0.5);

\draw (0,0) -- (2,2);
\draw (2,2) -- (2,0);
\draw[fill] (2.1,0) circle(0.05);
\draw[fill] (2.2,0) circle(0.05);
\draw[fill] (3,0) circle(0.05);
\draw[fill] (3.1,0) circle(0.05);
\draw[fill] (3.9,0) circle(0.05);
\draw[fill] (4,0) circle(0.05);
\draw[fill] (4.1,0) circle(0.05);
\draw[fill] (4.9,0) circle(0.05);
\draw[fill] (5,0) circle(0.05);
\draw[fill] (5.1,0) circle(0.05);

\end{tikzpicture}
\end{center}
\caption{$\Omega_*^{\Pin^-}(\B \PSU(2)\times \B ^2\Z_2)$}
\label{fig:Omega_*^{Pin^-}(BPSU(2)timesB^2Z_2)}
\end{figure}

\begin{theorem}
\begin{table}[!h]
\centering
\begin{tabular}{c c}
\hline
$i$ & $\Omega^{\Pin^-}_i(\B \PSU(2)\times \B ^2\Z_2)$\\
\hline
0& $\Z_2$\\
1& $\Z_2$\\
2& $\Z_2^2\times\Z_8$\\
3 & $\Z_2^2$\\
4 & $\Z_2^3$\\ 
5 & $\Z_2^3$\\
\hline
\end{tabular}
\end{table}
\end{theorem}

The bordism invariants of $\Omega^{\Pin^-}_2(\B \PSU(2)\times \B ^2\Z_2)$ are $w_2',x_2,\text{ABK}$.

The bordism invariants of $\Omega^{\Pin^-}_3(\B \PSU(2)\times \B ^2\Z_2)$ are $w_1w_2'=w_3',w_1x_2=x_3$.

The bordism invariants of $\Omega^{\Pin^-}_4(\B \PSU(2)\times \B ^2\Z_2)$ are $w_1^2w_2',w_1^2x_2,w_2'x_2$.

The bordism invariants of $\Omega^{\Pin^-}_5(\B \PSU(2)\times \B ^2\Z_2)$ are $x_2x_3,w_3'x_2,w_1w_2'x_2{(=w_2'x_3+w_3'x_2)}$.

\begin{theorem}
\begin{table}[!h]
\centering
\begin{tabular}{c c}
\hline
$i$ & $\TP_i(\Pin^-\times \PSU(2)\times \B \Z_2)$\\
\hline
0& $\Z_2$\\
1& $\Z_2$\\
2& $\Z_2^2\times\Z_8$\\
3 & $\Z_2^2$\\
4 & $\Z_2^3$\\ 
5 & $\Z_2^3$\\
\hline
\end{tabular}
\end{table}
\end{theorem}

The 2d topological terms are $w_2',x_2,\text{ABK}$.

The 3d topological terms are $w_1w_2'=w_3',w_1x_2=x_3$.

The 4d topological terms are $w_1^2w_2',w_1^2x_2,w_2'x_2$.

The 5d topological terms are $x_2x_3,w_3'x_2,w_1w_2'x_2{(=w_2'x_3+w_3'x_2)}$.

\subsubsection{$\Omega^{\tO}_d(\B \PSU(3)\times \B ^2\Z_3)$}
\bea
\Ext_{\A_3}^{s,t}(\H^*(M\tO\wedge(\B \PSU(3)\times \B ^2\Z_3)_+,\Z_3),\Z_3)\Rightarrow\Omega_{t-s}^{\tO}(\B \PSU(3)\times \B ^2\Z_3)_3^{\wedge}.
\eea

Since $\H^*(M\tO,\Z_3)=0$, $\Omega_d^{\tO}(\B \PSU(3)\times \B ^2\Z_3)_3^{\wedge}=0$.
\bea
\Ext_{\A_2}^{s,t}(\H^*(M\tO\wedge(\B \PSU(3)\times \B ^2\Z_3)_+,\Z_2),\Z_2)\Rightarrow\Omega_{t-s}^{\tO}(\B \PSU(3)\times \B ^2\Z_3)_2^{\wedge}.
\eea

Since $\H^*(\B \PSU(3)\times \B ^2\Z_3,\Z_2)=\H^*(\B \PSU(3),\Z_2)$, $\Omega_d^{\tO}(\B \PSU(3)\times \B ^2\Z_3)_2^{\wedge}=\Omega_d^{\tO}(\B \PSU(3))_2^{\wedge}$.

\begin{theorem}
\begin{table}[!h]
\centering
\begin{tabular}{c c}
\hline
$i$ & $\Omega^{\tO}_i(\B \PSU(3)\times \B ^2\Z_3)$\\
\hline
0& $\Z_2$\\
1& $0$\\
2& $\Z_2$\\
3 & $0$\\
4 & $\Z_2^3$\\ 
5 & $\Z_2$\\
\hline
\end{tabular}
\end{table}
\end{theorem}

The bordism invariant of $\Omega^{\tO}_2(\B \PSU(3)\times \B ^2\Z_3)$ is $w_1^2$.

The bordism invariants of $\Omega^{\tO}_4(\B \PSU(3)\times \B ^2\Z_3)$ are $w_1^4,w_2^2,c_2(\mod2)$.

The bordism invariant of $\Omega^{\tO}_5(\B \PSU(3)\times \B ^2\Z_3)$ is $w_2w_3$.

\begin{theorem}
\begin{table}[!h]
\centering
\begin{tabular}{c c}
\hline
$i$ & $\TP_i(\tO\times \PSU(3)\times \B \Z_3)$\\
\hline
0& $\Z_2$\\
1& $0$\\
2& $\Z_2$\\
3 & $0$\\
4 & $\Z_2^3$\\ 
5 & $\Z_2$\\
\hline
\end{tabular}
\end{table}
\end{theorem}

The 2d topological term is $w_1^2$.

The 4d topological terms are $w_1^4,w_2^2,c_2(\mod2)$.

The 5d topological term is $w_2w_3$.

\subsubsection{$\Omega^{\SO}_d(\B \PSU(3)\times \B ^2\Z_3)$}
\bea
\Ext_{\A_2}^{s,t}(\H^*(M\SO\wedge(\B \PSU(3)\times \B ^2\Z_3)_+,\Z_2),\Z_2)\Rightarrow\Omega_{t-s}^{\SO}(\B \PSU(3)\times \B ^2\Z_3)_2^{\wedge}.
\eea

Since $\H^*(\B \PSU(3)\times \B ^2\Z_3,\Z_2)=\H^*(\B \PSU(3),\Z_2)$, $\Omega_d^{\SO}(\B \PSU(3)\times \B ^2\Z_3)_2^{\wedge}=\Omega_d^{\SO}(\B \PSU(3))_2^{\wedge}$.
\bea
\Ext_{\A_3}^{s,t}(\H^*(M\SO\wedge(\B \PSU(3)\times \B ^2\Z_3)_+,\Z_3),\Z_3)\Rightarrow\Omega_{t-s}^{\SO}(\B \PSU(3)\times \B ^2\Z_3)_3^{\wedge}.
\eea

$\beta_{(3,3)} x_2'=x_3'$, $\beta_{(3,3)} z_2=z_3$, $\beta_{(3,3)} x_2'^2=2x_2'x_3'$, $\beta_{(3,3)}(x_2'z_2)=x_2'z_3+x_3'z_2$, $\beta_{(3,3)}(x_2'z_3)=x_3'z_3=-\beta_{(3,3)}(x_3'z_2)$.

The $E_2$ page is shown in Figure \ref{fig:Omega_*^{SO}(BPSU(3)timesB^2Z_3)_3}.

\begin{figure}[!h]
\begin{center}
\begin{tikzpicture}
\node at (0,-1) {0};
\node at (1,-1) {1};
\node at (2,-1) {2};
\node at (3,-1) {3};
\node at (4,-1) {4};
\node at (5,-1) {5};
\node at (6,-1) {$t-s$};
\node at (-1,0) {0};
\node at (-1,1) {1};
\node at (-1,2) {2};
\node at (-1,3) {3};
\node at (-1,4) {4};
\node at (-1,5) {5};
\node at (-1,6) {$s$};

\draw[->] (-0.5,-0.5) -- (-0.5,6);
\draw[->] (-0.5,-0.5) -- (6,-0.5);

\draw (0,0) -- (0,5);
\draw[fill] (1.9,0) circle(0.05);
\draw[fill] (2.1,0) circle(0.05);
\draw (4,1) -- (4,5);
\draw (4.1,0) -- (4.1,5);
\draw[fill] (3.9,0) circle(0.05);
\draw[fill] (4,0) circle(0.05);
\draw[fill] (5,0) circle(0.05);
\end{tikzpicture}
\end{center}
\caption{$\Omega_*^{\SO}(\B \PSU(3)\times \B ^2\Z_3)_3^{\wedge}$}
\label{fig:Omega_*^{SO}(BPSU(3)timesB^2Z_3)_3}
\end{figure}

\begin{theorem}
\begin{table}[!h]
\centering
\begin{tabular}{c c}
\hline
$i$ & $\Omega^{\SO}_i(\B \PSU(3)\times \B ^2\Z_3)$\\
\hline
0& $\Z$\\
1& $0$\\
2& $\Z_3^2$\\
3 & $0$\\
4 & $\Z^2\times\Z_3^2$\\ 
5 & $\Z_2\times\Z_3$\\
\hline
\end{tabular}
\end{table}
\end{theorem}

The bordism invariants of $\Omega^{\SO}_{2}(\B \PSU(3)\times \B ^2\Z_3)$ are $x_2',z_2$.

The bordism invariants of $\Omega^{\SO}_{4}(\B \PSU(3)\times \B ^2\Z_3)$ are $\sigma$, $c_2$, $x_2'^2$ and $x_2'z_2$.

The bordism invariants of $\Omega^{\SO}_5(\B \PSU(3)\times \B ^2\Z_3)$ are $w_2w_3,z_2x_3'(=-z_3x_2')$.

\begin{theorem}
\begin{table}[!h]
\centering
\begin{tabular}{c c}
\hline
$i$ & $\TP_i(\SO\times \PSU(3)\times \B \Z_3)$\\
\hline
0& $0$\\
1& $0$\\
2& $\Z_3^2$\\
3 & $\Z^2$\\
4 & $\Z_3^2$\\ 
5 & $\Z\times\Z_2\times\Z_3$\\
\hline
\end{tabular}
\end{table}
\end{theorem}

The 2d topological terms are $x_2',z_2$.

The 3d topological terms are $\frac{1}{3}\text{CS}_3^{(TM)}$, $\text{CS}_3^{(\PSU(3))}$.

The 4d topological terms are $x_2'^2$ and $x_2'z_2$.

The 5d topological terms are $\text{CS}_5^{(\PSU(3))},w_2w_3,z_2x_3'(=-z_3x_2')$.

\subsubsection{$\Omega^{\Spin}_d(\B \PSU(3)\times \B ^2\Z_3)$}
\bea
\Ext_{\A_2}^{s,t}(\H^*(M\Spin\wedge(\B \PSU(3)\times \B ^2\Z_3)_+,\Z_2),\Z_2)\Rightarrow\Omega_{t-s}^{\Spin}(\B \PSU(3)\times \B ^2\Z_3)_2^{\wedge}.
\eea

Since $\H^*(\B \PSU(3)\times \B ^2\Z_3,\Z_2)=\H^*(\B \PSU(3),\Z_2)$, $\Omega_d^{\Spin}(\B \PSU(3)\times \B ^2\Z_3)_2^{\wedge}=\Omega_d^{\Spin}(\B \PSU(3))_2^{\wedge}$.
\bea
\Ext_{\A_3}^{s,t}(\H^*(M\Spin\wedge(\B \PSU(3)\times \B ^2\Z_3)_+,\Z_3),\Z_3)\Rightarrow\Omega_{t-s}^{\Spin}(\B \PSU(3)\times \B ^2\Z_3)_3^{\wedge}.
\eea

Since $\H^*(M\SO,\Z_3)=\H^*(M\Spin,\Z_3)$, $\Omega_d^{\Spin}(\B \PSU(3)\times \B ^2\Z_3)_3^{\wedge}=\Omega_d^{\SO}(\B \PSU(3)\times \B ^2\Z_3)_3^{\wedge}$.

\begin{theorem}
\begin{table}[!h]
\centering
\begin{tabular}{c c}
\hline
$i$ & $\Omega^{\Spin}_i(\B \PSU(3)\times \B ^2\Z_3)$\\
\hline
0& $\Z$\\
1& $\Z_2$\\
2& $\Z_2\times\Z_3^2$\\
3 & $0$\\
4 & $\Z^2\times\Z_3^2$\\ 
5 & $\Z_3$\\
\hline
\end{tabular}
\end{table}
\end{theorem}

The bordism invariants of $\Omega^{\Spin}_{2}(\B \PSU(3)\times \B ^2\Z_3)$ are $\text{Arf},x_2',z_2$.

The bordism invariants of $\Omega^{\Spin}_{4}(\B \PSU(3)\times \B ^2\Z_3)$ are $\frac{\sigma}{16}$, $c_2$, $x_2'^2$ and $x_2'z_2$.

The bordism invariant of $\Omega^{\Spin}_5(\B \PSU(3)\times \B ^2\Z_3)$ is $z_2x_3'(=-z_3x_2')$.

\begin{theorem}
\begin{table}[!h]
\centering
\begin{tabular}{c c}
\hline
$i$ & $\TP_i(\Spin\times \PSU(3)\times \B \Z_3)$\\
\hline
0& $0$\\
1& $\Z_2$\\
2& $\Z_2\times\Z_3^2$\\
3 & $\Z^2$\\
4 & $\Z_3^2$\\ 
5 & $\Z\times\Z_3$\\
\hline
\end{tabular}
\end{table}
\end{theorem}

The 2d topological terms are $\text{Arf},x_2',z_2$.

The 3d topological terms are $\frac{1}{48}\text{CS}_3^{(TM)}$, $\text{CS}_3^{(\PSU(3))}$.

The 4d topological terms are $x_2'^2$ and $x_2'z_2$.

The 5d topological terms are $\frac{1}{2}\text{CS}_5^{(\PSU(3))},z_2x_3'(=-z_3x_2')$.

\subsubsection{$\Omega^{\Pin^+}_d(\B \PSU(3)\times \B ^2\Z_3)$}
\bea
\Ext_{\A_3}^{s,t}(\H^*(M\Pin^-\wedge(\B \PSU(3)\times \B ^2\Z_3)_+,\Z_3),\Z_3)\Rightarrow\Omega_{t-s}^{\Pin^+}(\B \PSU(3)\times \B ^2\Z_3)_3^{\wedge}.
\eea

Since $\H^*(M\Pin^-,\Z_3)=0$, $\Omega_d^{\Pin^+}(\B \PSU(3)\times \B ^2\Z_3)_3^{\wedge}=0$.
\bea
\Ext_{\A_2}^{s,t}(\H^*(M\Pin^-\wedge(\B \PSU(3)\times \B ^2\Z_3)_+,\Z_2),\Z_2)\Rightarrow\Omega_{t-s}^{\Pin^+}(\B \PSU(3)\times \B ^2\Z_3)_2^{\wedge}.
\eea

Since $\H^*(\B \PSU(3)\times \B ^2\Z_3,\Z_2)=\H^*(\B \PSU(3),\Z_2)$, $\Omega_d^{\Pin^+}(\B \PSU(3)\times \B ^2\Z_3)_2^{\wedge}=\Omega_d^{\Pin^+}(\B \PSU(3))_2^{\wedge}$.

\begin{theorem}
\begin{table}[!h]
\centering
\begin{tabular}{c c}
\hline
$i$ & $\Omega^{\Pin^+}_i(\B \PSU(3)\times \B ^2\Z_3)$\\
\hline
0& $\Z_2$\\
1& $0$\\
2& $\Z_2$\\
3 & $\Z_2$\\
4 & $\Z_2\times\Z_{16}$\\
5 & $0$\\
\hline
\end{tabular}
\end{table}
\end{theorem}

The bordism invariant of $\Omega^{\Pin^+}_2(\B \PSU(3)\times \B ^2\Z_3)$ is $w_1\tilde{\eta}$.

The bordism invariant of $\Omega^{\Pin^+}_3(\B \PSU(3)\times \B ^2\Z_3)$ is $w_1\text{Arf}$.

The bordism invariants of $\Omega^{\Pin^+}_4(\B \PSU(3)\times \B ^2\Z_3)$ are $c_2(\mod2)$ and $\eta$.

\begin{theorem}
\begin{table}[!h]
\centering
\begin{tabular}{c c}
\hline
$i$ & $\TP_i(\Pin^+\times \PSU(3)\times \B \Z_3)$\\
\hline
0& $\Z_2$\\
1& $0$\\
2& $\Z_2$\\
3 & $\Z_2$\\
4 & $\Z_2\times\Z_{16}$\\
5 & $0$\\
\hline
\end{tabular}
\end{table}
\end{theorem}

The 2d topological term is $w_1\tilde{\eta}$.

The 3d topological term is $w_1\text{Arf}$.

The 4d topological terms are $c_2(\mod2)$ and $\eta$.

\subsubsection{$\Omega^{\Pin^-}_d(\B \PSU(3)\times \B ^2\Z_3)$}
\bea
\Ext_{\A_3}^{s,t}(\H^*(M\Pin^+\wedge(\B \PSU(3)\times \B ^2\Z_3)_+,\Z_3),\Z_3)\Rightarrow\Omega_{t-s}^{\Pin^-}(\B \PSU(3)\times \B ^2\Z_3)_3^{\wedge}.
\eea

Since $\H^*(M\Pin^+,\Z_3)=0$, $\Omega_d^{\Pin^-}(\B \PSU(3)\times \B ^2\Z_3)_3^{\wedge}=0$.
\bea
\Ext_{\A_2}^{s,t}(\H^*(M\Pin^+\wedge(\B \PSU(3)\times \B ^2\Z_3)_+,\Z_2),\Z_2)\Rightarrow\Omega_{t-s}^{\Pin^-}(\B \PSU(3)\times \B ^2\Z_3)_2^{\wedge}.
\eea

Since $\H^*(\B \PSU(3)\times \B ^2\Z_3,\Z_2)=\H^*(\B \PSU(3),\Z_2)$, $\Omega_d^{\Pin^-}(\B \PSU(3)\times \B ^2\Z_3)_2^{\wedge}=\Omega_d^{\Pin^-}(\B \PSU(3))_2^{\wedge}$.

\begin{theorem}
\begin{table}[!h]
\centering
\begin{tabular}{c c}
\hline
$i$ & $\Omega^{\Pin^-}_i(\B \PSU(3)\times \B ^2\Z_3)$\\
\hline
0& $\Z_2$\\
1& $\Z_2$\\
2& $\Z_8$\\
3 & $0$\\
4 & $\Z_2$\\
5 & $0$\\
\hline
\end{tabular}
\end{table}
\end{theorem}

The bordism invariant of $\Omega^{\Pin^-}_2(\B \PSU(3)\times \B ^2\Z_3)$ is ABK.

The bordism invariant of $\Omega^{\Pin^-}_4(\B \PSU(3)\times \B ^2\Z_3)$ is $c_2(\mod2)$.

\begin{theorem}
\begin{table}[!h]
\centering
\begin{tabular}{c c}
\hline
$i$ & $\TP_i(\Pin^-\times \PSU(3)\times \B \Z_3)$\\
\hline
0& $\Z_2$\\
1& $\Z_2$\\
2& $\Z_8$\\
3 & $0$\\
4 & $\Z_2$\\
5 & $0$\\
\hline
\end{tabular}
\end{table}
\end{theorem}

The 2d topological term is ABK.

The 4d topological term is $c_2(\mod2)$.

\section{More computation of $\rm{O}/\SO$ bordism groups}

\label{sec:more-cobor}

\subsection{Summary}

\cred{Below we use the following notations, all cohomology class are pulled back to the $d$-manifold $M$ along the maps given in the definition of cobordism groups:\\
$\bullet$ $w_i$ is the Stiefel-Whitney class of the tangent bundle of $M$,\\
$\bullet$ $a$ is the generator of $\H^1(\B\Z_2,\Z_2)$,\\
$\bullet$ $a'$ is the generator of $\H^1(\B\Z_3,\Z_3)$, $b'=\beta_{(3,3)}a',$\\
$\bullet$ $x_2$ is the generator of $\H^2(\B^2\Z_2,\Z_2)$, $x_3=\Sq^1x_2$, $x_5=\Sq^2x_3$,\\
$\bullet$ $x_2'$ is the generator of $\H^2(\B^2\Z_3,\Z_3)$, $x_3'=\beta_{(3,3)}x_2'$,\\
$\bullet$ $x_2''$ is the generator of $\H^2(\B^2\Z_4,\Z_4)$, $x_3''=\beta_{(2,4)}x_2''$, $x_5''=\Sq^2x_3''$,\\
$\bullet$ $w_i'=w_i(\tO(n))\in\H^i(\B\tO(n),\Z_2)$ is the Stiefel-Whitney class of the principal $\tO(n)$ bundle,\\
$\bullet$ $p_1'=p_1(\tO(n))\in\H^4(\B\tO(n),\Z_2)$ is the first Pontryagin class of the principal $\tO(n)$ bundle,\\
$\bullet$ $z_2=w_2(\PSU(3))\in\H^2(\B\PSU(3),\Z_3)$ is the generalized Stiefel-Whitney class of the principal $\PSU(3)$ bundle, $z_3=\beta_{(3,3)}z_2$.\\
$\bullet$ $z_2'=w_2(\PSU(4))\in\H^2(\B\PSU(4),\Z_4)$ is the generalized Stiefel-Whitney class of the principal $\PSU(4)$ bundle, $z_3'=\beta_{(2,4)}z_2'$.\\
$\bullet$ For $n>1$, we also use the notation $a$ for the generator of $\H^1(\B\Z_{2^n},\Z_{2^n})$, $\tilde{a}=a\mod2$, $b$ is the generator of $\H^2(\B\Z_{2^n},\Z_{2^n})$, $\tilde{b}=b\mod2$ and $\tilde{b}=\beta_{(2,2^n)}a$.\\
$\bullet$ For $n>1$, we also use the notation $a'$ for the generator of $\H^1(\B\Z_{3^n},\Z_{3^n})$, $\tilde{a}'=a'\mod3$, $b'$ is the generator of $\H^2(\B\Z_{3^n},\Z_{3^n})$, $\tilde{b}'=b'\mod3$ and $\tilde{b}'=\beta_{(3,3^n)}a'$.\\
$\bullet$ $\cP_2$ is the Pontryagin square (see \ref{sec:plan-convention}).\\
$\bullet$ $\mathfrak{P}_3$ is the Postnikov square (see \ref{sec:plan-convention}).\\
Convention:
All product between cohomology classes are cup product.
}

\begin{table}[!h] %[!h] %[tb]
\centering
 \makebox[\textwidth][r]{
 \begin{tabular}{ |c| c | c|  c| c | c| c| c| c| c|}
\hline
$\Omega_d^H(-)$ & $\B \tO(3)$ &  $\B \tO(4)$ & $\B \tO(5)$  &   $\B (\Z_2\ltimes\PSU(3))$ &  $\B (\Z_2\ltimes\PSU(4))$
\\
\hline
$2$ SO& 
\begin{minipage}[c]{.5in}
$\Z_2:$\\
$w_2'$
\end{minipage}
&
\begin{minipage}[c]{.5in}
$\Z_2:$\\
$w_2'$
\end{minipage}
&
\begin{minipage}[c]{.5in}
$\Z_2:$\\
$w_2'$
\end{minipage}
&
\begin{minipage}[c]{.5in}
$\Z_3:$\\
$z_2$
\end{minipage}
&
\begin{minipage}[c]{.5in}
$\Z_4:$\\
$z_2'$
\end{minipage}

\\
\hline

$2$ O&
\begin{minipage}[c]{.5in}
$\Z_2^3:$\\
$w_1^2$,\\
$w_1'^2,w_2'$
\end{minipage}
&
\begin{minipage}[c]{.5in}
$\Z_2^3:$\\
$w_1^2$,\\
$w_1'^2,w_2'$
\end{minipage}
&
\begin{minipage}[c]{.5in}
$\Z_2^3:$\\
$w_1^2$,\\
$w_1'^2,w_2'$
\end{minipage}
&
\begin{minipage}[c]{.5in}
$\Z_2^2:$\\
$w_1^2,a^2$
\end{minipage}
&
\begin{minipage}[c]{.5in}
$\Z_2^2:$\\
$w_1^2,a^2$,\\
$\tilde{z}_2'$
\end{minipage}

\\
\hline
 \end{tabular}
 }%\hspace*{-10mm}
\caption{$2d$ bordism groups-1.}
 \label{2d bordism groups-1}
\end{table}

\begin{table}[!h] %[!h] %[tb]
\centering
 \makebox[\textwidth][r]{
 \begin{tabular}{ |c| c | c|  c| c | c| c| c| c| c|}
\hline
$\Omega_d^H(-)$ & $\B ^2\Z_4$ &  $\B\Z_4\times\B^2\Z_2$ & $\B\Z_6\times\B^2\Z_3$  &   $\B\Z_8\times\B^2\Z_2$ & $\B\Z_{18}\times \B^2\Z_3$
\\
\hline
$2$ SO& 
\begin{minipage}[c]{.5in}
$\Z_4:$\\
$x_2''$
\end{minipage}

&
\begin{minipage}[c]{.5in}
$\Z_2:$\\
$x_2$
\end{minipage}

&
\begin{minipage}[c]{.5in}
$\Z_3:$\\
$x_2'$
\end{minipage}
&
\begin{minipage}[c]{.5in}
$\Z_2:$\\
$x_2$
\end{minipage}
&
\begin{minipage}[c]{.5in}
$\Z_3:$\\
$x_2'$
\end{minipage}

\\
\hline

$2$ O&
\begin{minipage}[c]{.5in}
$\Z_2^2:$\\
$w_1^2,\tilde{x}_2''$
\end{minipage}
&
\begin{minipage}[c]{.5in}
$\Z_2^3:$\\
$w_1^2,\tilde{b},$\\
$x_2$
\end{minipage}
&
\begin{minipage}[c]{.5in}
$\Z_2^2:$\\
$w_1^2,a^2$
\end{minipage}
&
\begin{minipage}[c]{.5in}
$\Z_2^3:$\\
$w_1^2,\tilde{b},$\\
$x_2$
\end{minipage}

&
\begin{minipage}[c]{.5in}
$\Z_2^2:$\\
$w_1^2,a^2$
\end{minipage}

\\
\hline
 \end{tabular}
 }%\hspace*{-10mm}
\caption{$2d$ bordism groups-2.}
 \label{2d bordism groups-2}
\end{table}

\begin{table}[!h] %[!h] %[tb]
\centering
 \makebox[\textwidth][r]{
 \begin{tabular}{ |c| c | c|  c| c | c| c| c| c| c|}
\hline
$\Omega_d^H(-)$ & $\B \tO(3)$ &  $\B \tO(4)$ & $\B \tO(5)$  &   $\B (\Z_2\ltimes\PSU(3))$ &  $\B (\Z_2\ltimes\PSU(4))$
\\
\hline
$3$ SO& 
\begin{minipage}[c]{.7in}
$\Z_2^2:$\\
$w_1'^3$,\\
$w_1'w_2'=w_3'$
\end{minipage}
&
\begin{minipage}[c]{.7in}
$\Z_2^2:$\\
$w_1'^3$,\\
$w_1'w_2'=w_3'$
\end{minipage}
&
\begin{minipage}[c]{.7in}
$\Z_2^2:$\\
$w_1'^3$,\\
$w_1'w_2'=w_3'$
\end{minipage}
&
\begin{minipage}[c]{.5in}
$\Z_2:$\\
$a^3$
\end{minipage}
&
\begin{minipage}[c]{.5in}
$\Z_2^2:$\\
$a^3,a\tilde{z}_2'$
\end{minipage}

\\
\hline

$3$ O&
\begin{minipage}[c]{.7in}
$\Z_2^4:$\\
$w_1'w_1^2,w_1'^3$,\\
$w_1'w_2',w_3'$
\end{minipage}
&
\begin{minipage}[c]{.7in}
$\Z_2^4:$\\
$w_1'w_1^2,w_1'^3$,\\
$w_1'w_2',w_3'$
\end{minipage}
&
\begin{minipage}[c]{.7in}
$\Z_2^4:$\\
$w_1'w_1^2,w_1'^3$,\\
$w_1'w_2',w_3'$
\end{minipage}
&
\begin{minipage}[c]{.5in}
$\Z_2^2:$\\
$a^3,aw_1^2$
\end{minipage}
&
\begin{minipage}[c]{.5in}
$\Z_2^4:$\\
$a^3,aw_1^2$,\\
$z_3',a\tilde{z}_2'$
\end{minipage}

\\
\hline
 \end{tabular}
 }%\hspace*{-10mm}
\caption{$3d$ bordism groups-1.}
 \label{3d bordism groups-1}
\end{table}
%%%%%%%%%%
%%%%%%%%%%

\begin{table}[!h] %[!h] %[tb]
\centering
 \makebox[\textwidth][r]{
 \begin{tabular}{ |c| c | c|  c| c | c| c| c| c| c|}
\hline
$\Omega_d^H(-)$ & $\B ^2\Z_4$ &  $\B\Z_4\times\B^2\Z_2$ & $\B\Z_6\times\B^2\Z_3$  &   $\B\Z_8\times\B^2\Z_2$ & $\B\Z_{18}\times \B^2\Z_3$
\\
\hline
$3$ SO& 
\begin{minipage}[c]{.5in}
$0$
\end{minipage}

&
\begin{minipage}[c]{.5in}
$\Z_4\times\Z_2$:\\
$ab,\tilde{a}x_2$
\end{minipage}
&
\begin{minipage}[c]{.5in}
$\Z_3^2\times\Z_2$:\\
$a'b',a'x_2'$,\\
$a^3$
\end{minipage}
&
\begin{minipage}[c]{.5in}
$\Z_8\times\Z_2$:\\
$ab,\tilde{a}x_2$
\end{minipage}
&
\begin{minipage}[c]{.5in}
$\Z_9\times\Z_3\times\Z_2$:\\
$a'b',\tilde{a}'x_2'$,\\
$a^3$
\end{minipage}

\\
\hline

$3$ O&
\begin{minipage}[c]{.5in}
$\Z_2:$\\
$x_3''$
\end{minipage}
&
\begin{minipage}[c]{.5in}
$\Z_2^4$:\\
$\tilde{a}\tilde{b},x_3$,\\
$\tilde{a}x_2,\tilde{a}w_1^2$
\end{minipage}
&
\begin{minipage}[c]{.5in}
$\Z_2^2:$\\
$a^3,aw_1^2$
\end{minipage}
&
\begin{minipage}[c]{.5in}
$\Z_2^4$:\\
$\tilde{a}\tilde{b},x_3$,\\
$\tilde{a}x_2,\tilde{a}w_1^2$
\end{minipage}

&
\begin{minipage}[c]{.5in}
$\Z_2^2:$\\
$a^3,aw_1^2$
\end{minipage}

\\
\hline
 \end{tabular}
 }%\hspace*{-10mm}
\caption{$3d$ bordism groups-2.}
 \label{3d bordism groups-2}
\end{table}

%%%%%%%%%%
%%%%%%%%%%
\begin{table}[!h] %[!h] %[tb]
\centering
 \makebox[\textwidth][r]{
 \begin{tabular}{ |c| c | c|  c| c | c| c| c| c| c|}
\hline
$\Omega_d^H(-)$ & $\B \tO(3)$ &  $\B \tO(4)$ & $\B \tO(5)$ 

\\
\hline
$4$ SO& 
\begin{minipage}[c]{1.0in}
$\Z^2\times\Z_2:$\\
$\sigma,p_1'$,\\
$w_1'^2w_2'$
\end{minipage}
&
\begin{minipage}[c]{1.0in}
$\Z^2\times\Z_2^2:$\\
$\sigma,p_1'$,\\
$w_1'^2w_2',w_4'$
\end{minipage}

&
\begin{minipage}[c]{1.0in}
$\Z^2\times\Z_2^2:$\\
$\sigma,p_1'$,\\
$w_1'^2w_2',w_4'$
\end{minipage}

\\
\hline

$4$ O&
\begin{minipage}[c]{1.0in}
$\Z_2^8:$\\
$w_1^4,w_2^2$,\\
$w_1^2w_1'^2,w_1^2w_2'$,\\
$w_1'w_3',w_1'^2w_2'$,\\
$w_1'^4,w_2'^2$
\end{minipage}
&
\begin{minipage}[c]{1.0in}
$\Z_2^8:$\\
$w_1^4,w_2^2$,\\
$w_1^2w_1'^2,w_1^2w_2'$,\\
$w_1'w_3',w_1'^2w_2'$,\\
$w_1'^4,w_2'^2$,\\
$w_4'$
\end{minipage}
&

\begin{minipage}[c]{1.0in}
$\Z_2^8:$\\
$w_1^4,w_2^2$,\\
$w_1^2w_1'^2,w_1^2w_2'$,\\
$w_1'w_3',w_1'^2w_2'$,\\
$w_1'^4,w_2'^2$,\\
$w_4'$
\end{minipage}

\\
\hline
 \end{tabular}
 }%\hspace*{-15mm}
\caption{$4d$ bordism groups-1.}
 \label{4d bordism groups-1}
\end{table}

\begin{table}[!h] %[!h] %[tb]
\centering
 \makebox[\textwidth][r]{
 \begin{tabular}{ |c| c | c|  c| c | c| c| c| c| c|}
\hline
$\Omega_d^H(-)$ & $\B ^2\Z_4$ &  $\B\Z_4\times\B^2\Z_2$ & $\B\Z_6\times\B^2\Z_3$  &   $\B\Z_8\times\B^2\Z_2$ & $\B\Z_{18}\times \B^2\Z_3$

\\
\hline
$4$ SO& 
\begin{minipage}[c]{.7in}
$\Z\times\Z_8:$\\
$\sigma,\mathcal{P}_2(x_2'')$
\end{minipage}
&
\begin{minipage}[c]{.5in}
$\Z\times\Z_4\times\Z_2:$\\
$\sigma,\mathcal{P}_2(x_2)$,\\
$\tilde{b}x_2$
\end{minipage}
&
\begin{minipage}[c]{.5in}
$\Z\times\Z_3^2:$\\
$\sigma,a'x_3'=b'x_2'$,\\
$x_2'^2$
\end{minipage}
&
\begin{minipage}[c]{.5in}
$\Z\times\Z_4\times\Z_2:$\\
$\sigma,\mathcal{P}_2(x_2)$,\\
$\tilde{b}x_2$
\end{minipage}
&
\begin{minipage}[c]{.5in}
$\Z\times\Z_3^2:$\\
$\sigma,\tilde{b}'x_2'$,\\
$x_2'^2$
\end{minipage}

\\
\hline

$4$ O&
\begin{minipage}[c]{.7in}
$\Z_2^4:$\\
$w_1^4,w_2^2$,\\
$w_1^2\tilde{x}_2'',\tilde{x}_2''^2$
\end{minipage}
&
\begin{minipage}[c]{.5in}
$\Z_2^8:$\\
$\tilde{a}x_3,\tilde{b}x_2$,
$\tilde{b}^2,x_2^2$,
$w_1^4,w_2^2$,\\
$\tilde{b}w_1^2,x_2w_1^2$
\end{minipage}
&
\begin{minipage}[c]{.5in}
$\Z_2^4:$\\
$w_1^4,w_2^2$,\\
$a^4,a^2w_1^2$
\end{minipage}
&
\begin{minipage}[c]{.5in}
$\Z_2^8:$\\
$\tilde{a}x_3,\tilde{b}x_2$,
$\tilde{b}^2,x_2^2$,
$w_1^4,w_2^2$,\\
$\tilde{b}w_1^2,x_2w_1^2$
\end{minipage}
&
\begin{minipage}[c]{.5in}
$\Z_2^4:$\\
$w_1^4,w_2^2$,\\
$a^4,a^2w_1^2$
\end{minipage}

\\
\hline
 \end{tabular}
 }%\hspace*{-15mm}
\caption{$4d$ bordism groups-2.}
 \label{4d bordism groups-2}
\end{table}

%%%%%%%%%%
%%%%%%%%%%
\begin{table}[!h] %[!h] %[tb]
\centering
 \makebox[\textwidth][r]{
 \begin{tabular}{ |c| c | c|  c| c | c| c| c| c| c|}
\hline
$\Omega_d^H(-)$ & $\B \tO(3)$ &  $\B \tO(4)$ & $\B \tO(5)$ 
\\
\hline
$5$ SO&
\begin{minipage}[c]{1.0in}
$\Z_2^6:$\\
$w_2w_3,w_2^2w_1'$,\\
$w_2'w_3',w_1'w_2'^2$,\\
$w_1'^2w_3'=w_1'^3w_2',w_1'^5$
\end{minipage}

&
\begin{minipage}[c]{1.0in}
$\Z_2^6:$\\
$w_2w_3,w_2^2w_1'$,\\
$w_2'w_3',w_1'w_2'^2$,\\
$w_1'^2w_3'=w_1'^3w_2',w_1'^5$
\end{minipage}
&
\begin{minipage}[c]{1.0in}
$\Z_2^7:$\\
$w_2w_3,w_2^2w_1'$,\\
$w_2'w_3',w_1'w_2'^2$,\\
$w_1'^2w_3'=w_1'^3w_2',w_1'^5$,\\
$w_1'w_4'=w_5'$
\end{minipage}

\\
\hline

$5$ O&
\begin{minipage}[c]{1.0in}
$\Z_2^{11}:$\\
$w_2w_3,w_2^2w_1'$,\\
$w_1^4w_1',w_1^2w_1'^3$,\\
$w_1^2w_1'w_2',w_1^2w_3'$,\\
$w_2'w_3',w_1'w_2'^2$,\\
$w_1'^2w_3',w_1'^3w_2'$,\\
$w_1'^5$
\end{minipage}
&
\begin{minipage}[c]{1.0in}
$\Z_2^{12}:$\\
$w_2w_3,w_2^2w_1'$,\\
$w_1^4w_1',w_1^2w_1'^3$,\\
$w_1^2w_1'w_2',w_1^2w_3'$,\\
$w_2'w_3',w_1'w_2'^2$,\\
$w_1'^2w_3',w_1'^3w_2'$,\\
$w_1'^5,w_1'w_4'$
\end{minipage}
&
\begin{minipage}[c]{1.0in}
$\Z_2^{13}:$\\
$w_2w_3,w_2^2w_1'$,\\
$w_1^4w_1',w_1^2w_1'^3$,\\
$w_1^2w_1'w_2',w_1^2w_3'$,\\
$w_2'w_3',w_1'w_2'^2$,\\
$w_1'^2w_3',w_1'^3w_2'$,\\
$w_1'^5,w_1'w_4'$,\\
$w_5'$
\end{minipage}

\\
\hline
 \end{tabular}
 }%\hspace*{-15mm}
\caption{$5d$ bordism groups-1.}
 \label{5d bordism groups-1}
\end{table}

\begin{table}[!h] %[!h] %[tb]
\centering
 \makebox[\textwidth][r]{
 \begin{tabular}{ |c| c | c|  c| c | c| c| c| c| c|}
\hline
$\Omega_d^H(-)$ & $\B ^2\Z_4$ &  $\B\Z_4\times\B^2\Z_2$ & $\B\Z_6\times\B^2\Z_3$  &   $\B\Z_8\times\B^2\Z_2$ & $\B\Z_{18}\times \B^2\Z_3$\\
\hline
$5$ SO&
\begin{minipage}[c]{.8in}
$\Z_2^2:$\\
$w_2w_3,x_5''$
\end{minipage}
&
\begin{minipage}[c]{1.0in}
$\Z_4^3\times\Z_2^3:$\\
$a\mathcal{P}_2(x_2),ab^2$,\\
$a(\sigma\mod4),x_5=x_2x_3$,\\
$\tilde{a}\tilde{b}x_2,w_2w_3$
\end{minipage}
&
\begin{minipage}[c]{1.0in}
$\Z_2^3\times\Z_3^2\times\Z_9:$\\
$a^5,aw_2^2$,\\
$w_2w_3,a'b'x_2'$\\
$a'x_2'^2,\mathfrak{P}_3(b')$
\end{minipage}
&
\begin{minipage}[c]{1.2in}
$\Z_4\times\Z_8^2\times\Z_2^3:$\\
$(a\mod4)\mathcal{P}_2(x_2),ab^2$,\\
$a(\sigma\mod8),x_5=x_2x_3$,\\
$\tilde{a}\tilde{b}x_2,w_2w_3$
\end{minipage}
&
\begin{minipage}[c]{1.0in}
$\Z_2^3\times\Z_3^3\times\Z_{27}:$\\
$a^5,aw_2^2$,\\
$w_2w_3,\tilde{a}'(\sigma\mod3)$,\\
$\tilde{a}'\tilde{b}'x_2',\tilde{a}'x_2'^2$,\\
$\mathfrak{P}_3(b')$
\end{minipage}

\\
\hline

$5$ O&
\begin{minipage}[c]{.8in}
$\Z_2^4:$\\
$w_2w_3,w_1^2x_3''$,\\
$\tilde{x}_2''x_3'',x_5''$
\end{minipage}
&
\begin{minipage}[c]{1.0in}
$\Z_2^{12}:$\\
$\tilde{a}x_2^2,\tilde{b}x_3$,\\
$x_2x_3,\tilde{a}\tilde{b}^2$,\\
$x_5,\tilde{a}\tilde{b}x_2$,\\
$w_2w_3,\tilde{a}w_2^2$,\\
$\tilde{a}w_1^4,\tilde{a}\tilde{b}w_1^2$,\\
$x_3w_1^2=w_1^3x_2,\tilde{a}x_2w_1^2$
\end{minipage}
&
\begin{minipage}[c]{1.0in}
$\Z_2^5:$\\
$a^5,a^3w_1^2$,\\
$aw_1^4,aw_2^2$,\\
$w_2w_3$
\end{minipage}
& 
\begin{minipage}[c]{1.2in}
$\Z_2^{12}:$\\
$\tilde{a}x_2^2,\tilde{b}x_3$,\\
$x_2x_3,\tilde{a}\tilde{b}^2$,\\
$x_5,\tilde{a}\tilde{b}x_2$,\\
$w_2w_3,\tilde{a}w_2^2$,\\
$\tilde{a}w_1^4,\tilde{a}\tilde{b}w_1^2$,\\
$x_3w_1^2=w_1^3x_2,\tilde{a}x_2w_1^2$
\end{minipage}

&
\begin{minipage}[c]{1.0in}
$\Z_2^5:$\\
$a^5,a^3w_1^2$,\\
$aw_1^4,aw_2^2$,\\
$w_2w_3$
\end{minipage}

\\
\hline
 \end{tabular}
 }%\hspace*{-15mm}
\caption{$5d$ bordism groups-2.}
 \label{5d bordism groups-2}
\end{table}

\subsection{$\B ^2\Z_4$}

\subsubsection{$\Omega_d^{\tO}(\B ^2\Z_4)$}

\label{sec:OB2Z4}

\bea
\Ext_{\A_2}^{s,t}(\H^*(M\tO,\Z_2)\otimes\H^*(\B ^2\Z_4,\Z_2),\Z_2)\Rightarrow\Omega_{t-s}^{\tO}(\B ^2\Z_4)
\eea

\bea
\H^*(M\tO,\Z_2)=\A_2\otimes\Z_2[y_2,y_4,y_5,y_6,y_8,\dots]^*
\eea
where $y_2^*=w_1^2$, $(y_2^2)^*=w_2^2$, $y_4^*=w_1^4$, $y_5^*=w_2w_3$, etc.

\bea
\H^*(\B ^2\Z_4,\Z_2)=\Z_2[\tilde{x}_2'',x_3'',x_5'',x_9'',\dots]
\eea
where $\tilde{x}_2''=x_2''\mod 2$, $x_2''\in\H^2(\B ^2\Z_4,\Z_4)$, $x_3''=\beta_{(2,4)}x_2''$, $x_5''=\Sq^2x_3''$, $x_9''=\Sq^4x_5''$, etc.

\bea
\H^*(M\tO,\Z_2)\otimes\H^*(\B ^2\Z_4,\Z_2)&=&\A_2\otimes\Z_2[y_2,y_4,y_5,y_6,y_8,\dots]^*\otimes\Z_2[\tilde{x}_2'',x_3'',x_5'',x_9'',\dots]\\\notag
&=&\A_2\oplus2\Sigma^2\A_2\oplus\Sigma^3\A_2\oplus4\Sigma^4\A_2\oplus4\Sigma^5\A_2\oplus\cdots
\eea

Hence we have the following theorem
\begin{theorem}
\begin{table}[!h]
\centering
\begin{tabular}{c c}
\hline
$i$ & $\Omega^{\tO}_i(\B ^2\Z_4)$\\
\hline
0& $\Z_2$\\
1& $0$\\
2& $\Z_2^2$\\
3 & $\Z_2$\\
4 & $\Z_2^4$\\ 
5 & $\Z_2^4$\\
\hline
\end{tabular}
\end{table}
\end{theorem}

The 2d bordism invariants are $w_1^2,\tilde{x}_2''$.

The 3d bordism invariant is $x_3''$.

The 4d bordism invariants are $w_1^4,w_2^2,w_1^2\tilde{x}_2'',\tilde{x}_2''^2$.

The 5d bordism invariants are $w_2w_3,w_1^2x_3'',\tilde{x}_2''x_3'',x_5''$.

%$x_3=\frac{1}{2} w_1\tilde{x}_2$, $\Sq^1(\tilde{x}_2^2)=w_1\tilde{x}_2^2=2\tilde{x}_2\Sq^1\tilde{x}_2=4\tilde{x}_2x_3$, $x_2x_3=\frac{1}{4} w_1\tilde{x}_2^2$, $x_5=(w_2+w_1^2)x_3=\frac{1}{2}(w_3+w_1^3)\tilde{x}_2$.

\subsubsection{$\Omega_d^{\SO}(\B ^2\Z_4)$}
\label{sec:SOB2Z4}

\bea
\Ext_{\A_2}^{s,t}(\H^*(M\SO,\Z_2)\otimes\H^*(\B ^2\Z_4,\Z_2),\Z_2)\Rightarrow\Omega_{t-s}^{\SO}(\B ^2\Z_4)
\eea

\bea
\H^*(M\SO,\Z_2)=\A_2/\A_2\Sq^1\oplus\Sigma^4\A_2/\A_2\Sq^1\oplus\Sigma^5\A_2\oplus\cdots
\eea

\cred{
Note that
$\Sq^1\tilde{x}_2''=2\beta_{(2,4)}x_2''=0$, $\beta_{(2,4)}(x_2'')=\frac{1}{4}\delta x_2''=x_3''$, $\beta_{(2,4)}(x_2''^2)=2x_2''x_3''=2\tilde{x}_2''x_3''=0$, $\Sq^1(\tilde{x}_2''^2)=2\beta_{(2,4)}(x_2''^2)=0$, 
$\Sq^1x_3''=0$, $\Sq^1(\tilde{x}_2''x_3'')=0$,
$\Sq^1x_5''=\Sq^1\Sq^2\beta_{(2,4)}x_2''=\Sq^3\beta_{(2,4)}x_2''=(\beta_{(2,4)}x_2'')^2=x_3''^2$.
We have used the properties of Bockstein homomorphisms, \eqref{steenrel} and the Adem relations \eqref{Adem}.
}

\cred{Also note that
\bea
\beta_{(2,8)}\cP_2(x_2'')&=&\frac{1}{8}\delta\cP_2(x_2'')\mod2\nn\\
&=&\frac{1}{8}\delta(x_2''\cup x_2''+x_2''\hcup{1}\delta x_2'')\nn\\
&=&\frac{1}{8}(\delta x_2''\cup x_2''+x_2''\cup\delta x_2''+\delta(x_2''\hcup{1}\delta x_2''))\nn\\
&=&\frac{1}{8}(2x_2''\cup\delta x_2''+\delta x_2''\hcup{1}\delta x_2'')\nn\\
&=&x_2''\cup(\frac{1}{4}\delta x_2'')+2(\frac{1}{4}\delta x_2'')\hcup{1}(\frac{1}{4}\delta x_2'')\nn\\
&=&x_2''\beta_{(2,4)}x_2''+2\beta_{(2,4)}x_2''\hcup{1}\beta_{(2,4)}x_2''\nn\\
&=&x_2''\beta_{(2,4)}x_2''+2\Sq^2\beta_{(2,4)}x_2''\nn\\
&=&\tilde{x}_2''x_3''+2x_5''\nn\\
&=&\tilde{x}_2''x_3''
\eea
We have used $\beta_{(2,8)}=\frac{1}{8}\delta\mod2$, the Steenrod's formula \eqref{eq:Steenrod's}, $\beta_{(2,4)}=\frac{1}{4}\delta\mod2$, and the definition $\Sq^kx_n=x_n\hcup{n-k}x_n$.
}

\cred{
There is a differential $d_n$ corresponding to the Bockstein homomorphism  $\beta_{(2,2^n)}:\H^*(-,\Z_{2^n})\to\H^{*+1}(-,\Z_2)$ associated to $0\to\Z_2\to\Z_{2^{n+1}}\to\Z_{2^n}\to0$ \cite{may1981bockstein}. See \ref{Bockstein} for the definition of Bockstein homomorphisms.
}

\cred{
So there are differentials such that
$d_2(x_3'')=\tilde{x}_2''h_0^2$, $d_3(\tilde{x}_2''x_3'')=\tilde{x}_2''^2h_0^3$.
}

The $E_2$ page is shown in Figure \ref{fig:Omega_*^{SO}(B^2Z_4)}.

\begin{figure}[!h]
\begin{center}
\begin{tikzpicture}
\node at (0,-1) {0};
\node at (1,-1) {1};
\node at (2,-1) {2};
\node at (3,-1) {3};
\node at (4,-1) {4};
\node at (5,-1) {5};
\node at (6,-1) {$t-s$};
\node at (-1,0) {0};
\node at (-1,1) {1};
\node at (-1,2) {2};
\node at (-1,3) {3};
\node at (-1,4) {4};
\node at (-1,5) {5};
\node at (-1,6) {$s$};

\draw[->] (-0.5,-0.5) -- (-0.5,6);
\draw[->] (-0.5,-0.5) -- (6,-0.5);

\draw (0,0) -- (0,5);

\draw (2,0) -- (2,5);
\draw (3,0) -- (3,5);
\draw (4,0) -- (4,5);
\draw (4.1,0) -- (4.1,5);

\draw[fill] (5.1,0) circle(0.05);
\draw[fill] (5,0) circle(0.05);
\draw (4.9,0) -- (4.9,5);

\draw[color=red][->] (3,0) -- (2,2);
\draw[color=red][->] (3,1) -- (2,3);
\draw[color=red][->] (3,2) -- (2,4);
\draw[color=red][->] (3,3) -- (2,5);

\draw[color=red][->] (4.9,0) -- (4.1,3);
\draw[color=red][->] (4.9,1) -- (4.1,4);
\draw[color=red][->] (4.9,2) -- (4.1,5);

\end{tikzpicture}
\end{center}
\caption{$\Omega_*^{\SO}(\B ^2\Z_4)$}
\label{fig:Omega_*^{SO}(B^2Z_4)}
\end{figure}

Hence we have the following theorem
\begin{theorem}
\begin{table}[!h]
\centering
\begin{tabular}{c c}
\hline
$i$ & $\Omega^{\SO}_i(\B ^2\Z_4)$\\
\hline
0& $\Z$\\
1& $0$\\
2& $\Z_4$\\
3 & $0$\\
4 & $\Z\times\Z_8$\\ 
5 & $\Z_2^2$\\
\hline
\end{tabular}
\end{table}
\end{theorem}

The 2d bordism invariant is $x_2''$.

The 4d bordism invariants are $\sigma,\mathcal{P}_2(x_2'')$.

The 5d bordism invariants are $w_2w_3,x_5''$.

\subsection{$\B \tO(3)$}

\label{sec:BO3}

\subsubsection{$\Omega_d^{\tO}(\B \tO(3))$}

\bea
\Ext_{\A_2}^{s,t}(\H^*(M\tO,\Z_2)\otimes\H^*(\B \tO(3),\Z_2),\Z_2)\Rightarrow\Omega_{t-s}^{\tO}(\B \tO(3))
\eea

\bea
\H^*(M\tO,\Z_2)=\A_2\otimes\Z_2[y_2,y_4,y_5,y_6,y_8,\dots]^*
\eea
where $y_2^*=w_1^2$, $(y_2^2)^*=w_2^2$, $y_4^*=w_1^4$, $y_5^*=w_2w_3$, etc.

\bea
\H^*(\B \tO(3),\Z_2)=\Z_2[w_1',w_2',w_3']
\eea

\bea
\H^*(M\tO,\Z_2)\otimes\H^*(\B \tO(3),\Z_2)&=&\A_2\otimes\Z_2[y_2,y_4,y_5,y_6,y_8,\dots]^*\otimes\Z_2[w_1',w_2',w_3']\\\notag
&=&\A_2\oplus\Sigma\A_2\oplus3\Sigma^2\A_2\oplus4\Sigma^3\A_2\oplus8\Sigma^4\A_2\oplus11\Sigma^5\A_2\oplus\cdots
\eea

Hence we have the following theorem
\begin{theorem}
\begin{table}[!h]
\centering
\begin{tabular}{c c}
\hline
$i$ & $\Omega^{\tO}_i(\B \tO(3))$\\
\hline
0& $\Z_2$\\
1& $\Z_2$\\
2& $\Z_2^3$\\
3 & $\Z_2^4$\\
4 & $\Z_2^8$\\ 
5 & $\Z_2^{11}$\\
\hline
\end{tabular}
\end{table}
\end{theorem}

The 2d bordism invariants are $w_1^2,w_1'^2,w_2'$.

The 3d bordism invariant are $w_1'w_1^2,w_1'^3,w_1'w_2',w_3'$.

The 4d bordism invariants are $w_1^4,w_2^2,w_1^2w_1'^2,w_1^2w_2',w_1'w_3',w_1'^2w_2',w_1'^4,w_2'^2$.

The 5d bordism invariants are $w_2w_3,w_2^2w_1',w_1^4w_1',w_1^2w_1'^3,w_1^2w_1'w_2',w_1^2w_3',w_2'w_3',w_1'w_2'^2,w_1'^2w_3',w_1'^3w_2',w_1'^5$.

\subsubsection{$\Omega_d^{\SO}(\B \tO(3))$}

\bea
\Ext_{\A_2}^{s,t}(\H^*(M\SO,\Z_2)\otimes\H^*(\B \tO(3),\Z_2),\Z_2)\Rightarrow\Omega_{t-s}^{\SO}(\B \tO(3))
\eea

\bea
\H^*(M\SO,\Z_2)=\A_2/\A_2\Sq^1\oplus\Sigma^4\A_2/\A_2\Sq^1\oplus\Sigma^5\A_2\oplus\cdots
\eea

\bea
\H^*(\B \tO(3),\Z_2)=\Z_2[w_1',w_2',w_3']
\eea
where $\Sq^1w_2'=w_1'w_2'+w_3'$, $\Sq^1w_3'=w_1'w_3'$.

The $E_2$ page is shown in Figure \ref{fig:Omega_*^{SO}(BO(3))}.

\begin{figure}[!h]
\begin{center}
\begin{tikzpicture}
\node at (0,-1) {0};
\node at (1,-1) {1};
\node at (2,-1) {2};
\node at (3,-1) {3};
\node at (4,-1) {4};
\node at (5,-1) {5};
\node at (6,-1) {$t-s$};
\node at (-1,0) {0};
\node at (-1,1) {1};
\node at (-1,2) {2};
\node at (-1,3) {3};
\node at (-1,4) {4};
\node at (-1,5) {5};
\node at (-1,6) {$s$};

\draw[->] (-0.5,-0.5) -- (-0.5,6);
\draw[->] (-0.5,-0.5) -- (6,-0.5);

\draw (0,0) -- (0,5);
\draw[fill] (1,0) circle(0.05);
\draw[fill] (2,0) circle(0.05);
\draw[fill] (3,0) circle(0.05);
\draw[fill] (3.1,0) circle(0.05);
\draw (4,0) -- (4,5);
\draw (4.1,0) -- (4.1,5);
\draw[fill] (4.2,0) circle(0.05);
\draw[fill] (4.8,0) circle(0.05);
\draw[fill] (4.9,0) circle(0.05);
\draw[fill] (5,0) circle(0.05);
\draw[fill] (5.1,0) circle(0.05);
\draw[fill] (5.2,0) circle(0.05);
\draw[fill] (5.3,0) circle(0.05);

\end{tikzpicture}
\end{center}
\caption{$\Omega_*^{\SO}(\B \tO(3))$}
\label{fig:Omega_*^{SO}(BO(3))}
\end{figure}

Hence we have the following theorem
\begin{theorem}
\begin{table}[!h]
\centering
\begin{tabular}{c c}
\hline
$i$ & $\Omega^{\SO}_i(\B \tO(3))$\\
\hline
0& $\Z$\\
1& $\Z_2$\\
2& $\Z_2$\\
3 & $\Z_2^2$\\
4 & $\Z^2\times\Z_2$\\ 
5 & $\Z_2^6$\\
\hline
\end{tabular}
\end{table}
\end{theorem}

The 2d bordism invariant is $w_2'$.

The 3d bordism invariants are $w_1'^3,w_1'w_2'=w_3'$.

The 4d bordism invariants are $\sigma,p_1',w_1'^2w_2'$.

The 5d bordism invariants are $w_2w_3,w_2^2w_1',w_2'w_3',w_1'w_2'^2,w_1'^2w_3'=w_1'^3w_2',w_1'^5$.

%%%%%%%%%%%%%%%%%%%%%%%

\subsection{$\B \tO(4)$}

\subsubsection{$\Omega_d^{\tO}(\B \tO(4))$}

\bea
\Ext_{\A_2}^{s,t}(\H^*(M\tO,\Z_2)\otimes\H^*(\B \tO(4),\Z_2),\Z_2)\Rightarrow\Omega_{t-s}^{\tO}(\B \tO(4))
\eea

\bea
\H^*(M\tO,\Z_2)=\A_2\otimes\Z_2[y_2,y_4,y_5,y_6,y_8,\dots]^*
\eea
where $y_2^*=w_1^2$, $(y_2^2)^*=w_2^2$, $y_4^*=w_1^4$, $y_5^*=w_2w_3$, etc.

\bea
\H^*(\B \tO(4),\Z_2)=\Z_2[w_1',w_2',w_3',w_4']
\eea

\bea
\H^*(M\tO,\Z_2)\otimes\H^*(\B \tO(4),\Z_2)&=&\A_2\otimes\Z_2[y_2,y_4,y_5,y_6,y_8,\dots]^*\otimes\Z_2[w_1',w_2',w_3',w_4']\\\notag
&=&\A_2\oplus\Sigma\A_2\oplus3\Sigma^2\A_2\oplus4\Sigma^3\A_2\oplus9\Sigma^4\A_2\oplus12\Sigma^5\A_2\oplus\cdots
\eea

Hence we have the following theorem
\begin{theorem}
\begin{table}[!h]
\centering
\begin{tabular}{c c}
\hline
$i$ & $\Omega^{\tO}_i(\B \tO(4))$\\
\hline
0& $\Z_2$\\
1& $\Z_2$\\
2& $\Z_2^3$\\
3 & $\Z_2^4$\\
4 & $\Z_2^9$\\ 
5 & $\Z_2^{12}$\\
\hline
\end{tabular}
\end{table}
\end{theorem}

The 2d bordism invariants are $w_1^2,w_1'^2,w_2'$.

The 3d bordism invariant are $w_1'w_1^2,w_1'^3,w_1'w_2',w_3'$.

The 4d bordism invariants are $w_1^4,w_2^2,w_1^2w_1'^2,w_1^2w_2',w_1'w_3',w_1'^2w_2',w_1'^4,w_2'^2,w_4'$.

The 5d bordism invariants are 
$$w_2w_3,w_2^2w_1',w_1^4w_1',w_1^2w_1'^3,w_1^2w_1'w_2',w_1^2w_3',w_2'w_3',w_1'w_2'^2,w_1'^2w_3',w_1'^3w_2',w_1'^5,w_1'w_4'.$$

\subsubsection{$\Omega_d^{\SO}(\B \tO(4))$}

\bea
\Ext_{\A_2}^{s,t}(\H^*(M\SO,\Z_2)\otimes\H^*(\B \tO(4),\Z_2),\Z_2)\Rightarrow\Omega_{t-s}^{\SO}(\B \tO(4))
\eea

\bea
\H^*(M\SO,\Z_2)=\A_2/\A_2\Sq^1\oplus\Sigma^4\A_2/\A_2\Sq^1\oplus\Sigma^5\A_2\oplus\cdots
\eea

\bea
\H^*(\B \tO(4),\Z_2)=\Z_2[w_1',w_2',w_3',w_4']
\eea
where $\Sq^1w_2'=w_1'w_2'+w_3'$, $\Sq^1w_3'=w_1'w_3'$, $\Sq^1w_4'=w_1'w_4'$.

The $E_2$ page is shown in Figure \ref{fig:Omega_*^{SO}(BO(4))}.

\begin{figure}[!h]
\begin{center}
\begin{tikzpicture}
\node at (0,-1) {0};
\node at (1,-1) {1};
\node at (2,-1) {2};
\node at (3,-1) {3};
\node at (4,-1) {4};
\node at (5,-1) {5};
\node at (6,-1) {$t-s$};
\node at (-1,0) {0};
\node at (-1,1) {1};
\node at (-1,2) {2};
\node at (-1,3) {3};
\node at (-1,4) {4};
\node at (-1,5) {5};
\node at (-1,6) {$s$};

\draw[->] (-0.5,-0.5) -- (-0.5,6);
\draw[->] (-0.5,-0.5) -- (6,-0.5);

\draw (0,0) -- (0,5);
\draw[fill] (1,0) circle(0.05);
\draw[fill] (2,0) circle(0.05);
\draw[fill] (3,0) circle(0.05);
\draw[fill] (3.1,0) circle(0.05);

\draw[fill] (3.9,0) circle(0.05);
\draw (4,0) -- (4,5);
\draw (4.1,0) -- (4.1,5);
\draw[fill] (4.2,0) circle(0.05);
\draw[fill] (4.8,0) circle(0.05);
\draw[fill] (4.9,0) circle(0.05);
\draw[fill] (5,0) circle(0.05);
\draw[fill] (5.1,0) circle(0.05);
\draw[fill] (5.2,0) circle(0.05);
\draw[fill] (5.3,0) circle(0.05);

\end{tikzpicture}
\end{center}
\caption{$\Omega_*^{\SO}(\B \tO(4))$}
\label{fig:Omega_*^{SO}(BO(4))}
\end{figure}

Hence we have the following theorem
\begin{theorem}
\begin{table}[!h]
\centering
\begin{tabular}{c c}
\hline
$i$ & $\Omega^{\SO}_i(\B \tO(4))$\\
\hline
0& $\Z$\\
1& $\Z_2$\\
2& $\Z_2$\\
3 & $\Z_2^2$\\
4 & $\Z^2\times\Z_2^2$\\ 
5 & $\Z_2^6$\\
\hline
\end{tabular}
\end{table}
\end{theorem}

The 2d bordism invariant is $w_2'$.

The 3d bordism invariants are $w_1'^3,w_1'w_2'=w_3'$.

The 4d bordism invariants are $\sigma,p_1',w_1'^2w_2',w_4'$.

The 5d bordism invariants are $w_2w_3,w_2^2w_1',w_2'w_3',w_1'w_2'^2,w_1'^2w_3'=w_1'^3w_2',w_1'^5$.

%%%%%%%%%%%%%%%%%%%%
\subsection{$\B \tO(5)$}

\subsubsection{$\Omega_d^{\tO}(\B \tO(5))$}

\bea
\Ext_{\A_2}^{s,t}(\H^*(M\tO,\Z_2)\otimes\H^*(\B \tO(5),\Z_2),\Z_2)\Rightarrow\Omega_{t-s}^{\tO}(\B \tO(5))
\eea

\bea
\H^*(M\tO,\Z_2)=\A_2\otimes\Z_2[y_2,y_4,y_5,y_6,y_8,\dots]^*
\eea
where $y_2^*=w_1^2$, $(y_2^2)^*=w_2^2$, $y_4^*=w_1^4$, $y_5^*=w_2w_3$, etc.

\bea
\H^*(\B \tO(5),\Z_2)=\Z_2[w_1',w_2',w_3',w_4',w_5']
\eea

\bea
\H^*(M\tO,\Z_2)\otimes\H^*(\B \tO(5),\Z_2)&=&\A_2\otimes\Z_2[y_2,y_4,y_5,y_6,y_8,\dots]^*\otimes\Z_2[w_1',w_2',w_3',w_4',w_5']\\\notag
&=&\A_2\oplus\Sigma\A_2\oplus3\Sigma^2\A_2\oplus4\Sigma^3\A_2\oplus9\Sigma^4\A_2\oplus13\Sigma^5\A_2\oplus\cdots
\eea

Hence we have the following theorem
\begin{theorem}
\begin{table}[!h]
\centering
\begin{tabular}{c c}
\hline
$i$ & $\Omega^{\tO}_i(\B \tO(5))$\\
\hline
0& $\Z_2$\\
1& $\Z_2$\\
2& $\Z_2^3$\\
3 & $\Z_2^4$\\
4 & $\Z_2^9$\\ 
5 & $\Z_2^{13}$\\
\hline
\end{tabular}
\end{table}
\end{theorem}

The 2d bordism invariants are $w_1^2,w_1'^2,w_2'$.

The 3d bordism invariant are $w_1'w_1^2,w_1'^3,w_1'w_2',w_3'$.

The 4d bordism invariants are $w_1^4,w_2^2,w_1^2w_1'^2,w_1^2w_2',w_1'w_3',w_1'^2w_2',w_1'^4,w_2'^2,w_4'$.

The 5d bordism invariants are 
$$w_2w_3,w_2^2w_1',w_1^4w_1',w_1^2w_1'^3,w_1^2w_1'w_2',w_1^2w_3',w_2'w_3',w_1'w_2'^2,w_1'^2w_3',w_1'^3w_2',w_1'^5,w_1'w_4',w_5'.$$

\subsubsection{$\Omega_d^{\SO}(\B \tO(5))$}

\bea
\Ext_{\A_2}^{s,t}(\H^*(M\SO,\Z_2)\otimes\H^*(\B \tO(5),\Z_2),\Z_2)\Rightarrow\Omega_{t-s}^{\SO}(\B \tO(5))
\eea

\bea
\H^*(M\SO,\Z_2)=\A_2/\A_2\Sq^1\oplus\Sigma^4\A_2/\A_2\Sq^1\oplus\Sigma^5\A_2\oplus\cdots
\eea

\bea
\H^*(\B \tO(5),\Z_2)=\Z_2[w_1',w_2',w_3',w_4',w_5']
\eea
where $\Sq^1w_2'=w_1'w_2'+w_3'$, $\Sq^1w_3'=w_1'w_3'$, $\Sq^1w_4'=w_1'w_4'+w_5'$.

The $E_2$ page is shown in Figure \ref{fig:Omega_*^{SO}(BO(5))}.

\begin{figure}[!h]
\begin{center}
\begin{tikzpicture}
\node at (0,-1) {0};
\node at (1,-1) {1};
\node at (2,-1) {2};
\node at (3,-1) {3};
\node at (4,-1) {4};
\node at (5,-1) {5};
\node at (6,-1) {$t-s$};
\node at (-1,0) {0};
\node at (-1,1) {1};
\node at (-1,2) {2};
\node at (-1,3) {3};
\node at (-1,4) {4};
\node at (-1,5) {5};
\node at (-1,6) {$s$};

\draw[->] (-0.5,-0.5) -- (-0.5,6);
\draw[->] (-0.5,-0.5) -- (6,-0.5);

\draw (0,0) -- (0,5);
\draw[fill] (1,0) circle(0.05);
\draw[fill] (2,0) circle(0.05);
\draw[fill] (3,0) circle(0.05);
\draw[fill] (3.1,0) circle(0.05);

\draw[fill] (3.9,0) circle(0.05);
\draw (4,0) -- (4,5);
\draw (4.1,0) -- (4.1,5);
\draw[fill] (4.2,0) circle(0.05);
\draw[fill] (4.8,0) circle(0.05);
\draw[fill] (4.9,0) circle(0.05);
\draw[fill] (5,0) circle(0.05);
\draw[fill] (5.1,0) circle(0.05);
\draw[fill] (5.2,0) circle(0.05);
\draw[fill] (5.3,0) circle(0.05);
\draw[fill] (5.4,0) circle(0.05);

\end{tikzpicture}
\end{center}
\caption{$\Omega_*^{\SO}(\B \tO(5))$}
\label{fig:Omega_*^{SO}(BO(5))}
\end{figure}

Hence we have the following theorem
\begin{theorem}
\begin{table}[!h]
\centering
\begin{tabular}{c c}
\hline
$i$ & $\Omega^{\SO}_i(\B \tO(5))$\\
\hline
0& $\Z$\\
1& $\Z_2$\\
2& $\Z_2$\\
3 & $\Z_2^2$\\
4 & $\Z^2\times\Z_2^2$\\ 
5 & $\Z_2^7$\\
\hline
\end{tabular}
\end{table}
\end{theorem}

The 2d bordism invariant is $w_2'$.

The 3d bordism invariants are $w_1'^3,w_1'w_2'=w_3'$.

The 4d bordism invariants are $\sigma,p_1',w_1'^2w_2',w_4'$.

The 5d bordism invariants are $w_2w_3,w_2^2w_1',w_2'w_3',w_1'w_2'^2,w_1'^2w_3'=w_1'^3w_2',w_1'^5,w_1'w_4'=w_5'$.

\subsection{$\B \Z_{2n}\times \B ^2\Z_n$}

\label{sec:BZ2nB2Zn}

\subsubsection{$\Omega_d^{\tO}(\B \Z_4\times \B ^2\Z_2)$}

\bea
\H^*(\B \Z_4,\Z_4)=\Z_4[a,b]/(a^2=2b)
\eea
where $a\in\H^1(\B \Z_4,\Z_4)$, $b\in\H^2(\B \Z_4,\Z_4)$.

\bea
\H^*(\B \Z_4,\Z_2)=\Lambda_{\Z_2}(\tilde{a})\otimes\Z_2[\tilde{b}]
\eea
where $\tilde{a}=a\mod2\in\H^1(\B \Z_4,\Z_2)$, $\tilde{b}=b\mod2\in\H^2(\B \Z_4,\Z_2)$.

\bea
\H^*(\B ^2\Z_2,\Z_2)=\Z_2[x_2,x_3,x_5,x_9,\dots]
\eea

\bea
\Ext_{\A_2}^{s,t}(\H^*(M\tO,\Z_2)\otimes\H^*(\B \Z_4\times \B ^2\Z_2,\Z_2),\Z_2)\Rightarrow\Omega_{t-s}^{\tO}(\B \Z_4\times \B ^2\Z_2)
\eea

\bea
\H^*(M\tO,\Z_2)=\A_2\otimes\Z_2[y_2,y_4,y_5,y_6,y_8,\dots]^*
\eea
where $y_2^*=w_1^2$, $(y_2^2)^*=w_2^2$, $y_4^*=w_1^4$, $y_5^*=w_2w_3$, etc.

\bea
&&\H^*(M\tO,\Z_2)\otimes\H^*(\B \Z_4\times \B ^2\Z_2,\Z_2)\\\notag
&=&\A_2\otimes\Z_2[y_2,y_4,y_5,y_6,y_8,\dots]^*\otimes\Lambda_{\Z_2}(\tilde{a})\otimes\Z_2[\tilde{b}]\otimes\Z_2[\tilde{x}_2,x_3,x_5,x_9,\dots]\\\notag
&=&\A_2\oplus\Sigma\A_2\oplus3\Sigma^2\A_2\oplus4\Sigma^3\A_2\oplus8\Sigma^4\A_2\oplus12\Sigma^5\A_2\oplus\cdots
\eea

Hence we have the following theorem
\begin{theorem}

The bordism groups are
\begin{table}[!h]
\centering
\begin{tabular}{c c}
\hline
$i$ & $\Omega^{\tO}_i(\B \Z_4\times \B ^2\Z_2)$\\
\hline
0& $\Z_2$\\
1& $\Z_2$\\
2& $\Z_2^3$\\
3 & $\Z_2^4$\\
4 & $\Z_2^8$\\ 
5 & $\Z_2^{12}$\\
\hline
\end{tabular}
\end{table}

\end{theorem}

The 2d bordism invariants are $\tilde{b},x_2,w_1^2$.

The 3d bordism invariants are $\tilde{a}\tilde{b},x_3,\tilde{a}x_2,\tilde{a}w_1^2$.

The 4d bordism invariants are $\tilde{a}x_3,\tilde{b}x_2,\tilde{b}^2,x_2^2,w_1^4,w_2^2,\tilde{b}w_1^2,x_2w_1^2$.

The 5d bordism invariants are $\tilde{a}x_2^2,\tilde{b}x_3,x_2x_3,\tilde{a}\tilde{b}^2,x_5,\tilde{a}\tilde{b}x_2,w_2w_3,\tilde{a}w_2^2,\tilde{a}w_1^4,\tilde{a}\tilde{b}w_1^2,
x_3w_1^2\cblue{=w_1^3 x_2} ,\tilde{a}x_2w_1^2$.

%\cblue{
%Note
%\bea
%x_3=w_1 x_2
%\eea
%except $x_2x_3= \frac{1}{2} w_1 x_2^2$.
%}

\subsubsection{$\Omega_d^{\SO}(\B \Z_4\times \B ^2\Z_2)$}

\bea
\Ext_{\A_2}^{s,t}(\H^*(M\SO,\Z_2)\otimes\H^*(\B \Z_4\times \B ^2\Z_2,\Z_2),\Z_2)\Rightarrow\Omega_{t-s}^{\SO}(\B \Z_4\times \B ^2\Z_2)
\eea

\bea
\H^*(M\SO,\Z_2)=\A_2/\A_2\Sq^1\oplus\Sigma^4\A_2/\A_2\Sq^1\oplus\Sigma^5\A_2\oplus\cdots
\eea

\cred{
Note that
$\beta_{(2,4)}a=\tilde{b}$, $\Sq^1x_2=x_3$, $\Sq^1(\tilde{a}x_2)=\tilde{a}x_3$, $\Sq^1(\tilde{b}x_2)=\tilde{b}x_3$, $\beta_{(2,4)}(ab)=\tilde{b}^2$, $\beta_{(2,4)}(\mathcal{P}_2(x_2))=x_2x_3+x_5$, $\Sq^1(x_2x_3)=\Sq^1x_5=x_3^2$, $\Sq^1(\tilde{a}\tilde{b}x_2)=\tilde{a}\tilde{b}x_3$,
$\beta_{(2,4)}(a\mathcal{P}_2(x_2))=\tilde{b}x_2^2+\tilde{a}(x_2x_3+x_5)$, $\beta_{(2,4)}(ab^2)=\tilde{b}^3$, $\beta_{(2,4)}(a(\sigma\mod4))=\tilde{b}w_2^2$.
}

\cred{
There is a differential $d_2$ corresponding to the Bockstein homomorphism  $\beta_{(2,4)}:\H^*(-,\Z_{4})\to\H^{*+1}(-,\Z_2)$ associated to $0\to\Z_2\to\Z_{8}\to\Z_{4}\to0$ \cite{may1981bockstein}. See \ref{Bockstein} for the definition of Bockstein homomorphisms.
}

\cred{
So there are differentials such that
$d_2(\tilde{b})=\tilde{a}h_0^2$, $d_2(\tilde{b}^2)=\tilde{a}\tilde{b}h_0^2$, $d_2(x_2x_3+x_5)=x_2^2h_0^2$, $d_2(\tilde{b}x_2^2+\tilde{a}(x_2x_3+x_5))=\tilde{a}x_2^2h_0^2$, $d_2(\tilde{b}^3)=\tilde{a}\tilde{b}^2h_0^2$, $d_2(\tilde{b}w_2^2)=\tilde{a}w_2^2h_0^2$.
}

The $E_2$ page is shown in Figure \ref{fig:Omega_*^{SO}(BZ_4times B^2Z_2)}.

\begin{figure}[!h]
\begin{center}
\begin{tikzpicture}
\node at (0,-1) {0};
\node at (1,-1) {1};
\node at (2,-1) {2};
\node at (3,-1) {3};
\node at (4,-1) {4};
\node at (5,-1) {5};
\node at (6,-1) {6};
\node at (7,-1) {$t-s$};
\node at (-1,0) {0};
\node at (-1,1) {1};
\node at (-1,2) {2};
\node at (-1,3) {3};
\node at (-1,4) {4};
\node at (-1,5) {5};
\node at (-1,6) {$s$};

\draw[->] (-0.5,-0.5) -- (-0.5,6);
\draw[->] (-0.5,-0.5) -- (7,-0.5);

\draw (0,0) -- (0,5);
\draw (1,0) -- (1,5);

\draw (2,0) -- (2,5);
\draw[fill] (2.1,0) circle(0.05);

\draw (3,0) -- (3,5);
\draw[fill] (3.1,0) circle(0.05);

\draw[fill] (3.8,0) circle(0.05);
\draw (3.9,0) -- (3.9,5);
\draw (4,0) -- (4,5);
\draw (4.1,0) -- (4.1,5);

\draw[fill] (4.7,0) circle(0.05);
\draw (4.8,0) -- (4.8,5);
\draw (4.9,0) -- (4.9,5);
\draw (5,0) -- (5,5);
\draw (5.1,0) -- (5.1,5);
\draw[fill] (5.2,0) circle(0.05);
\draw[fill] (5.3,0) circle(0.05);

\draw (5.9,0) -- (5.9,5);
\draw (6,0) -- (6,5);
\draw (6.1,0) -- (6.1,5);

\draw[color=red][->] (2,0) -- (1,2);
\draw[color=red][->] (2,1) -- (1,3);
\draw[color=red][->] (2,2) -- (1,4);
\draw[color=red][->] (2,3) -- (1,5);

\draw[color=red][->] (3.9,0) -- (3,2);
\draw[color=red][->] (3.9,1) -- (3,3);
\draw[color=red][->] (3.9,2) -- (3,4);
\draw[color=red][->] (3.9,3) -- (3,5);

\draw[color=red][->] (4.8,0) -- (4.1,2);
\draw[color=red][->] (4.8,1) -- (4.1,3);
\draw[color=red][->] (4.8,2) -- (4.1,4);
\draw[color=red][->] (4.8,3) -- (4.1,5);

\draw[color=red][->] (5.9,0) -- (4.9,2);
\draw[color=red][->] (5.9,1) -- (4.9,3);
\draw[color=red][->] (5.9,2) -- (4.9,4);
\draw[color=red][->] (5.9,3) -- (4.9,5);

\draw[color=red][->] (6,0) -- (5,2);
\draw[color=red][->] (6,1) -- (5,3);
\draw[color=red][->] (6,2) -- (5,4);
\draw[color=red][->] (6,3) -- (5,5);

\draw[color=red][->] (6.1,0) -- (5.1,2);
\draw[color=red][->] (6.1,1) -- (5.1,3);
\draw[color=red][->] (6.1,2) -- (5.1,4);
\draw[color=red][->] (6.1,3) -- (5.1,5);

\end{tikzpicture}
\end{center}
\caption{$\Omega_*^{\SO}(\B \Z_4\times \B ^2\Z_2)$}
\label{fig:Omega_*^{SO}(BZ_4times B^2Z_2)}
\end{figure}

Hence we have the following theorem
\begin{theorem}

The bordism groups are
\begin{table}[!h]
\centering
\begin{tabular}{c c}
\hline
$i$ & $\Omega^{\SO}_i(\B \Z_4\times \B ^2\Z_2)$\\
\hline
0& $\Z$\\
1& $\Z_4$\\
2& $\Z_2$\\
3 & $\Z_4\times\Z_2$\\
4 & $\Z\times\Z_4\times\Z_2$\\ 
5 & $\Z_4^3\times\Z_2^3$\\
\hline
\end{tabular}
\end{table}

\end{theorem}

The 2d bordism invariant is $x_2$.

The 3d bordism invariants are $ab$ and $\tilde{a}x_2$.

The 4d bordism invariants are $\sigma$, $\mathcal{P}_2(x_2)$ and $\tilde{b}x_2$.

The 5d bordism invariants are $a\mathcal{P}_2(x_2)$, $ab^2$, $a(\sigma\mod4)$, $x_5=x_2x_3$, $\tilde{a}\tilde{b}x_2$ and $w_2w_3$.

\subsubsection{$\Omega_d^{\tO}(\B \Z_6\times \B ^2\Z_3)$}

\bea
\H^*(\B \Z_6\times \B ^2\Z_3,\Z_2)=\H^*(\B \Z_2,\Z_2)=\Z_2[a]
\eea
where $a\in\H^1(\B \Z_2,\Z_2)$.

\bea
\Ext_{\A_3}^{s,t}(\H^*(M\tO,\Z_3)\otimes\H^*(\B \Z_6\times \B ^2\Z_3,\Z_3),\Z_3)\Rightarrow\Omega_{t-s}^{\tO}(\B \Z_6\times \B ^2\Z_3)_3^{\wedge}
\eea

Since $\H^*(M\tO,\Z_3)=0$, we have $\Omega_d^{\tO}(\B \Z_6\times \B ^2\Z_3)_3^{\wedge}=0$.

\bea
\Ext_{\A_2}^{s,t}(\H^*(M\tO,\Z_2)\otimes\H^*(\B \Z_6\times \B ^2\Z_3,\Z_2),\Z_2)\Rightarrow\Omega_{t-s}^{\tO}(\B \Z_6\times \B ^2\Z_3)_2^{\wedge}
\eea

\bea
\H^*(M\tO,\Z_2)=\A_2\otimes\Z_2[y_2,y_4,y_5,y_6,y_8,\dots]^*
\eea
where $y_2^*=w_1^2$, $(y_2^2)^*=w_2^2$, $y_4^*=w_1^4$, $y_5^*=w_2w_3$, etc.

\bea
\H^*(M\tO,\Z_2)\otimes\H^*(\B \Z_6\times \B ^2\Z_3,\Z_2)&=&\A_2\otimes\Z_2[y_2,y_4,y_5,y_6,y_8,\dots]^*\otimes\Z_2[a]\\\notag
&=&\A_2\oplus\Sigma\A_2\oplus2\Sigma^2\A_2\oplus2\Sigma^3\A_2\oplus4\Sigma^4\A_2\oplus5\Sigma^5\A_2\oplus\cdots
\eea

Hence we have the following theorem
\begin{theorem}
The bordism groups are
\begin{table}[!h]
\centering
\begin{tabular}{c c}
\hline
$i$ & $\Omega^{\tO}_i(\B \Z_6\times \B ^2\Z_3)$\\
\hline
0& $\Z_2$\\
1& $\Z_2$\\
2& $\Z_2^2$\\
3 & $\Z_2^2$\\
4 & $\Z_2^4$\\ 
5 & $\Z_2^5$\\
\hline
\end{tabular}
\end{table}
\end{theorem}

The 2d bordism invariants are $w_1^2,a^2$.

The 3d bordism invariants are $a^3,aw_1^2$.

The 4d bordism invariants are $a^4,a^2w_1^2,w_1^4,w_2^2$.

The 5d bordism invariants are $a^5,a^3w_1^2,aw_1^4,aw_2^2,w_2w_3$.

\subsubsection{$\Omega_d^{\SO}(\B \Z_6\times \B ^2\Z_3)$}

\bea
\Ext_{\A_2}^{s,t}(\H^*(M\SO\wedge (\B \Z_6\times \B ^2\Z_3)_+,\Z_2),\Z_2)\Rightarrow\Omega_{t-s}^{\SO}(\B \Z_6\times \B ^2\Z_3)_2^{\wedge}.
\eea
Since $\H^*(\B \Z_6\times \B ^2\Z_3,\Z_2)=\H^*(\B \Z_2,\Z_2)$, we have $\Omega_d^{\SO}(\B \Z_6\times \B ^2\Z_3)_2^{\wedge}=\Omega_d^{\SO}(\B \Z_2)$.

The $E_2$ page is shown in Figure \ref{fig:Omega_*^{SO}(BZ_2)}.

\begin{figure}[!h]
\begin{center}
\begin{tikzpicture}
\node at (0,-1) {0};
\node at (1,-1) {1};
\node at (2,-1) {2};
\node at (3,-1) {3};
\node at (4,-1) {4};
\node at (5,-1) {5};
\node at (6,-1) {$t-s$};
\node at (-1,0) {0};
\node at (-1,1) {1};
\node at (-1,2) {2};
\node at (-1,3) {3};
\node at (-1,4) {4};
\node at (-1,5) {5};
\node at (-1,6) {$s$};

\draw[->] (-0.5,-0.5) -- (-0.5,6);
\draw[->] (-0.5,-0.5) -- (6,-0.5);

\draw (0,0) -- (0,5);
\draw[fill] (1,0) circle(0.05);
\draw[fill] (3,0) circle(0.05);

\draw (4,0) -- (4,5);

\draw[fill] (4.9,0) circle(0.05);

\draw[fill] (5,0) circle(0.05);
\draw[fill] (5.1,0) circle(0.05);

\end{tikzpicture}
\end{center}
\caption{$\Omega_*^{\SO}(\B \Z_2)$}
\label{fig:Omega_*^{SO}(BZ_2)}
\end{figure}

\bea
\Ext_{\A_3}^{s,t}(\H^*(M\SO\wedge (\B \Z_6\times \B ^2\Z_3)_+,\Z_3),\Z_3)\Rightarrow\Omega_{t-s}^{\SO}(\B \Z_6\times \B ^2\Z_3)_3^{\wedge}.
\eea

Since $\H^*(\B \Z_6\times \B ^2\Z_3,\Z_3)=\H^*(\B \Z_3\times \B ^2\Z_3,\Z_3)$, we have $\Omega_d^{\SO}(\B \Z_6\times \B ^2\Z_3)_3^{\wedge}=\Omega_d^{\SO}(\B \Z_3\times \B ^2\Z_3)_3^{\wedge}$.

Hence we have the following theorem
\begin{theorem}
The bordism groups are
\begin{table}[!h]
\centering
\begin{tabular}{c c}
\hline
$i$ & $\Omega^{\SO}_i(\B \Z_6\times \B ^2\Z_3)$\\
\hline
0& $\Z$\\
1& $\Z_3\times\Z_2$\\
2& $\Z_3$\\
3 & $\Z_3^2\times\Z_2$\\
4 & $\Z\times\Z_3^2$\\ 
5 & $\Z_2^3\times\Z_3^2\times\Z_9$\\
\hline
\end{tabular}
\end{table}
\end{theorem}

The 2d bordism invariant is $x_2'$.

The 3d bordism invariants are $a'b',a'x_2',a^3$.

The 4d bordism invariants are $\sigma$, $a'x_3'(=b'x_2')$ and $x_2'^2$.

The 5d bordism invariants are $a^5,aw_2^2,w_2w_3,a'b'x_2',a'x_2'^2,\mathfrak{P}_3(b')$.

Here $\mathfrak{P}_3$ is the Postnikov square.

\subsection{$\B \Z_{2n^2}\times \B ^2\Z_n$}

\label{sec:BZ2n2B2Zn}

\subsubsection{$\Omega_d^{\tO}(\B \Z_8\times \B ^2\Z_2)$}

\bea
\H^*(\B \Z_8,\Z_8)=\Z_8[a,b]/(a^2=4b)
\eea
where $a\in\H^1(\B \Z_8,\Z_8)$, $b\in\H^2(\B \Z_8,\Z_8)$.

\bea
\H^*(\B \Z_8,\Z_2)=\Lambda_{\Z_2}(\tilde{a})\otimes\Z_2[\tilde{b}]
\eea
where $\tilde{a}=a\mod2\in\H^1(\B \Z_8,\Z_2)$, $\tilde{b}=b\mod2\in\H^2(\B \Z_8,\Z_2)$.

\bea
\H^*(\B ^2\Z_2,\Z_2)=\Z_2[x_2,x_3,x_5,x_9,\dots]
\eea

\bea
\Ext_{\A_2}^{s,t}(\H^*(M\tO,\Z_2)\otimes\H^*(\B \Z_8\times \B ^2\Z_2,\Z_2),\Z_2)\Rightarrow\Omega_{t-s}^{\tO}(\B \Z_8\times \B ^2\Z_2)
\eea

\bea
\H^*(M\tO,\Z_2)=\A_2\otimes\Z_2[y_2,y_4,y_5,y_6,y_8,\dots]^*
\eea
where $y_2^*=w_1^2$, $(y_2^2)^*=w_2^2$, $y_4^*=w_1^4$, $y_5^*=w_2w_3$, etc.

\bea
&&\H^*(M\tO,\Z_2)\otimes\H^*(\B \Z_4\times \B ^2\Z_2,\Z_2)\\\notag
&=&\A_2\otimes\Z_2[y_2,y_4,y_5,y_6,y_8,\dots]^*\otimes\Lambda_{\Z_2}(\tilde{a})\otimes\Z_2[\tilde{b}]\otimes\Z_2[\tilde{x}_2,x_3,x_5,x_9,\dots]\\\notag
&=&\A_2\oplus\Sigma\A_2\oplus3\Sigma^2\A_2\oplus4\Sigma^3\A_2\oplus8\Sigma^4\A_2\oplus12\Sigma^5\A_2\oplus\cdots
\eea

Hence we have the following theorem
\begin{theorem}

The bordism groups are
\begin{table}[!h]
\centering
\begin{tabular}{c c}
\hline
$i$ & $\Omega^{\tO}_i(\B \Z_8\times \B ^2\Z_2)$\\
\hline
0& $\Z_2$\\
1& $\Z_2$\\
2& $\Z_2^3$\\
3 & $\Z_2^4$\\
4 & $\Z_2^8$\\ 
5 & $\Z_2^{12}$\\
\hline
\end{tabular}
\end{table}

\end{theorem}

The 2d bordism invariants are $\tilde{b},x_2,w_1^2$.

The 3d bordism invariants are $\tilde{a}\tilde{b},x_3,\tilde{a}x_2,\tilde{a}w_1^2$.

The 4d bordism invariants are $\tilde{a}x_3,\tilde{b}x_2,\tilde{b}^2,x_2^2,w_1^4,w_2^2,\tilde{b}w_1^2,x_2w_1^2$.

The 5d bordism invariants are $\tilde{a}x_2^2,\tilde{b}x_3,x_2x_3,\tilde{a}\tilde{b}^2,x_5,\tilde{a}\tilde{b}x_2,w_2w_3,\tilde{a}w_2^2,\tilde{a}w_1^4,\tilde{a}\tilde{b}w_1^2,
x_3w_1^2\cblue{=w_1^3 x_2} ,\tilde{a}x_2w_1^2$.

%\cblue{
%Note
%\bea
%x_3=w_1 x_2
%\eea
%except $x_2x_3= \frac{1}{2} w_1 x_2^2$.
%}

\subsubsection{$\Omega_d^{\SO}(\B \Z_8\times \B ^2\Z_2)$}

\bea
\Ext_{\A_2}^{s,t}(\H^*(M\SO,\Z_2)\otimes\H^*(\B \Z_8\times \B ^2\Z_2,\Z_2),\Z_2)\Rightarrow\Omega_{t-s}^{\SO}(\B \Z_8\times \B ^2\Z_2)
\eea

\bea
\H^*(M\SO,\Z_2)=\A_2/\A_2\Sq^1\oplus\Sigma^4\A_2/\A_2\Sq^1\oplus\Sigma^5\A_2\oplus\cdots
\eea

\cred{
Note that
$\beta_{(2,8)}a=\tilde{b}$, $\Sq^1x_2=x_3$, $\Sq^1(\tilde{a}x_2)=\tilde{a}x_3$, $\Sq^1(\tilde{b}x_2)=\tilde{b}x_3$, $\beta_{(2,8)}(ab)=\tilde{b}^2$, $\beta_{(2,4)}(\mathcal{P}_2(x_2))=x_2x_3+x_5$, $\Sq^1(x_2x_3)=\Sq^1x_5=x_3^2$, $\Sq^1(\tilde{a}\tilde{b}x_2)=\tilde{a}\tilde{b}x_3$,
$\beta_{(2,4)}((a\mod4)\mathcal{P}_2(x_2))=2\beta_{(2,8)}(a)x_2^2+\tilde{a}(x_2x_3+x_5)=2\tilde{b}x_2^2+\tilde{a}(x_2x_3+x_5)=\tilde{a}(x_2x_3+x_5)$, $\beta_{(2,8)}(ab^2)=\tilde{b}^3$, $\beta_{(2,8)}(a(\sigma\mod8))=\tilde{b}w_2^2$.
}

\cred{
There is a differential $d_n$ corresponding to the Bockstein homomorphism  $\beta_{(2,2^n)}:\H^*(-,\Z_{2^n})\to\H^{*+1}(-,\Z_2)$ associated to $0\to\Z_2\to\Z_{2^{n+1}}\to\Z_{2^n}\to0$ \cite{may1981bockstein}. See \ref{Bockstein} for the definition of Bockstein homomorphisms.
}

\cred{
So there are differentials such that
$d_3(\tilde{b})=\tilde{a}h_0^2$, $d_3(\tilde{b}^2)=\tilde{a}\tilde{b}h_0^2$, $d_2(x_2x_3+x_5)=x_2^2h_0^2$, $d_2(\tilde{a}(x_2x_3+x_5))=\tilde{a}x_2^2h_0^2$, $d_3(\tilde{b}^3)=\tilde{a}\tilde{b}^2h_0^2$, $d_3(\tilde{b}w_2^2)=\tilde{a}w_2^2h_0^2$.
}

The $E_2$ page is shown in Figure \ref{fig:Omega_*^{SO}(BZ_8times B^2Z_2)}.

\begin{figure}[!h]
\begin{center}
\begin{tikzpicture}
\node at (0,-1) {0};
\node at (1,-1) {1};
\node at (2,-1) {2};
\node at (3,-1) {3};
\node at (4,-1) {4};
\node at (5,-1) {5};
\node at (6,-1) {6};
\node at (7,-1) {$t-s$};
\node at (-1,0) {0};
\node at (-1,1) {1};
\node at (-1,2) {2};
\node at (-1,3) {3};
\node at (-1,4) {4};
\node at (-1,5) {5};
\node at (-1,6) {$s$};

\draw[->] (-0.5,-0.5) -- (-0.5,6);
\draw[->] (-0.5,-0.5) -- (7,-0.5);

\draw (0,0) -- (0,5);
\draw (1,0) -- (1,5);

\draw (2,0) -- (2,5);
\draw[fill] (2.1,0) circle(0.05);

\draw (3,0) -- (3,5);
\draw[fill] (3.1,0) circle(0.05);

\draw[fill] (3.8,0) circle(0.05);
\draw (3.9,0) -- (3.9,5);
\draw (4,0) -- (4,5);
\draw (4.1,0) -- (4.1,5);

\draw[fill] (4.7,0) circle(0.05);
\draw (4.8,0) -- (4.8,5);
\draw (4.9,0) -- (4.9,5);
\draw (5,0) -- (5,5);
\draw (5.1,0) -- (5.1,5);
\draw[fill] (5.2,0) circle(0.05);
\draw[fill] (5.3,0) circle(0.05);

\draw (5.9,0) -- (5.9,5);
\draw (6,0) -- (6,5);
\draw (6.1,0) -- (6.1,5);

\draw[color=red][->] (2,0) -- (1,3);
\draw[color=red][->] (2,1) -- (1,4);
\draw[color=red][->] (2,2) -- (1,5);

\draw[color=red][->] (3.9,0) -- (3,3);
\draw[color=red][->] (3.9,1) -- (3,4);
\draw[color=red][->] (3.9,2) -- (3,5);

\draw[color=red][->] (4.8,0) -- (4.1,2);
\draw[color=red][->] (4.8,1) -- (4.1,3);
\draw[color=red][->] (4.8,2) -- (4.1,4);
\draw[color=red][->] (4.8,3) -- (4.1,5);

\draw[color=red][->] (5.9,0) -- (4.9,2);
\draw[color=red][->] (5.9,1) -- (4.9,3);
\draw[color=red][->] (5.9,2) -- (4.9,4);
\draw[color=red][->] (5.9,3) -- (4.9,5);

\draw[color=red][->] (6,0) -- (5,3);
\draw[color=red][->] (6,1) -- (5,4);
\draw[color=red][->] (6,2) -- (5,5);

\draw[color=red][->] (6.1,0) -- (5.1,3);
\draw[color=red][->] (6.1,1) -- (5.1,4);
\draw[color=red][->] (6.1,2) -- (5.1,5);

\end{tikzpicture}
\end{center}
\caption{$\Omega_*^{\SO}(\B \Z_8\times \B ^2\Z_2)$}
\label{fig:Omega_*^{SO}(BZ_8times B^2Z_2)}
\end{figure}

Hence we have the following theorem
\begin{theorem}
The bordism groups are
\begin{table}[!h]
\centering
\begin{tabular}{c c}
\hline
$i$ & $\Omega^{\SO}_i(\B \Z_8\times \B ^2\Z_2)$\\
\hline
0& $\Z$\\
1& $\Z_8$\\
2& $\Z_2$\\
3 & $\Z_8\times\Z_2$\\
4 & $\Z\times\Z_4\times\Z_2$\\ 
5 & $\Z_4\times\Z_8^2\times\Z_2^3$\\
\hline
\end{tabular}
\end{table}
\end{theorem}

The 2d bordism invariant is $x_2$.

The 3d bordism invariants are $ab$ and $\tilde{a}x_2$.

The 4d bordism invariants are $\sigma$, $\mathcal{P}_2(x_2)$ and $\tilde{b}x_2$.

The 5d bordism invariants are $\cblue{(a\mod4)\mathcal{P}_2(x_2)}$, $ab^2$, $a(\sigma\mod8)$, $x_5=x_2x_3$, $\tilde{a}\tilde{b}x_2$ and $w_2w_3$.

\subsubsection{$\Omega_d^{\tO}(\B \Z_{18}\times \B ^2\Z_3)$}

\bea
\H^*(\B \Z_{18}\times \B ^2\Z_3,\Z_2)=\H^*(\B \Z_2,\Z_2)=\Z_2[a]
\eea
where $a\in\H^1(\B \Z_2,\Z_2)$.

\bea
\Ext_{\A_3}^{s,t}(\H^*(M\tO,\Z_3)\otimes\H^*(\B \Z_{18}\times \B ^2\Z_3,\Z_3),\Z_3)\Rightarrow\Omega_{t-s}^{\tO}(\B \Z_{18}\times \B ^2\Z_3)_3^{\wedge}
\eea

Since $\H^*(M\tO,\Z_3)=0$, we have $\Omega_d^{\tO}(\B \Z_{18}\times \B ^2\Z_3)_3^{\wedge}=0$.

\bea
\Ext_{\A_2}^{s,t}(\H^*(M\tO,\Z_2)\otimes\H^*(\B \Z_{18}\times \B ^2\Z_3,\Z_2),\Z_2)\Rightarrow\Omega_{t-s}^{\tO}(\B \Z_{18}\times \B ^2\Z_3)_2^{\wedge}
\eea

\bea
\H^*(M\tO,\Z_2)=\A_2\otimes\Z_2[y_2,y_4,y_5,y_6,y_8,\dots]^*
\eea
where $y_2^*=w_1^2$, $(y_2^2)^*=w_2^2$, $y_4^*=w_1^4$, $y_5^*=w_2w_3$, etc.

\bea
\H^*(M\tO,\Z_2)\otimes\H^*(\B \Z_{18}\times \B ^2\Z_3,\Z_2)&=&\A_2\otimes\Z_2[y_2,y_4,y_5,y_6,y_8,\dots]^*\otimes\Z_2[a]\\\notag
&=&\A_2\oplus\Sigma\A_2\oplus2\Sigma^2\A_2\oplus2\Sigma^3\A_2\oplus4\Sigma^4\A_2\oplus5\Sigma^5\A_2\oplus\cdots
\eea

Hence we have the following theorem
\begin{theorem}
The bordism groups are
\begin{table}[!h]
\centering
\begin{tabular}{c c}
\hline
$i$ & $\Omega^{\tO}_i(\B \Z_{18}\times \B ^2\Z_3)$\\
\hline
0& $\Z_2$\\
1& $\Z_2$\\
2& $\Z_2^2$\\
3 & $\Z_2^2$\\
4 & $\Z_2^4$\\ 
5 & $\Z_2^5$\\
\hline
\end{tabular}
\end{table}
\end{theorem}

The 2d bordism invariants are $w_1^2,a^2$.

The 3d bordism invariants are $a^3,aw_1^2$.

The 4d bordism invariants are $a^4,a^2w_1^2,w_1^4,w_2^2$.

The 5d bordism invariants are $a^5,a^3w_1^2,aw_1^4,aw_2^2,w_2w_3$.

\subsubsection{$\Omega_d^{\SO}(\B \Z_{18}\times \B ^2\Z_3)$}

\bea
\Ext_{\A_2}^{s,t}(\H^*(M\SO\wedge (\B \Z_{18}\times \B ^2\Z_3)_+,\Z_2),\Z_2)\Rightarrow\Omega_{t-s}^{\SO}(\B \Z_{18}\times \B ^2\Z_3)_2^{\wedge}.
\eea
Since $\H^*(\B \Z_{18}\times \B ^2\Z_3,\Z_2)=\H^*(\B \Z_2,\Z_2)$, we have $\Omega_d^{\SO}(\B \Z_{18}\times \B ^2\Z_3)_2^{\wedge}=\Omega_d^{\SO}(\B \Z_2)$.

\bea
\Ext_{\A_3}^{s,t}(\H^*(M\SO\wedge (\B \Z_{18}\times \B ^2\Z_3)_+,\Z_3),\Z_3)\Rightarrow\Omega_{t-s}^{\SO}(\B \Z_{18}\times \B ^2\Z_3)_3^{\wedge}.
\eea

Since $\H^*(\B \Z_{18}\times \B ^2\Z_3,\Z_3)=\H^*(\B \Z_9\times \B ^2\Z_3,\Z_3)$, we have $\Omega_d^{\SO}(\B \Z_{18}\times \B ^2\Z_3)_3^{\wedge}=\Omega_d^{\SO}(\B \Z_9\times \B ^2\Z_3)_3^{\wedge}$.

\bea
\H^*(\B \Z_9,\Z_9)=\Lambda_{\Z_9}(a')\otimes\Z_9[b'].
\eea
where $a'\in\H^1(\B \Z_9,\Z_9)$, $b'\in\H^2(\B \Z_9,\Z_9)$.

\bea
\H^*(\B \Z_9,\Z_3)=\Lambda_{\Z_3}(\tilde{a}')\otimes\Z_3[\tilde{b}'].
\eea
where $\tilde{a}'=a'\mod3$, $\tilde{b}'=b'\mod3$, $\tilde{b}'=\beta_{(3,9)}(a')$.

\bea
\H^*(\B ^2\Z_3,\Z_3)=\Z_3[x_2',x_8',\dots]\otimes\Lambda_{\Z_3}(x_3',x_7',\dots)
\eea

\cred{
Note that
$\beta_{(3,3)}(\tilde{a}')=3\beta_{(3,9)}(a')=3\tilde{b}'=0$, $\beta_{(3,3)}(x_2')=x_3'$, $\beta_{(3,3)}(x_2'^2)=2x_2'x_3'$, $\beta_{(3,9)}(a'b')=\tilde{b}'^2$, $\beta_{(3,9)}(a'b'^2)=\tilde{b}'^3$, $\beta_{(3,3)}(\tilde{a}'x_2')=\tilde{a}'x_3'$, $\beta_{(3,3)}(\tilde{b}'x_2')=\tilde{b}'x_3'$, 
$\beta_{(3,3)}(\tilde{a}'\tilde{b}'x_2')=\tilde{a}'\tilde{b}'x_3'$, $\beta_{(3,3)}(\tilde{a}'x_2'^2)=2\tilde{a}'x_2'x_3'$.
}

There is a differential $d_2$ corresponding to the $(3,9)$-Bockstein \cite{may1981bockstein}.

\cred{
So there are differentials such that
$d_2(\tilde{b}')=\tilde{a}'h_0'^2$, $d_2(\tilde{b}'^2)=\tilde{a}'\tilde{b}'h_0'^2$, $d_2(\tilde{b}'^3)=\tilde{a}'\tilde{b}'^2h_0'^2$.
}

The $E_2$ page is shown in Figure \ref{fig:Omega_*^{SO}(BZ_9timesB^2Z_3)_3}.

\begin{figure}[!h]
\begin{center}
\begin{tikzpicture}
\node at (0,-1) {0};
\node at (1,-1) {1};
\node at (2,-1) {2};
\node at (3,-1) {3};
\node at (4,-1) {4};
\node at (5,-1) {5};
\node at (6,-1) {6};
\node at (7,-1) {$t-s$};
\node at (-1,0) {0};
\node at (-1,1) {1};
\node at (-1,2) {2};
\node at (-1,3) {3};
\node at (-1,4) {4};
\node at (-1,5) {5};
\node at (-1,6) {$s$};

\draw[->] (-0.5,-0.5) -- (-0.5,6);
\draw[->] (-0.5,-0.5) -- (7,-0.5);

\draw (0,0) -- (0,5);
\draw (4,1) -- (4,5);

\draw (1,0) -- (1,5);
\draw (2,0) -- (2,5);
\draw[fill] (2.1,0) circle(0.05);
\draw (3,0) -- (3,5);
\draw[fill] (3.1,0) circle(0.05);
\draw (3.9,0) -- (3.9,5);
\draw[fill] (4,0) circle(0.05);
\draw[fill] (4.1,0) circle(0.05);

\draw (4.9,0) -- (4.9,5);
\draw (5,1) -- (5,5);
\draw[fill] (5,0) circle(0.05);
\draw[fill] (5.1,0) circle(0.05);
\draw (5.9,0) -- (5.9,5);
\draw (6,1) -- (6,5);

\draw[color=red][->] (2,0) -- (1,2);
\draw[color=red][->] (2,1) -- (1,3);
\draw[color=red][->] (2,2) -- (1,4);
\draw[color=red][->] (2,3) -- (1,5);

\draw[color=red][->] (3.9,0) -- (3,2);
\draw[color=red][->] (3.9,1) -- (3,3);
\draw[color=red][->] (3.9,2) -- (3,4);
\draw[color=red][->] (3.9,3) -- (3,5);

\draw[color=red][->] (5.9,0) -- (4.9,2);
\draw[color=red][->] (5.9,1) -- (4.9,3);
\draw[color=red][->] (5.9,2) -- (4.9,4);
\draw[color=red][->] (5.9,3) -- (4.9,5);

\draw[color=red][->] (6,1) -- (5,3);
\draw[color=red][->] (6,2) -- (5,4);
\draw[color=red][->] (6,3) -- (5,5);

\end{tikzpicture}
\end{center}
\caption{$\Omega_*^{\SO}(\B \Z_9\times \B ^2\Z_3)_3^{\wedge}$}
\label{fig:Omega_*^{SO}(BZ_9timesB^2Z_3)_3}
\end{figure}

Hence we have the following theorem
\begin{theorem}
The bordism groups are
\begin{table}[!h]
\centering
\begin{tabular}{c c}
\hline
$i$ & $\Omega^{\SO}_i(\B \Z_{18}\times \B ^2\Z_3)$\\
\hline
0& $\Z$\\
1& $\Z_9\times\Z_2$\\
2& $\Z_3$\\
3 & $\Z_9\times\Z_3\times\Z_2$\\
4 & $\Z\times\Z_3^2$\\ 
5 & $\Z_2^3\times\Z_3^3\times\Z_{27}$\\
\hline
\end{tabular}
\end{table}
\end{theorem}

The 2d bordism invariant is $x_2'$.

The 3d bordism invariants are $a'b',\tilde{a}'x_2',a^3$.

The 4d bordism invariants are $\sigma,\tilde{b}'x_2',x_2'^2$.

The 5d bordism invariants are $a^5,aw_2^2,w_2w_3,\tilde{a}'(\sigma\mod3),\tilde{a}'\tilde{b}'x_2',\tilde{a}'x_2'^2,\mathfrak{P}_3(b')$.

Here $\mathfrak{P}_3$ is the Postnikov square.

\subsection{$\B(\Z_2\ltimes\PSU(N))$}

\label{sec:BZ2PSUN}

For $N>2$, the outer automorphism group of $\PSU(N)$ is $\Z_2$ where
$\Z_2$ acts on $\PSU(N)$ via complex conjugation.

\subsubsection{$\Omega_3^{\tO}(\B(\Z_2\ltimes\PSU(3)))$}

\bea
\Ext_{\A_2}^{s,t}(\H^*(M\tO,\Z_2)\otimes\H^*(\B(\Z_2\ltimes\PSU(3)),\Z_2),\Z_2)\Rightarrow\Omega_{t-s}^{\tO}(\B(\Z_2\ltimes\PSU(3)))
\eea

\bea
\H^*(M\tO,\Z_2)=\A_2\otimes\Z_2[y_2,y_4,y_5,y_6,y_8,\dots]^*
\eea
where $y_2^*=w_1^2$, $(y_2^2)^*=w_2^2$, $y_4^*=w_1^4$, $y_5^*=w_2w_3$, etc.

We have a fibration
\bea
\B\PSU(3)\to \B (\Z_2\ltimes\PSU(3))\to \B\Z_2.
\eea
and
\bea
\H^*(\B\Z_2,\Z_2)=\Z_2[a]
\eea
\bea
\H^*(\B\PSU(3),\Z_2)=\Z_2[c_2,c_3]
\eea

By Serre spectral sequence, we have 
\bea
\H^p(\B\Z_2,\H^q(\B\PSU(3),\Z_2))\Rightarrow\H^{p+q}(\B(\Z_2\ltimes\PSU(3)),\Z_2).
\eea

The relevant piece is shown in Figure \ref{fig:SSS for (BZ_2,BPSU(3)) with coefficients Z_2}.

\begin{figure}[!h]
\centering
\begin{sseq}[grid=none,labelstep=1,entrysize=1.5cm]{0...3}{0...3}
\ssdrop{\Z_2}
\ssmoveto 1 0 
\ssdrop{\Z_2}
\ssmoveto 2 0
\ssdrop{\Z_2}
\ssmoveto 3 0
\ssdrop{\Z_2}
\ssmoveto 0 1
\ssdrop{0}
\ssmoveto 1 1
\ssdrop{0}
\ssmoveto 2 1
\ssdrop{0}
\ssmoveto 3 1
\ssdrop{0}
\ssmoveto 0 2
\ssdrop{0}
\ssmoveto 1 2
\ssdrop{0}
\ssmoveto 2 2
\ssdrop{0}
\ssmoveto 3 2
\ssdrop{0}
\ssmoveto 0 3
\ssdrop{0}
\ssmoveto 1 3
\ssdrop{0}
\ssmoveto 2 3
\ssdrop{0}
\ssmoveto 3 3
\ssdrop{0}
\end{sseq}
\centering
\caption{Serre spectral sequence for $(\B \Z_2,\B \PSU(3))$ with coefficients $\Z_2$}
\label{fig:SSS for (BZ_2,BPSU(3)) with coefficients Z_2}
\end{figure} 

Hence $\H^*(\B (\Z_2\ltimes\PSU(3)),\Z_2)=\H^*(\B\Z_2,\Z_2)$ for $*\le3$.

\bea
&&\H^*(M\tO,\Z_2)\otimes\H^*(\B\Z_2,\Z_2)\\\notag
&=&\A_2\otimes\Z_2[y_2,y_4,y_5,y_6,y_8,\dots]^*\otimes\Z_2[a]\\\notag
&=&\A_2\oplus\Sigma\A_2\oplus2\Sigma^2\A_2\oplus2\Sigma^3\A_2\oplus\cdots
\eea

Hence we have the following theorem
\begin{theorem}
The bordism groups are
\begin{table}[!h]
\centering
\begin{tabular}{c c}
\hline
$i$ & $\Omega_i^{\tO}(\B(\Z_2\ltimes\PSU(3)))$\\
\hline
0& $\Z_2$\\
1& $\Z_2$\\
2& $\Z_2^2$\\
3 & $\Z_2^2$\\
\hline
\end{tabular}
\end{table}
\end{theorem}

The 1d bordism invariant is $a$.

The 2d bordism invariants are $a^2,w_1^2$.

The 3d bordism invariants are $a^3,aw_1^2$.

\subsubsection{$\Omega_3^{\SO}(\B(\Z_2\ltimes\PSU(3)))$}

\bea
\Ext_{\A_2}^{s,t}(\H^*(M\SO,\Z_2)\otimes\H^*(\B(\Z_2\ltimes\PSU(3)),\Z_2),\Z_2)\Rightarrow\Omega_{t-s}^{\SO}(\B(\Z_2\ltimes\PSU(3)))_2^{\wedge}
\eea

\bea
\Ext_{\A_3}^{s,t}(\H^*(M\SO,\Z_3)\otimes\H^*(\B(\Z_2\ltimes\PSU(3)),\Z_3),\Z_3)\Rightarrow\Omega_{t-s}^{\SO}(\B(\Z_2\ltimes\PSU(3)))_3^{\wedge}
\eea

\bea
\H^*(\B\PSU(3),\Z_3)=(\Z_3[z_2,z_8,z_{12}]\otimes\Lambda_{\Z_3}(z_3,z_7))/(z_2z_3,z_2z_7,z_2z_8+z_3z_7)
\eea

By Serre spectral sequence, we have 
\bea
\H^p(\B\Z_2,\H^q(\B\PSU(3),\Z_3))\Rightarrow\H^{p+q}(\B(\Z_2\ltimes\PSU(3)),\Z_3).
\eea

The relevant piece is shown in Figure \ref{fig:SSS for (BZ_2,BPSU(3)) with coefficients Z_3}.

\begin{figure}[!h]
\centering
\begin{sseq}[grid=none,labelstep=1,entrysize=1.5cm]{0...3}{0...3}
\ssdrop{\Z_3}
\ssmoveto 1 0 
\ssdrop{0}
\ssmoveto 2 0
\ssdrop{0}
\ssmoveto 3 0
\ssdrop{0}
\ssmoveto 0 1
\ssdrop{0}
\ssmoveto 1 1
\ssdrop{0}
\ssmoveto 2 1
\ssdrop{0}
\ssmoveto 3 1
\ssdrop{0}
\ssmoveto 0 2
\ssdrop{\Z_3}
\ssmoveto 1 2
\ssdrop{0}
\ssmoveto 2 2
\ssdrop{0}
\ssmoveto 3 2
\ssdrop{0}
\ssmoveto 0 3
\ssdrop{\Z_3}
\ssmoveto 1 3
\ssdrop{0}
\ssmoveto 2 3
\ssdrop{0}
\ssmoveto 3 3
\ssdrop{0}
\end{sseq}
\centering
\caption{Serre spectral sequence for $(\B \Z_2,\B \PSU(3))$ with coefficients $\Z_3$}
\label{fig:SSS for (BZ_2,BPSU(3)) with coefficients Z_3}
\end{figure} 

Hence $\H^*(\B (\Z_2\ltimes\PSU(3)),\Z_3)=\H^*(\B\PSU(3),\Z_3)$ for $*\le3$.

Combining this with previous results, we have the following theorem

\begin{theorem}
The bordism groups are
\begin{table}[!h]
\centering
\begin{tabular}{c c}
\hline
$i$ & $\Omega_i^{\SO}(\B(\Z_2\ltimes\PSU(3)))$\\
\hline
0& $\Z$\\
1& $\Z_2$\\
2& $\Z_3$\\
3 & $\Z_2$\\
\hline
\end{tabular}
\end{table}
\end{theorem}

The 1d bordism invariant is $a$.

The 2d bordism invariant is $z_2$.

The 3d bordism invariant is $a^3$.

Here $z_2=w_2(\PSU(3))\in\H^2(\B\PSU(3),\Z_3)$ is the generalized Stiefel-Whitney class of the principal $\PSU(3)$ bundle.

\subsubsection{$\Omega_3^{\tO}(\B(\Z_2\ltimes\PSU(4)))$}

\bea
\Ext_{\A_2}^{s,t}(\H^*(M\tO,\Z_2)\otimes\H^*(\B(\Z_2\ltimes\PSU(4)),\Z_2),\Z_2)\Rightarrow\Omega_{t-s}^{\tO}(\B(\Z_2\ltimes\PSU(4)))
\eea

\bea
\H^*(M\tO,\Z_2)=\A_2\otimes\Z_2[y_2,y_4,y_5,y_6,y_8,\dots]^*
\eea
where $y_2^*=w_1^2$, $(y_2^2)^*=w_2^2$, $y_4^*=w_1^4$, $y_5^*=w_2w_3$, etc.

We have a fibration
\bea
\B\PSU(4)\to \B (\Z_2\ltimes\PSU(4))\to \B\Z_2.
\eea
and
\bea
\H^*(\B\Z_2,\Z_2)=\Z_2[a]
\eea

We also have a fibration
\bea
\B\SU(4)\to \B\PSU(4)\to \B^2\Z_4
\eea

and
\bea
\H^*(\B\SU(4),\Z_2)=\Z_2[c_2,c_3,c_4]
\eea
\bea
\H^*(\B^2\Z_4,\Z_2)=\Z_2[\tilde{x}_2'',x_3'',x_5'',x_9'',\dots]
\eea

By Serre spectral sequence, we have 

\bea
\H^p(\B^2\Z_4,\H^q(\B\SU(4),\Z_2))\Rightarrow\H^{p+q}(\B\PSU(4),\Z_2).
\eea

The relevant piece is shown in Figure \ref{fig:SSS for (B^2Z_4,BSU(4))}.

\begin{figure}[!h]
\centering
\begin{sseq}[grid=none,labelstep=1,entrysize=1.5cm]{0...3}{0...3}
\ssdrop{\Z_2}
\ssmoveto 1 0 
\ssdrop{0}
\ssmoveto 2 0
\ssdrop{\Z_2}
\ssmoveto 3 0
\ssdrop{\Z_2}
\ssmoveto 0 1
\ssdrop{0}
\ssmoveto 1 1
\ssdrop{0}
\ssmoveto 2 1
\ssdrop{0}
\ssmoveto 3 1
\ssdrop{0}
\ssmoveto 0 2
\ssdrop{0}
\ssmoveto 1 2
\ssdrop{0}
\ssmoveto 2 2
\ssdrop{0}
\ssmoveto 3 2
\ssdrop{0}
\ssmoveto 0 3
\ssdrop{0}
\ssmoveto 1 3
\ssdrop{0}
\ssmoveto 2 3
\ssdrop{0}
\ssmoveto 3 3
\ssdrop{0}

\end{sseq}
\centering
\caption{Serre spectral sequence for $(\B^2\Z_4,\B \SU(4))$}
\label{fig:SSS for (B^2Z_4,BSU(4))}
\end{figure} 

Hence $\H^*(\B\PSU(4),\Z_2)=\H^*(\B^2\Z_4,\Z_2)$ for $*\le3$.

Again by Serre spectral sequence, we have 
\bea
\H^p(\B\Z_2,\H^q(\B\PSU(4),\Z_2))\Rightarrow\H^{p+q}(\B(\Z_2\ltimes\PSU(4)),\Z_2).
\eea

The relevant piece is shown in Figure \ref{fig:SSS for (BZ_2,BPSU(4))}.

\begin{figure}[!h]
\centering
\begin{sseq}[grid=none,labelstep=1,entrysize=1.5cm]{0...4}{0...3}
\ssdrop{\Z_2}
\ssmoveto 1 0 
\ssdrop{\Z_2}
\ssmoveto 2 0
\ssdrop{\Z_2}
\ssmoveto 3 0
\ssdrop{\Z_2}
\ssmoveto 4 0
\ssdrop{\Z_2}
\ssmoveto 0 1
\ssdrop{0}
\ssmoveto 1 1
\ssdrop{0}
\ssmoveto 2 1
\ssdrop{0}
\ssmoveto 3 1
\ssdrop{0}
\ssmoveto 4 1
\ssdrop{0}
\ssmoveto 0 2
\ssdrop{\Z_2}
\ssmoveto 1 2
\ssdrop{\Z_2}
\ssmoveto 2 2
\ssdrop{\Z_2}
\ssmoveto 3 2
\ssdrop{\Z_2}
\ssmoveto 4 2
\ssdrop{\Z_2}
\ssmoveto 0 3
\ssdrop{\Z_2}
\ssmoveto 1 3
\ssdrop{\Z_2}
\ssmoveto 2 3
\ssdrop{\Z_2}
\ssmoveto 3 3
\ssdrop{\Z_2}
\ssmoveto 4 3
\ssdrop{\Z_2}

\end{sseq}
\centering
\caption{Serre spectral sequence for $(\B\Z_2,\B\PSU(4))$}
\label{fig:SSS for (BZ_2,BPSU(4))}
\end{figure} 

There are no differentials, 
\bea
\H^n(\B(\Z_2\ltimes\PSU(4)),\Z_2)=\left\{\begin{array}{llll}\Z_2&n=0\\\Z_2&n=1\\\Z_2^2&n=2\\\Z_2^3&n=3\end{array}\right.
\eea

\bea
&&\H^*(M\tO,\Z_2)\otimes\H^*(\B(\Z_2\ltimes\PSU(4)),\Z_2)\\\notag
&=&\A_2\oplus\Sigma\A_2\oplus3\Sigma^2\A_2\oplus4\Sigma^3\A_2\oplus\cdots
\eea

Hence we have the following theorem
\begin{theorem}
The bordism groups are
\begin{table}[!h]
\centering
\begin{tabular}{c c}
\hline
$i$ & $\Omega_i^{\tO}(\B(\Z_2\ltimes\PSU(4)))$\\
\hline
0& $\Z_2$\\
1& $\Z_2$\\
2& $\Z_2^3$\\
3 & $\Z_2^4$\\
\hline
\end{tabular}
\end{table}
\end{theorem}

The 1d bordism invariant is $a$.

The 2d bordism invariants are $a^2,\tilde{z}_2',w_1^2$.

The 3d bordism invariants are $a^3,z_3',a\tilde{z}_2',aw_1^2$.

Here $z_2'=w_2(\PSU(4))\in\H^2(\B\PSU(4),\Z_4)$ is the generalized Stiefel-Whitney class of the principal $\PSU(4)$ bundle, $\tilde{z}_2'=z_2'\mod2$, $z_3'=\beta_{(2,4)}z_2'$.

\subsubsection{$\Omega_3^{\SO}(\B(\Z_2\ltimes\PSU(4)))$}

\bea
\Ext_{\A_2}^{s,t}(\H^*(M\SO,\Z_2)\otimes\H^*(\B(\Z_2\ltimes\PSU(4)),\Z_2),\Z_2)\Rightarrow\Omega_{t-s}^{\SO}(\B(\Z_2\ltimes\PSU(4)))
\eea

\cred{
There is a differential $d_2$ corresponding to the Bockstein homomorphism  $\beta_{(2,4)}:\H^*(-,\Z_4)\to\H^{*+1}(-,\Z_2)$ associated to $0\to\Z_2\to\Z_8\to\Z_4\to0$ \cite{may1981bockstein}. See \ref{Bockstein} for the definition of Bockstein homomorphisms.
}

Since $\beta_{(2,4)}(z_2')=z_3'$, there is a differential such that
$d_2(z_3')=\tilde{z}_2'h_0^2$.

The $E_2$ page is shown in Figure \ref{Omega_*^{SO}(B(Z_2ltimesPSU(4)))}.

\begin{figure}[!h]
\begin{center}
\begin{tikzpicture}
\node at (0,-1) {0};
\node at (1,-1) {1};
\node at (2,-1) {2};
\node at (3,-1) {3};

\node at (4,-1) {$t-s$};
\node at (-1,0) {0};
\node at (-1,1) {1};
\node at (-1,2) {2};
\node at (-1,3) {3};

\node at (-1,4) {$s$};

\draw[->] (-0.5,-0.5) -- (-0.5,4);
\draw[->] (-0.5,-0.5) -- (4,-0.5);

\draw (0,0) -- (0,3);
\draw (2,0) -- (2,3);
\draw (3,0) -- (3,3);

\draw[fill] (1,0) circle(0.05);
\draw[fill] (2.9,0) circle(0.05);
\draw[fill] (3.1,0) circle(0.05);

\draw[->][color=red] (3,0) -- (2,2);
\draw[->][color=red] (3,1) -- (2,3);

\end{tikzpicture}
\end{center}
\caption{$\Omega_*^{\SO}(\B(\Z_2\ltimes\PSU(4)))$}
\label{Omega_*^{SO}(B(Z_2ltimesPSU(4)))}
\end{figure}

Hence we have the following theorem
\begin{theorem}
The bordism groups are
\begin{table}[!h]
\centering
\begin{tabular}{c c}
\hline
$i$ & $\Omega_i^{\SO}(\B(\Z_2\ltimes\PSU(4)))$\\
\hline
0& $\Z$\\
1& $\Z_2$\\
2& $\Z_4$\\
3 & $\Z_2^2$\\
\hline
\end{tabular}
\end{table}
\end{theorem}

The 1d bordism invariant is $a$.

The 2d bordism invariant is $z_2'$.

The 3d bordism invariants are $a^3,a\tilde{z}_2'$.

\section{Final Comments and Remarks} 

\subsection{Relations to Non-Abelian Gauge Theories and Sigma Models}

As we mentioned, this article is a companion Reference with further detailed calculations supporting other shorter articles \cite{Wan:2018zql,Wan:2018djl,Wan:2019oyr}.
Now we make final comments and remarks on how our cobordism group calculations in the preceding \Sec{sec:cobor} and \Sec{sec:more-cobor} are applied in these works 
\cite{Wan:2018zql,Wan:2018djl,Wan:2019oyr}.

\begin{enumerate} [leftmargin=6.mm] 

\item \emph{Pure SU(2) Yang-Mills theory's higher anomaly}: \Ref{Gaiotto2014kfa1412.5148} introduces the generalized global symmetries include higher symmetries 
(See a brief review in \Sec{sec:intro-anom}, Items \ref{1:charged} and \ref{2:charge}). The pure SU(N) Yang-Mills (YM) gauge theory has a higher-$1$-dimensional (1-form) 
electric symmetry, denoted as $\Z_{\rN,[1]}^e$ (previously known as the $\Z_{\rN}$-center symmetry). 
The pure SU(N) YM theory in 4d has the corresponding 1-form electric $\Z_{\rN,[1]}^e$ symmetry charged object: the 1-dimensional gauge-invariant Wilson line $W_e$:
\bea
W_e=\Tr_{\text{R}}( \text{P} \exp(\ii \oint a)).
\eea
and the 2-dimensional charge operator:  the 2-dimensional charge surface operator $U_e$.
\bea
U_e= \exp(\ii \frac{2\pi}{\rN} \oint \Lambda),
\eea
The spacetime path integral formulation of SU(N) YM higher symmetry becomes a relation:
 \bea \label{eq:link-We-Ue}
\langle W_e \;  U_e \rangle =\langle \Tr_{\text{R}}( \text{P} \exp(\ii \oint_{\gamma^1} a))\;  \exp(\ii \pi \oint_{\Sigma^2} \Lambda) \rangle = 
\exp(\frac{\ii 2 \pi}{\rN}{\text{Lk}({\gamma^1},{\Sigma^2})}),
 \eea
 with ${\text{R}}$ in fundamental representation.
The remarkable \Ref{Gaiotto2017TTT1703.00501} discovers the mixed higher 't Hooft anomaly of pure SU(N) YM theory at an even integer N
with a second Chern class topological term ($\pi \int\limits_{M^4} c_2 \simeq \int\limits_{M^4} \frac{  \theta}{8 \pi^2}  \text{Tr}\,F_a\wedge F_a$ at $\theta= \pi$ with 
the YM field strength curvature $F_a$)
between time-reversal $\Z_2^T$ symmetry (with a schematic background field $\cT$ or $w_1(TM)$) 
and the 1-form electric $\Z_{\rN,[1]}^e$ symmetry (with a schematic 2-form background field $B$), via a schematic 5d topological term:
\bea \label{eq:TBB}
\sim \exp(\ii \pi \int_{M^5} \cT B B). \label{eq:TBB}
\eea
%\Ref{1812.11968, 1904.00994} 
\Ref{Wan:2018zql,Wan:2019oyr}
shows that this precise 5d topological term written as a 5d bordism invariant (at N = 2) 
of a mod 2 class term
is:
	\begin{eqnarray}
	\label{Eq.TBBanomaly}
	\exp(\ii  \pi \int_{M^5} B \Sq^1 B+ \Sq^2 \Sq^1 B)= \exp(\ii  \frac{\pi}{2} \int_{M^5} \tilde{w}_1(TM)\cup \mathcal{P}(B)),
	\end{eqnarray}
based on the notation introduced earlier in
\Sec{sec:OB2Z2} and \Sec{sec:OB2Z2}, it can be written as:
\bea  \label{eq:w1PB}
\exp(\ii  \pi \int_{M^5}   x_2 \Sq^1 x_2+  \Sq^2 \Sq^1 x_2 )= \exp(\ii  \pi \int_{M^5}  x_2x_3 + x_5) = \exp(\ii  \pi \int_{M^5} \frac{1}{2}\tilde{w}_1\cP_2(x_2)),
\eea
where $x_2 = B$ is the generator of $\H^2(\B^2\Z_2,\Z_2)$.
Other than %\Ref{1812.11968, 1904.00994}
\Ref{Wan:2018zql,Wan:2019oyr}, the derivation of the relation of the topological invariant $x_2x_3 + x_5=\frac{1}{2}\tilde{w}_1\cP_2(x_2)$ has also been examined 
in an excellent note of Debray \cite{Arun2017}. 
For \eqn{eq:w1PB}, our relevant cobordism theory includes
the unoriented bordism group
$\Omega^{\tO}_d(\B ^2\Z_2)$ in
\Sec{sec:OB2Z2}
and the oriented bordism group
$\Omega^{\SO}_d(\B ^2\Z_2)$ in
\Sec{sec:SOB2Z2}.

\item  \emph{Pure SU(N) Yang-Mills theory's higher anomaly}: 
The above formulas \Eq{Eq.TBBanomaly} and \Eq{eq:w1PB}, are 5d topological invariants characterizing the 4d SU(2) YM at $\theta=\pi$'s higher anomaly. 
For a generic 4d SU(N) YM at $\theta=\pi$ of even integer $\rN = 2^n$,
\Ref{Wan:2018zql} %{1812.11968} 
proposes a precise 5d topological term written as a 5d bordism invariant (at $\rN = 2^n$) which includes at least a mod 2 class term:
\bea  \label{eq:SUNYM-5dSPT}
B \beta_{(2,\rN=2^n)}B+\frac{\rN}{2}\Sq^2\beta_{(2,\rN)}B = \frac{1}{\rN} \tilde w_1(TM) \cP(B), 
\eea
 characterizing (part of) the 4d SU(2) YM at $\theta=\pi$'s higher anomaly.
Pontryagin square is defined as
$\mathcal{P}:\H^2(-,\Z_{2^n})\to\H^4(-,\Z_{2^{n+1}})$.
For example, at $\rN=2$, we get  \eqn{eq:SUNYM-5dSPT}
coincides the same formula as \eqn{eq:w1PB}.
At $\rN=4$,  we get the formula 
$B \beta_{(2,\rN=4)}B = \frac{1}{4} \tilde w_1(TM) \cP(B).$
Our corresponding cobordism group calculations are presented in \Sec{sec:OB2Z4} and \Sec{sec:SOB2Z4}. 

\item \emph{More discrete symmetries (e.g. charge conjugation) and more higher anomalies}: 
SU(N) YM theory has charge conjugation symmetry $\Z_2^C$ when $\rN > 2$. Therefore,
%\Ref{1812.11968} 
\Ref{Wan:2018zql}
presents additional higher 't Hooft anomalies associated to the 
charge conjugation $\Z_2^C$  background field $A_{C}$.
The relevant cobordism group calculation involving additional $\Z_2^C$ symmetry requires 
adding a new $\B\Z_2$ sector into the previous classifying space.
Relevant cobordism group calculations are presented in \Ref{Wan:2018zql}, and also some trial toy-model examples 
in \Sec{sec:BZnB2Zn},
 \Sec{sec:BZ2nB2Zn},
 and  \Sec{sec:BZ2n2B2Zn} involving the classifying space $\B\Z_m$ and higher-classifying space $\B^2\Z_n$.
 The combined higher-classifying space includes the forms of $\B\Z_m \times \B^2\Z_n$ or 
 $\B\Z_m \ltimes \B^2\Z_n$ (in \Ref{Wan:2018zql}).

\item \emph{Non-linear sigma models and their anomalies}: Non-linear sigma models such as the $\CP^{\rN-1}$-sigma models (with the 
target space $\CP^{\rN-1}$) have a global symmetry of PSU(N). Therefore, the relevant 
cobordism group calculations presented in \Ref{Wan:2018zql} include the classifying space BPSU(N).
We include the pertinent cobordism group calculations also for BPSU(N) in \Sec{sec:BSUn},
BPSU(2)=BO(3) in \Sec{sec:BO3},
and $\B(\Z_2\ltimes\PSU(N))$ in \Sec{sec:BZ2PSUN}.
The time reversal symmetry $\Z_2^T$ of bosonic or fermionic version of sigma models corresponds to
O or Pin$^{\pm}$ structure respectively. 
The charge conjugation symmetry $\Z_2^C$ corresponds to the 
$\B\Z_2$ in $\B(\Z_2\ltimes\PSU(N))$ in \Sec{sec:BZ2PSUN}

\item \emph{Higher-symmetry extension, and the fate of gapped and gapless-ness of quantum phases}:\\
An SU(N) YM gauge theory coupled to SU(N) fundamental fermions break explicitly the  1-form $\Z_{\rN,[1]}^e$-symmetry
(thus does not have the  1-form $\Z_{\rN,[1]}^e$-symmetry).  
An SU(N) YM gauge theory coupled to SU(N) adjoint fermions can still possess a 1-form $\Z_{\rN,[1]}^e$-symmetry. 

The SU(N) adjoint fermion YM gauge theory is known as the adjoint QCD of SU(N) gauge group.
The relevant global symmetries of this adjoint QCD thus includes 
$\Z_{\rN,[1]}^e$ and a SU($m$) flavor chiral symmetry (say, if there is an $m$-flavor of Wely fermions in the adjoint representation of SU(N)).
Some trial toy-model examples of cobordism groups, involving these classifying spaces 
BSU(m), BPSU(m)  and B$^2\Z_{\rN}$, are presented 
in \Sec{sec:BPSU(m)B2ZN}.
For example, for the adjoint QCD with an SU(2) gauge group and $\rN_f=2$ adjoint Weyl fermions,
the pertinent symmetry groups are 
${\Spin \times_{{\Z_2^F}} \big(\frac{\SU(2) \times \Z_{8,\rm{A} }}{\Z_2^F} \big) \times  \Z_{2,[1]}^e}$
or
${\Pin^- \times_{{\Z_2^F}} \big(\frac{\SU(2) \times \Z_{8,\rm{A}}}{\Z_2^F} \big) \times  \Z_{2,[1]}^e}$ (including a time-reversal symmetry),
see their cobordism groups and higher-anomalies in \cite{Wan:2018zql}.\\

Along this development, the fate of relevant theories of the adjoint QCD
is explored recently using the modern language of higher-symmetries and higher-anomalies in
various other  
\Ref{Anber2018tcj1805.12290, 2018arXiv180609592C, Bi:2018xvr, SWW, Poppitz2019fnp1904.11640, Anber2019nfu1906.10315}, other than \cite{Wan:2018zql},
and References therein.

\Ref{Wan:2018zql} employs a generalization of a symmetry-extension method of \cite{Wang2017locWWW1705.06728} to 
a higher-symmetry-extension method, as a tool of constructing a fully-symmetry-preserving gapped phase saturating the higher 't Hooft anomalies.
It turns out that: 
\begin{itemize}
\item Certain higher 't Hooft anomalies \emph{cannot} be saturated by a fully-symmetry-preserving gapped phase (e.g. TQFT); which implies either the symmetry-breaking or gapless-ness
of the dynamical fate of the theories. Examples include $\cP(B)$ in $\H^4(M,\Z_4)$
and $A \cP(B)$ in $\H^5(M,\Z_4)$ where $M$ is the spacetime manifold \cite{Wan:2018zql}.
This higher-symmetry-extension approach \cite{Wan:2018zql} thus rules out some candidate low-energy infrared phases (as a dual phase of a high-energy QFT) proposed in \cite{Bi:2018xvr}.

\item Certain higher 't Hooft anomalies \emph{can} be saturated by a fully-symmetry-preserving gapped phase (e.g. TQFT); which implies a possible exotic dynamical fate
as \emph{the confinement with no chiral symmetry nor 1-form center symmetry breaking}.
Various examples of pure SU(N) YM gauge theories with $\theta=\pi$-topological term
indeed afford such an exotic confinement without any (ordinary or higher) symmetry-breaking, see
\Ref{Wan:2018zql,Wan:2019oyr}.

\end{itemize}

\end{enumerate}

\subsection{Relations to Bosonic/Fermionic Higher-Symmetry-Protected Topological states:
Beyond Generalized Super-Group Cohomology Theories}

\begin{enumerate} [leftmargin=6.mm] 

\item \emph{Bosonic higher-symmetry protected topological states} (b-higher-SPTs): \\[2mm]
(1) Bosonic symmetry-protected topological states (bSPTs) in $d+1$d of an internal (ordinary 0-form) global symmetry $G_{(0)}$ 
is proposed firstly in Chen-Gu-Liu-Wen \Ref{1106.4772} to be classified by a cohomology group 
\bea
\H^{d+1}(G_{(0)},\U(1))
\eea 
or the topological cohomology of
classifying space $\B G_0$ as 
\bea
\cH^{d+1}(\B G_{(0)},\U(1)).
\eea
(2) It is later proposed by Kapustin in 
\Ref{Kapustin2014tfa1403.1467}, for bosonic SPTs  of $G_{(0)}$ and for bosonic symmetric invertible topological order (denoted as b-iTO) of $G_{(0)}$, 
they are classify by a cobordism group classification, which is beyond the group cohomology framework. The torsion (finite group $\prod_j \Z_{n_j}$) part of cobordism group classification
contains:
\bea
\Hom(\Omega^H_{{d+1},\text{tors}}(\B G_{(0)}), \U(1))
\eea
where $H$ is an oriented $H=$ SO or an unoriented $H=$ O for the (co)bordism group.
To include the free part (the non-torsion part, infinite integer $\prod_j \Z$ classes), we need to include additional contribution: In physics,
this is related to the nontrivial thermal Hall response and gravitational Chern-Simons terms.\\[2mm]
(3) \Ref{Wen2014zga1410.8477} of Wen proposes the $\SO(\infty)$ version of bosonic cohomology group to classify the bSPTs beyond Chen-Gu-Liu-Wen's \Ref{1106.4772} via 
\bea
\H^{d+1}(\SO(\infty) \times G_{(0)},\U(1)), \quad \cH^{d+1}(\B (\SO(\infty) \times G_{(0)}),\U(1)).
\eea 
(4) \Ref{Freed2016} of Freed-Hopkins introduces this classification of topological phases (TP), 
including the torsion and the free parts, defined as a suitable new cobordism group denoted:
\bea
\TP_{{d+1}}(H\times G_{(0)}).
\eea
(5) In our work, we generalize the result of \Ref{Freed2016} of bosonic SPTs to bosonic higher-SPTs including the higher-symmetries (e.g. $G_{(1)}$), 
such as
\bea
\TP_{{d+1}}(H\times (G_{(0)} \times \B G_{(1)})),\quad  \TP_{{d+1}}(H \times (G_{(0)} \ltimes \B G_{(1)})), \quad \dots
\eea
and more general constructions in \Sec{sec:higher-G-cobordism}.

\item \emph{Fermionic higher-symmetry protected topological states} (f-higher-SPTs): \\
Fermionic symmetry-protected topological states (fSPTs) in $d+1$d of an internal (ordinary 0-form) global symmetry $G_{(0)}$ 
is proposed in \Ref{Gu2012ib1201.2648} by Gu-Wen to be classified by a super-cohomology group. 
A corrected modification of Gu-Wen model is presented by Gaiotto-Kapustin in
\Ref{Gaiotto2015zta1505.05856}. The  Gu-Wen model and Gaiotto-Kapustin model is more or less complete for
 the 3d (2+1D) fSPTs with a global symmetry of finite group $G_{(0)}$. They also provide lattice Hamiltonian or wavefunction model constructions.
 The relation between the full fermionic symmetry group $G_F$ and $G_{(0)}$ is based on a short exact sequence, extended by a normal subgroup
 fermionic parity $\Z_2^F$:
 \bea
 1\to \Z_2^F \to G_F \to G_{(0)} \to1.
 \eea

However, neither Gu-Wen nor Gaiotto-Kapustin models obtain a complete classification for 4d (3+1D) fSPTs, even for a global symmetry of finite group $G_{(0)}$.
Improvements are made via several different approaches:\\[2mm]

(1) Kitaev's in \Ref{Kitaev2015} proposes a homotopy-theoretic approach to SPT phases in action. This gives rise a correct $\Z_{16}$ classification of 3+1D topological superconductors, matching to the cobordism group classification. Kitaev's \Ref{Kitaev2015} 
can be regarded as the interaction version of SPT classification, improved from his previous K-theory approach for the topological phase classification of free-fermion systems \cite{Kitaevperiod}.
Kitaev's approach is reviewed, for example, in \Ref{Xiong2016deb1701.00004, Gaiotto2017zba1712.07950}.
\\[2mm]
(2) Kapustin-Thorngren-Turzillo-Wang \cite{Kapustin1406.7329} approaches is based on the $H=\Spin$ or $\Pin^{\pm}$ versions of cobordism group
$\Hom(\Omega^H_{{d+1},\text{tors}}(\B G_{(0)}), \U(1))$.\\[2mm]
(3) Freed-Hopkins \cite{Freed2016} introduces a cobordism group $\TP_{{d+1}}(H\times G_{(0)})$ whose effective computation is based on the Adams spectral sequence, with
$H=\Spin$ or $\Pin^{\pm}$ for a fermionic theory.\\[2mm]
(4)  Kapustin-Thorngren in \Ref{Kapustin2017jrc1701.08264} introduces the higher-dimensional bosonization to construct higher-dimensional fSPTs, 
mostly focusing on a finite symmetry group $G_F$.\\[2mm]
(5) Wang-Gu in \Ref{WangGu2017moj1703.10937, Wang2018pdc1811.00536} introduces a generalized group super-cohomology theory with multi-layers of group extension structures
of super-cohomology group, 
mostly focusing on a finite symmetry group $G_F$.
The computation of fSPTs classification based on the generalized super-cohomology group  is similar to the Atiyah-Hirzebruch spectral sequence method. See related discussions in
\Ref{ShiozakiKen2018yyj1810.00801, 1812.11959} on Atiyah-Hirzebruch spectral sequence for classifying fSPTs.
\\[2mm]
(6) \Ref{Montero2018arXiv180800009G, Hsieh2018ifc1808.02881} use a mixture of Dai-Freed theorem \cite{DaiFreed1994kq9405012} and 
Atiyah-Hirzebruch-like spectral sequence to determine fSPTs and their discrete anomalies on the boundaries.\\[2mm]

(7) \Ref{1812.11959} computes various finite-group fSPTs via Adams spectral sequence. Their methods and their derived fSPT terms can be regarded 
as the complementary approach to those derived via Atiyah-Hirzebruch-like spectral sequence \cite{WangGu2017moj1703.10937, Wang2018pdc1811.00536, ShiozakiKen2018yyj1810.00801}. \\[2mm]
(8) In our work, we generalize the result of \Ref{Freed2016} of fermionic SPTs to fermionic higher-SPTs including the higher-symmetries (e.g. $G_{(1)}$), 
such as
$$
\TP_{{d+1}}(H\times ( G_{(0)} \times \B G_{(1)})), \quad \TP_{{d+1}}(H\times (G_{(0)} \ltimes \B G_{(1)})), \quad \dots
$$
with $H=\Spin$ or $\Pin^{\pm}$. Or slightly more generally, consider the classification of fermionic higher-SPTs via:
$$
\TP_{{d+1}}(\mathbb{H}) 
$$ such that the $H$, $\mathbb{G}$
and $\mathbb{H}$ satisfy the following exact sequences:
\bea
\left\{\begin{array}{l} 
1 \to  \mathbb{G} \to \mathbb{H} \to \Spin({d+1}) \to 1,\\
1 \to \Z_2^F   \to  \Spin({d+1})  \to \SO({d+1})  \to 1,\\
  \B^2{G}_{(1)} \to  \B \mathbb{G} \to \B{G}_{(0)}  \to \B^3{G}_{(1)} \to   \dots.
\end{array}\right.
\eea
or
\bea
\left\{\begin{array}{l} 
1 \to  \mathbb{G} \to \mathbb{H} \to \Pin^{\pm}({d+1}) \to 1,\\
1 \to \Z_2^F   \to  \Pin^{\pm}({d+1})  \to \tO({d+1})  \to 1,\\
  \B^2{G}_{(1)} \to  \B \mathbb{G} \to \B{G}_{(0)}  \to \B^3{G}_{(1)} \to   \dots.
\end{array}\right.
\eea

Even more general constructions are explored in \Sec{sec:higher-G-cobordism}.

\item \emph{Braiding statistics and link invariants approach to characterize bosonic/fermionic SPTs and higher-SPTs}:
Another useful approach to classify SPTs is based on gauging the global symmetry group of SPTs, such that we obtain a gauge theory or TQFT at the end.
The braiding statistics of the fractionalized excitations of gauged SPTs can characterize the pre-gauged SPTs, the explicit method of 3d (2+1D) SPTs is outlined by Levin-Gu \cite{LG1202.3120}. 
Here we focus on the case of continuum field theory formulation of braiding statistics and link invariants approach to characterize these higher-dimensional SPTs.\\[2mm]
(1) 4d (3+1D) bSPTs: \Ref{CWangMLevin1412.1781, Putrov2016qdo1612.09298}\\
(2) 4d (3+1D) fSPTs: \Ref{Putrov2016qdo1612.09298, Kapustin2017jrc1701.08264, Cheng2017ftw1705.08911, 1812.11959}\\
(3) 5d (4+1D) SPTs or higher dimensions:  \Ref{Wan:2019oyr}.
\item 
\emph{A Generalized Cobordism Theory of higher-symmetry groups --- beyond
Higher-Group Super-Cohomology Theories}:\\[2mm]
It is known that the cobordism theory approach of 
Kapustin et al.~\cite{Kapustin2014tfa1403.1467, Kapustin1406.7329} and Freed-Hopkins \cite{Freed2016}
obtain the classification of fSPTs and bSPTs beyond Chen-Gu-Liu-Wen's group cohomology \cite{1106.4772} or Gu-Wen's group super-cohomology \cite{Gu2012ib1201.2648}.
A more refined version of generalized group super-cohomology \cite{WangGu2017moj1703.10937, Wang2018pdc1811.00536} 
can obtain some missing classes of \cite{Gu2012ib1201.2648} to match the cobordism classification.\\[2mm]
Therefore, we expect that the our approach, on a generalized cobordism theory including the higher-symmetry groups,
can classify higher-SPTs (including fSPTs and bSPTs) that may or may not be captured by 
higher-group super-cohomology theories.

For future work, it will be illuminating to understand the distinctions between the generalized higher-group cobordism theory approach and the generalized higher-group super-cohomology theories.
We expect the comparison between two approaches can be rephrased as 
a certain version of Adams spectral sequence method in contrast to 
a certain version of Atiyah-Hirzebruch spectral sequence method.

It will also be important to figure the possible lattice-regularization (e.g. lattice Hamiltonian on simplicial complex) of those higher-SPTs classified by our generalized cobordism theory.

\end{enumerate}

\section{Acknowledgements}

We thank Daniel Freed, Meng Guo, {Michael Hopkins}, Anton Kapustin, %Zohar Komargodski, 
Pavel Putrov, and Edward Witten for conversations.
JW thanks the collaborators for a previous collaboration on Ref.~\cite{2017arXiv171111587GPW}.
JW thanks the participants of Developments in Quantum Field Theory and Condensed Matter Physics (November 5-7, 2018) 
at Simons Center for Geometry and Physics at SUNY Stony Brook University
for giving valuable feedback where this work is reported \cite{SCGP}. 
JW thanks the feedback from the attendants of IAS seminar \cite{IAS}.  
ZW gratefully acknowledges the support from NSFC grants 11431010, 11571329.
%(PP gratefully acknowledges the support from Marvin L. Goldberger Fellowship and the DOE Grant DE-SC0009988.) 
JW gratefully acknowledges the Corning Glass Works Foundation Fellowship and NSF Grant PHY-1606531. %1314311. 
This work is also supported by NSF Grant DMS-1607871 ``Analysis, Geometry and Mathematical
Physics'' and Center for Mathematical Sciences and Applications at Harvard University.

%This work is also supported by the NSF Grant PHY-1306313, PHY-0937443, DMS-1308244, DMS-0804454, DMS-1159412 and Center for Mathematical Sciences and Applications at Harvard University.

%\cite{2018arXiv180107530B}

%\newpage

%\newpage
%\bibliographystyle{plain}
%\bibliographystyle{apsrev4} 
%\bibliographystyle{unsrt}
\bibliographystyle{Yang-Mills} %bst
%\bibliography{Yang-Mills-JW-2.bib,Yang-Mills.bib,Yang-Mills-ZW.bib} %bib
%\bibliography{YangMillsJW3,Yang-Mills.bib,Yang-Mills-ZW.bib}

\bibliography{YangMillsJW3,Yang-Mills-ZW,Yang-Mills}

%\bibliography{JCW-ref,all,mybib,Yang-Mills} %,all,

%{Yang-Mills.bib,Yang-Mills-JW.bib}

%%%%%%%%%%%%%%%%%%%%%%%%%%%% 
%%%%%%%%%%%%%%%%%%%%%%%%%%%% 
%%%%%%%%%%%%%%%%%%%%%%%%%%%% 
%%%%%%%%%%%%%%%%%%%%%%%%%%%% 

\end{document}